\documentclass[12pt,twoside,openright]{book}
\usepackage[utf8]{inputenc}
\usepackage[T1]{fontenc}
\usepackage{lmodern}
\usepackage{amssymb,amsmath,latexsym}
\usepackage{undertilde}
\usepackage{geometry}
\usepackage{epigraph}

\usepackage{amsfonts}
\usepackage{array}
\usepackage{tikz}
\newcommand{\dynkinradius}{.08cm}
\newcommand{\dynkinstep}{.65cm}
\newcommand{\dynkindot}[2]{\draw(\dynkinstep*#1,\dynkinstep*#2) circle (\dynkinradius);}
\newcommand{\dynkinddot}[2]{\fill(\dynkinstep*#1,\dynkinstep*#2) circle (\dynkinradius);}

\newcommand{\dynkinline}[4]{\draw[thin] (\dynkinstep*#1 + \dynkinradius,\dynkinstep*#2) -- (\dynkinstep*#3 - \dynkinradius,\dynkinstep*#4);}

\newcommand{\dynkindoubleline}[2]{\draw[double] (\dynkinstep*#1,\dynkinradius) -- (\dynkinstep*#2,\dynkinradius) (\dynkinstep*#1,-\dynkinradius) -- (\dynkinstep*#2,-\dynkinradius);}

\newenvironment{dynkin}{\begin{tikzpicture}}
{\end{tikzpicture}}

\newcommand{\C}{{\mathbb C}}

\newcommand{\im}{{\rm i }}
\newcommand{\D}{{\mathcal D }}

\newcommand\be{\begin{eqnarray}}
\newcommand\ee{\end{eqnarray}}

\usepackage{graphicx,ctable,booktabs}
\usepackage{float}
\usepackage{setspace}
	\usepackage{multicol}
	
	\usepackage{amsthm} 

\usepackage{fancyhdr}

\usepackage{cite}

\theoremstyle{plain}

\theoremstyle{definition}

\theoremstyle{remark}

\usepackage{mathtools}
\newcommand{\defeq}{\vcentcolon=}

\usepackage{multicol}

\usepackage{simplewick}	
	
\usepackage{slashed}

\makeatletter
\renewcommand*{\cleardoublepage}{\clearpage\if@twoside \ifodd\c@page\else
\hbox{}%
\thispagestyle{empty}%
\newpage%
\if@twocolumn\hbox{}\newpage\fi\fi\fi}
\makeatother

\usepackage{tikz}
\usetikzlibrary{calc}

\begin{document}
\pagenumbering{gobble}

\begin{titlepage}

%\begin{tikzpicture}[remember picture, overlay]
%  \draw[line width = 4pt] ($(current page.north west) + (1.23in,-1in)$) rectangle ($(current page.south east) + (-0.7in,1in)$);
%\end{tikzpicture}

%
%
%% Upper part of the page
%\includegraphics[width=0.45\linewidth]{University_of_Nottingham2.png}
%%\hspace{2cm} \includegraphics[width=0.23\linewidth]{erc.png}
%\\[1cm]    
%
%\textsc{\LARGE School of Mathematical Sciences}\\[4cm]
%
%%\textsc{\Large MAGIC080 Exam}\\[0.5cm]
%
%%\\[0.4cm]{\LARGE \bfseries Topics in Quantum Field Theory }
%% Title
%%\HRule \\[0.4cm]
%{ {\huge \bfseries Second-Order Fermions}\\[3cm]{\large \bf 2015}\\[3cm]
%\bigskip {\bfseries \Large PhD. in Mathematical Sciences}}\\[0.4cm]
%
%%\HRule \\[1.5cm]
%
%% Author and supervisor
%\begin{minipage}{0.4\textwidth}
%\begin{flushleft} \large
%\emph{Student:}\\
%Johnny \textsc{Espin}
%\end{flushleft}
%\end{minipage}
%\begin{minipage}{0.4\textwidth}
%\begin{flushright} \large
%\emph{Supervisor:} \\
%Prof.~Kirill  \textsc{Krasnov }
%\end{flushright}
%\end{minipage}
%
%\vfill
%% Bottom of the page
%
%
%\end{center}

\begin{center}

% Upper part of the page
\includegraphics[width=0.45\linewidth]{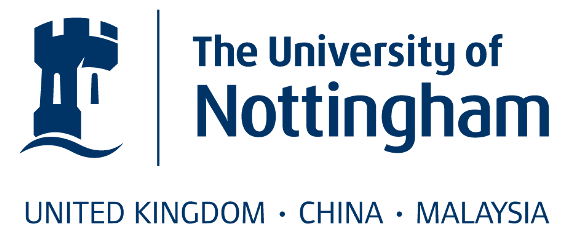}
%\hspace{2cm} \includegraphics[width=0.23\linewidth]{erc.png}
\\[1cm]    

\textsc{\LARGE School of Mathematical Sciences}\\[3.5cm]

%\textsc{\Large MAGIC080 Exam}\\[0.5cm]

%\\[0.4cm]{\LARGE \bfseries Topics in Quantum Field Theory }
% Title
%\HRule \\[0.4cm]
{ {\huge \bfseries Second-Order Fermions}\\[3cm]{\Large Johnny \textsc{Espin}, MSc.}\\[4cm]
\bigskip {\bfseries  Thesis submitted to the University of Nottingham for the degree of Doctor of Philosophy}}\\[1.5cm]

%\HRule \\[1.5cm]

{\large \bf August 2015}

\vfill
% Bottom of the page

\end{center}

\end{titlepage}

\cleardoublepage\thispagestyle{empty} \newpage

%  \begin{multicols}{2}
%\tableofcontents
%  \end{multicols}

\section*{Abstract}
\addcontentsline{toc}{chapter}{Abstract}
\epigraph{There is danger in reckless change; but greater danger in blind conservatism.}{\it Henry George, Social Problems}

It has been proposed several times in the past that one can obtain an equivalent, but in many aspects simpler description of fermions by first reformulating their first-order (Dirac) Lagrangian in terms of two-component spinors, and then integrating out the spinors of one chirality ($e.g.$ primed or dotted). The resulting new Lagrangian is second-order in derivatives, and contains two-component spinors of only one chirality. The new second-order formulation simplifies the fermion Feynman rules of the theory considerably, $e.g.$ the propagator becomes a multiple of an identity matrix in the field space. The aim of this thesis is to work out the details of this formulation for theories such as Quantum Electrodynamics, and the Standard Model of elementary particles. After having developed the tools necessary to establish the second-order formalism as an equivalent approach to spinor field theories, we proceed with some important consistency checks that the new formulation is required to pass, namely the presence or absence of anomalies in their perturbative and non-perturbative description, and the unitarity of the S-Matrix derived from their Lagrangian. Another aspect which is studied is unification, where we seek novel gauge-groups that can be used to embed all of the Standard Model content: forces and fermionic representations. Finally, we will explore the possibility to unify gravity and the Standard Model when the former is seen as a diffeomorphism invariant gauge-theory.

\cleardoublepage\thispagestyle{empty} \newpage

\section*{Acknoledgements}
\addcontentsline{toc}{chapter}{Acknoledgements}

I would like to thank my supervisor, Kirill Krasnov, for his support during these years in Nottingham; for his help and ideas that allowed this thesis to come into existence, and for all the additional knowledge in Physics and Mathematics he has imparted to me through many hours of discussions, seminars, and group meetings.

All the past and present members of his research group with whom I had the chance to interact also ought to be thanked: Gianluca Delfino, Kai Groh, Christian Steinwachs, Marco Cofano, Carlos Scarinci, Yannick Herfray, and Chih-Hao Fu.

I would also like to thank Mikhail Shaposhnikov and Yuri Shtanov for the suggestions that they have made which have led to new insights on the work accomplished in this thesis.

It was a pleasure to spend many hours in the office mainly thanks to all of B50 and other very good friends in the Department.

All my friends in Switzerland who visited me in Nottingham, and who kept me smiling thanks to never-ending hilarious conversations.

%Me gustar\'ia agradecer a mis padres, hermano y cu\~nada por su apoyo y paciencia.

Finally, I would like to thank my parents, my brother, and my sister-in-law, without whom the chance to write acknowledgements would never have happened in the first place. Muchas gracias!

This thesis was supported by the School of Mathematical Sciences at the University of Nottingham, and by the European Research Council.

\cleardoublepage\thispagestyle{empty} \newpage
\begingroup
\let\clearpage\relax
\tableofcontents
%\listoffigures
%\listoftables
\endgroup

\cleardoublepage\thispagestyle{empty} \newpage
%\newpage
\setlength\parindent{0pt}

\chapter*{Introduction}
\addcontentsline{toc}{chapter}{Introduction}
\pagenumbering{arabic}

\pagestyle{fancy}
%\addtolength{\headwidth}{\marginparsep} %these change header-rule width
%\addtolength{\headwidth}{\marginparwidth}
\lhead{ \bfseries \nouppercase{Introduction}}
\chead{} 
\rhead{} 
\lfoot{\small\scshape PhD Thesis} 
\cfoot{\thepage} 
\rfoot{\footnotesize Johnny Espin} 
\renewcommand{\headrulewidth}{.3pt} 
\renewcommand{\footrulewidth}{.3pt}
\setlength\voffset{-0.25in}
\setlength\textheight{648pt}

\section*{Why second-order?}

Since the emergence of Quantum Mechanics (QM) and Special Relativity (SR) at the beginning of the twentieth century, and after the success of Maxwell's unification of Electromagnetism, a lot of effort was put into merging the two theories. This led to a race, which aim was to find relativistic wave equations that would govern the dynamics of quantum-mechanical systems. Schr\"odinger and then Klein and Gordon formulated a second-order wave equation that was supposed to describe the relativistic evolution of the wavefunction. However, at that time it seemed that the nature of the latter violated some fundamental properties of mechanical systems: the Klein-Gordon solutions admitted both a positive and a negative energy mode. It must be emphasised that the theoretical framework that is nowadays called Quantum Field Theory was yet to be invented and understood. Nonetheless, British physicist Paul A.M. Dirac believed that the issue related to the presence of negative energy solutions relied on the second-order nature of the differential equation. Thus, he tried to construct a first-order differential equation that was compatible with the relativity principle. His theory was formulated in 1928 and the Dirac equation was later shown to describe relativistic spin $1/2$ particles: fermions. This was followed by the development of Quantum Electrodynamics (QED), the relativistic quantum theory of light and matter interactions which was then generalised into Yang-Mills (YM) theory, the theory of non-abelian $SU(n)$ gauge fields that describes the weak and the strong forces. This summarises the success of particle physics in the last century, success that culminated with the edification of the Standard Model (SM) of particle physics which is today the most accurate description of Nature that has been developed \cite{PhysRevLett.97.030801}.

Yet something can be seen as puzzling. Indeed, one of Dirac's reasons to construct a first-order wave equation was the misinterpretation of the negative energy solutions. However shortly after the discovery of his equation, it became clear that Nature admitted particles and antiparticles (positive and ``negative'' energy solutions). Nonetheless, fermions remained the only dynamical system that only admitted a first-order description.

Indeed, today it is well known that physical theories can be described by first- as well as by second-order Lagrangians. The classical example that every theoretical physicist has encountered is the relation between Hamiltonian and Lagrangian mechanics. The Hamiltonian formulation gives first-order evolution equations, but contains twice as many independent variables as the second-order Lagrangian formalism. Nevertheless, with a simple Legendre transform, it is possible to describe the system in whichever formalism is suitable. The Legendre transform amounts to inverting the relation between the momenta and the time derivatives of the generalised coordinates, and then ``plugging it back'' into the transform. In other words, it amounts to integrating out the momentum variables from the first-order Hamiltonian formulation to arrive to the second-order Lagrangian formalism (and vice-versa). The formulation that should be used for solving a given problem is a matter of convenience. Nonetheless, the community will generally side with the formulation that was developed first, unless the new approach brings unignorable advantages.

As a matter of fact, the most evident example is given by the first- and second-order formulations of General Relativity (GR). Physicists use the second-order metric formulation as the one in which GR was originally proposed. What is more, the first-order description of GR can seldom be found in textbooks. However, some aspects of the theory become more transparent in the first-order formalism. For example, in this formulation the Lagrangian of GR is polynomial (cubic) in the fields \cite{Herfray:2015rja,Deser:1987uk}, whereas the non-polynomiality of the second-order formulation makes it very cumbersome to work with. The availability of a simple polynomial Lagrangian, even though it contains more fields, is sometimes important. Another slightly less familiar example is the first order formalism for QCD \cite{Martellini:1997mu}. Thus, one can rewrite the Yang-Mills Lagrangian in the BF form, plus a term quadratic in the B-field. Again, the first-order Lagrangian is cubic in the fields.

As for fermions, their Lagrangian is first-order in derivatives. Hence, it is natural to ask whether a second-order formulation of fermions is possible.
\\
%could explain why dirac spinors have four components historically. why 4 DOF, etc
Let us make the last argument more precise. A Dirac spinor can be written as the sum of two unitary infinite dimensional representations of the Lorentz group $SO(1,3)$ (or its double cover $SL(2,\mathbb{C})$):
\begin{align*}
\Psi_D \in \left(1/2,0 \right) \oplus \left( 0,1/2 \right)
\end{align*}
which we call left-handed (unprimed) and right-handed (primed) respectively. The Dirac equation is then derived from the Dirac Lagrangian, here in $3+1$ dimensions with the metric $\eta_{\mu\nu}=(-,+,+,+)$:
\begin{align*}
\mathcal{L}_D= \bar\Psi_D \left(-i\slashed \partial  - m\right)\Psi_D
\end{align*}

with $\slashed \partial = \gamma^\mu \partial_\mu$, and $\{\gamma^\mu\}$ are the Dirac gamma matrices.
%and we have the algebra of (Dirac) gamma matrices:
%\begin{align}
%\{ \gamma^\mu, \gamma^\nu \}= -2\eta^{\mu\nu}, \quad \left( \gamma^\mu\right)^\dagger=\gamma^0 \gamma^\mu \gamma^0, \quad \gamma_5 = i\gamma^0\gamma^1\gamma^2\gamma^3,\quad \left( \gamma_5\right)^\dagger=\gamma_5
%\end{align}
This Lagrangian generalises in a straightforward way so as to include the interaction of fermions and photons (QED). We see that the Dirac equation
\begin{align*}
\left(-i\slashed \partial  - m\right)\Psi_D=0
\end{align*}

relates spinors of one chirality to the other through the off-diagonal entries of the Dirac matrices\footnote{In the case of Majorana fermions the spinor is linked to its hermitian conjugate through the Dirac equation.}. This is heuristically why a second order Lagrangian of the type
\begin{align*}
\mathcal{L}_D ~\stackrel{?}{=} ~ \bar\Psi_D \left(-\square +m^2\right)\Psi_D
\end{align*}
does not work since the Klein-Gordon operator is diagonal and hence we lose information contained in the Dirac equation. 

In this thesis, we will construct a set of second-order spinor field theories that can account for all of the information contained in the first-order Dirac equation.

%; those of a (Weyl-)Majorana and of a Dirac fermion. Then, we will apply the same construction to the Standard Model. 

\section*{Second-order fermions and the Standard Model}

In order to justify a second-order description of fermions, it is not only important but necessary to reformulate the whole SM in this formalism. In its usual form, it is a version of the first-order Dirac Lagrangian. As we argued above, there should also exist an associated second-order formulation that can be obtained by integrating out the ``momenta'' fields of the first-order formalism, and indeed, such a second order formulation exists and has been studied by many authors. The list of references that we are aware of is \cite{Feynman:1958ty,Brown:1958zz,Cortes:1992wr,Morgan:1995te,Veltman:1997am,Villasenor:2002kw,DelgadoAcosta:2010nx,Angeles:2011zz,DelgadoAcosta:2012yc,AngelesMartinez:2011nt,VaqueraAraujo:2012qa,Delgado-Acosta:2013kra,Vaquera-Araujo:2013bwa,DelgadoAcosta:2015ypa}, plus a few more works listed in \cite{Veltman:1997am}.  A lot of insight can be gained on the issue when one expresses all the quantities in terms of two-component spinors, and hence, our approach will be closest to that in \cite{Chalmers:1997ui}. 

In terms of two-component spinors it is straightforward to observe that for fermionic Lagrangians the ``momenta'' canonically conjugated to, say, unprimed spinors, are the primed spinors. In a path integral formulation of the the theory, these spinors are treated as independent degrees-of-freedom that have to be integrated over, and thus one can freely choose to integrate out the primed spinors only, arriving at the second-order Lagrangian for unprimed two-component spinors. A common aspect of second-order theories, is that the complexity of the first-order formalism is shifted from the propagator to the vertex, which in the second-order formalism contains a derivative operator. There is also now a new quartic vertex, absent in the first-order formulation. Thus, one obtains a formalism for fermions with Feynman rules very similar to those in QCD, with the familiar $(\partial A) A^2$ and $A^4$ vertices.  

Because the propagator of the second-order formulation is essentially a scalar-type propagator, and because we are working with two-component fermions, the spinor algebra calculations that are often cumbersome in usual Feynman diagrams are much simpler in this case. The second order formalism is also very ideally suited for computations using the spinor helicity methods, see also \cite{Chalmers:1997ui} for an emphasis of this point. Indeed, in computing Feynman diagrams, all that is left to do is proceeding with spinor contractions, and therefore projecting over helicity states becomes a trivial exercise. All in all, we will see that the second-order formalism is more efficient in perturbative calculations.

\section*{Checking the consistency of the theory}
It is important to note at this point, that the aim that we are trying to achieve here, is a completely equivalent description of spinor field theories. It is obviously possible to consider modifications of these as we will discuss later on, however, as a first check we would like to see whether we can reproduce all the basic properties of our usual well-known QFTs.

In the case of fermionic Lagrangians, a non-trivial consistency check is that of the presence (or absence) of anomalies. A lot of attention has been paid in the past to the treatment of anomalies in gauge theories (see \cite{ano1,ano2,ano3} for further reading). Indeed, although they would have catastrophic consequences if they affected a gauge symmetry, they are of evident use when affecting a global symmetry as in the effective field theory description of the pseudo-goldstone bosons of chiral transformations in QCD or for solving the $U(1)$ problem, again in QCD. They also lead to new phenomenological models such as the axion and appear not only in high-energy particle physics, but also in condensed matter physics when one is interested in an effective field theory description of the system through bosonisation of the fermions (see $e.g.$ \cite{weinberg2,altland}). In this thesis we will show that we can reproduce all of the non-trivial aforementioned results.

A reason why we do not learn about fermions directly in their more computationally superior second-order version is that there is a price to pay for going to the second-order formulation. Thus, having integrated out the primed spinors, which in the Dirac Lagrangian are Hermitian conjugates of the unprimed, we have lost manifest unitarity. As a matter of fact, when reformulated in a second-order formalism, the Lagrangian for a spin $1/2$ particle becomes non-hermitian. Although the theory is obtained from a first-order Lagrangian which is known to lead to a unitary S-matrix, an independent proof of unitarity in the former formalism is needed. In this work, we investigate how particular reality conditions, that we describe in the first part of the thesis, lead to a unitary theory in the context of perturbation theory, when imposed on the external states appearing in the S-matrix.
The unitarity of quantum field theories is a fundamental property required of any model aiming at describing Nature. For example, it leads to sensible probabilities when calculating the possible outcomes of a scattering experiment that can be measured in a laboratory. Here, we will only consider perturbative unitarity of the S-matrix, that is, we only require the latter to be unitary order-by-order in pertubation theory. We will follow an approach that was first developed by Veltman \cite{M-1963}, who used the decomposition of the Feynman propagator into forward and backward propagators to construct an equation that only depends on a combinatorics argument.

\section*{Novel aspects}
After having presented the main aspects of the formalism and checked its consistency, it is worth starting to look at novel aspects that are specific to our new formulation. As a matter of fact, one of the most striking aspects of the second-order theory is that it involves only half the number of fields. This has direct consequences on beyond SM physics (BSM), most particularly on Grand Unified Theories (GUTs) models. It is well know that the biggest successes of particle physics in the twentieth century have to do with the edification of the SM, but this relies on the important fact that the weak and electromagnetic forces have been unified into a Yang-Mills theory of electroweak interactions: $SU(2)_W\times U(1)_Y$. Once this was achieved and after evidence arose from high-energy Quantum Chromodynamics (QCD) that quarks come into three colours, leading to the establishment of the additional $SU(3)_c$ symmetry group, it was only a matter of time before physicists tried to unify further the SM. The most famous attempts were Georgi-Glashow's $SU(5)$ \cite{Georgi:1974sy} and Pati-Salam's $SU(4)\times SU(2)\times SU(2)$ \cite{Pati:1974yy}, which can be both further embedded into an $SO(10)$ gauge-group, see $e.g.$ \cite{Baez:2009dj}. The groups that are allowed in these GUTs are constrained by both the forces and matter content of the SM. As we mentioned above, the fact that we have a different number of fermionic representations in our model, directly influences the different GUT patterns that can be obtained. We will develop this in more details in Chapter \ref{sec:unific}. Further novel aspects of the theory that have not been covered in this thesis will be mentioned in the discussions throughout the chapters and in the final conclusion.

\section*{Plan for the thesis}
The aim of this work is to be as self-contained as possible, however, it is obviously unavoidable that a minimal set of concepts is assumed to be known. Nevertheless, we hope that this thesis can be thought of as a reference as far as second-order fermionic field theories are concerned, therefore, in Part I we construct explicitly the theory of Majorana-Weyl fermions, Chapter \ref{chapMaj}, and then generalise it to Dirac fermions, Chapter \ref{chapDirac}. In these two chapters, both a first- and second-order description can be found, their aim being the acquisition of a certain ease with the two-component spinor formalism.

Part II describes the construction of Quantum Electrodynamics, Chapter \ref{chapED}, and of the Standard Model, Chapter \ref{chapSM}. The former will be the framework with which we will be doing calculations in the rest of the thesis, whereas the latter has its obvious importance.

We then arrive to Part III, where in Chapter \ref{chaptertree} we deal with simple tree-level processes in order to get acquainted with the perturbative methods in their newly introduced second-order framework. Renormalisation problems in Chapter \ref{chapterreno} are the first step towards the non-trivial consistency checks that we derive in Part IV.

We start checking that the formalism can reproduce appropriately some non-trivial results such as the anomalies in Weyl and Dirac theories, Chapter \ref{chapterano}. In Chapter \ref{chapuni}, we prove that the theory we have been working with is indeed unitary, and finally in Chapter \ref{sec:unific} we present new possible unification patterns that are available due to the specificities of the second-order theory.

The thesis will end with a conclusion that will summarise what was achieved with this work, and a series of appendices follow in order to fill some gaps that the main text might have left.

\part{Free Field Theory}
\pagestyle{fancy}
%\addtolength{\headwidth}{\marginparsep} %these change header-rule width
%\addtolength{\headwidth}{\marginparwidth}
\lhead{ \bfseries \nouppercase{\leftmark}}
\chead{} 
\rhead{} 
\lfoot{\small\scshape PhD Thesis} 
\cfoot{\thepage} 
\rfoot{\footnotesize Johnny Espin} 
\renewcommand{\headrulewidth}{.3pt} 
\renewcommand{\footrulewidth}{.3pt}
\setlength\voffset{-0.25in}
\setlength\textheight{648pt}
\chapter{Majorana-Weyl Fermions}\label{chapMaj}

\section{Introduction}
When studying the representation theory of the Lorentz group in four dimensions, the simplest non-trivial representation that can be built is that of a Majorana(-Weyl) fermion. It amounts to taking one single copy of a state transforming under the $(1/2,0)$ or $(0,1/2)$ representation. A field theory can then be written for such a state and the latter is called a Majorana spinor if it is massive, and a Weyl spinor otherwise. In this chapter, we construct and quantise the field theory of such objects. First in the usual first-order formalism and later as a second-order theory. For a review on two-component spinors and for a guideline on the conventions that are used here, see Appendix \ref{appendixA}.

\section{First-order formalism}
\subsection{The Weyl (Majorana) Lagrangian}

Let us construct a free field theory of a single Grassmann-valued two-component spinor $\lambda_A$. The most general Lagrangian that is Hermitian and contains only terms of mass dimension four or lower is given by:
\begin{align}\label{L-weyl}
{\cal L}_{\rm Maj}=- i \sqrt{2} \,\lambda^\dagger_{A'} \theta^{\mu AA'} \partial_\mu \lambda_A - \frac{m}{2} \lambda^A\lambda_A - \frac{m}{2} \lambda^\dagger_{A'} \lambda^{\dagger\,A'}
\end{align}

%The fact that the first term is Hermitian (modulo a surface term) is an elementary computation. 
The first term can be rewritten as a combination of two terms that make the Lagrangian explicitly Hermitian, but this rewriting is equivalent to the above Lagrangian up to a surface term. Note that it is the Hermiticity that requires the presence of the imaginary unit in the kinetic term. Moreover, the sign in front of the latter is not arbitrary, it has to be chosen so that the Hamiltonian is positive definite, see below. Note that we could have taken the mass parameter $m$ to be complex as long as $m^*$ is used in the second mass term. However, the phase of $m$ can always be absorbed into $\lambda_A$, and is thus irrelevant. In particular, the sign in front of $m$ is arbitrary, and the sign as in (\ref{L-weyl}) can be achieved by a redefinition of the spinor fields. The factor of $\sqrt{2}$ in front of the first (kinetic) term is introduced for convenience. 

For $m=0$ one obtains the theory first considered by Weyl:
\begin{align}
{\cal L}_{\rm Weyl}=- i\sqrt{2} \,\lambda^\dagger_{A'} \theta^{\mu AA'} \partial_\mu \lambda_A
\end{align}

%\subsection{Index-free notation}

It is often convenient to rewrite formulas omitting the spinor indices. Using the index-free notation the above Lagrangian is rewritten as:
\begin{align}{\cal L}_{\rm Maj}=- i\sqrt{2} \,\lambda^\dagger \theta^{\mu} \partial_\mu \lambda - \frac{m}{2} \lambda\lambda - \frac{m}{2} \lambda^\dagger \lambda^\dagger
\end{align}

which is indeed more compact than (\ref{L-weyl}).

\subsection{Field equations and mode decomposition}

The field equations for (\ref{L-weyl}) are obtained by varying the action with respect to $\lambda_A$ and $\lambda^{\dagger\,A'}$, which for purposes of obtaining the field equations can be treated as independent variables. Note that special care needs to be taken when varying with respect to Grassmann-valued variables. Indeed, in the Grassmann case the left derivative is no longer the same as the right derivative. One has to decide which derivative is used. A good convention is that one varies with respect to unprimed spinors from the right, while with respect to primed spinors from the left. This gives, for the primed spinor equation:
\begin{align}\label{eqn-1} 
i\sqrt{2}\, \theta^{\mu AA'} \partial_\mu \lambda_A + m \lambda^{\dagger\,A'}=0
\end{align}

and for the unprimed spinor equation (note that we need to integrate by parts, hence an extra minus sign):
\begin{align}\label{eqn-2}
 i \sqrt{2} \,\theta^{\mu AA'} \partial_\mu \lambda^\dagger_{A'} - m\lambda^A = 0
\end{align}

Note that this is the Hermitian conjugate of (\ref{eqn-1}), as it should be. In order to see what these equations imply, we solve the first equation for $\lambda^\dagger$ and substitute the result into the second equation. Thus,
\begin{align}\lambda^{\dagger\,A'} = - \frac{i\sqrt{2}}{m} \theta^{\mu AA'} \partial_\mu \lambda_A
\end{align}

and then
\begin{align}i\sqrt{2} \,\theta^{\mu AA'} \partial_\mu \left(- \frac{i\sqrt{2}}{m} \theta^{\nu B}{}_{A'} \partial_\nu \lambda_B \right) - m\lambda^A = 0.\end{align}

Note that we have here two soldering forms with their primed spinor indices contracted. Moreover, their spacetime indices are contracted wit $\partial_\mu\partial_\nu$, which is symmetric in $\mu\nu$ (partial derivatives commute). Thus, we are interested in the object $\theta^{(\mu AA'} \theta^{\nu)B}{}_{A'}$, where the brackets denote the symmetrisation:
\begin{align} (\mu\nu) = \frac{1}{2}(\mu\nu + \nu\mu)\end{align}

This object can be computed explicitly from the formula for the soldering form given in (\ref{theta}). Alternatively, one may expect that this object must be proportional to the spacetime metric $\eta^{\mu\nu}$, and then compute the proportionality coefficient from the formula for the metric in terms of the soldering form. One gets:
\begin{align} \label{ident}
\theta^{(\mu AA'} \theta^{\nu)B}{}_{A'} = \frac{1}{2} \eta^{\mu\nu} \epsilon^{AB}
\end{align}

This formula is the simplest from a series of identities satisfied by the soldering forms. Many other useful identities can be derived. We now use this identity in the above equation for $\lambda^A$, which we multiply by $m$ to put it into the form:
\begin{align}\label{KG}
(\Box - m^2) \lambda^A = 0.
\end{align}

Thus, each component of our two-component spinor $\lambda^A$ satisfies the wave equation already familiar from the scalar field theory case. It is then clear that the parameter $m$ plays the role of the mass of our fermionic particles. This stems naturally from the group theory of the Poincar\'e group, where the momentum generator squared is a quadratic Casimir for any representation and can be used to define a differential equation for the states.

%\subsection{Field equations: Mode decomposition}

\medskip 

In deriving (\ref{KG}), we have used two first-order field equations for two-component spinors $\lambda_A$ and $\lambda^{\dagger\,A'}$ to obtain a second-order equation for $\lambda_A$. However, if we regard the latter as the defining equation of the system, it is clear that some information has been lost, and our original equations imply more than (\ref{KG}). In order to understand what the first-order differential equations imply for the theory, it is convenient to work in momentum space. Thus, we expand the spinor fields $\lambda_A$ and $\lambda^{\dagger\,A'}$ into Fourier modes. As it is usually done in field theory, a second-order wave equation will give rise to two linearly independent solutions: positive and negative energy modes. This leads to the following mode decomposition:
\begin{align}
\lambda_A(x)  &= \int \frac{d^3 k}{(2\pi)^3 2\omega_k} \left( a_A(k) e^{-i \omega_k  t + i \vec{k}\vec{x}} + b^\dagger_A(k) e^{i \omega_k  t - i\vec{k}\vec{x}}\right) \nonumber \\ & \equiv  \int d\Omega_k \left( a_A(k) e^{+ikx} + b^\dagger_A(k) e^{-ikx}\right) \label{decomp}
\end{align}

where $d\Omega_k $ is the Lorentz-invariant momentum-space measure, and $k_\mu = (-\omega_k, \vec k)$. As for the scalar field, we expect that the coefficient (operator) in front of the mode $e^{i\omega_k t}$ to be a creation operator, and this is why it was denoted by $b^\dagger_A(k)$. 
%However, since this is an unprimed spinor that is the Hermitian conjugate of some other spinor, this other spinor $b^{A'}$ is a primed one.
Similarly for the Hermitian conjugate spinor, we have:
\begin{align}
\lambda^{\dagger\,A'}(x) & = \int \frac{d^3 k}{(2\pi)^3 2\omega_k} \left( a^{\dagger\,A'}(k) e^{i \omega_k  t - i\vec{k}\vec{x}} + b^{A'}(k) e^{-i\omega_k  t + i\vec{k}\vec{x}}\right)\nonumber \\&\equiv  \int d\Omega_k \left(  a^{\dagger\,A'}(k) e^{-ikx} + b^{A'}(k) e^{+ikx}\right)
\label{decomp-herm}
\end{align}

We can then rewrite the first-order differential equations as algebraic equations for the modes. The equation (\ref{eqn-1}) then becomes:
\begin{align}\label{modes-1}
\sqrt{2} \theta^{\mu AA'} k_\mu a_A(k) - m b^{A'}(k)=0, \quad 
\sqrt{2} \theta^{\mu AA'} k_\mu b^\dagger_A(k)+ m a^{\dagger\,A'}(k)=0
\end{align}

The second equation (\ref{eqn-2}) is the complex conjugate of (\ref{eqn-1}), and so we get another pair of equations:
\begin{align}\label{modes-2}
\sqrt{2} \theta^{\mu AA'} k_\mu a^\dagger_{A'}(k) - m b^{\dagger\,A}(k)=0, \quad 
\sqrt{2} \theta^{\mu AA'} k_\mu b_{A'}(k)+ m a^{A}(k)=0
\end{align}

The above equations imply that the operators $a_A(k), b^{A'}(k)$ are not independent, they can be written as linear combinations of one another. Hence, the content of (\ref{eqn-1}), (\ref{eqn-2}) can be summarised by saying that they imply:
\begin{align}\label{ab}
b^{A'}(k) =\frac{\sqrt{2}}{m} \theta^{\mu AA'} k_\mu a_A(k)
\end{align}

as well as the on-shell condition $k^2+m^2=0$. This last condition, together with (\ref{ab}) is equivalent to the full set of first-order differential equations. In order to have a more compact notation, it is convenient to introduce:
\begin{align}\label{k-AA}
 k^{AA'} := \theta_\mu^{AA'} k^\mu
\end{align}

Using the above notation as well as (\ref{ab}), the mode decomposition (\ref{decomp}) can be rewritten as:
\begin{align}\label{weyl-mode}
 \lambda_A(x)  = \int d\Omega_k \left( a_A(k) e^{+ikx} -\frac{\sqrt{2} }{m} k_{AA'}a^{\dagger\,A'}(k) e^{-ikx}\right)
\end{align}

Thus, we see that in the case of a single Majorana fermion there is just one type of ladder operators, and therefore in a particle interpretation, a Majorana particle is its own anti-particle. This can be explained by noting that the Majorana equation (\ref{eqn-1}) can be interpreted as a {\it reality condition} for the fermion field $\lambda_A$. Indeed, we have a complex spinor field $\lambda_A$ satisfying the wave equation (\ref{KG}). In general, for a complex field we get two types of creation-annihilation operators, and thus particles and anti-particles. However, in this case the field satisfies an additional equation (\ref{eqn-1}) that can be interpreted as a (non-trivial) reality condition. This condition relates the anti-particle operators to the particle ones, and thus is the reason why there is only one type of operators in the mode expansion: the Majorana spinor is real in the sense of (\ref{eqn-1}). 

The above mode decomposition can be used as a starting point for the canonical quantisation, $i.e.$ the computation of the (anti-)commutators of $a_A(k)$ and $a^\dagger_{A'}(k)$, particle interpretation, and then the derivation of the LSZ formula needed for extracting the scattering amplitudes from the correlation functions. To do this we need the Hamiltonian formulation of the theory.

\subsection{Hamiltonian description of a single Majorana fermion}

We now proceed with a space-time split of our quantities in order to define the Hamiltonian of the theory. Necessary material for the understanding of what follows can be found in Appendix \ref{su2spinors}. The 3+1 decomposition of the Majorana Lagrangian (\ref{L-weyl}) is given by:
\begin{align}
{\cal L}_{\text{Maj}}=i \sqrt{2} \lambda^\dagger_{A'} \theta_{0}^{AA'} \partial_t \lambda_A -i \sqrt{2} \lambda^\dagger_{A'} \theta^{i\,AA'} \partial_i \lambda_A- \frac{m}{2} \lambda^A\lambda_A - \frac{m}{2} \lambda^\dagger_{A'} \lambda^{\dagger\,A'}
\end{align}

It readily follows that the canonically conjugated momentum to the spinor field $\lambda_A$ is given by:
\begin{align}\label{pi-weyl}
\pi^A=i \sqrt{2} \lambda^\dagger_{A'} \theta_0^{AA'} = i \lambda^{\star\,A}
\end{align}

Notice that the normalisation that we chose for the kinetic term involving a factor of $\sqrt{2}$ is needed precisely in order to have such a simple relation between the conjugate momentum $\pi^A$ and the $\star$-conjugate of $\lambda^A$. We can now rewrite our Lagrangian as:
\begin{align}\label{weyl-ham-form}
{\cal L}_{\text{Maj}}=\pi^A \partial_t \lambda_A - i \pi^A T^i_A{}^B \partial_i \lambda_B-\frac{m}{2}\lambda^A\lambda_A-\frac{m}{2}\pi^A\pi_A
\end{align}

where we have used the spatial soldering form in their version (\ref{T}). An alternative expression for the above Lagrangian involving the star-conjugation is given by:
\begin{align}
{\cal L}_{\text{Maj}}=i \lambda^{\star\,A} \partial_t \lambda_A +\lambda^{\star\,A} T^i_A{}^B \partial_i \lambda_B-\frac{m}{2}\lambda^A\lambda_A-\frac{m}{2}(\lambda^\star)_A\lambda^{\star\,A}
\end{align}

Using $(\lambda^A \eta^B)^\star = (\eta^\star)^{B} (\lambda^\star)^{A}$ as well as the fact that $\star^2=-1$ and that the quantities $T^i_A{}^B$ are $\star$-Hermitian, one can easily check this Lagrangian to be $\star$-Hermitian modulo a surface term.

A useful exercise for what follows is to find the field equations that follow from (\ref{weyl-ham-form}). Treating the fermionic fields $\lambda^A,~\pi^A$ as independent we get:
\begin{align}\label{majorana-eqs}
\dot{\lambda}_A-i T^i_A{}^B\partial_i \lambda_B -m\pi_A=0, \qquad
\dot{\pi}_A+i T^i_A{}^B\partial_i \pi_B +m\lambda_A =0
\end{align}

The second equation is the $\star$-conjugate of the first, as it should be. One can find the momentum $\pi_A$ from the first equation and substitute the result to the second. Using (\ref{T-algebra}) and multiplying the result by $m$ one gets:
\begin{align}\label{box-m-eqn}
(\partial_t \partial_t - \partial_i\partial^i +m^2 )\lambda_A =0
\end{align}

which is the desired massive wave equation for a two-component fermion.

\subsection{Momentum spinors and mode decomposition}

The mode decomposition (\ref{weyl-mode}) is not a good starting point for computations, because the operators $a^A(k)$ and $a^{\dagger\,A'}(k)$ are not the canonically normalised operators for creation and annihilation of particles. It is also clear that $a^A(k)$ contains in fact two operators when decomposed into some basis in the spinor space; it is a spinor-valued operator and we would like to have a mode decomposition where the ladder operators and the polarisation spinors appear explicitly. For all these reasons it is necessary to develop another tool --- that of momentum spinors. 

Consider a null real four-vector $k^{\mu}=(|k|,\vec{k})$ (not yet related to the momentum of any particle). As such, it can be written as a product of two spinors $k^\mu=-\theta^{\mu}_{AA'} k^A k^{A'}$, $i.e.$, $k^A k^{A'} = \theta_\mu^{AA'} k^\mu$. In the case of Lorentzian signature the spinors $k^A,~ k^{A'}$ must be complex conjugates of each other (for a real four-vector). It is then clear that $k^A$ is only defined modulo a phase. Moreover, as the vector $\vec{k}$ varies, thats is, as $\vec{n}=\vec{k}/|\vec{k}|$ varies over the sphere $S^2$, there is no continuous choice for the spinor $k^A$. We make the following choice for the momentum spinor:
\begin{align}\label{kA}
k^A \equiv k^A(\vec{k}) := 2^{1/4} \sqrt{|k|} \left( \cos(\theta/2) e^{i\phi/2} o^A + \sin(\theta/2) e^{-i\phi/2} \iota^A\right)
\end{align}

where $o^A,\iota^A$ is a basis in the space of unprimed spinors. Here $\theta,\phi$ are the usual coordinates on $S^2$ so that the momentum vector in the direction of the positive z-axis corresponds to $\theta=\phi=0$. We see that the corresponding spinor is $2^{1/4} \sqrt{|k|} o^A$. The formula (\ref{kA}) can be checked using the expression (\ref{theta}) for the soldering form.

We can now readily observe how the spinor changes as we rotate the vector $\vec k$. Consider, for example, what happens when the momentum direction gets reversed. This corresponds to $\theta\to \pi-\theta$ and $\phi\to \phi+\pi$. We get
\begin{align}
k^A(-\vec{k}) = 2^{1/4} i \sqrt{|k|}\left( \sin(\theta/2) e^{i\phi/2} o^A - \cos(\theta/2) e^{-i\phi/2} \iota^A\right)
\end{align}

Let us compare this to the effect of the $\star$-operation on the momentum spinor. We have
\begin{align}\label{k-k}
k^A(-\vec{k})=- i \, k^{\star A}(\vec{k})
\end{align}

We could have chosen a different phase factor in (\ref{kA}) so that there is no imaginary unit in this formula. However, in this case some formulas below become less symmetrical. 

An interesting consequence of (\ref{k-k}) can be obtained by taking the $\star$-conjugate of this formula. Using $\star^2=-1$ we get $k^{\star A}(-\vec{k}) = - i k^A(\vec{k})$. This means that flipping the sign of the momentum twice we get minus the original momentum spinor. In other words, $k^A$ takes values in a non-trivial spinor bundle over $S^2$. Let us now see how this formalism can be applied to the mode decomposition (\ref{weyl-mode}).

%
%\subsection{Mode decomposition at the Hamiltonian level}

As is usual in field theory, the Hamiltonian formulation of the theory allows for a standard derivation of the commutation relations between the ladder operators appearing in the mode decomposition of the fields. However, as was mentioned above, (\ref{weyl-mode}) is not suitable for this task as it contains a mixture of unprimed and primed spinors, whereas the Hamiltonian description only contains unprimed $SU(2)$ spinors. Therefore, we need an new adequate decomposition that will allow us to do so, and that additionally, will allows us to show that the Hamiltonian of the theory is indeed positive definite.

It is clear that momentum spinors $k^A(\vec{k})$ and $k^{\star A}(\vec{k})$ that we introduced above are linearly independent and can be used as a basis to decompose the operator-valued spinor $a^A(k)$. Thus, we introduce a pair of operators 
\begin{align}
a^A(\vec{k}) =  k^{\star A}(\vec{k}) \tilde{a}^+_k +  k^A(\vec{k}) \tilde{a}^-_k
\end{align}

where the interpretation of the new operators $\tilde{a}^\pm_k$ is to be clarified below and the momentum spinors $k^{\star A}$ and $k^A$ still have to be related to the four-momentum of the particle. We have denoted the new operators with an overtilde because they are still not canonically normalised to have a particle interpretation. 

The mode expansion convenient for the purposes of the Hamiltonian formulation is then obtained by either just expressing everything in terms of $\star$-conjugate spinors in (\ref{weyl-mode}), or alternatively by writing a general mode expansion and then using the field equations in their form (\ref{majorana-eqs}). For using this second method we note that the operator $T_A^i{}^B\partial_i$ becomes, when acting on the modes
\begin{align}
T_A^i{}^B\partial_i e^{i \vec{k}\vec{x}} =|k| \left( (o_A o^B  e^{i\phi} - \iota_A \iota^B  e^{-i\phi})\sin(\theta) - (o_A \iota^B +\iota_A o^B)\cos(\theta) \right) e^{i \vec{k}\vec{x}}
\end{align}

When acting on the momentum spinors $k^A, k^{\star A}$ this gives
\begin{align}
T_A^i{}^B\partial_i  k_B(\vec{k}) e^{i \vec{k}\vec{x}} =- |k| k_A(\vec{k}) e^{i \vec{k}\vec{x}}, \qquad
T_A^i{}^B\partial_i  k^\star_B(\vec{k}) e^{i \vec{k}\vec{x}} = |k| k_A^\star(\vec{k}) e^{i \vec{k}\vec{x}}
\end{align}

so these are eigenmodes of eigenvalues $\mp |k|$. Using this fact, we get the following mode expansion
\begin{align}
 \lambda^A(x)  = \int d\Omega_k &\Big( (k^{\star A} \tilde{a}^+_k +  k^A \tilde{a}^-_k)  e^{+ikx} \\ \nonumber &+\frac{1}{m}(k^{\star A}(\omega_k-|k|) \tilde{a}^{\dagger\,-}_k- k^A (\omega_k+|k|) \tilde{a}^{\dagger\,+}_k )e^{-ikx}\Big)
\end{align}

where we dropped the argument $\vec{k}$ from $k^A(\vec{k})$ and $k^{\star A}(\vec{k})$ for brevity. This coincides with what is obtained directly from (\ref{weyl-mode}).

\subsection{Quantisation and polarisation spinors}

Let us now compute the (anti-)commutational relations between the $\tilde{a}^{\pm},~\tilde{a}^{\dagger\,\pm}_k$ operators. With our conventions the anti-commutator between $\pi^A, ~\lambda^B$ is
\begin{align}
\{ \pi^A(x), \lambda^B(y)\} = - i \epsilon^{AB} \delta^3(x-y)
\end{align}

One then finds that the non-vanishing anti-commutators are
\begin{align}
\{ \tilde{a}^{\dagger\,\pm}_k, \tilde{a}^\pm_p\} = (2\pi)^3 2\omega_k \delta^3(k-p) \frac{m^2}{\sqrt{2}|k|(\omega_k\pm|k|)}
\end{align}

To obtain this result we have used the following relation for the momentum spinors
\begin{align}
k^A k^{\star B}- k^{\star A} k^B = \sqrt{2} |k| \epsilon^{AB}
\end{align}
which can be checked using the corresponding definitions. 

Thus, the operators we have introduced are not canonically normalised. Let us introduce new, canonically normalised operators via
\begin{align}\label{op-redef}
\tilde{a}^+_k = \sqrt{\frac{\omega_k-|k|}{\sqrt{2}|k|}} a^+_k, \quad \tilde{a}^-_k = \sqrt{\frac{\omega_k+|k|}{\sqrt{2}|k|}} a^-_k
\end{align}

The new operators satisfy 
\begin{align}
\{ \tilde{a}^{\dagger\,\pm}_k, a^\pm_p\} = (2\pi)^3 2\omega_k \delta^3(k-p)
\end{align}

The mode decomposition in terms of the canonically normalised operators is
\begin{align}\label{weyl-mode*}
 \lambda_A(x)  =  \int d\Omega_k\ \sqrt{m}\Big( ( \epsilon^+_A a^+_k + \epsilon^-_A a^-_k)  e^{+ikx} +( \epsilon^+_A a^{\dagger\,-}_k-\epsilon^-_A  a^{\dagger\,+}_k )e^{-ikx}\Big)
\end{align}

where we have introduced the polarisation spinors
\begin{align}
\epsilon^+_A =  \sqrt{\frac{\omega_k-|k|}{\sqrt{2}m |k|}} k^\star_A, \quad  \epsilon^-_A = \sqrt{\frac{\omega_k+|k|}{\sqrt{2}m |k|}} k_A 
\end{align}

which are normalised so that $\epsilon^{+ A} \epsilon^-_A=1$. 

For completeness, let us also give the expression for the momentum
\begin{align}
\pi_A(x) = \frac{i}{m}  \int d\Omega_k\sqrt{m}\Big( \left( -(\omega_k-|k|) \epsilon^{-}_A a^{\dagger\,+}_k + (\omega_k+|k|) \epsilon^{+}_A  a^{\dagger\,-}_k\right)  e^{-ikx} \\ \nonumber
+ \left(-(\omega_k-|k|) \epsilon^{-}_A a^-_k-(\omega_k+|k|) \epsilon^{+}_A a^+_k \right)e^{+ikx}\Big)
\end{align}

\subsubsection*{Alternative expression for the polarisations}

Here we motivate our choice for the normalisation of the operators (and the polarisation spinor $\epsilon^\pm_A$) by providing an alternative expression. We also relate the momentum spinors that we have been using to the four-momentum of each mode. Let us consider the quantity $k^{AA'}$ defined in (\ref{k-AA}) that corresponds to an excitation with the corresponding momentum. This is a massive quadrivector $k^A{}_{A'} k_A{}^{A'}=-m^2$. However, we can always represent it as a sum of two null vectors. In the spinor notation
\begin{align}
k^{AA'} = K^A K^{A'} - \frac{m^2 p^A p^{A'}}{2 (K^E p_E) (K_{E'} p^{E'})}\label{massivemomentum}
\end{align}

Here $p^{A} p^{A'}$ is a reference null vector (in the spinor form). The above decomposition of $k^{AA'}$ is defined once $p^{A}$ is chosen. Let us compute $K^A, p^A$ in the frame in which the spatial momentum vector $\vec{k}$ points along the $z$-axes. In this case we have from (\ref{theta-xyz})
\begin{align}
k^{AA'} = \frac{1}{\sqrt{2}} (\omega_k + |k|) o^A o^{A'} + \frac{1}{\sqrt{2}} (\omega_k - |k|) \iota^A \iota^{A'}
\end{align}

We would like our $K^A$ to be proportional to $k^A$, which in this case is a multiple of $o^A$. Thus, we take $p^A\sim \iota^A$ and get
\begin{align}
K^A = \frac{\sqrt{\omega_k+|k|}}{2^{1/4}} o^A = \sqrt{\frac{\omega_k + |k|}{2|k|}} k^A
\end{align}

Note that in the massless limit we get $K^A = k^A$ as one can expect. 

Using the spinor $K^A$ we can now write the polarisation spinors $\epsilon^\pm$ in the following convenient form convenient for calculations
\begin{align}
\epsilon^-_A = 2^{1/4} \frac{K_A}{\sqrt{m}}, \qquad \epsilon^+_A = \frac{\sqrt{m} \, p_A}{2^{1/4} (K^E p_E)}.\label{polarisations}
\end{align}

\subsection{Hamiltonian}

The total Hamiltonian is given by
\begin{align}\label{H-original}
H = \int d^3x \left(  i \pi^A T^i_A{}^B \partial_i \lambda_B+(m/2)\lambda^A\lambda_A+(m/2)\pi^A\pi_A\right)
\end{align}

Substituting (\ref{weyl-mode*}) one gets
\begin{align}
H=\int d\Omega_k \frac{m}{2} \Big( \frac{m}{\omega_k+|k|} a^{\dagger\,+}_k a_k^+  - \frac{\omega_k+|k|}{m} a_k^+ a^{\dagger\,+}_k \\ \nonumber + \frac{m}{\omega_k-|k|} a^{\dagger\,-}_k a_k^-  - \frac{\omega_k-|k|}{m} a_k^- a^{\dagger\,-}_k\Big)
\end{align}

where we have used
\begin{align}
k^{\star A} k_A = \sqrt{2}|k|
\end{align}

The normal ordered Hamiltonian ($i.e.$ ignoring the contribution from the zero modes) is then 
\begin{align}\label{weyl-H}
H=\int d\Omega_k \omega_k \left( (a_k^+)^\dagger a_k^+ + (a_k^-)^\dagger a_k^-\right)
\end{align}

which confirms the interpretation of $a_k^{\dagger\,\pm}$ and $a_k^\pm$ as the creation-annihilation operators of two species of particles of the same energy $\omega_k$.

\subsection{Massless limit}

It is now not hard to obtain the massless limit of the above theory. This is called a theory of a Weyl fermion, in contrast to the Majorana massive case considered so far. For small $m$ we have
\begin{align}
\omega_k = |k| + \frac{m^2}{2|k|} + O(m^4)
\end{align}

and two of the terms drop out from the expansion (\ref{weyl-mode*}). What remains is
\begin{align}
\lambda_A(x)  =  \int d\Omega_k 2^{1/4}k_A \left( a^-_k  e^{+ikx} -  a^{\dagger\,+}_k e^{-ikx}\right)
\end{align}

with the corresponding momentum given by
\begin{align}
\pi_A(x) = i \int  d\Omega_k 2^{1/4} k^\star_A \left( - a^+_k  e^{+ikx} +  a^{\dagger\,-}_k e^{-ikx}\right)
\end{align}

The main difference with the massive case above is that in the Weyl theory the field $\lambda(x)$ does not satisfy any equation that can be interpreted as a reality condition. It satisfies a constraint equation.

The full Hamiltonian in the Weyl case is
\begin{align}
H = \int d^3x \left(  i \pi^A T^i_A{}^B \partial_i \lambda_B\right) =\int d\Omega_k\omega_k \left( - a_k^+ a^{\dagger\,+}_k + a^{\dagger\,-}_k a_k^-\right)
\end{align}
reproducing of course the same normal ordered Hamiltonian (\ref{weyl-H}).

\subsection{Parity}

It is clear that both modes enter the Majorana theory symmetrically. Thus, there is a discrete symmetry with an action which changes the sign of the spatial momentum and exchanges the two modes. However, since for fermions the square of this operation can also take value minus one, we can have an additional phase in the definition of this transformation. %We choose the rule without any phase, but with a minus sign
\begin{align}
P^\dagger a^\pm_k P = \eta a^\mp_{-k}
\end{align}

as it leads to the most convenient transformation property. Let us determine what effect this has on the field operator $\lambda(x)$. It is clear that this operation must have something to do with the inversion of the spatial coordinate $\vec{x}\to -\vec{x}$. Thus, we compute
\begin{align}
P^\dagger \lambda(t,-\vec{x}) P = \frac{i}{m}  \int  d\Omega_k\sqrt{m}\Big( \eta \left(-(\omega_k-|k|) \epsilon^{-}_A a^-_k-(\omega_k+|k|) \epsilon^{+}_A a^+_k \right)e^{+ikx}\\ \nonumber
+ \eta^* \left( -(\omega_k-|k|) \epsilon^{-}_A a^{\dagger\,+}_k + (\omega_k+|k|) \epsilon^{+}_A  a^{\dagger\,-}_k \right) e^{-ikx}\Big)
\end{align}

where we have used $k_A(-\vec{k})=-i k^\star_A(\vec{k}),~ k_A^\star(-\vec{k})=-i k_A(\vec{k})$ and that
\begin{align}
\frac{\omega_k+|k|}{m} = \frac{m}{\omega_k-|k|}
\end{align}
We see that for the field to be a representation of parity we must have
\begin{align}
\eta =\eta^*
\end{align}

Then
\begin{align}
P^\dagger \lambda(t,-\vec{x}) P =\eta  \pi(t,\vec{x}) =\eta i \lambda^\star(t,\vec{x})
\end{align}

Using this transformation rule on (\ref{H-original}), we see that the theory is parity invariant.

\section{Second-order formulation of the Majorana theory}

We now turn to the main subject of this thesis, which is a second-order formulation of fermions. We first study it on the simplest example of Majorana theory. 

\subsection{Second-order Lagrangian}

As we have already mentioned in the introduction, a second-order formulation can be obtained by integrating out all fields of an ${SL}(2,\C)$ representation of one chirality. We choose to integrate out the primed spinors, so that the action depends only on an unprimed spinor. 

Let us carry out this simple exercise. We will keep all the spinor indices explicit to make the operation more transparent. The field equation that one gets for $\lambda^\dagger_{A'}$ is
\begin{align}
-i \sqrt{2} \partial^{ A'A} \lambda_A  - m\lambda^{\dagger\,A'}=0
\end{align}

where as before $\partial^{ A'A} \equiv\theta^{\mu\, A'A} \partial_\mu$.  This equation can be solved for the primed spinor, we find:
\begin{align}
\lambda^{\dagger\,A'} = -\frac{i\sqrt{2}}{m} \partial^{ A'A}\lambda_A
\end{align}

We now substitute this back into (\ref{L-weyl}) and get a second-order action involving only $\lambda_A$. We have
\begin{align}
{\cal L}_{\text{Maj}}=- \frac{1}{m}\partial_{A'}{}^{A}\lambda_A \partial^{ A'B} \lambda_B - \frac{m}{2}\lambda^A\lambda_A
\end{align}

Let us now integrate by parts to put both derivatives on the same spinor field; taking into account the fact that partial derivatives commute and using the simple identity:
\begin{align}\label{theta-ident}
\theta^{(\mu\, AA'}\theta^{\nu)}_{A'}{}^{B} = \frac{1}{2}\epsilon^{AB} \eta^{\mu\nu}
\end{align}
we get the following chiral Lagrangian
\begin{align}\label{weyl-chiral}
{\cal L}_{\text{Maj}}=\frac{1}{2m}\lambda^A \Box \lambda_A - \frac{m}{2}\lambda^A\lambda_A
\end{align}
which is just the obvious second-order Lagrangian leading to the Klein-Gordon equation as its field equation.

Finally, we rescale the field $\lambda_A$ to have the canonically normalised Lagrangian. Thus, we introduce
\begin{align}
\xi_A = \frac{1}{\sqrt{m}} \lambda_A
\end{align}

in terms of which the Lagrangian takes the form
\begin{align}\label{weyl-chiral*}
{\cal L}_{\text{Maj}}=-\partial_{A'}{}^{A}\xi_A \partial^{ A'B} \xi_B - \frac{m^2}{2}\xi^A\xi_A
\end{align}
where we have kept the unsymmetrised kinetic term for reasons that will become clear later on\footnote{As an advanced notice, when considering interactions for second-order fermions, it is the unsymmetrised version of the Lagrangian that allows for a simple minimal coupling. Nevertheless, whenever one is considering the free theory, the kinetic term can be rewritten as a Laplace operator without losing any information.}.

\subsection{Second-order Hamiltonian formulation}
Let us now derive the Hamiltonian density for the second-order Lagrangian (\ref{weyl-chiral*}).
The momentum conjugate to $\xi_A$ is
\begin{align}
p^A = \partial_t \xi^A
\end{align}

and then the Hamiltonian density is
\begin{align}\label{H-chiral}
{\cal H} = \frac{1}{2} p^A p_A + \frac{1}{2} \partial^i \xi^A \partial_i \xi_A + \frac{m^2}{2} \xi^A \xi_A
\end{align}
where we have again symmetrised the kinetic term as well as the spatial gradient. For each mode $e^{i \vec{k}\vec{x}}$ this is just a harmonic oscillator of frequency $\omega_k$. However, the field $\xi_A$ has 4 real components, and thus describes twice too many modes as compared to the original system. Thus, reality conditions need to be added to recover the original dynamics. The reality conditions are (\ref{majorana-eqs}), rescaled to make sense for $\xi$. We choose to write them in the form
\begin{align}\label{reality-maj}
\xi^\star_A= \frac{1}{i\, m}\left( \partial_t \xi_A - i T_A^{i\,\, B} \partial_i \xi_B\right)
\end{align}

which is of course just the original Majorana equation in its time plus space version. After the reality conditions are imposed, there are only two propagating modes left, and one recovers the original system. The reality conditions are quite non-trivial, and result in the following decomposition of the field $\xi$ into the canonically normalised modes
\begin{align}\label{xi-mode}
\xi_A(x)  =   \int d\Omega_k \Big( \left( \epsilon^+_A a^+_k + \epsilon^-_A a^-_k\right)  e^{+ikx} +\left( \epsilon^+_A a^{\dagger\,-}_k-\epsilon^-_A  a^{\dagger\,+}_k \right)e^{-ikx}\Big)
\end{align}

To write this we just took the mode decomposition (\ref{weyl-mode*}) and divided by $\sqrt{m}$ to get $\xi$. As we shall see below, this is completely legitimate because the passage between the first- and second-order descriptions is a canonical transformation. Thus, the commutational relations and the form of the Hamiltonian in terms of $a_k^\pm, a_k^{\dagger\,\pm}$ is as we found before. In particular, it is worth emphasising that the theory (\ref{weyl-chiral*}) that is in no sense Hermitian, after imposition on it of the reality condition (\ref{reality-maj}) becomes an ordinary theory with a Hermitian Hamiltonian. 

\subsubsection*{Canonical transformation}

A transformation between the original first-order description (\ref{H-original}) and (\ref{H-chiral}) is a canonical one. Indeed, starting from the second-order description, let us define new configuration and momentum variables via
\begin{align}
\lambda_A = \sqrt{m} \xi_A, \qquad \pi^A = \frac{1}{\sqrt{m}}\left( p^A - i T^{i\,AB} \partial_i \xi_B\right)
\end{align}

This is a canonical transformation because modulo a surface term
\begin{align}
\pi^A \partial_t \lambda_A = p^A \partial_t \xi_A
\end{align}

And in terms of the new variables the Hamiltonian density takes the form
\begin{align}
{\cal H}= i \pi^A T_A^{i\,\, B} \partial_i \lambda_B + \frac{m}{2} \pi^A \pi_A + \frac{m}{2} \lambda^A \lambda_A
\end{align}

which is the original Hamiltonian (\ref{H-original}). This implies, that the second-order formulation, even though it uses only one chirality of spinors, is still Hermitian and parity invariant.

\subsection{Working with the second-order formulation}

Let us now discuss if we could start with (\ref{weyl-chiral*}) and consistently work with the second-order formulation. The Lagrangian (\ref{weyl-chiral*}) is holomorphic in the field $\xi$, and is not Hermitian. It is clear that to lead to unitary physics it needs to be supplemented with a reality condition. In our case the required reality condition (\ref{reality-maj}) came from the original first-order formulation, but an interesting question is if it is possible to ``discover'' the condition (\ref{reality-maj}) without any prior knowledge of it. 

Let us first discuss this at the level of the Lagrangian. The reality condition is some equation that relates the field to its complex conjugate. In general it is the statement that some anti-linear operation $R$ applied to the field leaves it invariant. In the case of Majorana theory this anti-linear operation is the combination of, first, the application of the operation of Hermitian conjugation on the spinor $\lambda_A$, and then converting the resulting primed spinor to an unprimed one with the help of the Dirac operator
\begin{align}
(R \lambda)_A = \frac{\sqrt{2}}{i \, m} \theta^{\mu}_{AA'} \partial_\mu \lambda^{\dagger\,A'} \equiv \frac{1}{i \, m} \partial_{AA'} \lambda^{\dagger\,A'}
\end{align}

Thus, the anti-linear operation that gives the required reality condition can be schematically written as
\begin{align}\label{R}
R = \frac{1}{i \, m} \partial \circ \dagger
\end{align}

This operator maps unprimed spinors into unprimed spinors, and so can be used as imposing the reality via 
\begin{align}
R\lambda = \lambda
\end{align}

It is also a quite natural operator to be used for this purpose. Indeed, since $R$ must be anti-linear it must involve the operation of taking the Hermitian conjugation of $\lambda$. However, then the resulting spinor will be of a different type from the original $\lambda_A$, and thus cannot be compared to $\lambda$. But one does have an operator converting spinors of one type into those of the other - the Dirac operator. This is exactly what is used in (\ref{R}). The proportionality coefficient is then fixed from the requirement of compatibility with the dynamics of our theory. Indeed, the operator $R^2$ is a linear, second-order differential operator. Fields $\lambda$ satisfying $R\lambda=\lambda$ will also satisfy $R^2\lambda=\lambda$. This second-order differential equation for $\lambda$ must be compatible with the dynamics of the original complex theory. The easiest option is that it coincides with the field equation that one obtains for $\lambda$ from the original complex Lagrangian, and this is precisely what we have happening. 

To summarise, one could start with the theory (\ref{weyl-chiral*}), and then search for an anti-linear operator that can be used to impose the reality condition. Since Hermitian conjugation sends unprimed spinors into primed we have to use the Dirac operator to build up $R$. Then the linear operator $R^2$ must be compatible with the dynamics of the theory, which in our case fixes the choice of $R$ completely. After the reality condition $R\lambda=\lambda$ is imposed, the theory becomes an usual theory with a Hermitian Hamiltonian. There is however no guarantee that such a procedure of finding an appropriate $R$ works for an arbitrary complex Lagrangian. But the importance of the above example is in showing that the requirement of working with a Hermitian Lagrangian can be too restrictive, and that non-hermitian, holomorphic Lagrangians can also lead to the usual unitary dynamics provided an appropriate operator $R$ exists that can be used to impose the reality. As the example of Majorana theory shows, this operator can be quite non-trivial, and in fact be a differential operator. 

Let us now, in anticipation to later chapters, briefly discuss how to do computations with such a holomorphic, second-order in derivatives formulation. The key point is that one only needs to worry about the reality of the field (and resulting particle interpretation) on the external lines of all the diagrams. On the internal lines one can safely forget about any issues of reality, because in the fermionic path integral the spinors and their complex conjugate spinors are integrated over independently, and this takes care of the reality constraint: the second-order Lagrangian that arises from this integration automatically knows about the constraint, the interactions are built such as to respect the reality condition. So, the $R$-operation that we discussed here is only important to fix the mode decomposition (\ref{xi-mode}) and thus the particle interpretation. It is of no importance at all for computing the correlation functions of the field operator. It is only after these are computed, that one extracts the scattering amplitudes via an appropriate version of the LSZ formula that follows from (\ref{xi-mode}). Having these rules in mind simplifies computations significantly, because one can compute Feynman diagrams working with a holomorphic second-order theory (\ref{weyl-chiral*}), which has much simpler Feynman rules (even in the case in which we have interactions with external gauge fields) than the usual first order action. Below we shall apply these ideas to the Lagrangian of the Standard Model, after the Dirac spinors are discussed.

\chapter{Dirac Fermions}\label{chapDirac}

\section{More general fermionic Lagrangians}

We have considered the Lagrangian for a single two-component fermion. Let us now generalise this, and consider a collection of $N$ two-component fermions that we shall denote by $\lambda_A^i, i=1,\ldots, N$. It is clear that the following Lagrangian is the most general Hermitian Lagrangian of mass dimension four:
\begin{align}\label{gen-L}
{\cal L}= \sum_{i=1}^N -i\sqrt{2} \lambda^\dagger_i \partial \lambda^i - \frac{1}{2} M_{ij} \lambda^i\lambda^j - \frac{1}{2} M^{* ij} \lambda^\dagger_i \lambda^\dagger_j
\end{align}

Note that since $\lambda^i\lambda^j = \lambda^j \lambda^i$ the matrix $M_{ij}$ is symmetric. The matrix $M^{* ij}$ is then the complex conjugate of $M_{ij}$. 

Note now that the first kinetic term is invariant under the following ``flavour symmetry'':
\begin{align}
\lambda^i \to \mathcal{U}^i{}_j \lambda^j, \qquad \mathcal{U}\in {U}(N)
\end{align}

This symmetry mixes up the different fermionic species present. The mass terms are not invariant. However, we can use the chiral symmetry to diagonalise the mass matrix $M_{ij}$. Indeed, being a symmetric $N\times N$ matrix is is parametrized by $N(N+1)/2$ complex numbers, and thus by $N(N+1)$ real numbers. On the other hand, the (real) dimension of ${U}(N)$ is $N^2$. Thus, using the available ${ U}(N)$ freedom we can kill $N^2$ of the components of $M_{ij}$, which leaves us with only $N$ real diagonal entries. In fact, the transformation of absorbing the phase of the parameter $m$ of our original Majorana Lagrangian into the fermionic field $\lambda_A$, which allowed us to take $m$ to be real, was an example of a ${U}(1)$ transformation. Therefore, we can always go into the mass eigenstate basis and write down the most general fermionic Lagrangian (\ref{gen-L}) as follows:
\begin{align}
{\cal L}= \sum_{i=1}^N -i \sqrt{2} \lambda^\dagger_i \partial \lambda^i - \frac{1}{2} m_i \lambda^i\lambda^j - \frac{1}{2} m_i \lambda^\dagger_i \lambda^\dagger_j
\end{align}

where now $m_i=m_i^*$ are real mass parameters. These are, in general, all different. Thus, in general, the most general fermionic Lagrangian is just a collection of Majorana Lagrangians that we have considered before. 

\section{The Dirac Lagrangian}

A case that is very important for applications arises when two of the mass eigenstates of our collection of fermions have the same mass. Let us consider this case specifically. Thus, we now introduce different names to our two Majorana fermions, and call them $\eta$ and $\lambda$. We get the following Lagrangian:
\begin{align}
{\cal L}_{\rm D} = -i \sqrt{2} \lambda^\dagger \partial \lambda -\im \sqrt{2} \eta^\dagger  \partial\eta-\frac{m}{2} \lambda\lambda - \frac{m}{2} \eta\eta - \frac{m}{2} \lambda^\dagger \lambda^\dagger - \frac{m}{2} \eta^\dagger \eta^\dagger
\end{align}

With the mass spectrum being degenerate, this Lagrangian has a residual ${SO}(2)$ symmetry mixing the two fermions:
\begin{align}
\eta \to \cos(\alpha) \eta + \sin(\alpha) \lambda, \qquad  \lambda\to -\sin(\alpha) \eta + \cos(\alpha) \lambda
\end{align}
where $\alpha\in {\mathbb R}$. Let us rewrite this Lagrangian in the form that makes the above ${SO}(2)$ symmetry an ${ U}(1)$ symmetry. Hence, let us introduce the following complex linear combinations of our fermions:
\begin{align}
\chi:= \frac{1}{\sqrt{2}} (\eta+i \lambda), \qquad \xi :=  \frac{1}{\sqrt{2}} (\eta-i \lambda)
\end{align}

These can be seen to transform as
\begin{align}\label{u1}
\chi \to \chi e^{-i\alpha}, \qquad \xi \to \xi e^{i\alpha}
\end{align}

The Lagrangian can be written in terms of these new fermionic fields, and reads:
\begin{align}\label{Dirac-L}
{\cal L}_{\rm D} = -i \sqrt{2} \chi^\dagger \partial \chi -\im \sqrt{2} \xi^\dagger  \partial \xi - m(\chi \xi + \chi^\dagger \xi^\dagger)
\end{align}
In this form the Lagrangian is explicitly invariant under (\ref{u1}). We note that more generally, the mass matrix will break the chiral flavour ${ U}(N)$ symmetry down to some symmetry group $G$. Fermions of the theory can then be classified according to representations of $G$ that they realise. Thus, in the Dirac case the symmetry is broken down to ${ U}(1)$, and the two fermions that we have transform in complex conjugate representations of ${ U}(1)$.

\subsection{Dirac spinors}

As we shall soon see, the ``global'' ${ U}(1)$ symmetry of the Dirac Lagrangian can be made local by introducing a gauge field. The resulting theory is the one relevant for (quantum) electrodynamics. It was first discovered in a different form by Dirac. To motivate the original Dirac's version, let us note that the Feynman rules for the theory (\ref{Dirac-L}) are quite complicated. Indeed, the Lagrangian pairs fields $\chi$ with $\chi^\dagger$ and $\xi$ with $\xi^\dagger$ in the kinetic terms, as well as pairs $\chi\xi$ and $\chi^\dagger \xi^\dagger$ in the mass terms. Thus, there are 4 different propagators to be considered. This makes Feynman diagrams calculations with the above Lagrangian quite complex (because of the number of diagrams), see Appendix \ref{ApxRules}. This can be avoided if we put two two-component spinors into a single four-component (Dirac) spinor. It is only natural to clamp together fermionic fields that are in the same representation of the ${ U}(1)$ group. This is why we define:
\begin{align}
\Psi := \left( \begin{array}{c} \chi \\ \xi^\dagger \end{array} \right)
\end{align}

We also define the {\it Dirac conjugate} of $\Psi$ via:
\begin{align}
\bar{\Psi} := \left( \begin{array}{cc} \xi & \chi^\dagger \end{array} \right)
\end{align}

We now have:
\begin{align}
\bar{\Psi} \Psi = \xi\chi + \chi^\dagger \xi^\dagger
\end{align}

which is the correct mass term in (\ref{Dirac-L}). To rewrite the kinetic terms in terms of $\Psi$ we integrate by parts in the $\xi$ kinetic term to put the derivative onto $\xi^\dagger$. We then define
\begin{align}
\gamma^\mu := \left( \begin{array}{cc} 0 & \sqrt{2}\, \theta^{\mu} \\ \sqrt{2}\, \theta^{\mu} & 0 \end{array} \right)
\end{align}

and rewrite (\ref{Dirac-L}) as:
\begin{align}\label{Dirac}
{\cal L}_{\rm D} = -i\bar{\Psi} \gamma^\mu \partial_\mu \Psi - m\bar{\Psi} \Psi
\end{align}

which is the original form in which this Lagrangian was discovered by Dirac. As in the two-component version the main object was the soldering form $\theta^{\mu AA'}$, in the four-component version the object $\gamma^\mu$ plays the fundamental role. These $4\times 4$ matrices are called Dirac gamma-matrices. They satisfy the following basic algebra:
\begin{align}
\gamma^\mu \gamma^\nu + \gamma^\nu \gamma^\mu = -2\eta^{\mu\nu} {\bf 1}
\end{align}

which is the four-component analogue of (\ref{ident}). The Feynman rules of the Dirac version of the theory are much simpler in that there is only the pairing $\Psi \bar{\Psi}$ in the propagator.

\subsection{Hamiltonian formulation and mode decomposition}

The 3+1 decomposition of (\ref{Dirac-L}) is
\begin{align}
{\cal L}_{\text{D}}=\pi^A \partial_t \xi_A - i \pi^A T^i_A{}^B \partial_i \xi_B+\eta^A \partial_t \chi_A - i\eta^A T^i_A{}^B \partial_i \chi_B-m(\chi^A\xi_A +\pi^A \eta_A)
\end{align}

where $\pi^A:=i\xi^{\star\,A},~ \eta^A:=i\chi^{\star\,A}$ are the momenta conjugated to $\xi^A,~\chi^A$. The above Lagrangian leads us to the following equations of motion for the fields:
\begin{align}
\dot \xi_A - i T^i_A{}^B\partial_i \xi_B = m \eta_A, \quad \dot \chi_A - i T^i_A{}^B\partial_i \chi_B = m \pi_A
\end{align}

and similar ones for their respective momenta.

%\subsection{Canonical mode decomposition at the Hamiltonian level}
\newpage
In the exact same way as before, we can use the fact that both $\chi$ and $\xi$ satisfy the Klein-Gordon equation to expand the spinors in Fourier plane-waves:

\begin{align}
\xi_A = \int d\Omega_k \left( a_A(k)e^{-ikx} + b_A^\dagger(k)e^{ikx}\right),\quad \chi_A = \int d\Omega_k \left( c_A(k)e^{-ikx} + d_A^\dagger(k)e^{ikx}\right)
\end{align}

where $d\Omega_k \equiv \frac{d^3 k}{(2\pi)^3 2\omega_k}$ as before and the ladder operators $a,~b,~c,~d$ are not all independent. Indeed, using the equations of motion, we have for instance:
\begin{align}
-\sqrt{2} k^{ A A'} c^\dagger_{A'} - mb^{\dagger\,A} = 0, \quad -\sqrt{2} k^{ A A'}  a_A - md^{A'} = 0
\end{align}

and therefore there are only two sets of independent ladder operators. Expanded on the $(a,c)$ basis, the fields become:
\begin{align}
\xi_A &= \int d\Omega_k \left( a_A(k)e^{-ikx} +\frac{\sqrt{2}}{m} k_{AA'} c^{\dagger\,A'}(k) e^{ikx}\right) \nonumber \\ \chi_A &= \int d\Omega_k \left( c_A(k)e^{-ikx} +\frac{\sqrt{2}}{m} k_{AA'} a^{\dagger\,A'}(k) e^{ikx}\right)
\end{align}

Recall now that we can expand our ladder operators on the $(k,k^\star)$ basis:
\begin{align}
a_A(k) = k_A^\star \tilde{a}_k^+ + k_A \tilde{a}^-_k,\quad c_A(k) = k_A^\star \tilde{c}_k^+ + k_A \tilde{c}^-_k
\end{align}
where the tilde is meant to remind us that the modes are not yet canonically normalised. Then, using the equations of motion at the Hamiltonian level and recalling that our spinors $k_A$ and $k_A^\star$ are eigenvectors of the gradient energy, we obtain:
\begin{align}
 \xi^A(x)  &= \int d\Omega_k \Big( (k^{\star A} \tilde{a}^+_k +  k^A \tilde{a}^-_k)  e^{+ikx} \\ \nonumber &+\frac{1}{m}(k^{\star A}(\omega_k-|k|) \tilde{c}^{\dagger\,-}_k- k^A (\omega_k+|k|) \tilde{c}^{\dagger\,+}_k )e^{-ikx}\Big) \\
  \chi^A(x) & = \int d\Omega_k \Big( (k^{\star A} \tilde{c}^+_k +  k^A \tilde{c}^-_k)  e^{+ikx} \\ \nonumber &+\frac{1}{m}(k^{\star A}(\omega_k-|k|) \tilde{a}^{\dagger\,-}_k- k^A (\omega_k+|k|) \tilde{a}^{\dagger\,+}_k )e^{-ikx}\Big)
\end{align}

which is the same mode decomposition as for the Majorana fermion except that here we have two sets of ladder operators. It is worth noticing that under the exchange of ladder operators the fields simply are exchanged as well:
\begin{align}
\tilde{a}^\pm \leftrightarrow \tilde{c}^\pm ~~\Rightarrow~~ \xi \leftrightarrow \chi
\end{align}

Now, using the anti-commutational relations between the fields:
\begin{align}
\left\{\pi_A(\vec x), ~\xi_B(\vec y)\right\} = -i\epsilon_{AB} \delta^3(\vec x-\vec y),\quad \left\{\eta_A(\vec x), ~\chi_B(\vec y)\right\} = -i\epsilon_{AB} \delta^3(\vec x-\vec y)
\end{align}

and all others being zero, we can normalise our ladder operators so as to be canonical:
\begin{align}
\left\{ a^\pm_k, ~ {a}^{\dagger\,\pm}_p\right\} = (2\pi)^3 2\omega_k \delta^3(\vec k -\vec p) = \left\{ c^\pm_k, ~ c^{\dagger\,\pm}_p\right\}
\end{align}

This yields as before the canonical mode decomposition of the spinors:
\begin{align}
 \xi_A(x)  &=  \int d\Omega_k \sqrt{m}\Big( \left( \epsilon^+_A a^+_k + \epsilon^-_A a^-_k\right)  e^{+ikx} +\left( \epsilon^+_A c^{\dagger\,-}_k-\epsilon^-_A  c^{\dagger\,+}_k \right)e^{-ikx}\Big), \\
  \chi_A(x)  &=  \int d\Omega_k \sqrt{m}\Big( \left( \epsilon^+_A c^+_k + \epsilon^-_A c^-_k\right)  e^{+ikx} +\left( \epsilon^+_A a^{\dagger\,-}_k-\epsilon^-_A  a^{\dagger\,+}_k \right)e^{-ikx}\Big)
\end{align}

\subsection{Hamiltonian and CPT}

The total Hamiltonian is given by
\begin{align}
H = \int d^3x \left(  \im \pi^A T^i_A{}^B \partial_i \xi_B+\im \eta^A T^i_A{}^B \partial_i \chi_B+m\chi^A\xi_A+m\pi^A\eta_A\right).
\end{align}
Substituting the mode decomposition one gets for the normal ordered Hamiltonian:
\begin{align}
H=\int d \Omega_k  \omega_k \left( a^{\dagger\,+}_k a_k^+ + a^{\dagger\,-}_k a_k^- +c^{\dagger\,+}_k c_k^+ + c^{\dagger\,-}_k c_k^- \right),
\end{align}

which confirms the interpretation of $a^{\dagger\,\pm}_k, a_k^\pm$ and $c^{\dagger\,\pm}_k, c_k^\pm$ as the creation-annihilation operators of four species of particles of the same energy $\omega_k$.

Let us start considering the operation of parity acting on the ladder operators. As we recover Majorana theory when the two sets of ladder operators coincide we define parity transformations on the Fock space in the following way:
\begin{align}
P^\dagger a^\pm_k P = \varphi_p a^\mp_{-k}, \quad P^\dagger c^\pm_k P = \varphi_p c^\mp_{-k}
\end{align}

The action on the fields is then:
\begin{align}
P^\dagger \xi(t,-\vec{x}) P = \frac{i}{m}  \int d \Omega_k \sqrt{m}\Big( \varphi_p \left(-(\omega_k-|k|) \epsilon^{-}_A a^-_k-(\omega_k+|k|) \epsilon^{+}_A a^+_k \right)e^{+ikx}\\ \nonumber
+ \varphi_p^* \left( -(\omega_k-|k|) \epsilon^{-}_A c^{\dagger\,+}_k + (\omega_k+|k|) \epsilon^{+}_A  c^{\dagger\,-}_k \right)  e^{-ikx}\Big)
\end{align}

and just as in the Majorana case, we require the field to be a representation of parity, leading to:
\begin{align}
\varphi_p = \varphi_p^*, \quad \varphi_p^2=1
\end{align}

Then:
\begin{align}
P^\dagger \xi(t,-\vec{x}) P = \varphi_p \eta(t,\vec x),\quad P^\dagger \chi(t,-\vec{x}) P = \varphi_p \pi(t,\vec x)
\end{align}

Under charge conjugation, the ladder operators transform as:
\begin{align}
C^\dagger a^\pm_k C = \varphi_c c^\pm_{k}
\end{align}
with again $\varphi_c  \in \mathbb{R}$ and $\varphi_c^2=1$. Then:
\begin{align}
C^\dagger \xi C =\varphi_c \chi 
\end{align}

The Lagrangian is then $C$ invariant. Finally, for sake of completeness, time reversal is an anti-linear operator that flips both spin and momentum, thus:
\begin{align}
T^\dagger a^\pm_k T = \varphi_t a^\mp_{-k},\quad T^\dagger c^\pm_k T = \varphi_t c^\mp_{-k}
\end{align}

Then:
\begin{align}
T^\dagger \xi(-t,\vec{x}) T = \varphi_t \eta(t,\vec x),\quad T^\dagger \chi(-t,\vec{x}) T = \varphi_t \pi(t,\vec x)
\end{align}

with $\varphi_t^2=1$. Notice in particular:
\begin{align}
(PT)^\dagger \xi(-t,-\vec{x}) PT = \varphi_t \varphi_p \xi(t,\vec x),\quad (PT)^\dagger \chi(-t,-\vec{x}) PT = \varphi_t \varphi_p \chi(t,\vec x)
\end{align}

Thus, we see explicitly that the theory is $PT$ invariant (regardless of $C$ invariance).

\subsection{Second-order Dirac theory}

As for Majorana fermions considered above, at the level of the path integral we can integrate out the fermionic fields $\xi^\dagger, ~\chi^\dagger$ and obtain a chiral Lagrangian involving only unprimed spinors. From the field equations for the primed spinors we get:
\begin{align}
\xi^{\dagger\,A'} = -\frac{i\sqrt{2}}{m} \partial^{ A'A} \chi_A, \qquad \chi^{\dagger\,A'}= -\frac{i\sqrt{2}}{m} \partial^{ A'A} \xi_A.
\end{align}

Substituting this into the Lagrangian (\ref{dirac-a}) we get:
\begin{align}
{\cal L}_{D} = -\frac{2}{m}\partial_{A'}{}^{ A} \chi_A \partial^{A'B}  \xi_B - m \chi^A\xi_A\label{lagchiral}.
\end{align}
%We now use (\ref{theta-ident}) 

We now use:
\begin{align}
\theta_{A'}^{\mu\, A}  \theta^{\nu\,A'B}= -\frac{1}{2}\epsilon^{AB}\eta^{\mu\nu} + \Sigma^{\mu\nu\,AB}\label{metric_sigma}
\end{align}
where $\Sigma^{\mu\nu\,AB}$ is the self-dual two-form defined in (\ref{sd2f}), to rewrite this Lagrangian as:
\begin{align}
{\cal L}_{D} =-\frac{1}{m} \partial^\mu \chi^A \partial_\mu \xi_A - m \chi^A\xi_A
\end{align}

where we used the antisymmetry of $\Sigma^{\mu\nu\,AB}$ in its space time indices. In the following chapter we will explore more explicitly the theory of second-order Dirac fermions coupled to a gauge field. Notice that the Lagrangian that will be used for a straightforward minimal coupling is (\ref{lagchiral}).

\part{Interacting Field Theory}

\chapter{Electrodynamics}\label{chapED}

\section{First-order electrodynamics}
\subsection{Lagrangian and symmetries}

The Dirac Lagrangian (\ref{Dirac}) is invariant under the following {\it global} symmetry:
$$\Psi\to e^{-i \alpha} \Psi $$
where global means that the transformation parameter is a constant, not a function of spacetime coordinates. This is of course just the ${\rm U}(1)$ symmetry that we have discussed above, see (\ref{u1}). This symmetry can be promoted into a {\it local} symmetry if one introduces the so-called {\it gauge potential}. Thus, we introduce a new field $A_\mu$ which under local {\it gauge transformations} with a gauge parameter $\alpha(x)$ transforms as:
\begin{align}\label{gauge}
A_\mu \to A_\mu + (1/e) ~\partial_\mu \alpha
\end{align}

Now, with this local gauge transformation the original Lagrangian is not invariant anymore. Using Noether's method or alternatively, using the minimal coupling scheme, this non-invariance can be corrected. Let us now introduce the notion of {\it covariant} derivative of a spinor:
\begin{align}
D_\mu \Psi := (\partial_\mu +ie A_\mu) \Psi
\end{align}

It is easy to verify that 
\begin{align} D_\mu \Psi \to e^{-i\alpha} D_\mu \Psi
\end{align}

Thus, the following Lagrangian 
\begin{align}\label{Dirac-local}
{\cal L}_{\rm D} =- i \bar{\Psi} \gamma^\mu D_\mu \Psi - m\bar{\Psi} \Psi
\end{align}
is invariant under local ${ U}(1)$ transformations. When this Lagrangian is supplemented with a Lagrangian describing the dynamics of the gauge field (Maxwell Theory):
\begin{align}
{\cal L}_{\rm M} = -\frac{1}{4} F_{\mu\nu} F^{\mu\nu}, \qquad F_{\mu\nu} := \partial_\mu A_\nu - \partial_\nu A_\mu
\end{align}

one obtains the Lagrangian of (quantum) electrodynamics:
\begin{align}
{\cal L}= {\cal L}_{\rm D} + {\cal L}_{\rm M}
\end{align}

Note that, by construction, it is gauge invariant. Varying it with respect to the gauge field $A_\mu$ one gets the following field equation:
\begin{align}
\partial_\mu F^{\mu\nu} = -e\bar{\Psi} \gamma^\nu \Psi
\end{align}
This is just the Maxwell's (non-trivial) equations $\partial_\mu F^{\mu\nu}=-j^\nu$ with the current being equal to $j^\mu=e\bar{\Psi} \gamma^\mu \Psi$. Quantum field theory based on the above Lagrangian describes the quantum properties of electrons, their anti-particles positrons, as well as the mediators of interactions between them, photons.

\subsection{Two-component form}

We now promote the global ${U}(1)$ symmetry of (\ref{Dirac-L}) into local one at the level of the two-component Dirac Lagrangian (\ref{Dirac-L}). Thus, we introduce as above an ${U}(1)$ gauge field and convert the usual derivative into the covariant ones
\begin{align}
\partial \xi \to D \xi = (\partial - ie A)\xi, \qquad
\partial\chi \to D \chi = (\partial + ie A)\chi
\end{align}
where $D \equiv \theta^{\mu\,AA'}D_\mu$ and $A_\mu$ is the electromagnetic potential. Note that, since the fields $\xi$ and $\chi$ are charged in the opposite way, the expressions for the covariant derivatives on these fields differ by a sign in front of $A_\mu$. The gauge transformation rule for the electromagnetic potential is $A_\mu\to A_\mu+(1/e)~\partial_\mu \alpha$ as before. The Lagrangian becomes
\begin{align}\label{dirac-a}
{\cal L}_{\text{D}}= -i \sqrt{2} \xi^\dagger D \xi-i \sqrt{2} \chi^\dagger D \chi - m\chi\xi -m\xi^\dagger \chi^\dagger
\end{align}
This is the way that the two-component Dirac fermions couple to the electromagnetic potential.

\subsection{${\rm U(1)}$ charge, spin and Hamiltonian}
Let us describe the main quantities of the theory. The $U(1)$ current is given by:

\begin{align}
j^{\mu}= \sqrt{2}e \xi^\dagger \theta^\mu \xi - \sqrt{2}e \chi^\dagger \theta^\mu \chi
\end{align}
and hence, the $U(1)$ charge operator is given by:

\begin{align}
Q\equiv \int d^3x j^0 = e \int d^3x \left(-\xi^\star \xi + \chi^\star \chi\right)
\end{align}
In terms of the ladder operators, it becomes:

\begin{align}
Q= e \int d \Omega_k   \left[\left( a_k^{\dagger\,+} a_k^+ + a_k^{\dagger\,-} a_k^- \right)-\left(c_k^{\dagger\,+} c_k^+ + c_k^{\dagger\,-} c_k^- \right)\right]
\end{align}
which confirms the fact that we are dealing with two particles with opposite electromagnetic charges. One can also compute the spin-current and analyse how it acts on one-particle states created by either of the creation operators in its rest frame. If $a^{\dagger\,+}$ creates a spin up, negatively ($e<0$) charged particle, then Table \ref{zoo} encodes the different species and their quantum numbers.
%\vspace{-10pt}
%\begin{spacing}{1.8}
%\begin{align}
%\begin{array}{|c|c|c|}\hline
%\text{Species} & \text{Charge} & \text{Spin} \\ \hline
%a^{\dagger \,+}&  (-)  & (+)\\ \hline
%a^{\dagger \,-} &  (-)  & (-)\\ \hline
%c^{\dagger \,+} &   (+) & (+) \\ \hline
%c^{\dagger \,-} &  (+)  & (-) \\ \hline
%\end{array}
%\end{align}\end{spacing}

The free Hamiltonian is the same as in the non-interacting theory, we recall:
\begin{align}
H_0 = \int d^3x \left(  i \pi^A T^i_A{}^B \partial_i \xi_B+i \eta^A T^i_A{}^B \partial_i \chi_B+m\chi^A\xi_A+m\pi^A\eta_A\right)
\end{align}
or in terms of the ladder operators:
\begin{align}
H_0=\int d \Omega_k  \omega_k \left( a_k^{\dagger\,+} a_k^+ + a_k^{\dagger\,-} a_k^- +c_k^{\dagger\,+} c_k^+ + c_k^{\dagger\,-} c_k^- \right)
\end{align}
Whereas, the interactions Hamiltonian is given by:

\begin{align}
H_{int} = \int d^3 x \left( ie A_0(\eta^A\chi_A - \pi^A\xi_A) -e T^i_A{}^BA_i (\eta^A\chi_B - \pi^A\xi_B)\right)
\end{align}

\begin{spacing}{1.8}
\begin{table}[h]
\begin{center}
\begin{tabular}{|c|c|c|}\hline
\text{Species} & $Q$ & $S$ \\ \hline
$a^{\dagger \,+}$ &  $(-)$  & $(+)$\\ \hline
$a^{\dagger \,-}$ &  $(-)$  & $(-)$\\ \hline
$c^{\dagger \,+}$ &  $(+)$ & $(+)$ \\ \hline
$c^{\dagger \,-}$ &  $(+)$  & $(-)$ \\ \hline
\end{tabular}\caption{Fermions Zoology}\label{zoo}
\end{center}
\end{table}

\end{spacing}

\section{Second-order Electrodynamics}

\subsection{Lagrangian}
As we did before, at the level of the path integral we can integrate out the fermionic fields $\xi^\dagger$ and $\chi^\dagger$, and obtain a second-order Lagrangian involving only unprimed spinors. From the field equations for the primed spinors we get:
\begin{align}
\xi^{\dagger\,A'} = -\frac{i\sqrt{2}}{m} D^{ A'A} \chi_A, \qquad \chi^{\dagger\,A'}  = -\frac{i\sqrt{2}}{m} D^{ A'A} \xi_A.
\end{align}
Substituting this into the Lagrangian (\ref{dirac-a}) we get:
\begin{align}
{\cal L}_{D} = -\frac{2}{m}D_{A'}{}^{ A}  \chi_A D^{A'B}  \xi_B - m \chi^A\xi_A\label{lagchiral3}.
\end{align}
%We now use (\ref{theta-ident}) 

We now use:
\begin{align}
\theta_{A'}^{\mu\, A}  \theta^{\nu\,A'B}= -\frac{1}{2}\epsilon^{AB}\eta^{\mu\nu} + \Sigma^{\mu\nu\,AB}\label{metric_sigma2}
\end{align}
to rewrite this Lagrangian as:
\begin{align}
{\cal L}_{D} =-\frac{1}{m} D^\mu \chi^A D_\mu \xi_A - m \chi^A\xi_A - \frac{i}{m}e \Sigma^{\mu\nu\,AB}\chi_A  \xi_B F_{\mu\nu},
\end{align}
where we have integrated by parts to get the last term and $F_{\mu\nu}=2\partial_{[\mu} A_{\nu]}$. The last term describes interactions with the gauge field and can be seen to be essentially the spin to electromagnetic potential coupling term of Pauli's phenomenological description of spin. Note, however, that there are also interaction with the electromagnetic field vertices hidden in the first term. We can further simplify this Lagrangian by rescaling the fields. It is clear that in this formalism it is natural to introduce fermionic fields of mass dimension one via $\chi\to\sqrt{m}\chi, \xi\to\sqrt{m}\xi$. In terms of the rescaled fields the Lagrangian takes a particularly simple form:
\begin{align}
{\cal L}_{D} =-D^\mu \chi^A D_\mu \xi_A - m^2 \chi^A\xi_A - ie \Sigma^{\mu\nu\,AB} \chi_A  \xi_B F_{\mu\nu}.
\end{align}
However, for later convenience we will mainly work with 
\begin{align}
{\cal L}_{D} =-2D_{A'}{}^{ A} \chi_A D^{A'B} \xi_B - m^2 \chi^A\xi_A .\label{L-D-chiral}
\end{align}

\subsection{Parity at the Hamiltonian level}
It now arises the question on how parity and time reversal could be implemented in the Lagrangian formalism as they involve the canonically conjugated momenta. To answer this question, two points need to be recalled. First of all, in order to go from the first-order to the second-order formalism, one replaces the (first-order) canonically conjugated fields by their equation of motion. Second, as we previously saw this can be thought of as a canonical transformation between two sets of canonically conjugated variables. For an interacting Dirac field (with canonical pairs $(\chi,\eta),~(\xi,\pi))$, we have the four Dirac equations:
\vspace{-20pt}
\begin{spacing}{1.5}
\begin{align}\begin{array}{ll}
\dot \chi_A -i T^i_A{}^B \partial_i \chi_A -i\sqrt{2} e G_A{}^{A'}A_{A'}{}^B\chi_B&= m\pi_A \\
\dot \eta_A +i T^i_A{}^B \partial_i \eta_B +i\sqrt{2} e A_{AB'}G^{B'B}\eta_B   &= -m\xi_A \\
\dot \xi_A -i T^i_A{}^B \partial_i \xi_A +i\sqrt{2} e G_A{}^{A'}A_{A'}{}^B\xi_B  &= m\eta_A \\
\dot \pi_A +i T^i_A{}^B \partial_i \pi_B -i\sqrt{2} e A_{AB'}G^{B'B}\pi_B   &=- m\chi_A \\
\end{array}\end{align}\end{spacing}

with 
\begin{align}
\sqrt{2}G_A{}^{A'}A_{A'}{}^B = \delta_A{}^{B}A_0+T^i_A{}^BA_i
\end{align}

We define therefore in an analogous way to the Majorana fermions the following canonical transformations:
\begin{align}
\chi_A ~\rightarrow~\sqrt{m}\chi_A, \quad \eta_A ~\rightarrow \frac{1}{\sqrt{m}}\left(N_A - iT^i_A{}^{B}\partial_i\xi_B\right),\quad  N_A = \dot \xi_A  +i\sqrt{2} e G_A{}^{A'}A_{A'}{}^B\xi_B  \\
\xi_A ~\rightarrow~\sqrt{m}\xi_A, \quad \pi_A ~\rightarrow \frac{1}{\sqrt{m}}\left(\Pi_A - iT^i_A{}^{B}\partial_i\chi_B\right),\quad  \Pi_A = \dot \chi_A  -i\sqrt{2} e G_A{}^{A'}A_{A'}{}^B\chi_B 
\end{align}

In terms of the new variables, the Hamiltonian reads:
\begin{align}\begin{split}
\mathcal{H}=N^A\Pi_A + \partial_i\chi^A\partial_i\xi_A &- i\sqrt{2}e G_A{}^{A'}A_{A'}{}^{B}\left( \Pi^A\xi_B-N^A\chi_B \right)\\&-e \sqrt{2} T^{i\,AB}G_A{}^{A'}A_{A'}{}^{C}\left( \chi_C\partial_i\xi_B- \xi_C\partial_i\chi_B\right)
\end{split}\end{align}
For example, for the fields $N$ and $\xi$, it leads to the equations of motion:
\begin{align}
\dot \chi_A &= \Pi_A + i\sqrt{2} e G_A{}^{A'}A_{A'}{}^B\chi_B  \\
-\dot \Pi_A&= -\partial_i^2\chi_A-i\sqrt{2}eG^{BA'}A_{AA'}\Pi_B\nonumber \\&\qquad + 2ie \left( (\partial_i)_A{}^{A'} (A_{A'}{}^{B}\chi_B) + A_{AA'}(\partial_i)^{A'B}\chi_B\right)
\end{align}

with $(\partial_i)^{A'B}= \theta_i^{A'B}\partial_i$. Which leads finally to (using twice the field equation for $\Pi$):
\begin{align}
\square \chi_A+2ie\Big(\partial_A{}^{A'}(A_{A'}{}^{B}\chi_B) + A_{AA'} \partial^{A'B}\chi_B\Big) +2e^2 A_A{}^{A'}A_{A'}{}^B\chi_B = 0
\end{align}

This are the same field equations that follow from the second-order Lagrangian (\ref{lagchiral3}), showing that the above Hamiltonian properly describes QED.

Concerning parity, it is however easier to express the Hamiltonian as a function of the old conjugate fields (they are not anymore the conjugate momenta of the second-order fields). Nonetheless, as we just discussed, in terms of the old variables, the Hamiltonian simply resembles the first-order Hamiltonian with the fields rescaled, we recall:
\begin{align}\begin{split}
H = \int d^3x & \left(  i \pi^A T^i_A{}^B \partial_i \xi_B+i \eta^A T^i_A{}^B \partial_i \chi_B+m\chi^A\xi_A+m\pi^A\eta_A \right.  \\ &\left.+ ie A_0(\eta^A\chi_A - \pi^A\xi_A) -e T^i_A{}^BA_i (\eta^A\chi_B - \pi^A\xi_B)\right).
\end{split}\end{align}

As this is a canonical transformation and the above Hamiltonian is parity invariant, we conclude (trivially from the transformation properties of $A$ and the fermions) that the second-order formulation of QED is also parity invariant\footnote{Equivalently, one could rewrite the Lagrangian using the unprimed spinors and their $\star$-conjugate as was done for the free theory.}. This result about parity generalises straightforwardly to time invariance and charge conjugation.

\chapter{The Standard Model}\label{chapSM}

\section{Standard Model fields and Lagrangian}
\label{sec:SM}

In this chapter we will be working exclusively with two-component fermions. We follow \cite{Dreiner:2008tw}, with some differences in conventions. This chapter is based on \cite{Espin:2013wia,Espin:2015sxa}.

\subsection{Standard Model particles}
\subsubsection*{Fermions}

The SM fermions can be put together in the following table

\bigskip
\begin{center}
\begin{tabular}{c c c c c c}
Two-component fermions & $SU(3)$ & $SU(2)$ & $Y$ & $T_3$ & $Q=T_3+Y$ \\
\hline
$Q_i = \left(\begin{array}{c} u_i \\ d_i \end{array} \right)$ & $\begin{array}{c} 3  \\ 3 \end{array}$ & 2 & $\begin{array}{c} 1/6  \\ 1/6\end{array}$ & $\begin{array}{c} 1/2  \\ -1/2\end{array}$ & $\begin{array}{c} 2/3  \\ -1/3\end{array}$ \\
$\bar{u}_i$ & $\bar{3}$ & 1 & $-2/3$ & 0 & $-2/3$ \\
$\bar{d}_i$ & $\bar{3}$ & 1 & $1/3$ & 0 & $1/3$ \\
\hline
$L_i = \left(\begin{array}{c} \nu_i \\ e_i \end{array} \right)$ & $\begin{array}{c} 1  \\ 1 \end{array}$ & 2 & $\begin{array}{c} -1/2  \\ -1/2\end{array}$ & $\begin{array}{c} 1/2  \\ -1/2\end{array}$ & $\begin{array}{c} 0  \\ -1\end{array}$ \\
$\bar{e}_i$ & 1 & 1 & $1$ & 0 & $1$ \\
$\bar{\nu}_i$ & 1 & 1 & $0$ & 0 & $0$ \\
\hline
\end{tabular}
\end{center}
\bigskip
Where $3$ and $\bar{3}$ denote $SU(3)$ triplets, $2$ denotes an $SU(2)$ doublet and the 1 denotes the singlets. All fermionic fields here are unprimed two-component spinors. Hence, $\bar{u}_i$ denotes another set of fermions independent of $u_i$, whose Hermitian conjugate is denoted $u_i^\dagger$. The first half of the table corresponds to the quarks, whereas the second corresponds to the leptons. The last line is included here so as to complete the neutrino minimal standard model ($\nu$MSM, \cite{Asaka:2005pn}) that allows the presence of Majorana mass terms for the latter and enables different mechanisms that explain the neutrinos mass hierarchy and the baryon asymmetry of the Universe. Notice that as their name indicates, the ${SU}(3)$ triplet fields are a set of three two-component spinor fields that transform into each other under ${SU}(3)$ rotations. For example, if we were to be explicit with the index structure of $u_i$, we should write $u_{i\, A}^\alpha$ where $A=1,2$ is the usual spinor index, and $\alpha=1,2,3$ is the index on which ${SU}(3)$ acts. We only keep the index $i=1,2,3$ that denotes the spinor generation (flavour) as it is the only one that will play an important role in the construction to be carried out below. In total there are 16 two-component spinors for each generation of the Standard Model, plus their Hermitian conjugates.

\subsubsection*{Higgs field}

The Higgs field is the last ingredient necessary to the construction of a sensible theory that accommodates all the representations that we have mentioned above in a gauge-invariant fashion. It is a complex scalar field of ${U}(1)$ hypercharge $Y=1/2$. It is also a weak ${SU}(2)$ doublet, $i.e.$ :

\bigskip
\begin{center}
\begin{tabular}{c c c c c c}
Higgs & $SU(3)$ & $SU(2)$ & $Y$ & $T_3$ & $Q=T_3+Y$ \\
\hline
$\phi = \left(\begin{array}{c} \phi^+ \\ \phi^0 \end{array} \right)$ & 1 & 2 & $\begin{array}{c} 1/2  \\ 1/2\end{array}$ & $\begin{array}{c} 1/2  \\ -1/2\end{array}$ & $\begin{array}{c} 1  \\ 0\end{array}$ \\
\end{tabular}
\end{center}

\bigskip
Being a doublet, it is actually a collection of two complex scalar fields that are denoted by $\phi^+$ and $\phi^0$. We shall denote the weak ${SU}(2)$ index by $a, b, \ldots = 1,2$. Therefore, the Higgs field can be written as $\phi_a$, with $\phi_1=\phi^+$ and $\phi_2=\phi^0$.

\subsection{Fermionic sector of the Standard Model}

Using an index-free notation, the Lagrangian for the fermionic sector of the Standard Model is given by:
\begin{align}\begin{split}
\mathcal{L}_{ferm} = ~& -i\sqrt{2}Q^{\dagger i}{D} {Q_i} - i\sqrt{2}\bar u^{\dagger i}{D} \bar u_i- i\sqrt{2}\bar d^{\dagger i}{D} \bar d_i\\&- i\sqrt{2}L^{\dagger i}{D} {L_i}-i\sqrt{2} \bar e^{\dagger i}{D} {\bar e_i}- i\sqrt{2}\bar \nu^{\dagger i}{D} \bar \nu_i \\
&+ Y_u^{ij}  \phi^T \varepsilon Q_i \bar u_j - Y_d^{ij} \phi^\dagger Q_i \bar d_j +Y_\nu^{ij} \phi^T \varepsilon L_i \bar \nu_j - Y_e^{ij} \phi^\dagger L_i \bar e_j \\
&- (Y_u^\dagger)^{ij} \bar u_i^\dagger Q^\dagger_j\varepsilon  \phi^*   - (Y_d^\dagger)^{ij}\bar d_i ^\dagger Q_j  ^\dagger\phi -(Y_\nu^\dagger)^{ij} \bar \nu_i^\dagger L^\dagger_j\varepsilon  \phi^*   - (Y_e^\dagger)^{ij}\bar e_i ^\dagger L_j  ^\dagger\phi \\
& -\frac{1}{2}M^{ij}_{\bar\nu} \bar\nu_i\bar\nu_j - \frac{1}{2}(M_{\bar\nu}^\dagger)^{ij} \bar\nu_i^\dagger\bar\nu_j^\dagger 
\end{split}\label{Lferm}\end{align}

Here as before $D^{AA'} \equiv \theta^{\mu AA'}D_\mu$, where $D_\mu$ is a covariant derivative that acts on the fermions according to their SM representation. The quantities $Y^{ij}$ are arbitrary complex $3\times 3$ Yukawa matrices.

The above Lagrangian, is the most general that can be written using first-order kinetic terms, overall gauge-invariant and that contains operators of dimension up to four. This allows for the additional Majorana mass terms as mentioned earlier. To get the usual SM, any terms containing $\bar{\nu}$ or its Hermitian conjugate in (\ref{Lferm}) should be removed. As far as gauge-invariance is concerned, it is easier to understand the construction of the mass terms when the ${SU}(2)$ index structure is made explicit, all other implicit indices have straightforward contractions. As we have already mentioned, the Higgs field is a doublet $\phi_a$ with a single ${SU}(2)$ index in the lower position with $a=1,2$. Its transpose is then an object $(\phi^T)^a$. The complex conjugate field $(\phi^*)_a$ still carries a lower position index (for $SU(2)$, we have the well known $2\sim \bar{2}$), while the Hermitian conjugate is $(\phi^\dagger)^a$. Similarly, the quark doublet $Q_a$ has a lower position index. Its Hermitian conjugate is an object $(Q^\dagger)^a$. The quantity $\epsilon\equiv \epsilon_a{}^b$ is the matrix 
\begin{align}
\epsilon_a{}^b=\left(\begin{array}{cc} 0 & 1 \\-1 & 0 \end{array} \right)
\end{align}

Then the object $\phi^T \epsilon Q \equiv (\phi^T)^a \epsilon_a{}^b Q_a$ is invariant under the action of ${SL}(2,\C)$ via $Q\to gQ, \phi\to g\phi$ since $g^T \epsilon g=\epsilon$ and therefore $\epsilon$ can be taken as a metric over the fundamental representations of ${SL}(2,\C)$. In particular, $\phi^T \epsilon Q$ is ${SU}(2)$ invariant as it is a well defined scalar product. This property also applies to the Hermitian conjugate objects. It is then clear that all the mass terms in (\ref{Lferm}) are ${ SU}(2)$ invariant. The ${U}(1)$ and $SU(3)$ invariance is straightforwardly checked using the tables above.

\section{Second-order formulation of the Standard Model}
\label{sec:SOL}

As it was done antecedently for the Majorana and Dirac Lagrangians, we now proceed with the construction of the second-order Lagrangian for the SM.

\subsection{Quark sector}
We start with the quark sector as there is no Majorana mass term in this case. The equations of motion for the unprimed spinors are:
\vspace*{-30pt}
\begin{spacing}{2}
\begin{align}\begin{split}
\begin{array}{ccl}
 Q_i^{\dagger}: & i\sqrt{2}D Q^i =& -\left( \epsilon\phi^*\right) \bar{u}^\dagger_j (Y_u^\dagger)^{ji}- \phi ~\bar{d}^\dagger_j (Y_d^\dagger)^{ji} \\  
\bar u_i^{\dagger}: & i\sqrt{2}D\bar u^i =& - (Y_u^\dagger)^{ij}Q^\dagger_j\left( \epsilon\phi^*\right) \\
\bar{d} _i^\dagger: &i\sqrt{2} D\bar d^i =&- (Y_d^\dagger)^{ij}Q^\dagger_j~\phi \\ 
\end{array}\end{split}
\end{align}
\end{spacing}

Notice that some symmetry structure is appearing in the equations of motion. Indeed, let us combine the components of the Higgs field and of its Hermitian conjugate into the following $2\times 2$ matrix:
\begin{align}
\rho\Phi^\dagger \defeq\left( \epsilon\phi^*, \phi\right ) \equiv  \left( \begin{array}{cc}
(\phi^0)^* & \phi^+ \\ - \phi^- & \phi^0\\
\end{array}\right)
\end{align}

Under the weak $SU(2)$ the matrix $\Phi^\dagger$ transforms as:
\begin{align}
\Phi^\dagger  ~\mapsto~ \Omega\Phi^\dagger
\end{align}
while the field $\rho$ remains invariant. It is clear that $\rho^2= |\phi|^2$ is just the modulus squared of the Higgs field. Furthermore:
\begin{align}
\Phi^\dagger  \Phi = 1
\end{align}

so that $\Phi \in SU(2)$. This will become important in what follows. However, before using this fact, let us make the above equations look more transparent. We define new quark singlets as linear combinations of the old ones:
\begin{align}\label{new-bars}
\bar{u}^i ~\rightarrow~ (Y_u^\dagger)^{ij}\bar{u}_j, \quad \bar{d}^i ~\rightarrow~ (Y_d^\dagger)^{ij}\bar{d}_j
\end{align}

This constant reparametrisation of the fields makes the Yukawa matrices disappear from the last two equations of motion. Having done this, to further symmetrise the system, we combine the new quark singlets into a row
\begin{align}
\bar{Q} _i &\defeq \left( \bar{u}_i ~,~ \bar{d}_i\right)
\end{align}

In terms of the new quark singlets the equations of motion become:
\vspace*{-30pt}
\begin{spacing}{2}
\begin{align}\begin{split}
\begin{array}{ccl}
 Q _i^\dagger: & i\sqrt{2}D Q_i =& -\rho~\Phi^\dagger \left( \bar{Q}^\dagger\Lambda\right)_i\\  
\bar{Q} _i^\dagger: &i\sqrt{2} D\bar Q_i =&- \rho~ Q^\dagger_i \Phi^\dagger\\ 
\end{array}\end{split}
\end{align}
\end{spacing}
which is already much simpler than the previous system of equations. Here we introduced new hermitian Yukawa matrices
\begin{align}
\Lambda_q^{ij} &\defeq Y_q^{ik}(Y_q^\dagger)^{kj},
\end{align}

as well as a new column
\begin{align}\label{Q-Lambda}
\left(\bar{Q}^\dagger \Lambda\right)^i \equiv  \left(\begin{array}{c} \bar{u}^\dagger_j \Lambda_u^{ji} \\ \bar{d}^\dagger_j \Lambda_d^{ji}
\end{array}   \right)
\end{align}

While introducing the new pair of quarks $\bar{Q}_i$ has made the equations look more symmetric, there is no complete symmetry. Indeed, the doublet $Q_i$ transforms under the weak ${SU}(2)$ as before, and so does the Higgs matrix $\Phi^\dagger$ as we have seen, while $\bar{Q}_i$ does not transform, it is simply a rearrangement of the fields into a convenient structure. However, this suggests that we define a new set of ${SU}(2)$-invariant quark variables $\Phi Q_i$
\begin{align}\label{new-Q}
\Phi Q_i \defeq Q_i^{inv}
\end{align}
Notice that, as we mentioned earlier, $\Phi \in SU(2)$ and therefore this is a Higgs-field dependent ${SU}(2)$ gauge rotation of the original quark doublet. As such, after transforming the $SU(2)$ gauge fields accordingly, this transformation can be pulled through the covariant derivative as it is usually done. As we will work out in details below, the new gauge field will be an ${SU}(2)$-invariant object, the transformation that we have carried out effectively corresponds to fixing the gauge. It is similar to the well-known unitary gauge representation of a spontaneously broken gauge theory, therefore, the symmetry principle that allowed for the construction of $SU(2)$ invariant terms in the Lagrangian is still present, it is just rendered implicit. The new set of variables shall be called ``frozen''. Notice that the construction that we are implementing does not rely on this choice of gauge, the latter is merely a convenient choice. Now, keeping in mind this change in the covariant derivative operator we can write the field equations as:
\vspace*{-30pt}
\begin{spacing}{2}
\begin{align}\begin{split}
\begin{array}{ccl}
  Q _i^\dagger: & i\sqrt{2}D Q_i =& -\rho~\left(\bar{Q}^\dagger \Lambda\right)_i \\  
\bar{Q} _i^\dagger: &i\sqrt{2} D\bar Q_i =&- \rho~ Q^{\dagger}_i \\ 
\end{array}\end{split}\label{realquarks}
\end{align}
\end{spacing}

We have dropped the superscript $inv$ from the $Q_i$ as we will be dealing exclusively with these from here on. Notice that the equations become much simpler than in terms of the original variables. Let us now substitute the primed spinors obtained from the above field equations into the Lagrangian (\ref{Lferm}) and obtain the following second-order Lagrangian:
\begin{align}\label{L2-quarks}
\mathcal{L}_{quarks}= -\frac{2}{\rho}D\bar Q^i{D} {Q_i} - \rho \left( \Lambda \bar Q\right)^i Q_i 
\end{align}

where we have introduced a new row
\begin{align}
\left(\Lambda \bar{Q}\right)^i := \left( \Lambda_u^{ij} \bar{u}_j , \Lambda_d^{ij} \bar{d}_j \right)
\end{align}
which is the Hermitian conjugate of (\ref{Q-Lambda}) $\left(\Lambda \bar{Q}\right)^\dagger=\bar{Q}^\dagger \Lambda$. It is not difficult to see that in order to obtain (\ref{L2-quarks}) it is enough to note that half of the kinetic terms cancels the mass terms for the primed spinors, while the other half survives. This is the same phenomenon as before, in the simpler cases of Majorana and Dirac Lagrangians. Then the kinetic term in (\ref{L2-quarks}) is obtained from the first-order kinetic term $Q^{\dagger i} D Q_i$ by substituting the expression for $Q^{\dagger i}$. The mass term in (\ref{L2-quarks}) is easily obtained by combining the mass terms for the unprimed spinors in (\ref{Lferm}), and taking into account the definitions (\ref{new-bars}), (\ref{new-Q}) of the new fermionic variables. The covariant derivative acting on $Q_i$ in (\ref{L2-quarks}) takes into account the field redefinition (\ref{new-Q}). 

The Lagrangian (\ref{L2-quarks}) is much more compact than (\ref{Lferm}) from which it was obtained. However, it is evidently non-polynomial in the Higgs scalar field $\rho$, because of the presence of $1/\rho$ in the kinetic term. In the case of Dirac theory (\ref{L-D-chiral}), we simply needed a constant rescaling of the fields to bring the kinetic term into its canonical form. After doing this, the spinor fields effectively became fields of mass dimension one. The same, or rather an equivalent, procedure can be applied to (\ref{L2-quarks}). However, $\rho$ is now a dynamical field and absorbing it into the fermion fields thus changes the derivative operators acting on both $\bar{Q}_i, Q_i$. Indeed, when going through the derivative operator, we need to include an appropriate transformation of the latter. Otherwise stated $[ \rho^n , \partial ]\sim (\partial \rho)\rho^{n-1}$. Denoting the new Higgs-containing derivative operators by the curly ${\mathcal D}$ we finally write:
\begin{align}\label{L2-quarks*}
\mathcal{L}_{quarks} =  -2\mathcal{D}\bar Q^i\mathcal{D} {Q_i} - \rho^2 \left(\Lambda \bar Q\right)^i Q_i 
\end{align}
where $1/\sqrt{\rho}$ was absorbed into each spinor field. The new covariant derivative $\mathcal{D}$ contains non-polynomial Higgs-quarks interactions as well as the physical $SU(2)$-frozen gauge fields when acting on the unbarred doublet. In order to obtain the physical states of the theory, one should expand (\ref{L2-quarks*}) around the Higgs vacuum expectation value (vev) $\rho\to v+h(x)$, then one gets the free massive quarks with masses being multiples of the eigenvalues of the hermitian Yukawa matrices $\Lambda_q^{ij}$, together with quark interactions with the gauge fields as well as with the Higgs. We will give an example of the simplest interactions below. It is clear that interaction vertices with the Higgs can be of arbitrarily high valency (due to non-polynomiality in $\rho$). 

The field equations (\ref{realquarks}) for the new fermionic fields of mass dimension one read
\begin{align}\label{quarks-RC}
i\sqrt{2}{\mathcal D} Q_i = -\rho~\left(\bar{Q}^\dagger \Lambda\right)_i , \qquad i\sqrt{2} {\mathcal D}\bar Q_i =- \rho~ Q^{\dagger}_i
\end{align}

As in the previous chapters, these are now to be interpreted as the reality conditions, whose linearised versions are to be imposed on the external lines.

\subsection{Leptonic sector without the Majorana mass terms}

We first consider the case where all the Majorana mass terms are absent. The construction then follows exactly the one presented in the previous subsection. Hence, introducing the new barred lepton fields 
\begin{align}\label{new-bars-l}
\bar{\nu}^i ~\rightarrow~ (Y_\nu^\dagger)^{ij}\bar{\nu}_j, \quad \bar{e}^i ~\rightarrow~ (Y_e^\dagger)^{ij}\bar{e}_j
\end{align}

we gather the new fields together in a row
\begin{align}
\bar{L} _i \defeq \left( \bar{\nu}_i ~,~ \bar{e}_i\right)
\end{align}

and further define another row 
\begin{align}
\left(\Lambda \bar{L}\right)^i := \left( \Lambda_\nu^{ij} \bar{\nu}_j , \Lambda_e^{ij} \bar{e}_j \right)
\end{align}

where $\Lambda_l = Y_l Y_l^\dagger$ are the Hermitian Yukawa matrices. We also define the physical ${SU}(2)$-invariant unbarred leptonic doublet $L^{inv}_i = \Phi L_i$. Rewriting everything in terms of these quantities we get the following Lagrangian
\begin{align}\begin{split}
\mathcal{L}_{leptons} =  -i\sqrt{2}L^{\dagger i}{D} {L_i} - i\sqrt{2}\left( D \bar  L^i \right) \left(\bar L^\dagger \Lambda\right)_i - \rho \left( \Lambda \bar L\right)^i L_i -  \rho \, L^{\dagger i} \left( \bar L^\dagger \Lambda\right)_i 
\end{split}\end{align}

The resulting equations for the primed spinors are 
\vspace*{-30pt}
\begin{spacing}{2}
\begin{align}\begin{split}
\begin{array}{ccl}
 L _i^\dagger: & i\sqrt{2}D L_i =& -\rho~\left(\bar{L}^\dagger \Lambda\right)_{i}\\  
\bar{\nu} _i^\dagger: &i\sqrt{2} D\bar L_i =&- \rho~ L^{\dagger}_{i}  \\
\end{array}\end{split}\label{feqs-leptons}
\end{align}
\end{spacing}
And finally, substituting the resulting primed spinors into the Lagrangian we get
\begin{align}\begin{split}
\mathcal{L}_{leptons}=  -\frac{2}{\rho} D \bar L^i D {L_i} - \rho \left(\Lambda \bar L\right)^i L_i
\end{split}\label{L2-leptons}\end{align}
One can now rescale the lepton fields as we did with the quarks to give them mass dimension one and convert the kinetic terms into a canonical form. One obtains a Lagrangian as in (\ref{L2-quarks*}):
\begin{align}\begin{split}
\mathcal{L}_{leptons}=  -{2} \mathcal{D} \bar L^i \mathcal{D} {L_i} - \rho^2 \left(\Lambda \bar L\right)^i L_i
\end{split}\label{L2-leptons2}\end{align}

The sum of (\ref{L2-leptons2}) and (\ref{L2-quarks*}) is the Lagrangian that would describe the second-order SM without the inclusion of Majorana mass terms.

\subsection{Majorana mass terms included}

We now reinstate the Majorana mass terms. This leads to a more complicated analysis and not so simple final result, however, the construction is identical. Performing the same redefinitions of the fermionic variables as was done above, we can write the original Lagrangian in terms of the new spinor fields:
\begin{align}\begin{split}
\mathcal{L}_{leptons} = ~& -i\sqrt{2}L^{\dagger i}{D} {L_i} - i\sqrt{2}\left( D \bar  L^i \right) \left(\bar L^\dagger \Lambda\right)_i - \rho \left( \Lambda \bar L\right)^i L_i -  \rho \, L^{\dagger i} \left( \bar L^\dagger \Lambda\right)_i \\
&-\frac{1}{2}(Y_\nu^\dagger)^{ik}(Y_\nu^\dagger)^{jl}M^{ij}_{\bar\nu}\bar\nu_k\bar\nu_l - \frac{1}{2}Y_\nu^{ki}Y_\nu^{lj}(M_{\bar\nu}^\dagger)^{ij}\bar\nu_k^\dagger\bar\nu^\dagger_l 
\end{split}\end{align}

The structure of the last two terms suggests the following redefinition of the barred neutrino Yukawa-mass matrix
\begin{align}
 (Y_\nu^\dagger)^{ik}(Y_\nu^\dagger)^{jl}M^{ij}_{\bar\nu}~\rightarrow~M^{kl}_{\bar\nu}.
\end{align}
The new mass matrix is still symmetric. The first-order Lagrangian then becomes
\begin{align}\begin{split}
\mathcal{L}_{leptons} =  &-i\sqrt{2}L^{\dagger i}{D} {L_i} - i\sqrt{2}\left( D \bar  L^i \right) \left(\bar L^\dagger \Lambda\right)_i - \rho \left( \Lambda \bar L\right)^i L_i -  \rho \,L^{\dagger i}\left( \bar L^\dagger \Lambda\right)_i  \\
&-\frac{1}{2}M^{ij}_{\bar\nu}\bar\nu_i\bar\nu_j - \frac{1}{2}(M_{\bar\nu}^\dagger)^{ij}\bar\nu_i^\dagger\bar\nu^\dagger_j 
\end{split}\label{Lleptons}\end{align}

The resulting equations of motion for the primed spinors are as follows
\vspace*{-30pt}
\begin{spacing}{2}
\begin{align}\begin{split}
\begin{array}{ccl}
L _i^\dagger: & i\sqrt{2}D L_i =& -\rho~\left(\bar{L}^\dagger \Lambda\right)_{i}\\  
\bar{\nu} _i^\dagger: &i\sqrt{2} D\bar \nu^i =&- \rho~ \nu^{\dagger i} - \bar\nu^{\dagger j} (M_{\bar\nu}^\dagger)_{jk}(\Lambda^{-1})^{ki}\\ 
\bar{e} _i^\dagger: &i\sqrt{2} D\bar e_i =&- \rho~ e^\dagger_i  \\
\end{array}\end{split}\label{realleptons2}
\end{align}
\end{spacing}

We can now solve for the barred primed spinors using the first equation. The solution for $\bar{\nu}^\dagger$ is then substituted into the second equation. The last pair is then solved for the $L^\dagger$ fermions. It is not as easy as before to obtain the new Lagrangian after the solutions are substituted, however simplifications happen. For example, it is easy to note that the first and the last terms in the first line of (\ref{Lleptons}) cancel each other in view of the first equation in (\ref{realleptons2}). Indeed, we can combine these two terms as
\begin{align}
\mathcal{L}_{leptons} \supset L^\dagger_i\left(-i\sqrt{2}{D} {L_i} -  \rho  \left(\bar L^\dagger\Lambda\right)_i \right)
\end{align}

It is clear that this combination gives zero on the first equation in (\ref{realleptons2}). To eliminate the remaining primed spinors we again need just the first field equation that gives us $\bar{L}^\dagger$. Overall, this gives:
\begin{align}\begin{split}
\mathcal{L}_{leptons}= ~& -\frac{2}{\rho} D \bar L^i D {L_i} - \rho \left(\Lambda \bar L\right)^i L_i \\
&-\frac{1}{2}M_{\bar\nu}^{ij}\bar\nu_i\bar\nu_j + \frac{1}{\rho^2}(\Lambda_\nu^{-1}M_{\bar\nu}^\dagger\Lambda_\nu^{-1})^{ij} (D L_i)^\nu (D L_j)^\nu
\end{split}\label{Llep2nd}\end{align}

where $(DL_i)^\nu$ stands for the first $\nu$-component of the doublet $DL_i$. This Lagrangian is more complicated than the previous second-order Lagrangian due to the presence of the Majorana mass. There are two ways of understanding what is going on here. First of all, notice that the last term in (\ref{Llep2nd}) is simply equal to the Majorana mass term on the surface of the reality conditions (\ref{realleptons2}). Moreover, looking at the second field equation in (\ref{realleptons2}) and substituting the latter reality condition (the solution for $\bar{\nu}^\dagger_i$ from the first equation) one gets:
\begin{align}
i\sqrt{2}\left( D\bar{\nu}^i - \frac{1}{\rho} (\Lambda_\nu^{-1}M_{\bar\nu}^\dagger\Lambda_\nu^{-1})^{ij} (D L_j)^\nu\right) = -\rho \, \nu^{\dagger i}
\end{align}

If one expands all terms in this equation around the Higgs vev $\langle \rho\rangle = v$, the terms linear in the fields are
\begin{align}
i\sqrt{2} \, \partial \left( \bar{\nu}^i - \frac{1}{v} (\Lambda_\nu^{-1}M_{\bar\nu}^\dagger\Lambda_\nu^{-1})^{ij} \nu_j \right) = -v \, \nu^{\dagger i}.
\end{align}

This equation suggests that we should introduce a new barred neutrino field
\begin{align}\label{nubar-new}
\bar{\nu}^{new}_i := \bar{\nu}_i - \frac{1}{v} (\Lambda_\nu^{-1}M_{\bar\nu}^\dagger\Lambda_\nu^{-1})_{ij} \nu^j
\end{align}

as it is this field that satisfies reality conditions similar to those for all other two-component fermions present, otherwise stated, this field redefinition diagonalises the system of reality conditions that allow for a particle interpretation. However, rewriting the Lagrangian (\ref{Llep2nd}), as well as the interaction vertices in terms of $\bar{\nu}^{new}_i$, and then proceed with the usual perturbation theory calculations becomes rather cumbersome. It is clear that the result is complicated, as it relies on the field redefinition (\ref{nubar-new}) that in turn relies on the Higgs field assuming its vev. Therefore, the second-order Lagrangian with the Majorana mass terms added is not much simpler than its first-order counterpart. 

If we take the approach that the second-order formalism is more fundamental and we do not wish to accommodate the complicated structure of the Majorana mass terms, we have a very simple SM fermionic Lagrangian:
\begin{align}\label{SML}
\mathcal{L}_{SM,f} =  -2\mathcal{D}\bar Q^i\mathcal{D} {Q_i}-{2} \mathcal{D} \bar L^i \mathcal{D} {L_i} - \rho^2 \left(\Lambda_q \bar Q\right)^i Q_i  - \rho^2 \left(\Lambda_l \bar L\right)^i L_i
\end{align}

This Lagrangian explains neutrino oscillations by giving the neutrinos masses, but as previously stated, does not by itself explain the observed mass hierarchy or the baryon asymmetry of the Universe which relies on the lepton number violation introduced by the Majorana mass terms. 

The inclusion of Majorana masses or modifications of the SM can however be studied in the second-order framework. For example, a toy model that could be used would a model of a Dirac fermion with both Dirac and Majorana mass terms. This area of investigation has not been covered in this thesis, but could represent a future line of study.

\section{Bosonic sector revisited}
\label{sec:bosonic}

In this section we rewrite the bosonic sector of the Standard Model in terms of the same frozen gauge-fields that were used in the covariant derivatives of our fermionic Lagrangian. In doing so, we will have explicit expressions for the interactions between the bosons and the fermions. This approach allows for a more insightful perspective on the Higgs mechanism for spontaneously broken gauge-theories \cite{Vlasov:1987vt,Chernodub:2008rz,Faddeev:2008qc,Masson:2010vx}. However the presence of a vacuum breaking the larger symmetry group into its little group is not necessary. The material covered in this sections appears in \cite{Faddeev:2008qc} with minor differences in conventions, it is nonetheless interesting to spell it out for sake of completeness and in order to derive the interaction vertices for the physical states of the theory.

\subsection{Higgs sector}

We denote the gauge fields associated to the Standard Model group $SU(2)\times U(1)$\footnote{The $SU(3)$ part is omitted as it does not affect the construction that follows.} by $B_\mu$ and $Y_\mu$ respectively and their coupling constants by $g_2, ~g_1$.

First recall the construction of an ${SU}(2)$ matrix element out of the SM Higgs doublet. The Higgs field is an $SU(2)$ doublet with a $Y$-charge of $1/2$, therefore its covariant derivative is given by
\begin{align}
D_\mu \phi = \partial_\mu \phi + i g_2 B_\mu \phi + \frac{i g_1}{2} Y_\mu \phi
\end{align}

where $B_\mu = T^a B_\mu^a$ and $T^a=(1/2)\sigma^a$, with $\sigma^a$ the usual Pauli matrices. The transformation rules for the gauge-connections are of the standard form, so that the ${ SU}(2)$ connection $B_\mu$ transforms as
\begin{align}
B_\mu\to \Omega^\dagger B_\mu \Omega + (1/i g_2) \Omega^\dagger (\partial_\mu \Omega)
\end{align}

and trivially  the ${U}(1)$ connection as 
\begin{align}
Y_\mu\to Y_\mu + (1/g_1) \partial_\mu \xi
\end{align}

We now parametrize this doublet as
\begin{align}
\phi \equiv \rho \chi, \qquad \rho \in \mathbb{R}^+, ~\chi \in \mathbb{C}^2 ~{\rm with}~|\chi|^2 =1.
\end{align}
Using the $SL(2,\mathbb{C})$ metric $\varepsilon$, we can construct
\begin{align}
\Phi \equiv \left(\begin{array}{c} (\varepsilon \chi)^T \\ \chi^\dagger \end{array} \right)\in SU(2)
\end{align}

The transformation properties of the Higgs doublet under the SM group are then translated into the following transformations of the new object
\begin{align}
U(1)~&:~ \Phi ~\mapsto ~ e^{i\xi T^3}\Phi  \\
SU(2)~&:~ \Phi ~\mapsto ~ \Phi\Omega^\dagger
\end{align}
We can then define a covariant derivative operator such that $D_\mu \Phi$ transforms covariantly. This derivative operator is given by:
\begin{align}\label{d-Phi}
D_\mu \Phi := \partial_\mu \Phi - i g_2 \Phi B_\mu  +i g_1Y_\mu T^3 \Phi =  \left(\begin{array}{c} (\varepsilon D_\mu \chi)^T \\ (D_\mu \chi)^\dagger \end{array} \right)
\end{align}
where to obtain the last expression we have used $\epsilon B_\mu = B_\mu^T \epsilon^T$ which can be checked to hold for all 3 generators $T^a$. 

We can now use the object $\Phi$, as well as its covariant derivative (\ref{d-Phi}) to rewrite the Higgs kinetic term $|D_\mu \phi|^2$ as
\begin{align}
|D_\mu \phi |^2 = \left( \partial_\mu \rho \right)^2 + \rho^2 |D_\mu \chi |^2 = \left( \partial_\mu \rho \right)^2 + \frac{\rho^2}{2}{\rm Tr}  |D_\mu \Phi|^2
\end{align}
where we have defined as before
\begin{align}
|\phi|^2 = \rho^2.
\end{align}

Now let us recall the construction of ${SU}(2)$-invariant doublets. In (\ref{new-Q}) we have defined $Q_i^{inv}$ so that $Q_i = \Phi^\dagger Q_i^{inv}$. We then rewrote the Lagrangian in terms of $Q_i^{inv}$ and the gauge-transformed ${SU}(2)$ connection
\begin{align}
W_\mu \defeq \Phi B_\mu \Phi^\dagger + \frac{1}{ig_2}\Phi \partial_\mu \Phi^\dagger = 
\Phi B_\mu \Phi^\dagger - \frac{1}{ig_2}(\partial_\mu \Phi )\Phi^\dagger
\label{newconnection}
\end{align}
This connection is ${SU}(2)$-invariant. It however transforms under the ${U}(1)$ transformations 
\begin{align}
W_\mu ~\mapsto~ e^{i \xi T_3} W_\mu e^{-i\xi T_3}- \frac{1}{g_2} \left(\partial_\mu\xi\right) T_3 \label{u1sm}
\end{align}

Notice importantly that in the ${SU}(2)$-invariant connection $W_\mu$ a part of the quantity $D_\mu \Phi$ appears. Indeed, we have
\begin{align}
i (D_\mu \Phi )\Phi^\dagger = g_2 W_\mu - g_1 Y_\mu T^3
\end{align}

Therefore we have
\begin{align}
{\rm Tr}|D_\mu \Phi|^2 = \frac{1}{2}\left( g_2^2 \left( W_{\mu}^1 W^{1\,\mu} + W_{\mu}^2 W^{2\,\mu} \right) + \left( g_2 W_\mu^3 -g_1 Y_\mu \right)^2\right)\label{DH2}
\end{align}
where we decomposed $W_\mu = W_\mu^a T^a$. These are the ``will be'' mass terms\footnote{They become the mass terms for the gauge-fields only if the scalar field $\rho$ takes a non-zero value in the vacuum.} for the $W,Z$ bosons obtained from the kinetic term for the Higgs. Usually this is a consequence of choosing a vev for the Higgs field and therefore breaking the symmetry. Here, even though the cause is identical (perturbing the scalar field around its vev), we instead defined an ${SU}(2)$ invariant connection $W_\mu$, which appears in the covariant derivative acting on the ${SU}(2)$ invariant doublets. 

We can now define the usual linear combinations
\begin{align}
W^{\pm}_\mu \defeq \frac{1}{\sqrt{2}}\left( W_\mu^1 \mp i W_\mu^2\right),\qquad Z_\mu \defeq \frac{g_2W_\mu^3 - g_1 Y_\mu}{\sqrt{g_1^2+g_2^2}}
\end{align}
where the normalisation is a convention chosen for later convenience. By construction these fields are invariant under the weak $SU(2)$ and transform under $U(1)$ as
\begin{align}
W_\mu^\pm ~\mapsto~ e^{\pm i\xi(x)}W_\mu^\pm, \qquad Z_\mu ~\mapsto~ Z_\mu
\end{align}
These three gauge-fields are identified as the physical $SU(2)$ bosons which one can measure in an experiment. We can further define as usual the Weinberg angle $\theta_W$ so that
\begin{align}
Z_\mu \defeq \cos(\theta_W)W_\mu^3 - \sin(\theta_W)Y_\mu
\end{align}

From this equation one can deduce (we will further motivate this choice later on) that the second linear combination, the photon gauge field, will be given by:
\begin{align}
A_\mu \defeq \sin(\theta_W)W_\mu^3 + \cos(\theta_W)Y_\mu
\end{align}

Notice that the field redefinition from $W^3_\mu, Y_\mu$ to $Z^\mu, A_\mu$ is an $SO(2)\sim U(1)$ transformation. All in all, the Higgs sector Lagrangian can be rewritten in terms of physical quantities as follows:

\begin{align}\begin{split}
\mathcal{L}_{Higgs} &= -|D_\mu \phi|^2 - V\left( |\phi|^2 \right) \\
&=  -\left( \partial_\mu \rho \right)^2- \frac{(g_2\rho)^2}{2} \left( W^+ W^- +\frac{1}{2\cos^2(\theta_W)}Z_\mu Z^\mu\right) - V(\rho^2)\label{HiggsInter}
\end{split}\end{align} 

With this Lagrangian in hands, all that is needed to extract the mass terms is to set the Higgs field on its vev. Interactions can be obtained at cubic and quartic order, expanding everything with $\rho(x)=v+h(x)$. All that was done here is a reformulation of the Higgs sector in terms of the physical ${SU}(2)$-invariant degrees of freedom of the theory. More details on this reparametrisation of the Higgs field can be found in \cite{Faddeev:2008qc}.

\subsection{Yang-Mills sector}

We now perform the same change of variables in the Yang-Mills sector. The physical gauge-fields have already been defined, thus we only need to reconstruct their curvatures. Let us start with the following Lagrangian:
\begin{align}
\mathcal{L}_{YM} = -\frac{1}{8}{\rm Tr}\left(B_{\mu\nu}B^{\mu\nu} \right)- \frac{1}{4}Y_{\mu\nu}Y^{\mu\nu}
\end{align}
where the curvature tensors are defined according to (\ref{faraday}). Since the field redefinition (\ref{newconnection}) is a gauge transformation, we can immediately write
\begin{align}
\mathcal{L}_{YM} = -\frac{1}{8}{\rm Tr}\left(W_{\mu\nu}W^{\mu\nu} \right)- \frac{1}{4}Y_{\mu\nu}Y^{\mu\nu}
\end{align}

It is now convenient to define the following curvature combinations:
\begin{align}
W_{\mu\nu}^\pm \defeq \frac{1}{\sqrt{2}}\left( W_{\mu\nu}^1 \mp i W_{\mu\nu}^2\right) \equiv D_\mu W^\pm_\nu - D_\nu W^\pm_\mu,
\end{align}
where the covariant derivatives are
\begin{align}
D_\mu W^\pm_\nu \equiv \left(\partial_\mu \pm ig_2 W_\mu^3\right)W_\nu^\pm
\end{align}
We then have :
\begin{align}
{\rm Tr}\left(W_{\mu\nu}W^{\mu\nu} \right) = 2W^3_{\mu\nu}W^{3\,\mu\nu}+ 4 W^+_{\mu\nu}W^{-\,\mu\nu}
\end{align}

Recall now that the $W^3, Y$ connections can be expressed in terms of the physical $Z, A$ connections as
\begin{align}
\left(\begin{array}{c}  W^3 \\ Y \end{array} \right) = \left(\begin{array}{cc}  \cos(\theta_W) & \sin(\theta_W) \\ -\sin(\theta_W) & \cos(\theta_W) \end{array} \right)\left(\begin{array}{c}  Z \\ A \end{array} \right)
\end{align}
Hence,
\begin{align}
D_\mu W^\pm_\nu \equiv \left(\partial_\mu \pm ie A_\mu \pm i g_2 \cos(\theta_W)Z_\mu\right)W_\nu^\pm
\end{align}
where the electric charge $e$ is given by
\begin{align}
e \defeq g_2 \sin(\theta_W) = g_1 \cos(\theta_W)
\end{align}

Finally, another expression that we need is
\begin{align}
W^3_{\mu\nu} = \cos(\theta_W) Z_{\mu\nu} +\sin(\theta_W)F_{\mu\nu} + i g_2 \left(W_\mu^+ W_\nu^- - W^-_\mu W^+_\nu\right)
\end{align}

where $F_{\mu\nu}$ and $Z_{\mu\nu}$ are the curvature tensors of the photon and $Z$ boson. This gives the following final expression for the Yang-Mills sector Lagrangian:
\begin{align}\begin{split}
\mathcal{L}_{YM} = &-\frac{1}{4}F_{\mu\nu}F^{\mu\nu}-\frac{1}{4}Z_{\mu\nu}Z^{\mu\nu}-\frac{1}{2}W^+_{\mu\nu}W^{-\,\mu\nu} \\ & +\frac{g_2^2}{2}W_\mu^+W_\nu^-\left( W^{+\,\mu} W^{-\,\nu }-W^{-\,\mu} W^{+\,\nu} \right) \\& -ie \left( F_{\mu\nu} +\cot(\theta_W) Z_{\mu\nu} \right)W^{+\,\mu }W^{-\,\nu}
\end{split}\end{align}

\section{Interactions}
\label{sec:inter}

We have seen how to construct the Higgs and Yang-Mills sector Lagrangians in terms of the ${SU}(2)$-frozen variables. The physical components $W^\pm_\mu, Z_\mu$ of the connection $W_\mu$ are massive fields, with their mass determined by the vev of the scalar field $\rho$. We now explicitly show how the physical ${SU}(2)$ and $U(1)$ gauge fields interact with the physical fermions. The interaction vertices are obviously different in the second-order formulation. We will only look at the quark's vertices as the lepton's are similar.

\subsection{Weak interactions}

We first consider Higgsless interactions arising when the scalar field sits on its vev $\rho(x)=\nu$. Recall that the Lagrangian is given by (\ref{L2-quarks*}), the fermionic fields have canonical mass dimension one, and the covariant derivative contains some of the Higgs field interaction vertices. However, since we assume here that the scalar field sits on its vev, the covariant derivative $\D$ contains just the weak and electromagnetic connections. We have:
\begin{align}
\D_\mu Q_i = \left(\partial_\mu + i g_2 W_\mu + \frac{i g_1}{6} Y_\mu \right)Q_i,\quad \D_\mu \bar Q_i = \left(\partial_\mu + i g_1\mathcal{Q}Y_\mu   \right)\bar Q_i
\end{align}

where $\mathcal{Q}$ is the matrix of electric charges, which is in this case
\begin{align}
\mathcal{Q}\bar Q_i \equiv \left( \begin{array}{cc} -2/3&0\\ 0& 1/3\end{array}\right) \left( \begin{array}{c}\bar u_i\\\bar d_i \end{array}\right) 
\end{align}

The gauge fields that appear in the covariant derivative are the frozen fields, however, they have not yet been decomposed on the physical basis. Moreover, the kinetic term contains the term $\D^\mu \bar{Q}^i \D_\mu Q_i$, where the covariant derivatives acting on $\bar{Q}_i$ and $Q_i$ contain a different set of gauge fields. Nevertheless, as it is usually the case in the literature, we use the same symbol $\D_\mu$ to denote the covariant derivative acting on different representations. Indeed, it is assumed that the fermionic representation on which it acts is known and therefore, the choice of derivative operator is imposed. Furthermore, integration by parts also works as usual, since the application of $\D_\mu$ to $e.g.$ $Q_i$ maps this field into a different representation, namely the complex conjugate representation to the one describing $\bar{Q}_i$, as is clear from the reality conditions (\ref{quarks-RC}).
%
%
%Thus, the usage of the same symbol to denote the covariant derivative is only justified if one always keeps in mind the representation of the gauge group under which the corresponding fermionic field transforms. It is clear that $\bar{Q}_i, Q_i$ transform under two different representations. Moreover,  Keeping this in mind one can integrate in terms such as $\D^\mu \bar{Q}^i \D_\mu Q_i$ by parts without any inconsistencies. 

In terms of the physical field the covariant derivatives are rewritten as:
\begin{align}
 i g_2W_\mu+ \frac{i g_1}{6} Y_\mu = \frac{i g_2}{\sqrt{2}} \left( \begin{array}{cc} 0&W^+_\mu\\ W^-_\mu& 0\end{array}\right)
+ i e \mathcal{Q}A_\mu + \frac{i e}{s_Wc_W}Z_\mu\left( \frac{1}{2}T^3-s_W^2\mathcal{Q}\right)
\end{align}

where, as before, $ e \defeq g_1 \cos(\theta_W)$ and $s_W \equiv \sin(\theta_W),~c_W\equiv \cos(\theta_W)$. Whereas for the barred quarks we have:
\begin{align}
ig_1\mathcal{Q}Y_\mu = i e\mathcal{Q}A_\mu - i e\mathcal{Q} t_WZ_\mu 
\end{align}

where $t_W\equiv \tan(\theta_W)$. Notice that the matrix of electric charges acting on the unbarred quarks is the opposite to that of the barred ones. Hence, the quark fields interact with the electromagnetic field in the usual way. Let us now consider the interactions with the $W$-bosons. Recall, that due to the fact that these weak interactions are off-diagonal, expressing the quarks' free Lagrangian in terms of mass eigenstates brings up quark-mixing interactions. The relevant part of the Lagrangian (\ref{L2-quarks*}) becomes
\begin{align}\label{W-quarks}
-i \sqrt{2} g_2 \left( K^{ij} (\partial^A{}_{A'} \bar{u}_{i A}) W^{+ A'B} d_{j B} + (K^\dagger)^{ij} (\partial^A{}_{A'} \bar{d}_{i A}) W^{-A'B} u_{j B}\right),
\end{align}

where we reinstated the spinor indices for clarity. The unitary matrix $K^{ij}$ is the Cabibbo-Kobayashi-Maskawa (CKM) matrix that produces the aforementioned mixing. Terms in (\ref{W-quarks}) not only give the interactions responsible for the mixing between the generations, thus making the heavier generations unstable, but are also responsible for the $\beta$-decay. As was the case in Electrodynamics, in the SM second-order formalism there is a derivative present in the cubic vertex. Of course, this can be seen to be the standard vertex with no derivative present if one uses the ``reality condition'' (\ref{quarks-RC}) to express the derivative of the barred spinors in terms of the Hermitian conjugates of the unbarred. However, there is no need to introduce the primed spinors, and one can work with the Feynman rules that follow directly from (\ref{W-quarks}). We will explore this further in the simpler case of Electrodynamics.

The second-order formalism also introduces new quartic vertices that are quadratic in the gauge field in a similar way to scalar Electrodynamics. Indeed, we see that such vertices are present for both $A$ and $Z$ fields, but not for $W^\pm$ as they only appear in the unbarred quarks' covariant derivative. However, because the second-order Lagrangian was obtained after integrating out the primed two-component spinors, it is clear that the correlation functions of the unprimed fields are correctly reproduced. We will see how this fact can be proven perturbatively in the case of Electrodynamics.

\subsection{Interactions with the Higgs}

Although the construction of the interactions with the gauge fields is non-standard, something more interesting happens to the interactions of the fermions with the Higgs field. Indeed, due to the non-polynomiality of the Lagrangian, there exists vertices of arbitrarily high valency. In order to see this, consider fluctuations around the vev $\rho=\nu +h(x)$, where $h(x)$ stands for the physical Higgs field. The latter interacts polynomially with the gauge bosons and this can be read off from (\ref{HiggsInter}). The self-interactions of the Higgs are also as usual. As for the fermions, let us again consider only the quark sector; for leptons everything is analogous. 

We recall that in the form of the Lagrangian (\ref{L2-quarks*}) the covariant derivative was defined so that:
\begin{align}
\frac{1}{\sqrt{\rho}}D ~\leftrightarrow~\mathcal{D}\frac{1}{\sqrt{\rho}}.
\end{align}
Therefore:
\begin{align}\label{new-D}
\mathcal{D} = D +\frac{1}{2}\partial \ln \rho.
\end{align}

This logarithmic non-polynomiality suggests that we should parametrise the Higgs field in a different way: 
\begin{align}
\rho \equiv \nu e^{\phi(x)}.
\end{align}

This shifts the non-polynomiality from the covariant derivative to the mass terms, schematically
\begin{align}
\rho^2 (\Lambda \bar Q) Q ~\rightarrow~ m^2 e^{2\phi(x)} \bar Q Q,
\end{align}

where $m$ is the quark mass. At the same time, the covariant derivative is now simple:
\begin{align}
\mathcal{D}Q \equiv \left(D + \frac{1}{2}\partial \phi(x)\right) Q.
\end{align}

The exponential non-polynomiality also enters into the Higgs with gauge fields interaction vertices, as well as in the kinetic term for the Higgs that now becomes:
\begin{align}
(\partial_\mu \rho)^2 = \nu^2 (\partial_\mu \phi)^2 e^{2\phi}.
\end{align}

For practical purposes, one is only interested in terms involving a few external Higgs lines, and therefore, the exponentials can be expanded and the theory truncated. Hence, for calculations of this type it should not really matter which parameterisation of the field is used. However, one expects the theory to be renormalisable (as a resummation of all the vertices into the exponential) only when all valencies are considered.  Renormalisability of these modified theories has not been explored here and is left as a future investigation possibility.

\section{Curvature and covariant derivative conventions}

In this chapter we deal exclusively with unitary groups so that the inverse of a group object is its Hermitian conjugate. Let $A_\mu \defeq A_\mu^s T^s$ be a connection gauge field. $T^s$ ($s=1,\ldots, {\rm{dim}(G)}$) are the generators of the Lie group, which we take to be Hermitian, that satisfy:
\begin{align}
\left[ T^s, T^r \right] =i f^{srt}T^t
\end{align}
A vector $\phi$ in the fundamental representation transforms as:
\begin{align}
\phi~\mapsto~\phi^\Omega \equiv \Omega \phi, \quad \Omega \equiv \exp \left(i  q\xi^s(x)T^s \right),\quad \xi^s(x)\in \mathbb{R}
\end{align}
and $q$ stands for the charge of the field while $\xi^s(x)$ are coordinates that parametrise the transformation. The covariant derivative is constructed as follows
\begin{align}
D_\mu \phi \defeq \left(\partial_\mu +ig q A_\mu \right)\phi .
\end{align}
If we require that this transforms covariantly under the gauge transformations
\begin{align}
D_\mu \phi ~\mapsto ~\Omega D_\mu \phi
\end{align}
we deduce the transformation rule for the connection:
\begin{align}
A_\mu  ~\mapsto~ A_\mu^\Omega \equiv \Omega^\dagger A_\mu \Omega + \frac{1}{igq}\Omega^\dagger \left(\partial_\mu \Omega\right) ,\quad \Omega \in G
\end{align}
where $g$ denotes the coupling constant of the group. The field strength tensor or Yang-Mills curvature tensor is defined as:
\begin{align}
F_{\mu\nu} \defeq \partial_\mu A_\nu - \partial_\nu A_\mu + igq \left[A_\mu, A_\nu  \right]\label{faraday}
\end{align}
It transforms in the adjoint representation of the Lie group:
\begin{align}
F_{\mu\nu} ~\mapsto~ F_{\mu\nu}^\Omega \equiv \Omega^\dagger F_{\mu\nu} \Omega.
\end{align}

\part{Perturbative Calculations}

\chapter{Path-Integral Quantisation}

In this chapter we will derive the Feynman rules for second-order Majorana fermions minimally coupled to a vector field. In the massless limit, this gives the quantum theory of a Weyl fermion coupled to Electrodynamics when one treats the field as being charged. This particular example will be further discussed in the context of anomalies in Chapter \ref{chapterano}. The derivation then generalises straightforwardly to the case of Dirac Electrodynamics, for which we will simply state the rules. In Chapters \ref{chaptertree}, \ref{chapterreno}, we will be dealing exclusively with the latter.

\section{LSZ reduction formula}
The LSZ reduction formula is a mean to construct the appropriate initial and final states for scattering amplitudes starting from the correlation functions that can be calculated in quantum field theory. We will mainly follow the construction that can be found in \cite{Srednicki:2007qs}. Let us start recalling the mode decomposition of the Majorana field as described in the second-order formalism (\ref{xi-mode}):
\begin{align}
\xi_A(x)  &=   \int d\Omega_k \Big( \left( \epsilon^+_A a^+_k + \epsilon^-_A a^-_k\right)  e^{+ikx} +\left( \epsilon^+_A a^{\dagger\,-}_k-\epsilon^-_A  a^{\dagger\,+}_k \right)e^{-ikx}\Big) \\
\partial_t \xi_A(x)  &=   \int d\Omega_k (-i\omega_k) \Big( \left( \epsilon^+_A a^+_k + \epsilon^-_A a^-_k\right)  e^{+ikx}-\left( \epsilon^+_A a^{\dagger\,-}_k-\epsilon^-_A  a^{\dagger\,+}_k \right)e^{-ikx}\Big)
\end{align}

We now introduce a more compact notation as follows:
\begin{align}\label{xicompact}
\xi_A(x)  &=   \int d\Omega_k \Big(u_{k\,A}(s) a_k(s)  e^{+ikx} +  v_{k\,A}(s) a_k^\dagger(s) e^{-ikx}\Big)\\
\partial_t \xi_A(x)  &=   \int d\Omega_k (-i\omega_k) \Big(u_{k\,A}(s) a_k(s)   e^{+ikx}- v_{k\,A}(s) a_k^\dagger(s)e^{-ikx}\Big)
\end{align}

with $u_{k\,A}(s)= ( \epsilon^+_A(k) , \epsilon^-_A(k)),~v_{k\,A}(s)= ( -\epsilon^-_A(k) , \epsilon^+_A(k))$ and summation over the index $s=1,2$ is implicit. Let us look at the different states that can be constructed with this set of ladder operators. Consider a one particle state in the free theory:
\begin{align}
|{\bf p}, s  \rangle =  a_p^\dagger(s) |0\rangle
\end{align}

Also:
\begin{align}
\langle {\bf p}, s  | =   \langle 0|a_p(s)
\end{align}

and these are normalised so that:
\begin{align}
\langle {\bf p}, r |{\bf k}, s  \rangle = 2\omega_p (2\pi)^3 \delta^{(3)}(p-k)\delta_{rs}
\end{align}

%Whereas, the outgoing particle's state is given by:
%\begin{align}
%\langle{\bf p}, s |= \langle 0|  a_p(s) 
%\end{align}

Now, in order to relate physical scattering experiments to mathematical correlation functions calculated using the field theory, we need to express the creation operator as a function of the field. As a first step, the spatial Fourier transforms of (\ref{xicompact}) are:
\begin{align}
\int d^3x ~e^{ipx} \xi_A(x)&= \frac{1}{2\omega_p}\Big(u_{-p\,A}(s) a_{-p}(s)  e^{-2i\omega_p t} +  v_{p\,A}(s) a_p^\dagger(s)\Big) \\
\int d^3x~ e^{ipx} \partial_t\xi_A(x)&= \frac{-i}{2}\Big(u_{-p\,A}(s) a_{-p}(s)  e^{-2i\omega_p t} - v_{p\,A}(s) a_p^\dagger(s)\Big)
\end{align}

so that:
\begin{align}
\int d^3x ~e^{ipx} \overset\leftrightarrow{\partial_t}\xi_A(x)&= i v_{p\,A}(s) a_p^\dagger(s)
\end{align}

with $e^{ipx} \overset\leftrightarrow{\partial_t}\xi_A(x)= e^{ipx} {\partial_t}\xi_A - (\partial_t e^{ipx})\xi_A$. Finally, using $u^A_p(r)v_{p\,A}(s)= -\delta_{rs}$\footnote{Recall $\epsilon^{+\,A}\epsilon^-_A=1$.}, we obtain:
\begin{align}
i\int d^3x~ e^{ipx} \overset\leftrightarrow{\partial_t}\left(u^A_p(s)\xi_A(x)\right)&= a_p^\dagger(s)\label{creas}
\end{align}

Similarly, the annihilation operator can be constructed as (we do not want to take the Hermitian conjugate of the above expression to avoid introducing primed spinors):
\begin{align}
i\int d^3x~ e^{-ipx} \overset\leftrightarrow{\partial_t}\left(v^A_p(s)\xi_A(x)\right)&= a_p(s)\label{annis}
\end{align}

Notice the similarity of the construction to the scalar field case rather than to the first-order fermionic field case. Also note that the sign depends on the convention one chooses for the incoming and outgoing polarisations (whether, say, the incoming particle has polarisation $u_A$ or $u^A$), therefore the overall sign that we will obtain in the LSZ formula will depend on this choice.

For practical purposes, the ladder operators are not considered on their own (as they give an infinite spread to the particles they create), but rather smeared out by some wavepacket that localises the particle both in momentum and position space, $e.g.$ (dropping the spin index):
\begin{align}
a^\dagger_1 \defeq \int d^3k f_1(k) a^\dagger_k, \quad f_1(k) \propto e^{-(\vec{k}-\vec{k_1})^2/(4\sigma^2)}
\end{align}

where here $f_1(k)$ is a Gaussian wavepacket localised around the origin in position space and around $k_1$ with width $\sigma$ in momentum space. In the free theory, the states created by this smeared operator will time evolve in the Schr\"odinger picture; the wavepacket will propagate away from the origin and spread out. Similarly, if we consider a two-particle state, as $t\rightarrow \pm \infty$, the particles will effectively become widely separated. Let us now consider the case of an interacting theory, and more specifically a 2-2 scattering (it generalises straightforwardly). In this case, both the operators and the states evolve with time. We can however assume that we can construct a well defined multi-particle state in the far past (or far future) following the principles we just discussed. Therefore, we assume that our initial state is of the form:
\begin{align}
|i\rangle = \lim_{t\rightarrow - \infty} a_1^\dagger(t) a_2^\dagger(t)|0\rangle, \quad k_1\neq k_2
\end{align}

with $\langle i| i\rangle =1$ and $\langle i|$ is defined using (\ref{annis}). Similarly our final state will be given by:
\begin{align}
|f\rangle = \lim_{t\rightarrow + \infty} a_3^\dagger(t) a_4^\dagger(t)|0\rangle, \quad k_3\neq k_4
\end{align}
 with $\langle f| f\rangle =1$ and $\langle f| $ is defined in a similar way to $\langle i| $. The physical scattering experiment will measure the amplitude $\langle f |i\rangle$, and it is this quantity that needs to be related to the fields present in the theory. In order to obtain the LSZ formula the following trick is used:
\begin{align}\begin{split}
a_i^\dagger(+\infty) - a_i^\dagger(-\infty) &=\int_{-\infty}^{+\infty} dt ~ \partial_t a_i^\dagger(t)  \\
 &=i\int d^3kf_i(k) \int d^4x ~\partial_t \left(e^{ikx} \overset\leftrightarrow{\partial_t}\left(u_k\xi(x)\right) \right)  \\
  &=i\int d^3kf_i(k) \int d^4x ~\left(e^{ikx} (\partial_t ^2+\omega_k^2)\left(u_k\xi(x)\right) \right)  \\
  &=i\int d^3kf_i(k) \int d^4x ~\left(e^{ikx} (-\square+m^2)\left(u_k\xi(x)\right) \right)  \\
\end{split}\end{align}
where in the last line we used the on-shell condition $\omega_k^2 = |\vec{k}|^2+m^2$ and integration by parts after converting $\vec{k}$ into a derivative operator acting on the Fourier mode. Notice that for a free second-order spinor field, we have $(-\square+m^2)\xi(x)=0$ (as this is the free theory equation of motion). However, when interactions are included, this is no longer true. Similarly for the annihilation operators, we have:
\begin{align}\begin{split}
a_i(+\infty) - a_i(-\infty)  &=i\int d^3kf_i(k) \int d^4x ~\left(e^{-ikx} (-\square+m^2)\left(v_k\xi(x)\right) \right)  \\
\end{split}\end{align}

In the following, we will use:
\begin{align}\begin{split}
a_i^\dagger(-\infty) &= a_i^\dagger(+\infty)+ i\int d^3kf_i(k) \int d^4x ~\left(e^{ikx} (-\square+m^2)\left(u_k\xi(x)\right) \right) \\
a_i(+\infty) &= a_i(-\infty)+ i\int d^3kf_i(k) \int d^4x ~\left(e^{-ikx} (-\square+m^2)\left(v_k\xi(x)\right) \right)  \label{linkladder}
\end{split}\end{align}

So that if we look at the scattering amplitude in which we are interested, we have (dropping the wavepacket factors):
\begin{align}\begin{split}
\langle f |i\rangle &= \langle 0| a_3(+\infty)a_4(+\infty)a_1^\dagger(-\infty)a_2^\dagger(-\infty)|0\rangle \\
&= \langle 0|T a_3(+\infty)a_4(+\infty)a_1^\dagger(-\infty)a_2^\dagger(-\infty)|0\rangle \\
&= i^2(-i)^2\int d^4x_1~e^{ik_1x_1}u_{k_1}^A (-\square +m^2) \cdots \\
&\quad \int d^4x_3~e^{-ik_3x_3}v_{k_3}^C (-\square +m^2) \cdots\langle 0|T \xi_A(x_1)\xi_B(x_2)\xi_C(x_3)\xi_D(x_4)|0\rangle 
\end{split}\end{align}
where in the second line we used the fact the the operators are time-ordered in the first line, and in the third line we replaced the ladder operators using (\ref{linkladder}) and that the annihilation operators acting on the vacuum give zero. This generalises easily to $n_i$ incoming and $n_f$ outgoing particles:
\begin{align}\begin{split}
\langle f |i\rangle &= i^{n_i}(-i)^{n_f}\int\prod_{i=1}^{n_i} d^4x_i~e^{ik_ix_i}u_{k_i}^{A_i} (-\square +m^2)\prod_{j=1}^{n_f} d^4 y_j~e^{-ik_jy_j}v_{k_j}^{B_j} (-\square +m^2)\\
&\qquad  \times\langle 0|T \xi_{A_1}(x_1)\cdots \xi_{A_{n_i}}(x_{n_i})\xi_{B_1}(y_1)\cdots \xi_{B_{n_f}}(y_{n_f})|0\rangle \label{LSZ}
\end{split}\end{align}

This equation is known as the Lehmann-Symanzik-Zimmerman (LSZ) reduction formula. It links the correlation functions on the RHS that are computed by means of Feynman diagrams to the physical scattering amplitudes on the LHS. Note that, as usual, the LSZ reduction formula holds provided the vacuum expectation value of the field as well as the matrix element for the creation of a one-particle state from the vacuum satisfy some constraints:
\begin{align}
\langle 0| \xi_A(x) |0\rangle &=0 \\
\langle {\bf p},s| \xi_A(x) |0\rangle &=v_{p\,A}(s)e^{-ipx} \\
\langle 0| \xi_A(x) |{\bf p},s\rangle &=u_{p\,A}(s)e^{ipx} 
\end{align} 
This implies that the Lagrangian ought to be modified to satisfy the quantum theory constraints: this modification is usually encoded in the $Z_i$ factors that renormalise the field, the mass, and other couplings present in the theory.

\section{The path-integral for second-order fermions}

We now develop the tools necessary to compute the correlation functions that appear on the RHS of the LSZ reduction formula (\ref{LSZ}). The main quantity that is used is the path-integral (partition function, or generating functional), which for the free-field theory is given by:
\begin{align}
Z_0[J] \defeq \langle 0|0\rangle_J = \int \mathcal{D}\xi ~e^{i\int d^4x \left( \mathcal{L}_0 + J^A\xi_A\right)}, \quad Z_0[0]=1\label{origpart}
\end{align}

where
\begin{align}
\mathcal{L}_0 = -\frac{1}{2}\xi^A\left( -\square+m^2\right) \xi_A
\end{align}

is the free Lagrangian, $J^A$ is a grassmann-valued source for the spinor field $\xi_A$, and $\mathcal{D}\xi$ is an appropriate functional measure. The path-integral can be rewritten using a Fourier transform of the fields:
\begin{align}
\xi_A(x)= \int \frac{d^4k}{(2\pi)^4}~e^{ikx}\xi_A(k)
\end{align}
The action then becomes:
\begin{align}
S_0 =-\frac{i}{2}\int \frac{d^4k}{(2\pi)^4} ~\xi^A(k) (k^2+m^2)\xi_A(-k)
\end{align}

and the exponent of the path integral:
\begin{align}
S_0[J] = -\frac{i}{2}\int \frac{d^4k}{(2\pi)^4} ~\left(\xi^A(k) (k^2+m^2)\xi_A(-k) + J^A(k)\xi_A(-k)+J^A(-k)\xi_A(k)\right)
\end{align}
This can be rewritten after a constant shift of the variables (which leaves the measure unchanged):
\begin{align}
\xi^A(k) \rightarrow \xi^A(k) - \frac{J^A(k)}{k^2+m^2}
\end{align}
Then:
\begin{align}
S_0[J] = -\frac{i}{2}\int \frac{d^4k}{(2\pi)^4} ~\left(\xi^A(k) (k^2+m^2)\xi_A(-k) - \frac{J^A(k)J_A(-k)}{k^2+m^2}\right)
\end{align}

So that the partition function becomes:
\begin{align}
Z_0[J]= e^{\frac{i}{2}\int \frac{d^4k}{(2\pi)^4} ~ \frac{J^A(k)J_A(-k)}{k^2+m^2}}\equiv e^{\frac{1}{2}\int \frac{d^4k}{(2\pi)^4} ~ {J^A(k)(S_F)_{AB}(k)J^B(-k)}}
\end{align}
where we used $Z_0[0]=1$. In position space it becomes:
\begin{align}
Z_0[J]= e^{\frac{1}{2}\int d^4x d^4y ~ {J^A(x)(S_F)_{AB}(x-y)J^B(y)}}
\end{align}
The quantity $S_F(x-y)$ is the Feynman propagator, it is the Green function of the field equation for the free field:
\begin{align}
(-\square +m^2)S_F(x-y)= i\delta^{(4)}(x-y)
\end{align}
It is given by:
\begin{align}
(S_F)_{AB}(k)= \frac{-i\epsilon_{AB}}{k^2+m^2}, \quad (S_F)_{AB}(x-y) = \int \frac{d^4k}{(2\pi)^4}\frac{-i\epsilon_{AB}}{k^2+m^2-i\varepsilon}e^{ik(x-y)}
\end{align}
where in position space the contour to be chosen is dictated by the $\varepsilon>0$ regularisation. More will be said about the position space propagator in Chapter \ref{chapuni}.

The question that arises now is, how do we relate this partition function to the correlation functions in which we are interested? Let us go back to (\ref{origpart}) and observe that:
\begin{align}
\langle 0| T \xi_A(x_1) \cdots |0\rangle = \left(\frac{1}{i}\frac{\delta}{\delta J^A(x_1)}\right)\cdots \left.Z_0[J]\right|_{J=0}
\end{align}
We have, for example:
\begin{align}
\langle 0| T \xi_A(x)\xi_B(y) |0\rangle =  \left(\frac{1}{i}\frac{\delta}{\delta J^A(x)}\right)\left(\frac{1}{i}\frac{\delta}{\delta J^B(y)}\right)\left.Z_0[J]\right|_{J=0} = (S_F)_{AB}(x-y)
\end{align}

or in momentum space:
\begin{align}
\langle 0| T \xi_A(k)\xi_B(-k) |0\rangle =  \left(\frac{1}{i}\frac{\delta}{\delta J^A(-k)}\right)\left(\frac{1}{i}\frac{\delta}{\delta J^B(k)}\right)\left.Z_0[J]\right|_{J=0} = (S_F)_{AB}(k)
\end{align}
We see that correlation function in the free theory are given by products of propagators. However, we are interested in the correlation functions for the interacting theory. In order to obtain the latter, we define the path-integral for this theory:
\begin{align}
Z[J] \defeq\langle 0|0\rangle_J &= \int \mathcal{D}\xi ~e^{i\int d^4x \left( \mathcal{L}_0 + \mathcal{L}_{int}+ J^A\xi_A\right)}, \quad Z[0]=1\label{origpart2}
\end{align}

where $\mathcal{L}_{int}$ is the interaction part of the Lagrangian, which depends on $\xi$ but which can also depend on other fields ($e.g.$ a vector field, see below). For now, let us suppose that it depends only on the spinor field. We can then rewrite:
\begin{align}\begin{split}
Z[J]  &= \int \mathcal{D}\xi ~e^{i\int d^4x \left( \mathcal{L}_0 + \mathcal{L}_{int}+ J^A\xi_A\right)}\\
&= e^{i\int d^4x \mathcal{L}_{int}[\delta_x]}\int \mathcal{D}\xi ~e^{i\int d^4x \left( \mathcal{L}_0 + J^A\xi_A\right)}\\
&= e^{i\int d^4x \mathcal{L}_{int}[\delta_x]}Z_0[J], \quad Z[0]=1\label{origpart3}
\end{split}\end{align}

where we introduced the notation $\delta_x \equiv \frac{1}{i}{\delta}/{\delta J^A(x)}$. Note that in general $Z[0]\neq 1$, so that the condition has to be imposed by hand. However, for sake of simplicity (it will not affect the following statements), we will assume that it has been done. The generating functional for the interaction theory generates, as its name suggests, the correlation functions of the latter. As before, we have:
\begin{align}
\langle 0| T \xi_A(x_1) \cdots |0\rangle = \left(\frac{1}{i}\frac{\delta}{\delta J^A(x_1)}\right)\cdots \left.Z[J]\right|_{J=0}
\end{align}

Which in this case becomes:
\begin{align}
\langle 0| T \xi_A(x_1) \cdots |0\rangle = \left(\frac{1}{i}\frac{\delta}{\delta J^A(x_1)}\right)\cdots \left.e^{i\int d^4x \mathcal{L}_{int}[\delta_x]}Z_0[J]\right|_{J=0}
\end{align}

As it is well known, these correlation functions are generally calculated in perturbation theory after expanding the exponential containing the interactions Lagrangian in a power series in the (assumed small) coupling constant. Since the interactions are typically higher than quadratic, the two-point functions (or propagators) remain unchanged at tree-level. Similarly, after having specified the interactions, the (tree level) vertices of the theory can be derived. The set of expressions containing the propagators and the vertices form the Feynman rules of the theory. We now derive these for two simple models of second-order fermions.

\section{Feynman rules for Majorana-Weyl theory}\label{WeylMajoRules}

Let us now look more particularly at the case discussed at the beginning of the chapter. We consider a massive Majorana fermion minimally coupled to a vector field. In the massless limit, the now called Weyl fermion can be considered as a charged field and then the vector field is the appropriate gauge field under whose symmetry group the fermion transforms. The Lagrangian is given by:
\begin{align}
\mathcal{L}= -(D_{A'}{}^A\xi_A)(D^{A'B}\xi_B) - \frac{m^2}{2}\xi^A\xi_A
\end{align}
with reality condition:
\begin{align}
\xi^{\dagger\,A'}= -\frac{i \sqrt{2}}{m} D^{ A'A} \xi_A , \quad D^{ A'A}\xi_A = (\partial -ieA)^{ A'A}\xi_A
\end{align}

which leads to the mode decomposition (\ref{xicompact}):
\begin{align}
\xi_A(x)  &=     \int d\Omega_k \Big(u_{k\,A}(s) a_k(s)  e^{+ikx} +  v_{k\,A}(s) a_k^\dagger(s) e^{-ikx}\Big)
\end{align}

The Lagrangian splits into its free part (upon integration by parts):
\begin{align}
\mathcal{L}_0= -\frac{1}{2}\xi^A(-\square +m^2)\xi_A
\end{align}
and interacting part:
\begin{align}
\mathcal{L}_{int} = ie A_{A'}{}^A\left( \xi_A\partial^{A'B}\xi_B - (\partial^{A'B}\xi_B)\xi_A\right) +\frac{e^2}{2}A^2 \xi^A\xi_A
\end{align}
where we chose to write the terms in the brackets in a way that will mimic Dirac theory to be discussed below, and $A^2 \equiv A_{A'}{}^A A_{A}{}^{A'}= A_\mu A^\mu$. We are first of all interested in the cubic vertex in momentum space. After Fourier transforming all the fields (all momenta incoming by convention), we have:
\begin{align}\begin{split}
iS^{(3)}= ie\int \frac{d^4k_1}{(2\pi)^4}\frac{d^4k_2}{(2\pi)^4}\frac{d^4k_3}{(2\pi)^4}(2\pi)^4 \delta^{(4)}(k_1+&k_2+k_3)A_{A'}{}^A(k_1)\xi^B(k_2)\xi^C(k_3)\\ &\times\left[\epsilon_{BA}k_3^{A'}{}_C-\epsilon_{CA}k_2^{A'}{}_B \right]
\end{split}\end{align}

The Feynman rule is related to the three-point correlation function:
\begin{align}\begin{split}
\langle 0 | T A_{A'}{}^A(k_1)\xi^B(k_2)\xi^C(k_3)|0\rangle \equiv &\\ \left(\frac{1}{i}\frac{\delta}{\delta B^{A'}_A(-k_1)}\right)&\left(\frac{1}{i}\frac{\delta}{\delta J_B(-k_2)}\right)\left(\frac{1}{i}\frac{\delta}{\delta J_C(-k_3)}\right)\left.Z[B,J]\right|_{B,J=0}\label{3ptA}
\end{split}\end{align}
where $B_{A'}^A$ is the source current associated to the vector field, and the generating functional has to be expanded to first order. In the following we will assume that the free part of the generating functional corresponding to the vector field is given by:
\begin{align}
Z_0[B]= \exp\left(-\frac{1}{2}\int \frac{d^4q}{(2\pi)^4} B_{A'}{}^{A}(q)D^{A'B'}_{AB}(q)B_{B'}{}^{B}(-q)\right)
\end{align}
with $D(q)$ an appropriate propagator and the extra minus sign in the exponent corresponds to the Bose symmetry of the vector field. Let us introduce some notation. For a generic $N$-particle interaction, the integration measure is:
\begin{align}
\int dK^{(N)}\equiv \int\prod_{i=1}^{N} \frac{d^4k_i}{(2\pi)^4}(2\pi)^4 \delta^{(4)}(\sum_{j=1}^N k_j)  
\end{align}
Furthermore, the functional derivatives will be denoted as:
\begin{align}
\Delta_{A'}^A(k)\equiv \left(\frac{1}{i}\frac{\delta}{\delta B^{A'}_A(k)}\right),\quad \delta^{A}(k)\equiv\left(\frac{1}{i}\frac{\delta}{\delta J_A(k)}\right)
\end{align}

We then have, at order $e$:
\begin{align}
Z[B,J]= ie\int dK^{(3)}\left[\epsilon_{BA}k_3^{A'}{}_C-\epsilon_{CA}k_2^{A'}{}_B \right]\Delta_{A'}^A(-k_1) \delta^{B}(-k_2) \delta^{C}(-k_3)Z_0[B]Z_0[J]
\end{align}

Together with (\ref{3ptA}), we see that we need to expand the exponential containing the vector propagator once, and the exponential containing the fermion propagator twice. This yields an overall factor of $1/(2!)$. Once the functional derivatives have been taken care of, we are left with:
\begin{align}\begin{split}
\langle 0 | T A_{A'}{}^A(k_1)\xi^B(k_2)&\xi^C(k_3)|0\rangle = (2\pi)^4\delta^{(4)}(k_1+k_2+k_3)\\&\times 2ie\left[\epsilon_{FD}k_2^{D'}{}_E-\epsilon_{ED}k_3^{D'}{}_F\right]S^{BE}(k_2)S^{CF}(k_3)D^{AD}_{A'D'}(k_1)
\end{split}\end{align}

Let us now recall, that the correlation functions are linked to the scattering amplitudes through the LSZ reduction formula (\ref{LSZ}). In momentum space, the latter is given by:
\begin{align}\begin{split}
\langle f |i\rangle &= i^{n_i}(-i)^{n_f}\prod_{i=1}^{n_i} u_{k_i}^{A_i} (k_i^2 +m^2)\prod_{i=1}^{n_f}v_{p_j}^{B_j} (p_j^2 +m^2)\\
&\qquad  \times\langle 0|T \xi_{A_1}(k_1)\cdots \xi_{A_{n_i}}(k_{n_i})\xi_{B_1}(-p_1)\cdots \xi_{B_{n_f}}(-p_{n_f})|0\rangle \label{LSZm}
\end{split}\end{align}
and similarly, a formula can be derived for vector fields. Notice that:
\begin{align}
i(p^2+m^2)S_F^{AB}(p) = \epsilon^{AB}
\end{align}
So that, what the LSZ effectively does is to amputate the external propagators from the correlation functions! Therefore, we have schematically (up to signs and before projecting on polarisations):
\begin{align}
\langle f |i\rangle \sim \left.\langle 0|T \xi_{A_1}(k_1)\cdots \xi_{A_{n_i}}(k_{n_i})\xi_{B_1}(-p_1)\cdots \xi_{B_{n_f}}(-p_{n_f})|0\rangle \right|_{amputated}
\end{align}
For the case of the cubic vertex Feynman rule, we are not interested in projecting the correlation function on external polarisations (as this is only done for external states that are on-shell). Moreover, it is usual to define the transition matrix $\mathcal{T}$:
\begin{align}
\langle f |i\rangle \equiv  i (2\pi)^4\delta^{(4)}(K_{in}-K_{out})\mathcal{T}
\end{align}
%where on the RHS ``1'' stands for the free-theory contribution as is generally disregarded.
So that the overall conservation of momentum is factored out. All in all, the off-shell cubic vertex Feynman rules is given by (all particles incoming):
\begin{align}
2ie\left[\epsilon_{CA}k_2^{A'}{}_B-\epsilon_{BA}k_3^{A'}{}_C\right]
\end{align}

In a similar way, the Feynman rule for the quartic vertex can be derived, one obtains:
\begin{align}
2ie^2\epsilon^{A'B'}\epsilon_{AB}\epsilon_{CD}
\end{align}

\begin{figure}[H]
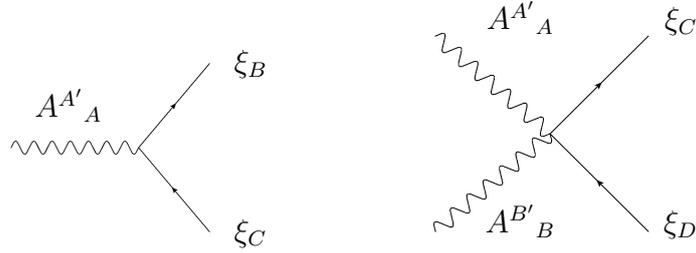
\begin{center}
\input{cubicmajo.pdf_tex} \quad \input{quarticmajo.pdf_tex}
\caption{Cubic and quartic vertices for the Majorana-Vector interactions}\label{vertmajo}
\end{center}\end{figure}

We can now summarise the Feynman rules for this theory as it is usually done in any quantum field theory textbook.
\begin{itemize}
\item Draw all amputated connected diagrams at a given order in the coupling constant (topologically inequivalent).
\item Each internal line corresponds to a propagator: \begin{align}
S_F^{AB} = \frac{-i\epsilon^{AB}}{p^2+m^2}
\end{align}
\item Enforce momentum conservation at each vertex. The latter are either cubic or quartic (Fig.\ref{vertmajo}) and are given by:
\begin{align}
2ie\left[\epsilon_{CA}k_2^{A'}{}_B-\epsilon_{BA}k_3^{A'}{}_C\right],\quad 2ie^2\epsilon^{A'B'}\epsilon_{AB}\epsilon_{CD}
\end{align}
respectively, with all momenta incoming.
\item The rules for the signs are as follows: we write spinor arrows at each vertex as we previously drew charge arrows. In this case, an outgoing arrow denotes the spinor that sits to the left in the vertex interaction and an incoming arrow denotes the spinor sitting to the right. The sign in the momentum is positive if the spinor arrow and the momentum flow arrow are antiparallel. The indices in the propagator and in the quartic vertex correspond to the order that ``climbs up'' the spinor arrow.
\item For loops, an extra minus sign arises as in the usual first-order formalism, and the momentum running into each of them has to be integrated over. 
\item External lines are contracted with polarisation spinors\footnote{Notice that in the LSZ formula (\ref{LSZm}), there is an extra $(-1)^{n_f}$ coming from the outgoing particles. Later on, we will develop an index free notation where the contraction of the polarisation spinor of the outgoing particles is reversed compared to the incoming particles. This will cancel this overall sign.}
\begin{align}
{\rm incoming: }~~ u_k(s), \quad {\rm outgoing: } ~~v_k(s)
\end{align}
\item Symmetry factors need to be accounted for.
\end{itemize}
Using these rules, it is possible to carry on with perturbation theory in this formalism, but before doing so, we will give the same rules for the case of Dirac Electrodynamics.

%\vspace*{-20pt}

\section{Feynman rules for Dirac theory}\label{DiracRules}

The Lagrangian for second-order Quantum Electrodynamics is given by (we simply consider the fermionic part):
\begin{align}
{\cal L} =-{2} D_{A'}{}^{ A} \chi_A  D^{A'B} \xi_B - m^2 \chi^A\xi_A\label{lagchiral2PI}
\end{align}
with 
\begin{align}
D_\mu \xi = (\partial_\mu - i e A_\mu)\xi, \qquad
 D_\mu \chi = (\partial_\mu + i eA_\mu)\chi,
\end{align}
where we included the electromagnetic coupling $|e|\ll 1$, and the fields transform under the $U(1)$ symmetry group as:
\begin{align}
\delta\xi = +ie\alpha \xi, \quad \delta\chi = -ie\alpha \chi
\end{align}

The mode decomposition that follows is given by:
\begin{align}
\xi_A(x)  &=     \int d\Omega_k \Big(u_{k\,A}(s) a_k(s)  e^{+ikx} +  v_{k\,A}(s) c_k^\dagger(s) e^{-ikx}\Big)\\
\chi_A(x)  &=     \int d\Omega_k \Big(u_{k\,A}(s) c_k(s)  e^{+ikx} +  v_{k\,A}(s) a_k^\dagger(s) e^{-ikx}\Big)
\end{align}
where summation over the index $s=1,2$ is implicit. Recalling Table \ref{zoo}, $a^\dagger$ creates electrons, while $c^\dagger$ creates positrons.
Being not Hermitian, the theory is supplemented with reality conditions:
\begin{align}
\xi^{\dagger\,A'} = -\frac{i\sqrt{2}}{m}  D^{ A'A}\chi_A, \qquad \chi^{\dagger\,A'} = -\frac{i\sqrt{2}}{m}  D^{ A'A}\xi_A. \label{realD}
\end{align}

The Lagrangian can be expanded so that:
\begin{align}
{\cal L}= \mathcal{L}_0 + \mathcal{L}_{int}
\end{align}

with
\begin{align}
 \mathcal{L}_0 = -\partial^\mu \chi^A \partial_\mu \xi_A - m^2 \chi^A\xi_A, 
\end{align}
and
\begin{align}
 \mathcal{L}_{int} = 2ieA^{AA'}\left( \chi_A (\partial_{A'}{}^{B}\xi_B) +  (\partial_{A'}{}^{B}\chi_B ) \xi_A \right) - e^2 A^B{}_{B'}A^{B'}{}_{B}\chi^A\xi_A
\end{align}
Because there are now two distinct fermionic fields, both the propagator and vertices are oriented. Using the same method as above, one arrives to the following rules for the propagator:
\begin{align}
\langle 0| T\{\xi_A(p)\chi_B(-p)\}|0 \rangle \equiv S_F(p)_{AB} = \frac{-i}{p^2+m^2}\epsilon_{AB}
\end{align}
where, the field $\xi_A$ sits at the end of the directed line. Similarly, taking all our particles to be incoming, the vertices are (cubic and quartic resp.): 
\begin{align}
2ie\left[\epsilon_{CA}k_2^{A'}{}_B+\epsilon_{BA}k_3^{A'}{}_C\right],\quad 
 -2i e^2 \epsilon^{A'B'}\epsilon_{AB}\epsilon_{CD}
\end{align}
\vspace*{-20pt}
\begin{figure}[H]\begin{center}
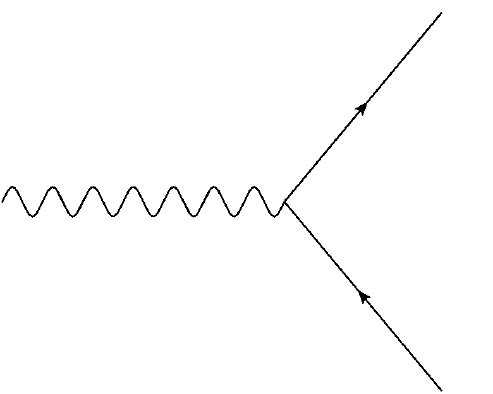 \quad 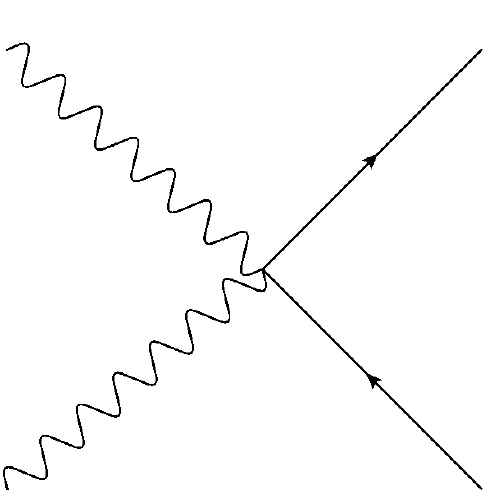
\caption{Cubic and quartic vertices for Dirac Electrodynamics interactions}\label{vertdira}
\end{center}\end{figure}

As for the external polarisations to be used, let us for example consider an incoming electron. Its state is given by:
\begin{align}
|{\bf p}, s ,e \rangle =  a_p^\dagger(s) |0\rangle
\end{align}

We need to consider the following Wick contraction:
\begin{align}
  \acontraction{}{\xi_A(x)}{}{|{\bf p}, s,e\rangle} \xi_A(x)|{\bf p}, s,e\rangle  =e^{\im px} u_A(s,p)|0\rangle
\end{align}

Similarly an incoming positron will be given by:
\begin{align}
|{\bf p}, s ,-e\rangle = c_p^\dagger(s) |0\rangle
\end{align}

and then:
\begin{align}
  \acontraction{}{\chi_A(x)}{}{|{\bf p}, s,-e\rangle} \chi_A(x)|{\bf p}, s,-e\rangle  =e^{\im px}  u_A(s,p)|0\rangle
\end{align}
As for the outgoing particles, an outgoing electron state is given by:
\begin{align}
\langle{\bf p}, s ,e|= \langle 0|  a_p(s) 
\end{align}
So that the Wick contraction to consider is:
\begin{align}
  \acontraction{}{\langle{\bf p}, s ,e|}{}{\chi_A(x)} \langle{\bf p}, s,e |\chi_A(x) =\langle 0| e^{-\im px} v_A(s)
\end{align}

Finally, an outgoing positron:
\begin{align}
\langle{\bf p}, s,-e |= \langle 0|  c_p(s) 
\end{align}
Hence:
\begin{align}
  \acontraction{}{\langle{\bf p},s ,-e|}{}{\xi_A(x)} \langle{\bf p}, s ,-e|\xi_A(x) =\langle 0| e^{-\im px}  v_A(s)
\end{align}

We see that electrons and positrons are described by the same polarisation spinor. This is due to the charge symmetry invariance of the theory. Finally, let us summarise the Feynman rules:

\begin{itemize}
\item Draw all amputated connected diagrams at a given order in the coupling constant (topologically inequivalent).
\item Each, oriented, internal line corresponds to a propagator: \begin{align}
S_F^{AB} = \frac{-i\epsilon^{AB}}{p^2+m^2}
\end{align}
The field $\xi_A$ sits at the end of the directed line.
\item Enforce momentum conservation at each vertex. The latter are either cubic or quartic (Fig.\ref{vertmajo}) and are given by:
\begin{align}
2ie\left[\epsilon_{CA}k_2^{A'}{}_B+\epsilon_{BA}k_3^{A'}{}_C\right],\quad -2ie^2\epsilon^{A'B'}\epsilon_{AB}\epsilon_{CD}
\end{align}
respectively, with all momenta incoming. The spinor field $\xi$ has an incoming directed line.
\item For loops, an extra minus sign arises as in the usual first-order formalism, and the momentum running into each of them as to be integrated over. Notice that due to the orientation of the lines, there is in general more than one loop orientation.
\item External lines are contracted with polarisation spinors (electrons and positrons share the same polarisation spinors):
\begin{align}
{\rm incoming: }~~ u_k(s), \quad {\rm outgoing: } ~~v_k(s)
\end{align}
\item Symmetry factors need to be accounted for.
\end{itemize}

We can now proceed with some basic Perturbation Theory calculations. There, we will see that these Feynman rules (although sufficient) can be improved to simplify calculations.

\chapter{Tree-level Processes}\label{chaptertree}

We start by computing some of the most typical QED amplitudes. If not otherwise stated, we will be dealing exclusively with Dirac Electrodynamics from now on. In this chapter we will be quite explicit in deriving the amplitudes in order to get acquainted with the two-component spinor formalism.

\section{On-shell formalism for the three-valent vertex}\label{onshellsect}
Before carrying out any calculation, it is interesting to construct the Berends-Giele currents \cite{Berends:1987me} for the cubic vertex that allow for simpler calculations later on. This exercise also serves as an introduction to the research area of ``Scattering Amplitudes'' using a spinor-helicity formalism.
Our three-valent vertex with incoming (fermions) momenta $k_1$ and $k_2$ is:
\begin{align}
\mathcal{V}_3(k_1,k_2)_{ABC}{}^{C'} =  2\im e \left[k_{1A}{}^{C'}\epsilon_{BC} + k_{2B}{}^{C'}\epsilon_{AC} \right]
\end{align}
with momentum conservation imposed. When computing scattering amplitudes, this vertex will be projected on polarisation spinors for incoming fermions. We therefore compute the following ``on-shell'' amplitudes:
\begin{align}
\mathcal{M}(h_1,h_2)_C{}^{C'} = \epsilon^A(k,h_1)\epsilon^B(p,h_2)\mathcal{V}_3(k_1,k_2)_{ABC}{}^{C'}
\end{align}

where we split the previously used spinors $u_A$ and $v_A$ into their components $\epsilon_A \equiv \epsilon^\pm_A$. In order to do so, recall that a massive momentum admits the following spinor decomposition:
\begin{align}
k^{AA'}= K^{A}K^{A'} - \frac{m^2 p^{A}p^{A'}}{2(Kp)[Kp]}
\end{align}
where both $K$ and $p$ are null spinors. In this formula, $p$ is a reference spinor from which any physical amplitude can not depend and which can be chosen freely. For clarity we also recall:
\begin{align}
\epsilon^-_A(k)= \frac{2^{1/4}K_A}{\sqrt{m}}, \quad \epsilon^+_A(k) = \frac{\sqrt{m}p_A}{2^{1/4}(Kp)}
\end{align}

Let us then compute the partially on-shell amplitudes (currents) for two incoming particles with momentum $k_1$ and $k_2$ respectively:
\begin{align}
\mathcal{M}(-,-)_C{}^{C'}&= -\sqrt{2}iem \left(  \frac{p_1^{C'}K_{2C}}{[1p_1]}+ \frac{p_2^{C'}K_{1C}}{[2p_2]}\right) \\
\mathcal{M}(+,+)_C{}^{C'}&= -\sqrt{2}iem \left(   \frac{p_{1C}K_{2}^{C'}}{(1p_1)}+ \frac{p_{2C}K_1^{C'}}{(2p_2)}\right) \\
\mathcal{M}(-,+)_C{}^{C'}&= -2ie\left(K_{1C}K_{2}^{C'} + \frac{m^2}{2}\frac{ p_1^{C'}p_{2C}}{[1p_1](2p_2)} \right) \\
\mathcal{M}(+,-)_C{}^{C'}&= -2ie\left(K_{1}^{C'}K_{2C} + \frac{m^2}{2}\frac{ p_{1C}p_{2}^{C'}}{(1p_1)[2p_2]}\right)
\end{align}

where $(1p_1) \defeq K_1^A p_{1A}$ and similarly $[1p_1] \defeq K_{1A'} p_{1}^{A'}$. Notice that for real momenta, we have the following identity:
\begin{align}
\mathcal{M}(h_1,h_2)_C{}^{C'} &= (-1)^{(h_1+h_2)}\epsilon_{CD}\epsilon^{C'D'}\mathcal{M^*}(-h_1,-h_2)_{D'}{}^{D}\label{complexamp}
\end{align}
with $h_i = \pm 1/2$. From the above formulas, the computation of different diagrams for external fermions becomes much simpler, specially at tree level, as one simply needs to contract them with internal propagators. We will not consider here the amplitudes for on-shell photon and fermions as for those cases we will have to consider quartic interactions to be treated later on.
It is worth noticing here that when dealing with massless incoming/outgoing fermions, there are only two currents that contribute to the process. In this formalism one immediately sees that only fermions with opposite helicity (if both incoming or outgoing) contribute, with:
\begin{align}
\mathcal{M}(-,+)_C{}^{C'}&~\stackrel{m=0}{=}~ -2ieK_{1C}K_{2}^{C'}\label{m0amp1} \\
\mathcal{M}(+,-)_C{}^{C'}&~\stackrel{m=0}{=}~ -2ieK_{1}^{C'}K_{2C}\label{m0amp2} 
\end{align}

Lastly, if we consider one outgoing particle or both of them outgoing, there is a change in sign in the three-valent vertex. Moreover, a positive helicity outgoing particle will be represented by $\epsilon^-$ as we previously saw. Accordingly, there will be either a global sign change in the on-shell amplitudes if both particles are outgoing, or a relative sign if there is one incoming particle and one outgoing, as well as a change of helicity state. For example, for two outgoing particles we have:
\begin{align}
\mathcal{M}^{out}(-,+) = - \mathcal{M}^{in}(+,-)
\end{align}

\section{Sum rules: spin averaged probabilities}

When specific helicity combinations are not of interest, one is led to consider unpolarised cross-sections. As it is done in the usual first-order formalism, we develop now similar tools for an efficient computation of averaged probabilities. When we sum (or average) over photon polarisation states, one can make use of a Ward identity to obtain:
\begin{align}
\sum_{\rm pol.}\epsilon_\mu\epsilon^*_\nu ~\rightarrow ~ \eta_{\mu\nu} 
\end{align}

In our case, this will become:
\begin{align}
\sum_{\rm pol.}\epsilon_{AA'}\epsilon^*_{BB'} ~\rightarrow ~ -\epsilon_{AB}\epsilon_{A'B'}  
\end{align}

As for the fermions, we need to compute:
\begin{align}
\epsilon_A^+\epsilon_{A'}^{*+} + \epsilon_A^-\epsilon_{A'}^{*-} 
\end{align}

Using (\ref{massivemomentum}) and (\ref{polarisations}), we obtain:
\begin{align}
\epsilon_A^+\epsilon_{A'}^{*+}(k) + \epsilon_A^-\epsilon_{A'}^{*-}(k) = \sqrt{\frac{2}{m^2}}k_{AA'}
\end{align}

\section{Unpolarised processes: $e^-\mu^- \rightarrow e^-\mu^-$ scattering}

\begin{figure}[H]
	\begin{center}
	 	\includegraphics[width=0.3\linewidth]{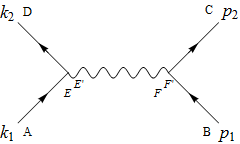}
	 	\caption{$e^-\mu^- \rightarrow e^-\mu^-$}\label{emuemu}
	\end{center}
\end{figure}

We now have everything we need to work with our formalism. Hence, we start considering the simplest QED process: electron-muon scattering at tree level in the limit $m_e \ll m_\mu$, Fig.\ref{emuemu}.

Let us first compute the amputated amplitude $\mathcal{M}_{ABCD}$ for an incoming electron with momentum $k_1$ scattered off an incoming muon with momentum $p_1$. We have:
\begin{align}\begin{split}
\mathcal{M}_{ABCD}&= 4\frac{(ie)^2(-i)}{q^2}\left(k_{1A}{}^{E'}\epsilon_{DE}-k_{2D}{}^{E'}\epsilon_{AE}\right) \epsilon^{EF}\epsilon_{E'F'}\left(p_{1B}{}^{F'}\epsilon_{CF}-p_{2C}{}^{F'}\epsilon_{BF}\right)\\ &= \frac{4ie^2}{q^2}\left[\left(k_1\cdot p_1\right)_{AB}\epsilon_{CD} +\left(k_2 \cdot p_1\right)_{DB}\epsilon_{AC} \right. \\ & \hspace*{3cm} \qquad\left.-\left(k_1\cdot  p_2\right)_{AC}\epsilon_{BD} -\left(k_2\cdot  p_2\right)_{DC}\epsilon_{AB}   \right]
\end{split}\end{align}

where we defined:
\begin{align}
(k\cdot p)_{AB} \defeq k_A{}^{C'}p_{BC'}
\end{align}

and $q^2=(k_1-k_2)^2=(p_1-p_2)^2 = t$. The complex conjugate amplitude is simply obtained after replacing every unprimed spinor by a primed one and vice versa, so that (taking into account the extra minus sign from the imaginary unit):
\begin{align}\begin{split}
\mathcal{M}^*_{A'B'C'D'}&= \frac{4ie^2}{q^2}\left[\left(k_1\cdot  p_1\right)_{A'B'}\epsilon_{C'D'} +\left(k_2\cdot  p_1\right)_{D'B'}\epsilon_{A'C'}  \right. \\ & \hspace*{3cm} \qquad\left. -\left(k_1\cdot  p_2\right)_{A'C'}\epsilon_{B'D'} -\left(k_2\cdot  p_2\right)_{C'D'}\epsilon_{A'B'}   \right]\label{ampprime}
\end{split}\end{align}

and we defined:
\begin{align}
(k\cdot  p)_{A'B'} \defeq k_{A'C}p_{B}{}^{C}
\end{align}

Now, one generally needs to compute an unpolarised cross-section. In order to do so, we project our amplitude on external polarisation states, average over incoming particles' spin and sum over outgoing particles'. In the end, we have:
\begin{align}\begin{split}
\overline{|\mathcal{M}|^2} &= \frac{1}{4}\sum_{\rm pol.} \mathcal{M}_{ABCD}\mathcal{M}^*_{A'B'C'D'}\epsilon^A\epsilon^{*A'}(k_1)\epsilon^B\epsilon^{*B'}(p_1)\epsilon^C\epsilon^{*C'}(p_2)\epsilon^D\epsilon^{*D'}(k_2) \\ &=\frac{1}{4} \mathcal{M}_{ABCD}\mathcal{M}^*_{A'B'C'D'}k_1^{AA'}p_1^{BB'}p_2^{CC'}k_2^{DD'}\frac{2}{m_e^2}\frac{2}{m_\mu^2}
\end{split}\end{align}

Let us consider the following quantity:
\begin{align}
\mathcal{M}^*_{A'B'C'D'}k_1^{AA'}p_1^{BB'}p_2^{CC'}k_2^{DD'}\frac{2}{m_e^2}\frac{2}{m_\mu^2}
\end{align}

Either using (\ref{ampprime}) and
\begin{align}
k^{AA'}k_{A'}{}^{B}= -\frac{m^2}{2}\epsilon^{AB}
\end{align}
or similarly noticing:
\begin{align}
\frac{2}{m_k^2}k^{AA'}p^{BB'} \mathcal{V}^*_3(k,p)_{A'B'C'}{}^C = -  \mathcal{V}_3(k,p)_{ABC'}{}^C
\end{align}

It is easy to derive the following equality\footnote{In Section \ref{realityampli} we derive the general formula for an arbitrarily high number of external particles in order to prove unitarity.}
\begin{align}
\mathcal{M}^*_{A'B'C'D'}k_1^{AA'}p_1^{BB'}p_2^{CC'}k_2^{DD'}\frac{2}{m_e^2}\frac{2}{m_\mu^2} = -\mathcal{M}^{ABCD}
\end{align}

So that:
\begin{align}\begin{split}
\overline{|\mathcal{M}|^2}  &=- \frac{1}{4} \mathcal{M}_{ABCD}\mathcal{M}^{ABCD}
\end{split}\end{align}

In the above formula, we only need to compute three different expressions. Consider four momenta $k,~p,~q,~l$ describing massive particles. We have:\begin{spacing}{1.5}
\begin{align}
(k\cdot  p)_{AB}\epsilon_{CD} \left\lbrace \begin{array}{ll} (k\cdot  p)^{AB}\epsilon^{CD} &= m_k^2 m_p^2 \\ (k \cdot q)^{AC}\epsilon^{BD} &= -\frac{1}{2}m_k^2 (p\cdot q) \\ (q\cdot l)^{CD}\epsilon^{AB} &= (k\cdot p)(q\cdot l) \end{array} \right.
\end{align}
\end{spacing}

where
\begin{align}
k\cdot p = k^{A}{}_{A'}p^{A'}{}_{A}= k_\mu p^\mu
\end{align}

Using this and neglecting terms proportional to the electron mass, we obtain:
\begin{align}
\overline{|\mathcal{M}|^2}  &=\frac{8e^4}{q^4}\Big[(k_1\cdot p_1)(k_2\cdot p_2) + (k_1\cdot p_2)(k_2 \cdot p_1) + m_\mu^2 (k_1\cdot k_2) \Big]
\end{align}

which is the well know squared amplitude for the unpolarised process.

\section{Helicity structure: muon pair production $e^-e^+ \rightarrow \mu^-\mu^+$}

It is important to understand what happens physically in scattering processes and, even though unpolarised cross-sections are easier to compute, they often do not provide any insight about what is really going on. One could instead consider individual physical processes, that is, different helicity structures for a given process. Once this has been done, one is still free to sum over all helicity channels to recover an averaged cross-section.
\begin{figure}[H]
	\begin{center}
	 	\includegraphics[width=0.17\linewidth]{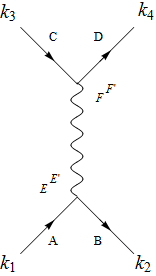}
	 	\caption{$e^-e^+ \rightarrow \mu^-\mu^+$}\label{eemumu}
	\end{center}
\end{figure}
We will consider the high-energy behaviour of the muon pair production from an electron-positron annihilation (Fig.\ref{eemumu}), $i.e.$ its massless limit. The latter makes the computation much easier and at the same time is insightful toward exhibiting the simplicity of our formalism in the helicity basis.

Using the results of (\ref{onshellsect}), we only have to consider four amplitudes. We need to connect (\ref{m0amp1}) and (\ref{m0amp2}) for, on one side, the electron/positron pair and, on the other side, the muon/antimuon pair, with a photon propagator in between. Moreover, from (\ref{complexamp}), we have:
\begin{align}
\mathcal{M}(+,-)_C{}^{C'} &= \epsilon_{CD}\epsilon^{C'D'}\mathcal{M^*}(-,+)_{D'}{}^{D}
\end{align}

so that there are effectively only two distinct amplitudes. These are:
\begin{align}
\mathcal{M}(+,- ; +,-) &= \mathcal{M}_e^{in}(+,-)_E{}^{E'} ~\frac{-i }{q^2}\epsilon^{EF}\epsilon_{E'F'}~\mathcal{M}^{out}_\mu(+,-)_{F}{}^{F'} \\
\mathcal{M}(+,- ; -,+) &= \mathcal{M}^{in}_e(+,-)_E{}^{E'} ~\frac{-i }{q^2}\epsilon^{EF}\epsilon_{E'F'}~ \mathcal{M}^{out}_\mu(-,+)_{F}{}^{F'}
\end{align}

We can relate them to the amplitudes with opposite helicities as follows:
\begin{align}
\mathcal{M}(+,- ; +,-) =- \mathcal{M}^*(-,+ ; -,+)\\
\mathcal{M}(+,- ; -,+) =- \mathcal{M}^*(-,+ ; +,-)
\end{align}

Let us then compute these amplitudes. We label the particles with their momentum $k_i$ and use the abbreviation $k_i^A k_{jA} \defeq (ij)$ and $k_{iA'} k_{j}^{A'} \defeq [ij]$. We then have:
\begin{align}
\mathcal{M}(+,- ; -,+) =\frac{4ie^2}{q^2}(13)[24]\\
\mathcal{M}(-,+ ; +,-) =\frac{4ie^2}{q^2}[13](24)\\
\mathcal{M}(+,- ; +,-) =\frac{4ie^2}{q^2}(14)[23]\\
 \mathcal{M}(-,+ ; -,+)= \frac{4ie^2}{q^2}[14](23)
\end{align}

where we used $(ij)^* = -[ij] $. Finally, using
\begin{align}
(ij)[ij] = (k_i\cdot k_j)
\end{align}

we obtain:
\begin{align}
|\mathcal{M}(+,- ; +,-)|^2 =\frac{16e^4}{q^4}(k_1\cdot k_4)(k_2\cdot k_3)= |\mathcal{M}(-,+ ; -,+)|^2\\
|\mathcal{M}(+,- ; -,+)|^2 =\frac{16e^4}{q^4}(k_1\cdot k_3)(k_2\cdot k_4) = |\mathcal{M}(-,+ ; +,-) |^2
\end{align}

And trivially:
\begin{align}
\overline{|\mathcal{M}|^2} = \frac{8e^4}{q^4}\Big[ (k_1\cdot k_3)(k_2\cdot k_4) + (k_1\cdot k_4)(k_2\cdot k_3)  \Big]
\end{align}

We can qualitatively analyse the physics of the process. Consider the above process in the centre-of-mass frame of the incoming particles and assume without loss of generality that the particle flies along the positive direction of the z-axis (the antiparticle flies along the negative direction). There are only two distinct amplitudes, as for an incoming pair of particle and antiparticle with opposite polarisations\footnote{By polarisation, we mean helicity (information carried by the polarisation spinors), which has to be distinguished from the spin (eigenvalue of the spin operator defined in some reference frame). The helicity is the projection of the spin onto the direction of the momentum.}, say +/- respectively, both particles will have spin up along the z-axis, summing up to a spin 1 state. After they decay and their product creates the other pair of particles, the latter will either have opposite helicities to the original pair, or carry the same helicities as their predecessors. In any case they could not have the same helicity, as it would correspond to a spin 0 state. Another way to see this is to recall that the cubic vertex vanishes (in the massless limit) for particles with identical helicities. Therefore, one can physically only distinguish those two different states if one does not know about the initial state of the incoming particles.

\section{Compton scattering $e^-\gamma \rightarrow e^-\gamma$ and the quartic vertex}\label{seccompton}
\begin{figure}[H]  \centering\begin{center}
\includegraphics[width=0.5\linewidth]{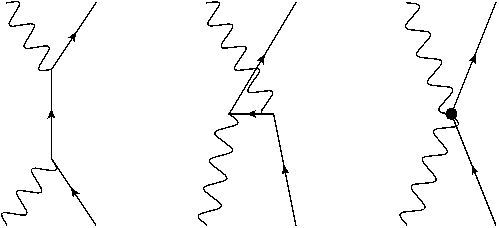}\caption{The three tree-level diagrams entering the Compton scattering calculation.}\label{compton1}
\end{center}\end{figure}

We consider now Compton-scattering, a very well known process, but more specifically in our case, the first tree-level process in which the new quartic vertex comes into play. We label as before the particles by their momenta $k_i$ and their spinor indices. The process is described by three diagrams (Fig.\ref{compton1}): the s-channel diagram with the momentum flowing in the internal propagator is given by $(k_1+k_2)=q=(k_3+k_4)$ and $q^2=s$; the u-channel with $(k_2-k_3)=p=(k_4-k_1)$ and $p^2=u$; and the new quartic vertex.

The three amputated amplitudes are given by:
\begin{align}\begin{split}
\mathcal{M}_1^{A'}{}_{A}{}_{B}{}^{C'}{}_{C}{}_{D}&= 2 i e\left(-k_{4D}{}^{C'}\epsilon_{FC}+q_{F}{}^{C'}\epsilon_{DC}\right)\frac{-i\epsilon^{FE}}{s+m^2}2 i e\left(k_{2B}{}^{A'}\epsilon_{EA}-q_{E}{}^{A'}\epsilon_{BA}\right) \\ 
&=-\frac{4e^2i}{s+m^2}\left( -\epsilon_{AC}k_{2B}{}^{A'}k_{4D}{}^{C'}+\epsilon_{AB}q_{C}{}^{A'}k_{4D}{}^{C'}  \right. \\ & \hspace*{3cm} \qquad\left.-\epsilon_{CD}q_{A}{}^{C'}k_{2B}{}^{A'}+\frac{1}{2}q^2\epsilon^{A'C'}\epsilon_{AB}\epsilon_{CD}\right)   \end{split}\\
\begin{split}\mathcal{M}_2^{A'}{}_{A}{}_{B}{}^{C'}{}_{C}{}_{D}&= 2 i e\left(-k_{4D}{}^{A'}\epsilon_{FA}+p_{F}{}^{A'}\epsilon_{DA}\right)\frac{-i\epsilon^{FE}}{u+m^2}2 i e\left(k_{2B}{}^{C'}\epsilon_{EC}-p_{E}{}^{C'}\epsilon_{BC}\right) \\ 
&=-\frac{4e^2i}{u+m^2}\left( +\epsilon_{AC}k_{2B}{}^{C'}k_{4D}{}^{A'}-\epsilon_{BC}p_{A}{}^{C'}k_{4D}{}^{A'} \right. \\ & \hspace*{3cm} \qquad\left.-\epsilon_{AD}p_{C}{}^{A'}k_{2B}{}^{C'}+\frac{1}{2}p^2\epsilon^{A'C'}\epsilon_{BC}\epsilon_{AD}\right)   \end{split}\\
\begin{split}\mathcal{M}_3^{A'}{}_{A}{}_{B}{}^{C'}{}_{C}{}_{D}&=-2 i e^2\epsilon^{A'C'}\epsilon_{AC}\epsilon_{DB}= 2 i e^2\epsilon^{A'C'}\epsilon_{AC}\epsilon_{BD}  \end{split}
\end{align}

Let us make a few comments. First of all, notice that in the propagators the Levi-Civita symbol has to be contracted from the outgoing fermion onto the incoming fermion (this is the usual ``climb up the fermions arrows'' rule). If in the Dirac formalism, this extra sign does not matter, it is simply because consistency between all diagrams is sufficient to ensure the right sign. However, here, relative signs matter as the channel amplitudes have interferences with the quartic vertex, which does not have any propagator. Instead, when stating the four-valent amplitude, one has to take into account that the spinor indices of the fermions also have to be placed in the same order (from the outgoing state inwards).\\

Second, while the quartic vertex describes the ``identity'' amplitude, we see how each of the eight terms entering the channel amplitudes describe all possible ways of mixing spinor indices among the particles. Indeed, if one is to ``scatter'' the spinor index of the incoming fermion with the incoming photon, one is considering an s-channel process for which the indices $A$ and $B$, on one side, $C$ and $D$ on the other side, mix. Indices belonging to the same spinor representation mix naturally with a Levi-Civita symbol, however, if one is to mix an unprimed index with a primed index, one understands naturally the appearance of the momenta in this formalism. Similary, if one wishes to mix, say A and D indices, one is considering a u-channel subprocess. Finally, the extra factor of two in the last term of each channel amplitude is understood as a symmetry factor of the consequent subprocess.
Last but not least, crossing symmetry is already apparent at this level of the computation. Indeed, under the exchange of the two photons indices\footnote{It is understood that by ``photon index $A$'', what is meant is ``the pair of photon indices $AA'$''.} ($A$ and $C$) as well as $q$ and $p$, the two channel amplitudes are mapped into each other. This symmetry will greatly shorten the computation of the amplitude squared since as a consequence one only needs to compute half of the terms.\\

We now proceed with the computation. As before, the sum over all polarisations is carried out and we have:
\begin{align}
\sum_{pol.} ~\rightarrow~\frac{2}{m^2}\epsilon^{AE}\epsilon^{CF}\epsilon_{A'E'}\epsilon_{C'F'}k_2^{BB'}k_4^{DD'}
\end{align}

The dual amplitudes are this time not equal to the original ones as the fermion in the propagator is off-shell (recall that the sum over polarisations amounts to contracting the complex amplitude with the Dirac operator in momentum space). However, it is possible to regroup the amplitudes in the physical scattering channels. Indeed we can simplify the calculation if we look more carefully at the four-valent vertex. Recall:
\begin{align}
\begin{split}\mathcal{M}_3^{A'}{}_{A}{}_{B}{}^{C'}{}_{C}{}_{D}&=-2 i e^2\epsilon^{A'C'}\epsilon_{AC}\epsilon_{DB}= 2 i e^2\epsilon^{A'C'}\epsilon_{AC}\epsilon_{BD}  \end{split}
\end{align}

Using the Shouten identity
\begin{align}
\epsilon_{AB}\epsilon_{CD}=\epsilon_{AC}\epsilon_{BD}-\epsilon_{AD}\epsilon_{BC}
\end{align}

We can rewrite the amplitude as:
\begin{align}
\begin{split}\mathcal{M}_3^{A'}{}_{A}{}_{B}{}^{C'}{}_{C}{}_{D}&= -4 i e^2\epsilon^{A'C'}\left(  -\frac{1}{2}\frac{s+m^2}{s+m^2}\epsilon_{AB}\epsilon_{CD}-\frac{1}{2}\frac{u+m^2}{u+m^2}\epsilon_{AD}\epsilon_{BC}\right)  \end{split}
\end{align}

Then, one can split this amplitude within the $s-$ and $u-$channel amplitudes to have:
\begin{align}\begin{split}
\mathcal{M}_s^{A'}{}_{A}{}_{B}{}^{C'}{}_{C}{}_{D}&=-\frac{4e^2i}{s+m^2}\left( -\epsilon_{AC}k_{2B}{}^{A'}k_{4D}{}^{C'}+\epsilon_{AB}q_{C}{}^{A'}k_{4D}{}^{C'}\right. \\ & \hspace*{3cm} \qquad\left.-\epsilon_{CD}q_{A}{}^{C'}k_{2B}{}^{A'}-\frac{1}{2}m^2\epsilon^{A'C'}\epsilon_{AB}\epsilon_{CD}\right)   \end{split}
\end{align}\begin{align}
\begin{split}\mathcal{M}_u^{A'}{}_{A}{}_{B}{}^{C'}{}_{C}{}_{D}&=-\frac{4e^2i}{u+m^2}\left( +\epsilon_{AC}k_{2B}{}^{C'}k_{4D}{}^{A'}-\epsilon_{BC}p_{A}{}^{C'}k_{4D}{}^{A'}\right. \\ & \hspace*{3cm} \qquad\left. -\epsilon_{AD}p_{C}{}^{A'}k_{2B}{}^{C'}-\frac{1}{2}m^2\epsilon^{A'C'}\epsilon_{BC}\epsilon_{AD}\right)   \end{split} \\
\mathcal{M}&=\mathcal{M}_s + \mathcal{M}_u
\end{align}

With this trick the dual amplitudes become equal to minus the bare amplitudes! The four valent vertex comes to help us to relate the complex amplitudes to the original ones through the Dirac equation\footnote{See Section \ref{realityampli} and Appendix \ref{ApxRules}.}. 

%We then have:
%\begin{align}
%\overline{|\mathcal{M}|^2} = -\frac{1}{4}\mathcal{M}\cdot \mathcal{M}
%\end{align}

Let us now carry on with our computation. The spin averaged squared amplitude will be given by:
\begin{align}
\overline{|\mathcal{M}|^2} = \frac{1}{4}\sum_{ij}\mathcal{M}_i^{A'}{}_{A}{}_{B}{}^{C'}{}_{C}{}_{D}\tilde{\mathcal{M}}_{jA'}{}^{AB}{}_{C'}{}^{C}{}_{D} =-\frac{1}{4}\mathcal{M}\cdot \mathcal{M}
\end{align}
%Thanks to crossing symmetry, there are only four quantities to compute:
%\begin{align}
%|\mathcal{M}_1|^2,\quad |\mathcal{M}_3|^2,\quad \mathcal{M}_1\tilde{\mathcal{M}_2},\quad \mathcal{M}_1\tilde{\mathcal{M}_3}
%\end{align}

Thanks to crossing symmetry, we only need to compute two squared amplitudes:
\begin{align}
|\mathcal{M}_s|^2, \quad \mathcal{M}_s\cdot \mathcal{M}_u
\end{align}

The task of computing Compton scattering became at least as simple as in the usual Dirac formalism, likely easier as one needs only to plug together the Feynman rules and contract spinors. In computing these amplitudes, we will express the momenta dot products in terms of the Mandelstam variables here defined:
\begin{align}
s&= (k_1+k_2)^2 = 2 (k_1\cdot k_2) -m^2 = 2(k_3\cdot k_4)-m^2 \\
t&= (k_1-k_3)^2 = -2 (k_1\cdot k_3) = -2(k_2\cdot k_4)-2m^2 \\
u&= (k_2-k_3)^2 = -2 (k_2\cdot k_3) -m^2 = -2(k_1\cdot k_4)-m^2 \\
0&=s+t+u + 2m^2
\end{align}
We then have the following identities:
\begin{align}
(k_2\cdot q) &= \frac{s-m^2}{2}= (k_4\cdot q)\\
(k_2\cdot p) &= \frac{u-m^2}{2}= (k_4\cdot p)\\
(k_2\cdot k_4) &= \frac{s+u}{2}\\
(q\cdot p) &= -m^2
\end{align}
And finally, we need:
\begin{align}
\Sigma^{\mu\nu\,AB}\Sigma^{\alpha\beta}{}_{AB} = \frac{1}{2}\left(  \eta^{\mu\alpha}\eta^{\nu\beta}- \eta^{\mu\beta}\eta^{\nu\alpha} - i \epsilon^{\mu\nu\alpha\beta} \right)
\end{align}
So that we have the following identity:
\begin{align}
k^{AA'}p_A{}^{B'}p_{A'}{}^{B}q_{BB'} = (k\cdot p)(q\cdot p) - \frac{1}{2}p^2 (k\cdot q)
\end{align}

We obtain:
\begin{spacing}{1.5}
\begin{align}
|\mathcal{M}_s|^2 &= -8 e^4 \left[  \frac{(s+u)}{(s+m^2)}-\frac{(s-m^2)^2}{(s+m^2)^2}  \right] \\
|\mathcal{M}_u|^2 &= -8 e^4 \left[  \frac{(s+u)}{(u+m^2)}-\frac{(u-m^2)^2}{(u+m^2)^2}  \right]= |\mathcal{M}_s|^2(s\leftrightarrow u)\\
\mathcal{M}_s\mathcal{M}_u&= -16e^4\frac{m^2}{(s+m^2)(u+m^2)}\left[m^2-\frac{1}{2}(s+u)\right]=\mathcal{M}_u\mathcal{M}_s
\end{align}
\end{spacing}

So that:
\begin{align}\begin{split}
\overline{|\mathcal{M}|^2} &= -2e^4\left[ \left( \frac{u+m^2}{s+m^2} +\frac{s+m^2}{u+m^2} \right) + 4m^2\left( \frac{1}{s+m^2}+\frac{1}{u+m^2} \right)\right. \\ & \hspace*{3cm} \qquad\left.- 4m^4\left( \frac{1}{s+m^2}+\frac{1}{u+m^2} \right)^2 \right]\end{split}
\end{align}

As we will see in Section \ref{realityampli}, the four vertex can be excluded of any tree level calculation if we define a set of rules to be followed. This is very similar to what is done in Yang-Mills theory to reconstruct all tree-level diagrams from a basis of three-valent vertices. Firstly, define the reduced channel amplitudes:
\begin{align}
\mathfrak{M}_{s_i} \defeq (s_i+m^2)\mathcal{M}_{s_i}
\end{align}
where $m$ is the mass of the fermion in the channel. Then, the amputated amplitude for a two-fermions-two-photons process with momenta $k_i$ is:
\begin{align}\begin{split}
\mathcal{M}(s_1,s_2, \{k_i\})&= \frac{\mathfrak{M}_{s_1}(s_1,\{k_i\})}{(s_1+m^2)}+\frac{\mathfrak{M}_{s_2}(s_2,\{k_i\})}{(s_2+m^2)}+\mathcal{V}_4\\ &=  \frac{\mathfrak{M}_{s_1}(s_1=-m^2,\{k_i\})}{(s_1+m^2)}+\frac{\mathfrak{M}_{s_2}(s_2=-m^2,\{k_i\})}{(s_2+m^2)}\end{split} \label{rule1}
\end{align}
This fixes the rules for the cases in which the four-valent vertex appears as a tree. In the following we will see how this rule also works for loops involving the four-vertex; as long as the amplitudes describes a physical process.

\chapter{Renormalisation}\label{chapterreno}

After having explored a few basic tree-level processes, it is important to check that both formalisms (first- and second-order) are equivalent at loop level. Indeed, QED has been experimentally shown to be the most accurate theory of Nature; it would be a huge blow to the new formalism if it were not able to predict the same results as its first-order counterpart. In this chapter we explore the simplest loop processes, for which an analytical calculation is both tractable and pedagogical. In order to check full consistency, higher loops calculations would be needed, however this is not covered in this thesis and is left as a future possible line of investigation. For further explanations about the Physics behind these processes, we refer the reader to \cite{Srednicki:2007qs,Peskin:1995ev}.

\section{Using dimensional regularisation}

We use dimensional regularisation, which is a natural choice for gauge theories, to deal with the divergent one-loop integrals. However, one must be careful when using soldering form identities in this scheme. Indeed, even though its algebraic properties are retained, many identities only hold in four dimensions (see \cite{Dreiner:2008tw} Appendix B.2). We shortly summarise here the main identities in $D\neq 4$. As we mentioned, the algebraic equation:
\begin{align}
\theta^\mu_{AA'}\theta^{\nu\,A'}_B &=-\frac{1}{2}\eta^{\mu\nu}\epsilon_{AB}+\Sigma^{\mu\nu}_{AB}
\end{align}

is still valid in dimensional regularisation. The trace identity is then
\begin{align}
\theta^\mu_{AA'}\theta^{\nu\,AA'} &=-{\eta^{\mu\nu}},\quad \epsilon_{AB}\epsilon^{AB} =2
\end{align}

where the internal spinor space is two-dimensional even in our regularisation scheme. Similarly, another identity that follows from the algebra is:
\begin{align}
\theta^\mu_{AA'}\theta_{\mu\,B}^{A'} &=-\frac{D}{2}\epsilon_{AB}
\end{align}

Now, any identity that involves the Levi-Civita tensor cannot be valid in $D\neq 4$ dimensions, for example:
\begin{align}
\Sigma^{\mu\nu}_{AB}\Sigma^{\alpha\beta\, AB} = \frac{1}{2}\left( \eta^{\mu\alpha}\eta^{\nu\beta} -\eta^{\mu\beta}\eta^{\nu\alpha}-i\epsilon^{\mu\nu\alpha\beta}\right)
\end{align}
does not hold any longer. However, for any symmetric tensor $T_{(\mu\alpha)}$, the identity:
\begin{align}
\Sigma^{\mu\nu}_{AB}\Sigma^{\alpha\beta\, AB}T_{(\mu\alpha)} = \frac{1}{2}\left( \eta^{\mu\alpha}\eta^{\nu\beta} -\eta^{\mu\beta}\eta^{\nu\alpha}\right)T_{(\mu\alpha)}
\end{align}
remains valid. After having set the rules for using dimensional regularisation in our formalism, we can finally start calculating some simple amplitudes.

\section{More on the quartic vertex: charge renormalisation}\label{seccharge}

\begin{figure}[H]  \centering\begin{center}
\includegraphics[width=0.3\linewidth]{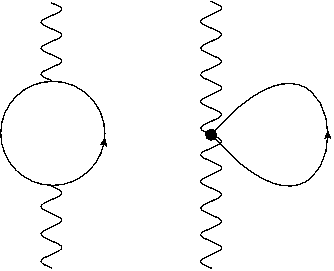}\caption{Photon two-point function diagrams at one loop}\label{charge}
\end{center}\end{figure}

We will, first of all, compute the amputated two-point photon amplitude at one loop in second-order QED. We begin our calculation by considering the two diagrams that contribute to the one-loop amplitude, (Fig.\ref{charge}). We then have, using the Feynman rules listed in Section \ref{DiracRules}:
\begin{align}\begin{split}
i&\Pi^{(1)}(k)^{A'}{}_{A}{}^{B'}{}_{B} = \\ &~(-1)4e^2 \int\frac{d^4 p}{(2\pi)^4}\frac{\left[ p^{A'}{}_{B}(p+k)^{B'}{}_{A} + (p+k)^{A'}{}_{B}p^{B'}{}_{A} - \frac{1}{2}\left( (p+k)^2+p^2\right) \epsilon^{A'B'}\epsilon_{AB}  \right]}{\left[p^2+m^2 \right]\left[(p+k)^2+m^2 \right]}\\
&~+ (-1)4e^2 \int\frac{d^4 p}{(2\pi)^4}\frac{\epsilon^{A'B'}\epsilon_{AB} }{\left[p^2+m^2 \right]}
\end{split}
\end{align}

%\begin{figure}[H]  \centering\begin{center}
%\input{charge_s4.pdf_tex}\caption{One-loop photon two-point amplitude}\label{charge}
%\end{center}\end{figure}
%\begin{figure}[H]  \centering\begin{center}
%$\vcenter{\hbox{\input{charge_s.pdf_tex}}}$
%  \hspace*{.2in}
%  $\vcenter{\hbox{ \input{charge_4.pdf_tex}}}$
%    	\caption{One-loop photon two-point amplitude}\label{charge}
%\end{center}\end{figure}
In dimensional regularisation, we have $D=4-\varepsilon$. This allows us to freely shift momenta and to rewrite the second integral as:
\begin{align}\begin{split}
\int\frac{d^D p}{(2\pi)^D}\frac{1}{\left[p^2+m^2 \right]}&=\int\frac{d^D p}{(2\pi)^D}\left( \frac{1/2}{\left[p^2+m^2 \right]}+\frac{1/2}{\left[(p+k)^2+m^2 \right]}\right)\\&= \int\frac{d^D p}{(2\pi)^D}\frac{\left[ \frac{1}{2}\left( (p+k)^2+p^2\right) +m^2 \right]}{\left[p^2+m^2 \right]\left[(p+k)^2+m^2 \right]}\end{split}
\end{align}
Then:
\begin{align}\begin{split}
i\Pi^{(1)}&(k)^{A'}{}_{A}{}^{B'}{}_{B} \\&= (-1)4e^2 \int\frac{d^D p}{(2\pi)^D}\frac{\left[ p^{A'}{}_{B}(p+k)^{B'}{}_{A} + (p+k)^{A'}{}_{B}p^{B'}{}_{A} +m^2 \epsilon^{A'B'}\epsilon_{AB}  \right]}{\left[p^2+m^2 \right]\left[(p+k)^2+m^2 \right]}
\end{split}
\end{align}

We expect the amplitude to be proportional to the transverse projector so as to satisfy the Ward-Takahashi identity. Therefore, it should depend on $k^{A'}{}_A$ and $k^{B'}{}_B$ with this index structure. This is not the case in our integrand, we thus use the following identity:
\begin{align}
 \theta^{(\mu\,A'}{}_{B}\theta^{\nu)\,B'}{}_{A} =  \theta^{(\mu\,A'}{}_{A}\theta^{\nu)\,B'}{}_{B} +\frac{1}{2}\eta^{\mu\nu}\epsilon^{A'B'}\epsilon_{AB}
\end{align}

in order to rewrite the numerator. Once this has been done, Feynman parameters are introduced to rewrite this integral (keeping terms involving even powers of the loop momentum $\ell$ only) and we Wick rotate the time component of our loop momentum ($\ell^0 = i\ell^0_E$). We then have for $i\Pi^{(1)}{}^{A'}{}_{A}{}^{B'}{}_{B}$:
\begin{align}\begin{split}
& (-1)4e^2i \int_0^1 dx \int\frac{d^D \ell_E}{(2\pi)^D}\frac{1}{\left[{\ell_E}^2+\Delta \right]^2}\left[ 2{\ell_E}^{A'}{}_{A}{\ell_E}^{B'}{}_{B} +{\ell_E}^2 \epsilon^{A'B'}\epsilon_{AB}\right.\\ &\hspace*{1cm}\left.-x(1-x)k^2\epsilon^{A'B'}\epsilon_{AB} -2 x(1-x)k^{A'}{}_{A}k^{B'}{}_{B} +m^2 \epsilon^{A'B'}\epsilon_{AB}  \right]\\
=& (-1)4e^2 \int_0^1 dx \int\frac{d^D {\ell_E}}{(2\pi)^D}\frac{1}{\left[{\ell_E}^2+\Delta \right]^2}\left[ \left(-\frac{2}{D}+1\right){\ell_E}^2 \epsilon^{A'B'}\epsilon_{AB}\right.\\ &\hspace*{1cm} \left. -x(1-x)k^2\epsilon^{A'B'}\epsilon_{AB}- 2x(1-x)k^{A'}{}_{A}k^{B'}{}_{B} +m^2 \epsilon^{A'B'}\epsilon_{AB}  \right]
\end{split}
\end{align}

with
\begin{align}
\ell = p+xk, \quad \Delta= x(1-x)k^2+m^2
\end{align}

and we replaced
\begin{align}
{\ell_E}^{A'}{}_{A}{\ell_E}^{B'}{}_{B} \rightarrow -\frac{1}{D}{\ell_E}^2 \epsilon^{A'B'}\epsilon_{AB}
\end{align}
Finally, using:
\begin{align}
\left(-\frac{2}{D}+1\right) \int\frac{d^D\ell_E}{(2\pi)^D}\frac{\ell_E^2}{\left(\ell_E^2+\Delta\right)^2} &= \frac{1}{(4\pi)^{D/2}}\Gamma(2-{D}/{2}) \left(\frac{1}{\Delta}\right)^{2-\frac{D}{2}} (-\Delta) \\
 \int\frac{d^D\ell_E}{(2\pi)^D}\frac{1}{\left(\ell_E^2+\Delta\right)^2} &= \frac{1}{(4\pi)^{D/2}}\Gamma(2-{D}/{2}) \left(\frac{1}{\Delta}\right)^{2-\frac{D}{2}} 
\end{align}

We obtain:
\begin{align}
i\Pi^{(1)}{}^{A'}{}_{A}{}^{B'}{}_{B}= 4e^2i\frac{\Gamma(2-{D}/{2})}{(4\pi)^{D/2}}\left(k^2\epsilon^{A'B'}\epsilon_{AB}+k^{A'}{}_{A}k^{B'}{}_{B}\right)\int_0^1 dx  \frac{2x(1-x)}{\Delta^{2-\frac{D}{2}}}\label{2pt1loop}
\end{align}

The two point function is usually written as:
\begin{align}
i\Pi^{(1)}{}^{A'}{}_{A}{}^{B'}{}_{B}= -\left(k^2\epsilon^{A'B'}\epsilon_{AB}+k^{A'}{}_{A}k^{B'}{}_{B}\right)\cdot i\Pi^{(1)}(k^2)
\end{align}

We have here :
\begin{align}
\Pi^{(1)}(k^2)= -8e^2\frac{\Gamma(2-{D}/{2})}{(4\pi)^{D/2}}\int_0^1 dx  \frac{x(1-x)}{\Delta^{2-\frac{D}{2}}}
\end{align}

We shall now use our regularisation and take the $D\rightarrow 4$ limit:
\begin{align}
\Pi^{(1)}(k^2)~\substack{\longrightarrow \\ D\rightarrow 4} ~- \frac{2\alpha}{\pi}\int_0^1 dx ~ x(1-x)\left(\frac{2}{\varepsilon}-\gamma -\log\frac{\Delta}{4\pi} + \mathcal{O}(\varepsilon)\right)
\end{align}

Let us now shortly recall that the quantity we have just computed describes the renormalisation of the electromagnetic coupling constant $\alpha$. Indeed, because of the Ward-Takahashi identities, we have that the photon two-point function is given by:
\begin{align}
-\frac{i\epsilon^{A'B'}\epsilon_{AB}}{k^2(1-\Pi(k^2))}
\end{align}

Therefore, as long as $\Pi(k^2)$ is regular for on-shell momenta, the propagator always has a simple pole at $k^2=0$ and the photon remains massless (this is for example not the case in 2D massless QED where the photon acquires a mass at the one-loop level\footnote{To see this, consider (\ref{2pt1loop}) at $m^2=0$ and $D=2$. Including a factor of $1/2$ as a dimensional correction for the loop, the two-point function has the structure of a transverse propagator for a massive photon with mass $e^2/\pi$. Notice however that the isomorphism between the Lorentz group in two dimensions and $SU(2)$ spinors is no longer valid. Therefore, the limit should be understood in the following way: compute the quantity in 4D, use the isomorphism to go back to spacetime indices, take the limit $D\rightarrow2$ and multiply by one half for each fermion loop as a dimensional correction.}). The residue of the pole is related to the wave-function renormalisation:
\begin{align}
\frac{1}{1-\Pi(0)}= Z_3
\end{align}

Referring to the quantity that multiplies the vector-current interaction in the Lagrangian as the bare charge $e_0$, after renormalisation we have:
\begin{align}
e= \sqrt{Z_3}e_0 \label{chargeren}
\end{align}

with $e$ the physical renormalised charge. Notice that if one looks at the counterterms in the Lagrangian (computed as usual, see $e.g.$ \cite{Peskin:1995ev}), one has
\begin{align}
Z_1=\frac{e_0}{e}Z_2\sqrt{Z_3},\quad \delta_i=Z_i-1
\end{align}

with $\delta_1$ the counterterm corresponding to the three-valent vertex and $Z_2$ the fermion wavefunction renormalisation. Equation (\ref{chargeren}) is a first indication that 
\begin{align}
Z_1=Z_2
\end{align}

Furthermore, in our case, we introduce a fourth counterterm for the four-valent vertex such that:
\begin{align}
\delta_4=Z_4-1 = \left(\frac{e_0}{e}\right)^2 Z_3Z_2-1 = \frac{Z_1^2}{Z_2}-1
\end{align}

Then, we have:
\begin{align}
Z_1=Z_2=Z_4
\end{align}

This follows from gauge invariance (or equivalently from the Ward identities).

Coming back to the charge renormalisation, when one computes a scattering process at non-zero $k^2$ at one-loop, one deals with the quantity:
\begin{align}\begin{split}
-\frac{i\epsilon^{A'B'}\epsilon_{AB}}{k^2}\frac{e_0^2}{1-\Pi(k^2)}&=  -\frac{i\epsilon^{A'B'}\epsilon_{AB}}{k^2}e^2(1+\Pi(k^2)-\Pi(0)) \\&=  -\frac{i\epsilon^{A'B'}\epsilon_{AB}}{k^2}\frac{e^2}{1-(\Pi(k^2)-\Pi(0))}\label{effectivecoupling}
\end{split}\end{align}

Then, although the first order shift in the electric charge is divergent, the effective electromagnetic coupling that appears in (\ref{effectivecoupling}) is well defined:
\begin{align}
\alpha_{eff}(k^2)= \frac{\alpha}{1-(\Pi(k^2)-\Pi(0))}
\end{align}

with, at one-loop
\begin{align}
\Pi(k^2)-\Pi(0)= -\frac{2\alpha}{\pi}\int_0^1dx~x(1-x)\log\left(\frac{m^2}{m^2+x(1-x)k^2}\right)
\end{align}

\section{Charge renormalisation using the Passarino-Veltman reduction}
The Passarino-Veltman (PV) reduction is a useful tool to simplify the calculation of one-loop integrals. Its principle relies on the fact that one can expand any kind of one-loop tensor integral in a basis of scalar integrals (see \cite{Ellis:2011cr} for a comprehensive review). We give here a short overview of the technique in both first- and second-order formalisms. Although the method is the same, we will see that it is easier to obtain the decomposition on basis integrals in the latter.

\subsection{Within the Dirac formalism}
Let us first look at the workings of the reduction in the usual case for the specific example that we have treated above. We write the expression for the one-loop polarisation tensor given by usual Dirac first-order Feynman rules:
\begin{align}
i\Pi^{{(1)}\,\mu\nu}= (-1)(ie)^2(-i)^2\int \frac{d^D\ell}{(2\pi)^D}{\rm{Tr}}\left(S_{\ell+k} \gamma^\mu S_\ell \gamma^\nu  \right), \quad S_p = \frac{-\slashed{p}+m}{p^2+m^2}
\end{align}
with $\slashed p \defeq \gamma^\mu p_\mu$ and the algebra of Dirac gamma matrices is defined in (\ref{diracgammarules}). Using the definition of the one- and two-point scalar integrals:
\begin{align}
A_0(i) &= \frac{1}{i\pi^{D/2}}\int d^D\ell \frac{1}{D_i}, \quad D_i = (\ell +k_i)^2+m^2 \\
B_0; B^\mu; B^{\mu\nu} &= \frac{1}{i\pi^{D/2}}\int d^D\ell \frac{1;\ell^\mu;\ell^{\mu}\ell^\nu}{D_0D_1},\quad k_0=0,~ k_1=k
\end{align}

we can rewrite it as (we drop from now on the superscript (1) as we are exclusively dealing with the one-loop quantity):
\begin{align}
i\Pi^{\mu\nu}= -\frac{ie^2}{(4\pi)^{D/2}} \Bigg[ {\rm Tr}\left(\gamma^\alpha\gamma^\mu\gamma^\beta\gamma^\nu\right) \left(B_{\alpha\beta}+k_\alpha B_\beta\right)+m^2{\rm Tr}\left(\gamma^\mu\gamma^\nu\right)B_0 \Bigg]
\end{align}

Our aim is to scalarise the integrals. In order to do so, we give a reminder of the Passarino-Veltman reduction.
\begin{center}
\line(1,0){250}
\end{center}

We make explicit the Passarino-Veltman reduction algorithm for the case of two-point tensor integrals. By Lorentz invariance the integrals can be rewritten as:
\begin{align}
B_\mu &= k_\mu B_1 \\
B_{\mu\nu} &= g_{\mu\nu} B_{00} + k_\mu k_\nu B_{11}
\end{align}

where $B_1,B_{00},B_{11}$ are form factors. Dotting the first equation with the external momentum and using\footnote{We may use a shift in the integral to prove the second equality. The latter is allowed since the integrals are convergent in dimensional regularisation.}:
\begin{align}
2(k\cdot \ell)= D_1-D_0 -k^2, \quad A_0(1)=A_0(0)
\end{align}

for $m_0=m_1=m$. We have:
\begin{align}
B_1 = -\frac{B_0}{2}
\end{align}

Let us now define the transverse projector:
\begin{align}
P^{(T)}_{\mu\nu} \defeq g_{\mu\nu} - \frac{k_\mu k_\nu}{k^2}\label{tproj}
\end{align}

We then have:
\begin{align}
\frac{1}{D-1}P^{(T)}_{\mu\nu} B^{\mu\nu} &= B_{00} \\
-\frac{1}{D-1}P^{(T)}_{\mu\nu} B^{\mu\nu} + \frac{k_\mu k_\nu}{k^2}B^{\mu\nu} &= k^2B_{11}
\end{align}

Using loop-momentum shifts and rewriting numerators as inverse propagators, we obtain:
\begin{align}
(D-1)B_{00} &= \frac{A_0(1)}{2}- m^2 B_0 - \frac{k^2}{4}B_0\\
k^2 B_{11}& = -B_{00} + \frac{k^2}{4}B_0+\frac{A_0(1)}{2}
\end{align}
\begin{center}
\line(1,0){250}
\end{center}

Let us now go back to our calculation. The photon self-energy can be written as:
\begin{align}
i\Pi_{\mu\nu} = P^{(T)}_{\mu\nu}\Pi_T(k) + \frac{k_\mu k_\nu}{k^2}\Pi_L(k)
\end{align}

where $P^{(T)}_{\mu\nu}$ is the projector defined in (\ref{tproj}) and $\Pi_{T/L}$ are form factors. We then have:
\begin{align}
ik^\mu k^\nu\Pi_{\mu\nu} &= k^2 \Pi_L \label{form1}\\
ig^{\mu\nu}\Pi_{\mu\nu}&= (D-1)\Pi_T + \Pi_L\label{form2}
\end{align}

Using the gamma matrices algebra and the decomposition of the tensor integrals in scalar form factors, it is straightforward, however cumbersome, to show that $\Pi_L=0$ and that:
\begin{align}
(D-1)\Pi_T = \frac{4ie^2}{(4\pi)^{D/2}}\Bigg[(D-2)A_0(1) + 2m^2 B_0 -\frac{k^2}{2}(D-2)B_0 \Bigg]
\end{align}

So that:
\begin{align}
i\Pi_{\mu\nu} = iP^{(T)}_{\mu\nu}\frac{4e^2}{(4\pi)^{D/2}}\frac{1}{D-1}\Bigg[(D-2)A_0(1) + 2m^2 B_0 -\frac{k^2}{2}(D-2)B_0 \Bigg]
\end{align}

\subsection{Within the second-order formalism}
In this formalism, things work in a similar fashion. We want to scalarise the following integral:
\begin{align}\begin{split}
i\Pi(k)^{A'B'}_{AB}&= \int\frac{d^D \ell}{(2\pi)^D}\frac{(-1)4e^2 }{D_0D_k}\left[ \ell^{A'}{}_{B}(\ell+k)^{B'}{}_{A}+ (\ell+k)^{A'}{}_{B}\ell^{B'}{}_{A} +m^2 \epsilon^{A'B'}\epsilon_{AB}  \right]
\end{split}
\end{align}
%{\left[\ell^2+m^2 \right]\left[(\ell+k)^2+m^2 \right]}
where as before
\begin{align}
D_k = (\ell+k)^2+m^2
\end{align}

Again, the amplitude should be proportional to the transverse projector so that to satisfy the Ward-Takahashi identity. Using the same identity as before in order to shuffle the indices of the soldering forms, we can rewrite the numerator so as to match the external (physical) index structure:
\begin{align}\begin{split}
i\Pi(k)^{A'B'}_{AB} &= (-1)\frac{4ie^2}{(4\pi)^{D/2}} \int\frac{d^D \ell}{i\pi^{D/2}}\frac{1}{D_0D_k}\left[ 2\ell^{A'}{}_{A}\ell^{B'}{}_{B} + \ell^{A'}{}_{A}k^{B'}{}_{B} + \ell^{B'}{}_{B}k^{A'}{}_{A}\right. \\ &\hspace*{4cm} \left. +(\ell \cdot (\ell +k) +m^2) \epsilon^{A'B'}\epsilon_{AB}  \right]\\
&= (-1)\frac{4ie^2}{(4\pi)^{D/2}} \left[ 2B^{A'}_{A}{}^{B'}_{B} + B^{A'}_{A}k^{B'}{}_{B} + B^{B'}_{B}k^{A'}{}_{A} \right. \\ &\hspace*{4cm} \left.+(A_0(1)+k^2B_1) \epsilon^{A'B'}\epsilon_{AB}  \right]
\end{split}
\end{align}

where $B^{A'}_{A}{}^{B'}_{B}$ and $B^{A'}_{A}$ are given as before by their form-factor decomposition\footnote{Recall that $g_{\mu\nu} \rightarrow - \epsilon^{A'B'}\epsilon_{AB} $}. Compared to the Dirac-formalism's expression, the latter is much simpler as there are no gamma matrices to worry about. The scalarisation in itself is the same, but equations (\ref{form1},\ref{form2}) are calculated almost straightforwardly using as before $\epsilon^{A'B'}\epsilon_{AB}\epsilon_{A'B'}\epsilon^{AB}=D$.

\section{Fermion self-energy}

\begin{figure}[H]  \centering\begin{center}
\includegraphics[width=0.25\linewidth]{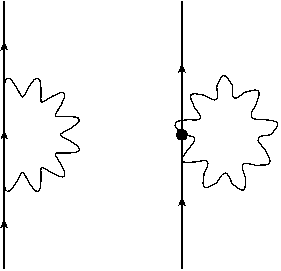}\caption{Fermion self-energy diagrams at one loop}\label{fermself}
\end{center}\end{figure}

We compute now the fermion self energy at one-loop order (Fig.\ref{fermself}). We are mainly interested in extracting the counterterms corresponding to the mass operator and the fermion wave function. We have:
\begin{align}
-i\Sigma(p^2)= -i\Sigma^{(1)}(p^2)- i(p^2+m^2)\delta_2 -i\delta_m
\end{align}

With the renormalisation conditions:
\begin{align}
\Sigma(p^2+m^2=0)= 0, \quad \left.\frac{\partial\Sigma}{{\partial p^2}}\right|_{p^2+m^2=0}=0
\end{align}
Leading to:
\begin{align}
\delta_m= -\Sigma^{(1)}(m^2), \quad \delta_2 = - \left.\frac{\partial\Sigma^{(1)}}{{\partial p^2}}\right|_{p^2+m^2=0}
\end{align}
Let us then compute these quantities. The tadpole diagram vanishes in dimensional regularisation due to the photon being massless. We use a non-zero photon mass to regularise the infrared divergence of the remaining diagram, we then have:
\begin{align}\begin{split}
-i\Sigma^{(1)}(p^2)&=\lim_{m_\gamma\rightarrow 0}2e^2\epsilon_{AB}\int \frac{d^Dq}{(2\pi)^D}\frac{(m^2-p^2)-q^2}{(q^2+m_\gamma^2)((q+p)^2+m^2)}\\
&=\lim_{m_\gamma\rightarrow 0}2e^2i\epsilon_{AB}\left[(m^2-p^2)\int_0^1 dx\int \frac{d^D\ell_E}{(2\pi)^D}\frac{1}{(\ell_E^2+\Delta)^2}\right. \\ &\hspace*{4cm} \left.-\int \frac{d^D\ell_E}{(2\pi)^D}\frac{1}{(\ell_E^2+m^2)}\right]
\end{split}\end{align}
%&=\lim_{m_\gamma\rightarrow 0}\int \frac{d^Dq}{(2\pi)^D}(2ei)^2 \left[ -p^{D'}{}_{B}\epsilon_{FD}+(p+q)^{D'}{}_{F}\epsilon_{BD}\right] \frac{-i\epsilon^{FE}}{(q+p)^2+m^2}\frac{-i\epsilon_{CD}\epsilon^{C'D'}}{q^2+m_\gamma^2} \\
%&{}\quad \times \left[  p^{C'}{}_{A}\epsilon_{EC}-(p+q)^{C'}{}_{E}\epsilon_{AC} \right] \\

with 
\begin{align}
\Delta = p^2x(1-x)+(1-x)m^2+ xm_\gamma^2
\end{align}

and we used
\begin{align}
\int \frac{d^Dq}{(2\pi)^D}\frac{(q+p)^2}{q^2((q+p)^2+m^2)} = \int \frac{d^Dq}{(2\pi)^D}\frac{-m^2}{q^2((q+p)^2+m^2)}
\end{align}

Now, the second integral is straightforwardly evaluated and yields:
\begin{align}
i\epsilon_{AB}\frac{\alpha}{2\pi}m^2\left(\frac{2}{\epsilon}-\gamma +\log4\pi-\log\frac{ m^2}{\mu^2}+1\right)
\end{align}

where $\mu$ is a UV cutoff scale. The first integral is less trivial because of the integration over the Feynman parameter $x$:
\begin{align}
i\epsilon_{AB}\frac{\alpha}{2\pi}\left(\frac{2}{\epsilon}-\gamma+\log4\pi\right)(m^2-p^2)\left(1-\frac{\epsilon}{2}\int_0^1dx\log \Delta \right)
\end{align}

As we previously said, we are interested in the counterterms. The first renormalisation condition is then given by the value of the self-energy on-shell. In that case we have:
\begin{align}
\int_0^1dx\log \Delta(p^2+m^2=0) = -2+\log\frac{m^2}{\mu^2}
\end{align}

So that:
\begin{align}
\delta_m = -\frac{\alpha}{2\pi}3m^2\left(\frac{2}{\epsilon}-\gamma+\log4\pi  -\log\frac{m^2}{\mu^2} +5/3\right)
\end{align}
where we removed the spinor metric $\epsilon_{BA}$ from the definition. The remaining counterterm is given by the derivative of the self-energy set on-shell:
\begin{align}
- i\left.\frac{\partial\Sigma^{(1)}}{{\partial p^2}}\right|_{p^2+m^2=0}=i\epsilon_{AB}\frac{\alpha}{2\pi}\left(\frac{2}{\epsilon}-\gamma+\log4\pi -\log\frac{m^2}{\mu^2}  -\log\frac{m_\gamma^2}{m^2}+1 \right)
\end{align}

So that:
\begin{align}
\delta_2=  -\frac{\alpha}{2\pi}\left(\frac{2}{\epsilon}-\gamma+\log4\pi -\log\frac{m^2}{\mu^2}  -\log\frac{m_\gamma^2}{m^2}+1 \right)
\end{align}

These are the usual counterterms obtained in the first-order formalism. We will not cover the analysis of their UV and IR divergences in this thesis, which is a typical exercise to be found in any QFT textbook.

\section{Three-valent vertex renormalisation}

We will now give the UV divergent part of the one-loop correction to the three valent vertex.
\begin{figure}[H]  \centering\begin{center}
\includegraphics[width=0.7\linewidth]{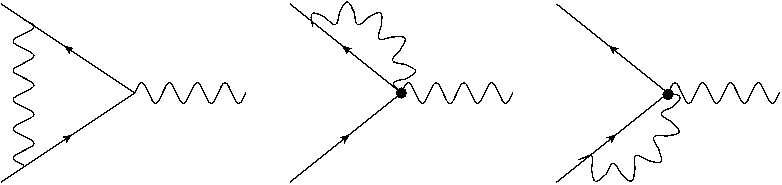}\caption{Three-valent vertex renormalisation at one loop}\label{vertrenorm}
\end{center}\end{figure}

There are three diagrams (Fig.\ref{vertrenorm}), whose amplitudes are given (resp.) by:
\begin{align}\begin{split}
8e^3\int\frac{d^4q}{(2\pi)^4}&D_F(k_1-q) D_F (k_2+q) D_\gamma(q) \Bigg[ \left({k_1}_{BN'} {k_2}^{N'}{}_{C} \right)\left(k_1-k_2-2q\right)^{A'}{}_{A}\\
& +\left(k_2+q\right)^{A'}{}_B {k_2}_C{}^{N'} \left(k_1-q \right)_{N'A}   +\left(k_1-q\right)^{A'}{}_C {k_1}_B{}^{N'}\left(k_2+q \right)_{N'A} \\& +\frac{1}{2}(k_1-q)^2\left({k_2}^{A'}{}_C \epsilon_{BA} - (k_2+q)^{A'}{}_A\epsilon_{BC} \right) \\& +\frac{1}{2}(k_2+q)^2\left({k_1}^{A'}{}_B \epsilon_{CA} + (k_1-q)^{A'}{}_A\epsilon_{BC}\right)   \Bigg]
\end{split}\end{align}
\begin{align}
4e^3\int\frac{d^4q}{(2\pi)^4}D_F(k_1-q)  D_\gamma(q) \Bigg[  {k_1}^{A'}{}_{B}\epsilon_{CA}-(k_1-q)^{A'}{}_{C}\epsilon_{BA} \Bigg]
\end{align}
\begin{align}
4e^3\int\frac{d^4q}{(2\pi)^4} D_F (k_2+q) D_\gamma(q) \Bigg[ {k_2}^{A'}{}_{C}\epsilon_{BA}-(k_2+q)^{A'}{}_{B}\epsilon_{CA} \Bigg]
\end{align}

where $D_F (p) =1/(p^2+m^2)$ and $D_\gamma (p) =1/p^2$. We now only keep the UV divergent pieces as we are concerned by the (in dimensional regularisation) $1/\epsilon$ part of the counterterm needed to renormalise the 3-vertex:
\begin{align}\begin{split}
\mathcal{V}^{(1)}_3(k_1,k_2)_{ABC}{}^{A'} = &
8e^3\int\frac{d^4q}{(2\pi)^4}D_F(k_1-q) D_F (k_2+q) D_\gamma(q) \\ \times &\Bigg[ -q^{A'}{}_{B}q_{N'A}{k_2}^{N'}{}_C  -q^{A'}{}_{C}q_{N'A}{k_1}^{N'}{}_B  \\&+\frac{1}{2}(k_1-q)^2 (k_2-q)^{A'}{}_C\epsilon_{BA} + \frac{1}{2}(k_2+q)^2 (k_1+q)^{A'}{}_B\epsilon_{CA}\Bigg]
\end{split}\end{align}

where $\mathcal{V}^{(1)}_3(k_1,k_2)$ denotes the one loop contribution to the 3-vertex, and we have gathered all the terms under the same integral. We now use Feynman parameters, so that $\ell = q-xk_1+yk_2$, $\Delta= xyq^2+ (1-z)^2m^2$ and $\ell^{A'}_A\ell^{B'}_B \rightarrow -\frac{\epsilon^{A'B'}\epsilon_{AB}}{D}\ell^2$ to obtain (keeping only UV divergent pieces):
\begin{align}\begin{split}
\mathcal{V}^{(1)}_3(k_1,k_2)_{ABC}{}^{A'} = 
16e^3\int & DF\int\frac{d^D\ell}{(2\pi)^D}\frac{\ell^2}{(\ell^2+\Delta)^3}\\ &\times \Bigg[ {k_1}^{A'}{}_B\epsilon_{CA}\left( \frac{1+x}{2}-\frac{1-x}{D}\right)\\&+{k_2}^{A'}{}_C\epsilon_{BA}\left( \frac{1+y}{2}-\frac{1-y}{D}\right)  {k_1}^{A'}{}_C\epsilon_{BA}\left( \frac{1-x}{D}-\frac{x}{2}\right)\\&+{k_2}^{A'}{}_B\epsilon_{CA}\left( \frac{1-y}{D}-\frac{y}{2}\right)\Bigg]
\end{split}\end{align}

where $ \int DF =\int dxdydz\,\delta(x+y+z-1)$. Using
\begin{align}
\int\frac{d^D\ell}{(2\pi)^D}&\frac{\ell^2}{(\ell^2+\Delta)^3}  = \frac{i}{(4\pi)^2} \left(\frac{2}{\epsilon}\right) +\mathcal{O}(1), \quad {\rm as~}\epsilon\rightarrow 0
\end{align}

We are left with the straightforward Feynman parameters integrations:
\begin{align}
\int DF\left(\frac{1+x}{2}-\frac{1-x}{4}\right) =\frac{1}{4} \\
\int DF\left( \frac{1-x}{4}-\frac{x}{2}\right) = 0
\end{align}
So that we obtain:
\begin{align}\begin{split}
\mathcal{V}^{(1)}_3(k_1,k_2)_{ABC}{}^{A'} &= 2ei\left({k_1}^{A'}{}_B\epsilon_{CA}+{k_2}^{A'}_C\epsilon_{BA} \right)\times \left( \frac{\alpha}{2\pi}\right)\left( \frac{2}{\epsilon}\right)\\& =  \mathcal{V}_3(k_1,k_2)_{ABC}{}^{A'} \times (1~{\rm loop})
\end{split}\end{align}
Finally, the renormalised vertex is:
\begin{align}\begin{split}
\mathcal{V}^{R}_3(k_1,k_2) = \mathcal{V}_3(k_1,k_2)\left(1+  (1~{\rm loop})+\delta_1\right)
\end{split}\end{align}

where $\delta_1$ is the vertex counterterm. Using $\delta_1 = -\left.(1~{\rm loop})\right|_{\rm on-shell}$, we obtain:
\begin{align}
\delta_1 = -\left( \frac{\alpha}{2\pi}\right)\left( \frac{2}{\epsilon}\right) +\mathcal{O}(1) \sim \delta_2
\end{align}

In order to check the equality of the counterterms, we would have had to calculate accurately the finite parts contribution of the integrals. We refer the reader to \cite{Peskin:1995ev}.

We now move on to calculations and properties in/of the second-order formalism that are not straightforwardly equivalent to its first-order counterpart.

\part{Advanced Aspects}

\chapter{Anomalies}\label{chapterano}

\section{Introduction}

We will see how the anomaly can be calculated in pertubation theory using Feynman diagrams, and then how its non-perturbative nature can be demonstrated using a path-integral derivation. In both cases, we review first how the anomaly is computed in the usual Dirac formalism and thereafter study its construction in a theory with second-order fermions.

\section{Fermion number anomaly in perturbation theory}

\subsection{First-order perturbative calculation}
The anomaly can be computed in perturbation theory by means of Feynman diagrams. Indeed, one shows that the divergence of the current has a non-zero matrix element to create two photons:
\begin{align}
\langle k_1,k_2 | j^{ A'}_{A}(p)|0\rangle &= \epsilon^{*\, B}_{B'}(k_1)\epsilon^{*\, C}_{C'}(k_2)\mathcal{M}^{ABC}_{A'B'C'}(k_1,k_2)\nonumber\\
\langle k_1,k_2 | p\cdot j(p)|0\rangle &\neq 0
\end{align}

\begin{figure}[H]  \centering\begin{center}
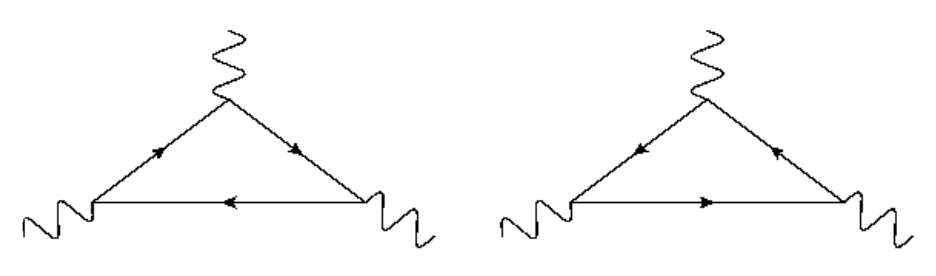\caption{First-order Feynman diagrams for the fermion number anomaly in Weyl theory.}\label{anofirst}
\end{center}\end{figure}

We briefly translate the two-component anomaly calculation of \cite{Dreiner:2008tw} into our notation. The Feynman rules are as follows:
\begin{align}
\langle 0| \bar\lambda_{A'}(p)\lambda_B(-p)|0 \rangle & = \frac{-i\sqrt{2}p_{A'B}}{p^2+m^2}\nonumber \\
\mathcal{V}_3(q,p,k) &= i\sqrt{2}e~\epsilon^{A'B'}\delta_{A}{}^{C}
\end{align}
%\langle 0| A^{A'}{}_{A}(q) \bar\lambda^{B'}(p)\lambda^C(k)|0 \rangle &=i\sqrt{2}e~\epsilon^{A'B'}\delta_{A}{}^{C}~\delta^3(q+p+k)
Taking into account the two orientations for the triangle diagram (Fig. \ref{anofirst}), and denoting the incoming photons by spinor indices $(AA'),~(BB'),~(CC')$ and their massless momenta $k_1,~k_2,~k_3$:
\begin{align}\begin{split}
i\mathcal{M}^{(1)}(k_2,k_3)= 8e^3\int&\frac{d^4q}{(2\pi)^4}\Bigg[q^{B'}{}_{C}(q-k_2)^{A'}{}_{B}(q+k_3)^{C'}{}_{A} \\& - q^{C'}{}_{B}(q-k_2)^{B'}{}_{A}(q+k_3)^{A'}{}_{C}\Bigg]\times \frac{1}{(q-k_2)^2q^2(q+k_3)^2}\label{amplitudeweyl1st}
\end{split}\end{align}

From here, one can show that the amplitude is shift dependent ($q\rightarrow q+a$) and that the divergence of the amplitude with respect to the currents is given by:
\begin{align}
ik_1 \cdot i\mathcal{M}^{(1)}(k_2,k_3;c)&= \frac{ie^3}{8\pi^2}(2c)\epsilon^{\mu\nu\alpha\beta}{k_2}_\mu{k_3}_\beta \label{an1}\\
ik_2 \cdot i\mathcal{M}^{(1)}(k_2,k_3;c)&= -\frac{ie^3}{8\pi^2}(1+c)\epsilon^{\mu\nu\alpha\beta}{k_2}_\nu{k_3}_\beta\label{an2} \\
ik_3 \cdot i\mathcal{M}^{(1)}(k_2,k_3;c)&= -\frac{ie^3}{8\pi^2}(1+c)\epsilon^{\mu\nu\alpha\beta}{k_2}_\alpha{k_3}_\beta\label{an3}
\end{align}
where we have used $a^\beta= c(k_2-k_3)^\beta$.

\subsection{Second-order Lagrangian}

We will now carry out the calculation for a Weyl fermion in a second-order formalism. The Lagrangian in this case is given by:
\begin{align}
{\cal L} = - D_{A'}{}^{ A} \lambda_A  D^{A'B} \lambda_B - \frac{m^2}{2} \lambda^A\lambda_A\label{lagweyl2}.
\end{align}

This should be supplemented with the reality conditions:
\begin{align}
m\lambda^{\dagger\, A'} = -i\sqrt{2}D^{ A'A}\lambda_A \label{realityweyl}
\end{align}

The field equations that result from the above Lagrangian are
\begin{align}
{2} D_{A'}{}^{ A} D^{A'B} \lambda_B + m^2 \lambda^A=0
\end{align}

In what follows we will consider only Weyl theory, which amounts to setting the mass to zero in the above equations. We see that the Lagrangian is not invariant under the usual $U(1)$ transformations
\begin{align}
 \delta\lambda= +ie\alpha \lambda
\end{align}

However, the field equations and the reality condition are. Furthermore, the current given by:
\begin{align}
j_{A}{}^{A'}=ie\left( \lambda_A  D^{A'B} \lambda_B -  D^{A'B} \lambda_B\lambda_A \right) = 2ie \lambda_A  D^{A'B} \lambda_B
\end{align}

is conserved on-shell:
\begin{align}
D_{A'}{}^{A}j_{A}{}^{A'} = 0
\end{align} 
In order to see if the transformations are a symmetry of the theory, we need to check the above equation in the quantum theory. We will therefore compute pertubatively the anomaly corresponding to $U(1)$ transformations. As we have seen, the current conservation equation is anomalous in the presence of only one Weyl fermion. However, the anomaly should cancel out if several Weyl fermions are present and their charges sum up to zero.

\subsection{Perturbative calculation in the second-order formalism}

In order to proceed, we need Feynman rules to compute the diagrams (\ref{WeylMajoRules}) that we recall here (all incoming, and where the order of the fields in the vertices is $\gamma ff$ and $\gamma\gamma ff$):
\begin{align}
\langle 0| \lambda_A(p)\lambda_B(-p)|0 \rangle & = \frac{-i}{p^2+m^2}\epsilon_{AB} \\
\mathcal{V}_3(q,p,k)&=  2ie( p_B{}^{A'}\epsilon_{CA}-k_C{}^{A'}\epsilon_{BA})\\
\mathcal{V}_4(q_1,q_2,p,k)&= 2\im e^2 \epsilon^{A'B'}\epsilon_{AB}\epsilon_{CD}
\end{align}

With these rules, we give the amplitude for the process where we denote the incoming photons by spinor indices $(AA'),~(BB'),~(CC')$ and they are labelled by their massless momenta $k_1,~k_2,~k_3$. We split the amplitude into different contributions:
\begin{align}\begin{split}
i\mathcal{M}(k_2,k_3;s)= 8e^3\int&\frac{d^4q}{(2\pi)^4}\Bigg[ \frac{\mathcal{I} + \mathcal{J} }{D(-k_2)D(k_3)D(0)  } + \frac{\mathcal{A}}{D(-k_2)D(k_3)} \\& + \frac{\mathcal{B}}{D(-k_2)D(0)}+\frac{\mathcal{C}}{D(0)D(k_3)}\Bigg] \label{amplitudeweyl}
\end{split}\end{align}

\begin{figure}[H]  \centering\begin{center}
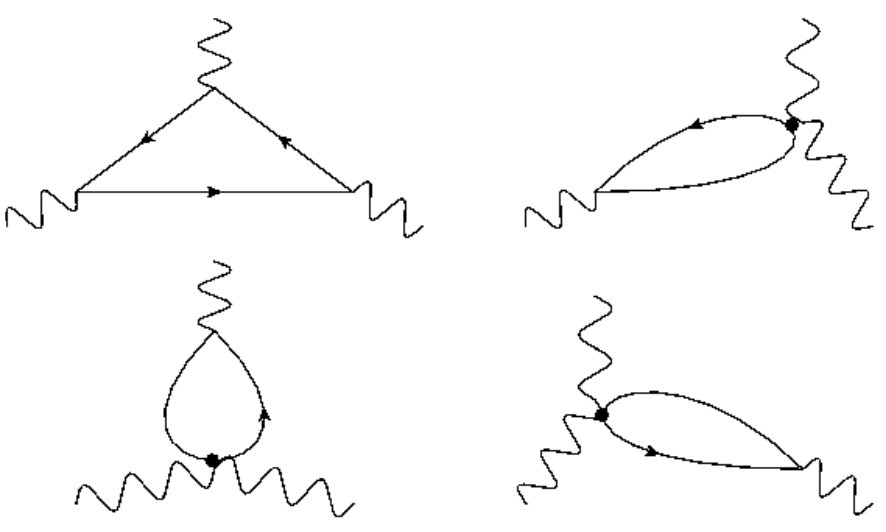\caption{Second-order Feynman diagrams for the fermion number anomaly in Weyl theory. There is only one triangle diagram, however, the quartic vertex now also contributes.}\label{anosecond}
\end{center}\end{figure}
where $D(k) = (q_s+k)^2$ and we allowed the amplitude to depend on a shifted loop momentum $q_s\defeq q+s$, and where $\mathcal{I} + \mathcal{J}$ are contributions from the diagram made of cubic vertices only and $\mathcal{A},~\mathcal{B},~\mathcal{C}$ are contributions from the diagrams with a quartic vertex (Fig.{\ref{anosecond}). For sake of clarity, we omitted the external indices on all the integrands and on the full amplitude $\mathcal{M}^{A}{}_{A'}{}^{B}{}_{B'}{}^{C}{}_{C'}(k_2,k_3;s)\equiv \mathcal{M}(k_2,k_3;s)$. Let us now make explicit the different terms. First of all:
\begin{align}\begin{split}
\mathcal{I}= q_s^{B'}{}_{C}(q_s-k_2)^{A'}{}_{B}(q_s+k_3)^{C'}{}_{A}- q_s^{C'}{}_{B}(q_s-k_2)^{B'}{}_{A}(q_s+k_3)^{A'}{}_{C}
\end{split}\end{align}

They correspond to $i\mathcal{M}^{(1)}(k_2,k_3)$ in (\ref{amplitudeweyl1st}). In our case, the triangle diagram yields an extra contribution:
\begin{align}\begin{split}
\mathcal{J}=&\frac{1}{2}q_s^2\epsilon^{B'C'}\left(\epsilon_{AB}(q_s+k_3)^{A'}{}_{C} +\epsilon_{AC}(q_s-k_2)^{A'}{}_{B}\right) \\
&+\frac{1}{2}(q_s+k_3)^2\epsilon^{A'C'}\left(\epsilon_{AB}q_s^{B'}{}_{C}+\epsilon_{CB}(q_s-k_2)^{B'}{}_{A}\right)\\
&+\frac{1}{2}(q_s-k_2)^2\epsilon^{A'B'}\left(-\epsilon_{AC}q_s^{C'}{}_{B}-\epsilon_{BC}(q_s+k_3)^{C'}{}_{A}\right)\end{split}
\end{align}

These terms arise from the contractions of momenta by propagators as it was the case when we computed the photon two-point function. They are expected to cancel out with terms arising from the quartic vertex, let us then look at these:
\begin{align}
\mathcal{A}&= \frac{1}{4}\epsilon^{B'C'}\epsilon_{BC}(k_2+k_3)^{A'}{}_{A} \\
\mathcal{B}&=-\frac{1}{4} \epsilon^{A'C'}\epsilon_{AC}{k_2}^{B'}{}_{B} \\
\mathcal{C}&=-\frac{1}{4}\epsilon^{A'B'}\epsilon_{AB}{k_3}^{C'}{}_{C}
\end{align}

We can combine these four terms in the following way: in the $\mathcal{J}$ term, one can cancel one propagator and add to $\mathcal{A},~\mathcal{B}$ or $\mathcal{C}$ the correct term in terms of scalar propagators remaining uncancelled. We have:
\begin{align}
\mathcal{\tilde A}&= \frac{1}{2}\epsilon^{B'C'}\left( \frac{1}{2}\epsilon_{BC}(k_2+k_3)^{A'}{}_{A} +\epsilon_{AB}(q_s+k_3)^{A'}{}_{C} +\epsilon_{AC}(q_s-k_2)^{A'}{}_{B}\right)\\
\mathcal{\tilde B}&=\frac{1}{2} \epsilon^{A'C'}\left(-\frac{1}{2}\epsilon_{AC}{k_2}^{B'}{}_{B}+\epsilon_{AB}q_s^{B'}{}_{C}+\epsilon_{CB}(q_s-k_2)^{B'}{}_{A}\right) \\
\mathcal{\tilde C}&=\frac{1}{2}\epsilon^{A'B'}\left(-\frac{1}{2}\epsilon_{AB}{k_3}^{C'}{}_{C}-\epsilon_{AC}q_s^{C'}{}_{B}-\epsilon_{BC}(q_s+k_3)^{C'}{}_{A}\right)
\end{align}

To summarise what we have done so far, the amplitude can be rewritten as:
\begin{align}\begin{split}
i\mathcal{M}(k_2,k_3; s)= i\mathcal{M}^{(1)}(k_2,k_3; s) ~~+~~ ( {\rm Bubbles})
\end{split}\end{align}

where the bubbles are given by $\mathcal{\tilde A},~\mathcal{\tilde B}$ or $\mathcal{\tilde C}$. We want to show that the shift dependence of the integral allows us to reproduce the same result as in the first order case. We therefore compute the overall shift dependence of the amplitude (and by extension of the bubbles):
\begin{align}
i\delta_s\mathcal{M} \equiv i\mathcal{M}(k_2,k_3; s)-i\mathcal{M}(k_2,k_3)
\end{align}

Then, we have:
\begin{align}
i\delta_s\mathcal{M} =8e^3i s^{D}_{D'} \lim_{q\rightarrow \infty} \int \frac{d\Omega_4}{(2\pi)^4}q^{D'}_{D} q^2 I(q)
\end{align}
where $q$ has been Wick rotated and $I(q)$ is the whole integrand in (\ref{amplitudeweyl}). The only contribution comes from the order $q^3$ part of the triangle diagrams, and from the order $q$ in the bubbles. Hence:
\begin{align}\begin{split}
i\delta_s\mathcal{M} = 4e^3i s^{D}_{D'}& \lim_{q\rightarrow \infty} \int \frac{d\Omega_4}{(2\pi)^4}\frac{q^{D'}_{D}}{q^2}  \Bigg[\epsilon^{B'C'}\left( \epsilon_{AB}q^{A'}_{C} +\epsilon_{AC}q^{A'}_{B}\right) \\
&+\epsilon^{A'C'}\left(\epsilon_{AB}q^{B'}_{C}+\epsilon_{CB}q^{B'}_{A}\right)+\epsilon^{A'B'}\left(-\epsilon_{AC}q^{C'}_{B}-\epsilon_{BC}q^{C'}_{A}\right) \\
&\hspace*{4cm}+2\left(q^{B'}_{C}q^{A'}_{B}q^{C'}_{A}- q^{C'}_{B}q^{B'}_{A}q^{A'}_{C}\right)\Bigg]
\end{split}\end{align}

Using the usual replacements:
\begin{align}
q^{A'}_{A}q^{B'}_{B}& \rightarrow -\frac{q^2}{4}\epsilon^{A'B'}\epsilon_{AB} \\
q^{A'}_{A}q^{B'}_{B}q^{C'}_{C}q^{D'}_{D}&\rightarrow \frac{q^4}{24}\left( \epsilon^{A'B'}\epsilon_{AB} \epsilon^{C'D'}\epsilon_{CD} +\epsilon^{A'C'}\epsilon_{AC} \epsilon^{B'D'}\epsilon_{BD} +\epsilon^{A'D'}\epsilon_{AD} \epsilon^{B'C'}\epsilon_{BC}  \right)
\end{align}

we obtain
\begin{align}
i\delta_s\mathcal{M} &= \frac{ie^3}{3\pi^2}s^{D}_{D'} \Bigg[ \epsilon^{D'A'}\epsilon^{B'C'}\epsilon_{A(B}\epsilon_{C)D} - \epsilon^{D'B'}\epsilon^{A'C'}\epsilon_{D(A}\epsilon_{C)B} + \epsilon^{D'C'}\epsilon^{A'B'}\epsilon_{D(A}\epsilon_{B)C} \Bigg] \label{epseps}\\
&= \frac{ie^3}{3\pi^2}s^{D}_{D'} \Bigg[ -\Sigma^{D'}{}_{D}^{A'}{}_{A} \cdot\Sigma^{B'}{}_{B}^{C'}{}_{C}+\Sigma^{D'}{}_{D}^{B'}{}_{B} \cdot\Sigma^{A'}{}_{A}^{C'}{}_{C}-\Sigma^{D'}{}_{D}^{C'}{}_{C} \cdot\Sigma^{A'}{}_{A}^{B'}{}_{B}\Bigg]
\end{align}

where 
\begin{align}
\Sigma^{D'}{}_{D}^{A'}{}_{A} \cdot\Sigma^{B'}{}_{B}^{C'}{}_{C} \equiv \left(\Sigma^{EF}\right)^{D'A'}_{DA}\left(\Sigma_{EF}\right)^{B'C'}_{BC}
\end{align}

and we have used the expression for the self-dual two-forms in spinorial notation:
\begin{align}
\left( \Sigma^{A'B'}_{AB}\right)^{EF} \equiv 2\epsilon^{A'B'}\delta^E_{(A}\delta^{F}_{B)}
\end{align}

We can now replace the pairs of spinor indices by their Minkowski correspondence:
\begin{align}
i\delta_s\mathcal{M} &= \frac{ie^3}{3\pi^2}s_\beta \Bigg[ -\Sigma^{\beta\mu} \cdot\Sigma^{\nu\alpha}+\Sigma^{\beta\nu} \cdot\Sigma^{\mu\alpha}-\Sigma^{\beta\alpha} \cdot\Sigma^{\mu\nu}\Bigg] \\
&= \frac{e^3}{2\pi^2}\epsilon^{\mu\nu\alpha\beta}s_\beta
\end{align}

where we used\footnote{Using the self-dual properties of the $\Sigma$s, it is also possible to show that the quantity in square braquets (written as in (\ref{epseps})) is equal to $\frac{3}{2i}\epsilon^{\mu\nu\alpha\beta}$. }
\begin{align}
\Sigma^{\mu\nu}\cdot \Sigma^{\alpha\beta} = \frac{1}{2}\left(\eta^{\mu\alpha}\eta^{\nu\beta}-\eta^{\mu\beta}\eta^{\nu\alpha} -i\epsilon^{\mu\nu\alpha\beta}\right)
\end{align}

The dependence of the shift is on the momenta $k_2,~k_3$. We consider a symmetric interchange of momenta and indices. We therefore write $s_\mu = c(k_2-k_3)_\mu$, so that:
\begin{align}
i\delta_s\mathcal{M} &= c\frac{e^3}{2\pi^2}\epsilon^{\mu\nu\alpha\beta}(k_2-k_3)_\beta
\end{align}

Because we already know how the first-order amplitude depends on the shift we can extract the shift dependence of the bubbles (it can also be read off the calculation as an intermediate step):
\begin{align}
i\delta_s \Delta = i\delta_s\mathcal{M}-i\delta_s\mathcal{M}^{(1)} = \frac{e^3}{2\pi^2}\epsilon^{\mu\nu\alpha\beta}s_\beta-\frac{e^3}{8\pi^2}\epsilon^{\mu\nu\alpha\beta}s_\beta = \frac{3e^3}{8\pi^2}\epsilon^{\mu\nu\alpha\beta}s_\beta
\end{align}

where we denoted by $\Delta$ the contribution from the bubbles. Now, in order to conclude, we notice that one can explicitly compute the value of $\Delta(s=0)$ (see Appendix \ref{apxtransition} and \ref{apxexplicit}):
\begin{align}\begin{split}
i\Delta(0) = \frac{ie^3}{16\pi^2}\Bigg[ \left( \eta^{\mu\nu}\eta^{\alpha\beta} - \eta^{\mu\beta}\eta^{\nu\alpha} - 2i \epsilon^{\mu\nu\alpha\beta}\right)&{k_3}_\beta\\& \hspace*{-1cm}+\left( \eta^{\mu\alpha}\eta^{\nu\beta} - \eta^{\mu\beta}\eta^{\nu\alpha} + 2i \epsilon^{\mu\nu\alpha\beta}\right){k_2}_\beta\Bigg]
\end{split}\end{align}

The contractions with the metric will vanish on-shell, we can therefore write:
\begin{align}
i\Delta(0) = -\frac{e^3}{8\pi^2} \epsilon^{\mu\nu\alpha\beta}({k_2}-{k_3})_\beta 
\end{align}

where the equality is valid on-shell. We can now finally write the expression for the anomaly in the second-order formalism. Since the expressions for the anomaly are given in the first-order case by Eqs.(\ref{an1}-\ref{an3}), we fix the shift dependence of these to be $c_1(k_2-k_3)$ and write:
\begin{align}
i\mathcal{M}(c_2)&= i\mathcal{M}^{(1)}(c_2) + i\Delta(c_2) \nonumber \\
&= i\mathcal{M}^{(1)}(c_1) + i\delta_{21}\mathcal{M}^{(1)}+ i\Delta(0)+ i\delta_{c_2} \Delta
\end{align}

where $i\delta_{21}\mathcal{M}^{(1)}$ is given by $\frac{e^3}{8\pi^2} \epsilon^{\mu\nu\alpha\beta}({k_2}-{k_3})_\beta( c_2-c_1)$. Therefore, the second-order amplitude at shift $c_2$ will be equal to the first-order amplitude at shift $c_1$ when 
\begin{align}
0 &= i\delta_{21}\mathcal{M}^{(1)}+ i\Delta(0)+ i\delta_{c_2} \Delta \nonumber \\
&=\frac{e^3}{8\pi^2}\epsilon^{\mu\nu\alpha\beta}(k_2-k_3)_\beta \Big[ 4c_2-c_1 -1 \Big]
\end{align}

In the end, the anomalies are equal in both formalisms, when the shifts are related by:
\begin{align}
\boxed{4c_2=c_1 +1 }\label{anomalyequiv}
\end{align}

This means that there is always a shift $c_2$ so that the anomaly is given by Eqs.(\ref{an1}-\ref{an3}). In the appendix, we compute explicitly the anomaly in the second-order formalism. There we obtain:
\begin{align}
ik_1 \cdot i\mathcal{M}(k_2,k_3;c_2)&= -\frac{ie^3}{4\pi^2}(1-4c_2)\epsilon^{\mu\nu\alpha\beta}{k_2}_\mu{k_3}_\beta \\
ik_2 \cdot i\mathcal{M}(k_2,k_3;c_2)&= -\frac{ie^3}{4\pi^2}(2c_2)\epsilon^{\mu\nu\alpha\beta}{k_2}_\nu{k_3}_\beta \\
ik_3 \cdot i\mathcal{M}(k_2,k_3;c_2)&= -\frac{ie^3}{4\pi^2}(2c_2)\epsilon^{\mu\nu\alpha\beta}{k_2}_\alpha{k_3}_\beta
\end{align}
One readily sees that using (\ref{anomalyequiv}), one obtains exactly Eqs.(\ref{an1}-\ref{an3}), hence proving the result consistent.

\subsection{Generalisation to $N$ Weyl fermions}
The result generalises easily to the case in which we are dealing with $N$ Weyl spinors with charges $eQ_i$. Choosing $c_2=1/6$ (symmetric in all channels), the anomaly is given by:
\begin{align}
ik_1 \cdot i\mathcal{M}(k_2,k_3;1/6)&= -{\rm Tr}(Q_i^3)\frac{ie^3}{12\pi^2}\epsilon^{\mu\nu\alpha\beta}{k_2}_\mu{k_3}_\beta 
\end{align}

So that the theory is anomaly-free if, for example, we are dealing with a Dirac fermion for which there are two Weyl spinors of opposite charge. The condition for an anomaly-free theory reads:
\begin{align}
{\rm Tr}(Q_i^3) = 0
\end{align}

It also generalises to the case of the axial anomaly in Electrodynamics. Indeed, the axial current arises from what can be seen as two equally charged Weyl fermions, whereas the vector current is constructed out of two oppositely charged spinors:
\begin{align}
{j_{a}}^{A'}_{A} = -2i \lambda _A D^{A'B}\lambda_B - 2i \chi _A D^{A'B}\chi_B,\quad {j_{v}}^{A'}_{A} = 2ie \lambda _A D^{A'B}\lambda_B - 2i e\chi _A D^{A'B}\chi_B
\end{align}

So that defining:
\begin{align}
Q_A = \left( \begin{array}{cc} -1 & 0 \\ 0 & -1
\end{array}\right)\quad Q_V =\left( \begin{array}{cc} 1 & 0 \\ 0 & -1
\end{array}\right)
\end{align}

The anomaly of the axial current (where the vector currents are gauge and therefore $c_2=0$ above) is given by\footnote{There is an extra minus sign coming from the fact both photons are now outgoing, so that $k_A = k_2+k_3$.}:
\begin{align}
ik_A \cdot i\mathcal{M}(k_2,k_3;0)&= {\rm Tr}(Q_A Q_V^2)\frac{ie^2}{4\pi^2}\epsilon^{\mu\nu\alpha\beta}{k_2}_\mu{k_3}_\beta = -\frac{ie^2}{2\pi^2}\epsilon^{\mu\nu\alpha\beta}{k_2}_\mu{k_3}_\beta
\end{align}

Which leads to 
\begin{align}
\langle k_2, k_3 | \partial_\mu j^\mu_a |0\rangle = - \frac{e^2}{(4\pi)^2}\epsilon^{\mu\nu\alpha\beta} \langle k_2, k_3 | F_{\mu\nu}F_{\alpha\beta} |0\rangle
\end{align}

where there is a factor of $1/2$ for each antisymmetrisation and an extra factor of $1/2$ for the 2 different contractions of the operator on the final bosonic state. This is the same result as we obtained above in the case of Dirac QED.

\section{Path-integral methods}
We now follow a non-perturbative approach to the calculation of the anomaly. In order to do so, we first look at the chiral anomaly in Dirac theory and then we will see how the fermion number anomaly can also be constructed similarly. The first three subsections are a review of the calculation that is usually carried out using non-perturbative methods. We then repeat the calculation using two-component spinors in both first- and second-order formalisms.

\subsection{First-order Dirac Lagrangian and chiral symmetry}\label{diracgammarules}
We work with the Dirac Lagrangian coupled to Electrodynamics in $3+1$ dimensions with metric $\eta_{\mu\nu}=(-,+,+,+)$. We have:
\begin{align}
\mathcal{L}_D= -i\bar\Psi \slashed D \Psi - m\bar\Psi\Psi,\quad \slashed D \Psi = (\slashed \partial +ie\slashed A)\Psi\label{dirac}
\end{align}

with $\slashed D = \gamma^\mu D_\mu$ and we have:
\begin{align}
\{ \gamma^\mu, \gamma^\nu \}= -2\eta^{\mu\nu}, \quad \left( \gamma^\mu\right)^\dagger=\gamma^0 \gamma^\mu \gamma^0, \quad \gamma_5 = i\gamma^0\gamma^1\gamma^2\gamma^3,\quad \left( \gamma_5\right)^\dagger=\gamma_5
\end{align}

The Lagrangian has the usual $U(1)$ gauge symmetry:
\begin{align}
\Psi~\mapsto~ e^{-ie\alpha(x)}\Psi, \quad A_\mu ~\mapsto~ A_\mu+\partial_\mu \alpha
\end{align}

If the Dirac fermion were massless, the above Lagrangian would be invariant under an additional global symmetry:
\begin{align}
\Psi~\mapsto~ e^{i\theta \gamma_5}\Psi
\end{align}
which is the so-called axial symmetry. The respective currents are:

\begin{align}
j^\mu = -e\bar\Psi\gamma^\mu \Psi, \quad  j^\mu_5 = \bar\Psi\gamma^\mu\gamma_5 \Psi
\end{align}
Using the equations of motion we have:
\begin{align}
\partial_\mu j^\mu = 0 , \quad \partial_\mu j^\mu_5 = -2  mi \bar \Psi \gamma_5 \Psi\label{laws}
\end{align}

Therefore, we see that, at the classical level, the gauge current is conserved and the axial current is also conserved if the fermions are massless. As we have seen, anomalies arise when the classical conservation of a current is broken by quantum corrections. For gauge theories involving Dirac fermions, the gauge symmetry is never anomalous, however we will see that the axial symmetry is.

\subsection{Euclidean path integral and chiral Jacobian}\label{anodiraceuclid}
We describe here the method developed by Fujikawa \cite{Fujikawa:1979ay} to compute the anomaly due to the chiral transformations. We continue analytically our quantities into Euclidean space such that $x_4= ix^0,~\partial_4= -i\partial_0,~\gamma^4= i\gamma^0$ and $A_4= iA^0$. We now have:
\begin{align}
\{ \gamma^\mu, \gamma^\nu \}= -2\delta^{\mu\nu}, \quad \left( \gamma^\mu\right)^\dagger=-\gamma^\mu , \quad \gamma_5 = -\gamma^1\gamma^2\gamma^3\gamma^4,\quad \left( \gamma_5\right)^\dagger=\gamma_5
\end{align}

The Euclidean path integral of the Dirac action becomes:
\begin{align}
\int \mathcal{D}\bar\Psi \mathcal{D}\Psi \exp \left[ -\int d^4x\bar\Psi (i\slashed{D}+m) \Psi \right]
\end{align}

where $\slashed{D}$ is a Hermitian operator. In order to analyse the Jacobian for the chiral transformation, we expand the Dirac fields into a basis of eigenfunctions of the latter:
\begin{align}
\Psi(x) &= \sum_n a_n \varphi_n(x) = \sum_n \langle x| n\rangle a_n, \\
\bar\Psi(x) &= \sum_n \bar{b}_n \varphi^\dagger_n(x) = \sum_n \bar{b}_n \langle n| x\rangle 
\end{align}
 
with:
\begin{align}
\slashed{D} \varphi_n(x)= \lambda_n \varphi_n(x),\\
\int d^4x \varphi^\dagger_n(x)\varphi_m(x) = \delta_{nm}
\end{align}

This basis formally diagonalises the Dirac action:
\begin{align}
\int d^4x\bar\Psi (i\slashed{D}+m) \Psi  = \lim_{N\rightarrow\infty} \sum_{n=1}^N (i\lambda_n+m)\bar{b}_na_n,
\end{align}

where the sum runs over the non-vanishing eigenvalues. Similarly, the measure transforms into:
\begin{align}\begin{split}
\mathcal{D}{\bar\Psi}\mathcal{D}\Psi &= \left[ {\rm det}\langle n| x\rangle {\rm det} \langle x| n\rangle \right]^{-1} \lim_{N\rightarrow\infty}\prod_{n=1}^{N} d\bar{b}_n da_n \\
&= \left( {\rm det}\int d^4 x \langle n| x\rangle \langle x| m\rangle \right)^{-1} 
\lim_{N\rightarrow\infty}\prod_{n=1}^{N} d\bar{b}_n da_n \\
&=\left( {\rm det}\,\delta_{nm} \right)^{-1} \lim_{N\rightarrow\infty}\prod_{n=1}^{N} d\bar{b}_n da_n \\
&=  \lim_{N\rightarrow\infty}\prod_{n=1}^{N} d\bar{b}_n da_n 
\end{split}\end{align}

We can use this definition of the path integral measure to carry out our calculations. Let us start by considering the Jacobian for a local infinitesimal chiral transformation:
\begin{align}
\Psi(x)~&\mapsto~ \Psi'(x)= e^{i\theta(x) \gamma_5}\Psi(x) = (1+i\theta(x)\gamma_5) \Psi(x), \\
\bar\Psi(x)~&\mapsto~ \bar\Psi'(x)=\bar\Psi(x)e^{i\theta(x) \gamma_5} =  \bar\Psi(x)(1+i\theta(x)\gamma_5)
\end{align}

In order to see how the coefficients transform, we expand the fields in the above equation in terms of the eigenfunctions:
\begin{align}\begin{split}
\Psi'(x)&= \sum_n a'_n \varphi_n(x) \\
&=\sum_n(1+i\theta(x)\gamma_5) a_n\varphi_n(x)
\end{split}\end{align}

So that 
\begin{align}\begin{split}
a'_n = \sum_m(\delta_{nm}+i\int d^4x \varphi_n^\dagger(x)\theta(x)\gamma_5\varphi_m(x)) a_m
\end{split}\end{align}

and similarly:
\begin{align}\begin{split}
\bar{b}'_n = \sum_m \bar{b}_m (\delta_{mn}+i\int d^4x \varphi_m^\dagger(x)\theta(x)\gamma_5\varphi_n(x)) 
\end{split}\end{align}

From these expressions, we obtain that the measure transforms as (keeping $N$ finite as a mode cut-off):
\begin{align}
\begin{split}
\prod_{n=1}^{N} d\bar{b}'_n da'_n& = {\rm det}\left[\delta_{mn} +i\int d^4x \varphi_m^\dagger(x)\theta(x)\gamma_5\varphi_n(x)\right] ^{-1}   \prod_{n=1}^{N} d\bar{b}_n  \\
& \times {\rm det}\left[\delta_{nm} +i\int d^4x \varphi_n^\dagger(x)\theta(x)\gamma_5\varphi_m(x)\right] ^{-1}  \prod_{n=1}^{N} da_n 
\end{split}
\end{align}

Using the relation for an infinitesimal matrix $\delta M$:
\begin{align}
{\rm det} (1+\delta M) = \exp {\rm tr} \ln (1+\delta M) = \exp {\rm tr}\, \delta M
\end{align}

we have:
\begin{align}
\begin{split}
\prod_{n=1}^{N} d\bar{b}'_n da'_n& =\exp \left[ -2i \sum _{n=1}^N\int d^4x \varphi_n^\dagger(x)\theta(x)\gamma_5\varphi_n(x)\right]  \prod_{n=1}^{N} d\bar{b}_n da_n 
\end{split}
\end{align}

We want to replace the mode cut-off ($N\rightarrow\infty$) by a cut-off in terms of eigenvalues. This proceeds as follows:
\begin{align}\begin{split}
\lim_{N\rightarrow\infty}\sum _{n=1}^N\int d^4x \varphi_n^\dagger(x)\theta(x)\gamma_5\varphi_n(x) &= \lim_{M\rightarrow\infty}\sum _{n=1}^\infty\int d^4x \varphi_n^\dagger(x)\theta(x)\gamma_5\mathcal{F}(\lambda_n^2/M^2)\varphi_n(x)\\
&= \lim_{M\rightarrow\infty}\sum _{n=1}^\infty\int d^4x \varphi_n^\dagger(x)\theta(x)\gamma_5\mathcal{F}(\slashed{D}^2/M^2)\varphi_n(x) \\
&= \lim_{M\rightarrow\infty}{\rm Tr}\theta(x)\gamma_5\mathcal{F}(\slashed{D}^2/M^2)
\end{split}\end{align}

where $\mathcal{F}(x)$ is an arbitrary smooth regulator that rapidly approaches 0 as $x\rightarrow \infty$ and $\mathcal{F}(0)=1$. For the choice $\mathcal{F}(x)= e^{-x}$, we obtain results related to the heat-kernel or $\zeta$-function regularisation. In terms of the regulator, the Jacobian for the chiral transformations is given by:
\begin{align}
J= \exp \left[-2i \lim_{M\rightarrow\infty}{\rm Tr}\theta(x)\gamma_5\mathcal{F}(\slashed{D}^2/M^2)\right] \equiv \exp \left[{i\int d^4x\theta(x)\mathcal{A}(x)} \right]
\end{align}

where the function $\mathcal{A}(x)$ is the anomaly.

The anomaly produces, here in a quantum regime, the same modification of theory at the level of the action as would a Lagrangian that is not classically invariant but is instead shifted by the same anomaly function when performing an axial transformation of its content. Thereby, when we use an effective Lagrangian with the fermions integrated out ($i.e.$ when their path-integral has been solved), one should add to the Lagrangian a term transforming accordingly to take into account the anomaly:
\begin{align}
\mathcal{L}_{eff} ~\rightarrow \mathcal{L}_{eff} + \theta(x)\mathcal{A}(x)
\end{align}
This is, for example, what happens in the effective theory description of the pions. 

Before proceeding with the computation of the anomaly, let us come back to our conservations laws (\ref{laws}). An infinitesimal axial transformation will change the partition function of the theory as:
\begin{align}
\delta_\theta(\mathcal{Z}) ~\mapsto ~ \int \mathcal{D}\bar\Psi \mathcal{D}\Psi \left[ i\int d^4x \left(\theta(x)\mathcal{A}(x) + j^\mu_5 \partial_\mu \theta(x)\right)\right] e^{-S_E}
\end{align}
And therefore we have the anomalous conservation of the axial current:
\begin{align}
\partial_\mu \langle j^\mu_5\rangle = \mathcal{A}(x)
\end{align}
Notice that this equation does not depend on any loop-expansion or perturbative definition. It is therefore clearly a consequence that depends on the non-perturbative aspects of the theory.

\subsection{Heat-kernel regularisation}
We now explicitly calculate the anomaly using a decreasing exponential as regulator. We then have:
\begin{align}\begin{split}
\mathcal{A}(x) &=\lim_{M^2\rightarrow\infty}-2{\rm tr}\left[ \sum_n e^{-\lambda^2_n/M^2} \left(\varphi_n^\dagger(x)\gamma_5\varphi_n(x)\right)\right] \\
&=\lim_{M^2\rightarrow\infty}-2{\rm tr}\left[ \sum_n  \left(\varphi^\dagger_n(x)\gamma_5 e^{-\slashed{D}^2/M^2} \varphi_n(x)\right)\right] 
\end{split}\label{tracegamma5}
\end{align}

Fujikawa's method proceeds now as follows: as the eigenfunctions of the Dirac operator also satisfy the Klein-Gordon equation, they admit a plane-wave decomposition; then expand the Dirac operator such as to obtain:
\begin{align}
\slashed{D}^2 = -D^2 + E,\quad {D_\mu}= \partial_\mu + ieA_\mu, \quad E= ie\sigma^{\mu\nu}F_{\mu\nu}
\end{align}
with $\sigma_{\mu\nu}= \frac{1}{4}\left[\gamma_\mu,\gamma_\nu \right]$ (this convention is different from the one usually used, but will become clear later on). We then have:
\begin{align}\begin{split}
\mathcal{A} (x) &= \lim_{M^2\rightarrow\infty}-2\int\frac{d^4k}{(2\pi)^4}{\rm tr}\left[\gamma_5 e^{-ikx}e^{({D}^2 - E)/M^2}e^{ikx}\right]\\
&= \lim_{M^2\rightarrow\infty}-2\int\frac{d^4k}{(2\pi)^4}{\rm tr}\left[\gamma_5 e^{(({D}+ik)^2 - E)/M^2}\right]
\end{split}
\end{align}

We now rescale the momentum by a factor of $M$ such that $k\mapsto Mk$. Then, in the limit where $M$ goes to infinity, one can Taylor expand the exponential around $k^2$ and the only terms that will contribute will be of order $\mathcal{O}(M^{-4})$ or less. Moreover, since we are tracing the matrices in the exponent with $\gamma_5$, only traces involving at least four extra gamma matrices will be non-vanishing. This leaves us with one single term:
\begin{align}\begin{split}
\mathcal{A}(x) 
&= -\int\frac{d^4k}{(2\pi)^4}e^{-k^2}{\rm tr}\left[\gamma_5E^2\right]
\end{split}
\end{align}

Using that, in Euclidean signature
\begin{align}
{\rm tr} \Big[ \gamma_5 \Big[\gamma^\mu,\gamma^\nu \Big]\Big[\gamma^\alpha,\gamma^\beta \Big] \Big] = -16 \epsilon^{\mu\nu\alpha\beta}
\end{align}

with $\epsilon^{1234}=+1$. We have finally:
\begin{align}
\begin{split}
\mathcal{A} (x) &= -\frac{e^2}{16\pi^2}\epsilon^{\mu\nu\alpha\beta}F_{\mu\nu}F_{\alpha\beta}
\end{split}
\end{align}

To obtain the anomaly in Minkowski space, we need to analytically continue each step of the derivation. This will bring up a factor of $i$ from the Wick rotation of the momentum integral, a second factor of $i$ from the trace of the gamma matrices and a minus sign for changing the orientation to $\epsilon^{0123}=+1$. In the end, the axial anomaly is given by:
\begin{align}
\begin{split}
\mathcal{A}_M (x) &= -\frac{e^2}{16\pi^2}\epsilon^{\mu\nu\alpha\beta}F_{\mu\nu}F_{\alpha\beta}
\end{split}
\end{align}

where we denoted by a subscript ``M'' the anomaly continued to Minkowski spacetime. Let us now shortly come back to (\ref{tracegamma5}). Since $\gamma_5$ anticommutes with the Dirac operator, for each eigenfunction $\varphi_n$ with eigenvalue $\lambda_n\not=0$, there will be an eigenfunction $\varphi_{-n}$ with eigenvalue $-\lambda_n$ given by $\varphi_{-n}=\gamma_5\varphi_n$.
Since these eigenfunctions have different eigenvalues, they are orthonormal. In particular they can not be eigenfunctions of $\gamma_5$ because of the relation $\varphi_{-n}=\gamma_5\varphi_n$. Nevertheless, it is possible to decompose them such that:
\begin{align}
\varphi_n^\pm = \frac{1}{2}(1\pm\gamma_5) \varphi_n,\quad \gamma_5\varphi_n^\pm=\pm\varphi_n^\pm,\quad \slashed{D}^2 \varphi_n^\pm = \lambda_n^2\varphi_n^\pm
\end{align}

Then, both $\varphi_n^+$ and $\varphi_{n}^-$ will appear in the sum with the same exponential prefactor but with different chirality eigenvalues, thus cancelling in the sum. What remains are the eigenfunctions of the Dirac operator with null eigenvalue. In this case, since both $\varphi_0$ and $\gamma_5\varphi_0$ live in the same eigensubspace we can diagonalise $\gamma_5$ in the latter such that:
\begin{align}
\gamma_5\varphi_i^\pm = \pm \varphi_i^\pm 
\end{align}

Notice however that for $\lambda=0$, eigenfunctions of opposite chirality do not necessarily come in pairs (indeed for a general Lagrangian, we could have several left-handed Weyl fermions and no right-handed partner). If $n_\pm$ counts the number of eingenfunctions with positive/negative eigenvalues respectively, we have:
\begin{align}\begin{split}
\mathcal{A}(x) &=\lim_{M^2\rightarrow\infty}-2{\rm tr}\left[ \sum_n e^{-\lambda^2_n/M^2} \left(\varphi_n(x)^\dagger\gamma_5\varphi_n(x)\right)\right] 
\end{split} \\
&=-2\left( \sum_{i=1}^{n^+} (\varphi_i^+)^\dagger\varphi_i^+- \sum_{i=1}^{n^-} (\varphi_i^-)^\dagger\varphi_i^-\right)
\end{align}

And therefore:
\begin{align}
\int d^4 x \mathcal{A}(x) = -2(n^+ -n^-) = -2 ~{\rm index} (\slashed{D})
\end{align}
This is an example of the Atiyah-Singer index theorem, the anomaly is proportional to the index of the Dirac operator.\\

Last but not least, we give here a similar way to compute the anomaly. Recall:
\begin{align}\begin{split}
\mathcal{A}(x) &= \lim_{M^2\rightarrow\infty}-2{\rm tr}\left[\gamma_5 e^{-\slashed{D}^2/M^2}\right]
\end{split}
\end{align}
Now, this quantity can be related to the heat kernel coefficients of the effective action derived from the Dirac Lagrangian, \cite{Vassilevich:2003xt}:
\begin{align}
{\rm Tr} \left[f e^{-t\slashed{D}^2}  \right] \simeq \sum_{k\geq 0}t^{(k-D)/2}a_k(f,\slashed{D}^2)
\end{align}

where the trace denotes also the integration over the Euclidean space. By making the replacement $1/M^2 \rightarrow t$, we have:
\begin{align}
\int d^4 x\mathcal{A}(x) = \lim_{t\rightarrow 0}-2{\rm Tr}\left[\gamma_5 e^{-t\slashed{D}^2}\right] = a_D(-2\gamma_5,\slashed{D}^2)
\end{align}

In four dimensions, we have:
\begin{align}
a_4(f,\slashed{D}^2)= \frac{1}{(4\pi)^2}{\rm Tr}\left[ f\cdot\left( \frac{D^2E}{6}+ \frac{E^2}{2}+\frac{\Omega_{\mu\nu}\Omega^{\mu\nu}}{12}\right)\right]
\end{align}

where $\Omega_{\mu\nu}= ieF_{\mu\nu}$ here. So that, keeping only the non-vanishing trace:
\begin{align}
\int d^4 x\mathcal{A}(x) = \frac{1}{(4\pi)^2}{\rm Tr}\left[ -2\gamma_5 \cdot \frac{E^2}{2}\right] = -\frac{e^2}{16\pi^2}\int d^4 x~ \epsilon^{\mu\nu\alpha\beta}F_{\mu\nu}F_{\alpha\beta}\label{anofirstPI}
\end{align}

This way of computing the anomaly will turn out to be useful when dealing with two component spinors.

\subsection{First-order Lagrangian and symmetries}

We now show briefly how the calculation proceeds when dealing with two-component spinors. The Dirac Lagrangian is:
\begin{align}\label{Dirac-L2}\begin{split}
{\cal L}_{D} &= \im \chi^\dagger  \bar D \chi +\im \xi^\dagger  D \xi - m(\chi \xi+ \chi^\dagger \xi^\dagger)\\
 &= \im \sqrt{2} \chi^\dagger_{A'}\bar{D}^{A'}{}_{A}\chi^A +\im \sqrt{2} \xi^{\dagger A'}D_{A'}{}^{A}\xi_A - m(\chi^A \xi_A + \chi^\dagger_{A'} \xi^{\dagger A'})
 \end{split}\end{align}
The electromagnetic $U(1)$ transformations are given by
\begin{alignat}{4}
 & \delta\xi= +ie\alpha \xi, && \delta\chi= -ie\alpha \chi\\ 
  &D^{A'A}\xi_A= (\partial^{A'A} - ieA^{A'A})\xi_A, &\qquad &  \bar{D}^{A'A}\chi_A= (\partial^{A'A} + ieA^{A'A})\chi_A
\end{alignat}

and we have\begin{spacing}{1.2}
\begin{align}
\Psi = \left( \begin{array}{c} \chi \\ \xi^\dagger \end{array}\right)
\end{align}\end{spacing}

On the other hand, chiral transformations are given by
\begin{align}
\delta\xi= -i\alpha \xi, \quad   \delta\chi= -i\alpha \chi\label{chiralt}
\end{align}
We recall the Dirac equations for the four spinors\footnote{Recall that the adjoint of a differential operator is defined through integration by parts and complex conjugation.}:
\begin{alignat}{3}
i D\xi &=-m\chi^\dagger ,\quad i{D}^\dagger \chi^\dagger = -m\xi \\
i\bar{D} \chi &= m\xi^\dagger ,\quad   i\bar{D}^\dagger \xi^\dagger = m\chi
\end{alignat}

Using the Euclidean conjugation defined in the Appendix \ref{appendix_euclidean}, there are four gauge-covariant, self-adjoint operators that we can define:
\begin{align}\begin{split}
{D}^\dagger D =-2(\partial - ieA)_{A}{}^{A'} (\partial - ieA)_{A'}{}^{B},\quad \bar{D}^\dagger \bar{D} =-2(\partial + ieA)^{A}{}_{A'} (\partial + ieA)^{A'}{}_B \\
D {D}^\dagger =-2(\partial - ieA)_{A'}{}^{A}(\partial - ieA)_{A}{}^{B'} ,\quad   \bar{D} \bar{D}^\dagger  =-2(\partial + ieA)^{A'}{}_{A} (\partial + ieA)^A{}_{B'}
\end{split}\end{align}

They are self-adjoint in the sense that, $e.g.$:
\begin{align}
\langle \lambda_1 |D^\dagger D \lambda_2 \rangle = \int d^4 x \hat \lambda_1 D^\dagger D \lambda_2 =  \int d^4 x \widehat{\left(D^\dagger D  \lambda_1\right)} \lambda_2 = \langle D^\dagger D \lambda_1 |\lambda_2 \rangle
\end{align}

Where we used integration by parts and the antilinearity of the conjugation. We then define four complete orthonormal basis of eigenfunctions:
\begin{align}
\begin{split}
D^\dagger D\phi_{L\,n}(x) = \lambda_n^2 \phi_{L\,n}(x), \quad DD^\dagger \phi_{R\,n}(x) = \lambda_n^2 \phi_{R\,n}(x) \\
\bar{D}^\dagger \bar{D}\varphi_{L\,n}(x) = \kappa_n^2 \varphi_{L\,n}(x), \quad \bar{D}\bar{D}^\dagger \varphi_{R\,n}(x) = \kappa_n^2 \varphi_{R\,n}(x)
\end{split}
\end{align}

The relation between the eigenfunctions is chosen to be:
\begin{align}
\begin{split}
D\phi_{L\,n}(x) = {\lambda_n} \phi_{R\,n}(x),\quad \bar{D}\varphi_{L\,n}(x) = {\kappa_n}\varphi_{R\,n}(x)
\end{split}
\end{align}
with $\lambda_n,~\kappa_n >0$ for $\lambda_n,~\kappa_n \neq 0$, so that $D^\dagger D $ and $DD^\dagger$ (as well as the other pair) share the same number of non-vanishing eigenvalues. We can also normalise the functions so that $\hat \phi \sim \varphi$ and $\kappa_n = \lambda_n$. We then expand our fields as\footnote{$\xi^\dagger$ can be decomposed in the dual basis of eigenfuntions of $DD^\dagger$ as the terms $\int d^4x\xi^\dagger (x)DD^\dagger \phi_{R}(x)$ are then $SU(2)$ invariant and we still have $\int d^4x\xi^\dagger (x)DD^\dagger \phi_{R}(x)=\int d^4x (\bar{D}\bar{D}^\dagger\xi^\dagger (x)) \phi_{R}(x)\sim  \sum\int d^4 x (\bar{D}\bar{D}^\dagger\hat \phi_R (x)) \phi_{R\,n}(x)\sim\sum\int d^4 x\widehat{({D}{D}^\dagger\phi_R (x))} \phi_{R}(x)  $.}\footnote{Note that we could also have expanded $\xi^\dagger \sim \sum \varphi_R$ and $\chi^\dagger \sim \sum \phi_R$ as they are eigenfunctions of the corresponding operators. Similarly for the unprimed spinors, we could have chosen the basis corresponding to the dual of the basis of their operator integrated by parts: $\xi \sim \sum \varphi_{L}^\dagger$ and $ \chi \sim \sum \phi_{L}^\dagger$.}:
\begin{align}
\begin{split}
\xi(x) = \sum_n a_n \phi_{L\,n}(x), \quad \chi(x) = \sum_n b_n \varphi_{L\,n}(x) \\
\xi^\dagger(x) = \sum_n \bar{a}_n \hat \phi_{R\,n}(x), \quad \chi^\dagger(x) = \sum_n \bar{b}_n \hat\varphi_{R\,n}(x)
\end{split}
\end{align}

As before, we can compute the Jacobian for the transformation of the measure:
\begin{align}
\begin{split}
\prod_{n=1}^{N} d\bar{b}'_n d\bar{a}'_n db'_nda'_n& =\exp \Big[ i \sum _{n=1}^N\int d^4x \alpha(x) \left(\hat \phi_{L\,n}(x)\phi_{L\,n}(x)+ \hat \varphi_{L\,n}(x)\varphi_{L\,n}(x)\right.  \\
 &- \left. \hat \phi_{R\,n}(x)\phi_{R\,n}(x)- \hat \varphi_{R\,n}(x)\varphi_{R\,n}(x) \right) \Big]  \prod_{n=1}^{N} d\bar{b}_n d\bar{a}_n db_n da_n
\end{split}
\end{align}

Using as before the appropriate regulator for each sum over modes, we have for the anomaly:
\begin{align}\begin{split}
\mathcal{A}(x) &= \lim_{M^2\rightarrow\infty}\left(+{\rm tr}\left[ e^{-D^\dagger D /M^2}\right]+{\rm tr}\left[ e^{-\bar {D}^\dagger \bar D /M^2}\right]-{\rm tr}\left[ e^{-D {D}^\dagger /M^2}\right]-{\rm tr}\left[ e^{-\bar{D}\bar{D}^\dagger /M^2}\right]\right)
\end{split}
\end{align}

where the operators have been appropriately continued to Euclidean space as before. We then rewrite our operators as:
\begin{align}
D^\dagger D = -{D}^2  \underbrace{-ie\Sigma^{\mu\nu}F_{\mu\nu}}_{E},\quad D {D}^\dagger = -D^2  \underbrace{-ie\bar{\Sigma}^{\mu\nu}F_{\mu\nu}}_{\bar E}
\end{align}
\begin{align}
\bar{D}^\dagger \bar{D} = -{\bar{D}}^2  -E,\quad \bar{D} \bar{D}^\dagger = -\bar{D}^2  -{\bar E}
\end{align}
where $\Sigma$ and $\bar\Sigma$ can be identified as the 't Hooft symbols once properly rescaled (in Minkowski spacetime they also correspond to the basis of self-dual and anti-self-dual two forms). We then have:
\begin{align}
\int d^4 x\mathcal{A}(x) = \frac{1}{(4\pi)^2}\left({\rm Tr}\left[ +2 \cdot \left(\frac{E^2}{2}+\frac{\Omega_{\mu\nu}\Omega^{\mu\nu}}{12}\right)\right] +{\rm Tr}\left[ -2 \cdot \left(\frac{\bar{E}^2}{2}+\frac{\bar{\Omega}_{\mu\nu}\bar{\Omega}^{\mu\nu}}{12}\right)\right]\right)
\end{align}

Using (in Euclidean signature):
\begin{align}
{\rm tr} E^2 &= -\frac{e^2}{2}\left(\eta^{\mu\alpha}\eta^{\nu\beta}-\eta^{\mu\beta}\eta^{\nu\alpha} +\epsilon^{\mu\nu\alpha\beta}\right)F_{\mu\nu}F_{\alpha\beta}\\
{\rm tr} \bar{E}^2 &= -\frac{e^2}{2}\left(\eta^{\mu\alpha}\eta^{\nu\beta}-\eta^{\mu\beta}\eta^{\nu\alpha} -\epsilon^{\mu\nu\alpha\beta}\right)F_{\mu\nu}F_{\alpha\beta}
\end{align}
and:
\begin{align}
\Omega_{\mu\nu} = \pm ieF_{\mu\nu},\quad \bar{\Omega}_{\mu\nu} = \pm ieF_{\mu\nu}
\end{align}
We obtain finally:
\begin{align}
\int d^4 x\mathcal{A}(x) = -\frac{e^2}{(4\pi)^2}\int d^4 x~\epsilon^{\mu\nu\alpha\beta}F_{\mu\nu}F_{\alpha\beta} 
\end{align}

which is the same result as (\ref{anofirstPI}), as expected.

\subsection{Axial anomaly in the second-order formalism}

%When one is dealing with the anomalous conservation of the global axial symmetry, the heat kernel method, where one treats the photon field as a background upon which the rest of the fields fluctuate, can be used. This is the case of second-order QED which consists of two EM coupled Weyl fermions. 
We now proceed with the same calculation, but in the second-order formalism. Recall that the Lagrangian for Dirac fermions in the second-order formalism is given by:
\begin{align}\begin{split}
{\cal L} &=  -(\bar{D} \chi )^T( D \xi) - m^2 \chi\xi \\
&= -2(\bar{D}^{A'}{}_{ A}\chi^A )(  D_{A'}{}^{B} \xi_B ) - m^2 \chi^A\xi_A \label{lagchiral2}.
\end{split}\end{align}

This should be supplemented with the reality conditions:
\begin{align}
m\xi^{\dagger} = \im (\bar{D}\chi)^T , \qquad m\chi^{\dagger} = \im D\xi. \label{reality}
\end{align}

Noticing that $\bar{D}$ integrated by parts gives $D^\dagger$, the field equations that result from the above Lagrangian are
\begin{align}\label{feqs-2nd-order}
(D^\dagger D- m^2) \xi=0,\quad (\bar {D}^\dagger \bar D - m^2) \chi=0.
\end{align}
We see that the Lagrangian is invariant under the usual $U(1)$ transformations
\begin{align}
 \delta\xi= +ie\alpha \xi, \qquad \delta\chi= -ie\alpha \chi, \qquad \delta A_{AA'}=\partial_{AA'} \alpha.
\end{align}
However, the Lagrangian is explicitly {\it not} invariant under the (local or global) chiral transformations
\begin{align}
 \delta_{chiral}\xi=-i\alpha \xi, \qquad \delta_{chiral}\chi= -i\alpha \chi
 \end{align}
 that act on both spinor fields in the same way. Note that the electromagnetic potential is not transformed. We have:
\begin{align}
\delta_{chiral} \mathcal{L}  =  -2i\alpha (x) \mathcal{L}  + j^{5\,\mu} \partial_{\mu} \alpha(x),
\end{align}
where\footnote{Recall $D= \sqrt{2}\theta^\mu D_\mu$.}
\begin{align}
j^{5\,\mu} := \sqrt{2}i\left( (\theta^\mu\chi)^T D \xi +  (\bar{D}\chi)^T (\theta^\mu\xi) \right).
\end{align}
This expression should be compared to the usual $U(1)$ current
\begin{align}
j^\mu = \sqrt{2}ie\left( (\theta^\mu\chi)^T D \xi -  (\bar{D}\chi) ^T(\theta^\mu\xi) \right)
\end{align}
At the classical level, using the equations of motion, the gauge current is conserved $D_\mu j^\mu=0$, whereas the axial current is not:
\begin{align}
D_\mu j^{5\,\mu}=-2i ~ {\cal L}. \label{classicalcons}
\end{align}

In the massless limit the field equations (\ref{feqs-2nd-order}) are invariant under the global chiral transformations. So are the reality conditions (\ref{reality}). In this limit the right-hand-side of (\ref{classicalcons}) becomes
\begin{align}
D_\mu j^{5\,\mu}=2i(\bar{D} \chi )^T( D \xi),
\end{align}
which vanishes on the surface of the reality conditions (\ref{reality}). It is only in this sense that the massless theory is invariant under the chiral transformations. 

To see what becomes of the axial current conservation in the quantum theory, let us consider the effect of the local chiral transformation on the path integral. We have:
\begin{align}
Z= \int \mathcal{D}\xi \mathcal{D}\chi \exp \left[ i\int d^4x \left( \mathcal{L}+ \alpha(x)\left\{ \mathcal{A}^{chiral}_M-2i\mathcal{L} -\partial_{\mu}j^{5\,\mu}\right\} \right) \right],
\end{align}
where $\mathcal{A}^{chiral}_M$ is the coming from non-invariance of the (chiral half of the) integration measure. We thus see that the usual quantum non-conservation of the axial current is replaced in our case by:
\begin{align}
\partial^\mu\langle j^5_\mu\rangle =  \mathcal{A}^{chiral}_M-2i\langle \mathcal{L} \rangle\equiv \mathcal{A}_M,
\end{align}
where we have introduced the notation $\mathcal{A}_M$ for the full anomaly. We will proceed as before to compute the anomaly. We have the same four complete basis of eigenfunctions:
\begin{align}
\begin{split}
D^\dagger D\phi_{L\,n}(x) = \lambda_n^2 \phi_{L\,n}(x), \quad DD^\dagger \phi_{R\,n}(x) = \lambda_n^2 \phi_{R\,n}(x) \\
\bar{D}^\dagger \bar{D}\varphi_{L\,n}(x) = {\lambda_n^2} \varphi_{L\,n}(x), \quad \bar{D}\bar{D}^\dagger \varphi_{R\,n}(x) = {\lambda_n^2} \varphi_{R\,n}(x)
\end{split}
\end{align}
with the relation between the eigenfunctions:
\begin{align}
\begin{split}
D\phi_{L\,n}(x) = {\lambda_n} \phi_{R\,n}(x),\quad \bar{D}\varphi_{L\,n}(x) = {\lambda_n}\varphi_{R\,n}(x)
\end{split}
\end{align}
Together with the equations of motion, that we recall:
\begin{align}
(D^\dagger D- m^2) \xi=0,\quad (\bar {D}^\dagger \bar D - m^2) \chi=0.
\end{align}

and the reality conditions:
\begin{align}\label{realDano}
m\xi^{\dagger} = \im (\bar{D}\chi)^T , \qquad m\chi^{\dagger} = \im D\xi
\end{align}

%which together, imply:
%
%
%
%
%
%\begin{align}
%( DD^\dagger- m^2) \chi^\dagger=0,\quad (\bar D\bar {D}^\dagger  - m^2) \xi^\dagger=0.
%\end{align}

We expand our fields as follows:
\begin{align}
\begin{split}
\xi(x) = \sum_n a_n \phi_{L\,n}(x), \quad \chi(x) = \sum_n b_n  \hat\phi_{L\,n}(x)
\end{split}
\end{align}
%\\ \xi^\dagger(x) = \sum_n \bar{a}_n \varphi_{R\,n}(x), \quad \chi^\dagger(x) = \sum_n \bar{b}_n \hat\varphi_{R\,n}(x)

where the hat denotes the dual basis and we have the relation:
\begin{align}
\begin{split}
\bar{D}\hat\phi_{L\,n}(x) = {\lambda_n}\hat\phi_{R\,n}(x)
\end{split}
\end{align}
%,\quad {D}\hat\varphi_{L\,n}(x) = {\lambda_n}\hat\varphi_{R\,n}(x)

This formally diagonalises the Lagrangian:
\begin{align}
\int d^4x \mathcal{L} = -\sum_{\lambda_n \neq 0}^\infty b_n a_n\lambda_n^2
\end{align}

Now, for the anomaly, we can calculate the change in the measure as we have previously done, expanding the chiral transformations on the basis of eigenfunctions and extracting the transformation of the coefficients, we obtain in the end:
\begin{align}
\begin{split}
\prod_{n=0}^{N}  db'_nda'_n& =\exp \Big[ 2i \sum _{n=0}^N\int d^4x \alpha(x) \left(\hat\phi_{L\,n}(x)\phi_{L\,n}(x) \right) \Big]  \prod_{n=1}^{N} db_n da_n\label{changemeas1}
\end{split}
\end{align}

Similarly, it is straightforward to extract the contribution coming from the Lagrangian, since it is proportional to the latter and is given by\footnote{Intuitively, in the LHS, only the $(N-1)^{th}$ term of the Taylor expansion of the exponential contributes to the integral, while in the RHS, we used the prefactor as the $N^{th}$ term, and then, after an $N/N$ normalisation, we rewrote the whole as an exponential, thereby leaving the sum over modes.}:
\begin{align}
\begin{split}
\int \prod_{n=0}^{N}  db_nda_n ~\Big[ (-2i\alpha) \sum _{\lambda_n\neq 0 }^N b_n a_n\lambda_n^2 \Big] ~ \exp\left[\sum_{\lambda_n\neq 0 }^N b_n a_n\lambda_n^2 \right]  \\
=\int \prod_{n=0}^{N}  db_nda_n ~\Big[ (-2i\alpha)  \sum _{\lambda_n\neq 0 }^N 1 \Big] ~ \exp\left[\sum_{\lambda_n\neq 0 }^N b_n a_n\lambda_n^2 \right] 
\end{split}
\end{align}

We have to carefully think about the sum over modes that we have on the RHS. If we put together the variation of the measure and the contribution coming from the Lagrangian, we have:
\begin{align}
\begin{split}
\delta Z &= \int \prod_{n=0}^{N}  db_nda_n ~\Big[ (2i\alpha) \left(  \sum _{n=0}^N -\sum _{\lambda_n\neq 0 }^N \right)\Big] ~ \exp\left[\sum_{\lambda_n\neq 0 }^N b_n a_n\lambda_n^2 \right]  \\
&=\int \prod_{n=0}^{N}  db_nda_n ~\Big[ (2i\alpha)n^0_L \Big] ~ \exp\left[\sum_{\lambda_n\neq 0 }^N b_n a_n\lambda_n^2 \right] 
\end{split}
\end{align}

where $n^0_L$ is the number of zero modes of the unprimed quadratic operator. This is a piece of the Atiyah-Singer index theorem which states (we recall):
\begin{align}
\int d^4x \mathcal{A}(x) = 2(n^0_L - n^0_R) =2( \dim\ker D - \dim\ker D^\dagger ) = 2 {\rm ind}~ D
\end{align}

It seems that we are missing the bit coming from $\dim\ker D^\dagger $. To understand what happens, we must go back to the first-order formulation in terms of grassmann coefficients. Recall the Lagrangian:
\begin{align}\begin{split}
{\cal L}_{D} &= \im \chi^\dagger  (\bar D \chi)^T +\im \xi^\dagger  D \xi - m(\chi \xi+ \chi^\dagger \xi^\dagger)
 \end{split}\end{align}
 
with the expansion in modes
 \begin{align}
\begin{split}
\xi(x) = \sum_n a_n \hat \varphi_{L\,n}(x), \quad \chi(x) = \sum_n b_n \varphi_{L\,n}(x) \\
\xi^\dagger(x) = \sum_n \bar{a}_n \varphi_{R\,n}(x), \quad \chi^\dagger(x) = \sum_n \bar{b}_n \hat\varphi_{R\,n}(x)
\end{split}
\end{align}
 
The Lagrangian is then:
\begin{align}
\int d^4x \mathcal{L} = \sum_{\lambda_n\neq 0 } \Big[\left(\bar{a}_n a_n + \bar{b}_n b_n \right)i\lambda_n -\left( b_n a_n + \bar{b}_n \bar{a}_n \right) m \Big]  - m\left(\sum_{i= 1}^{n_L^0}b^0_i a^0_i +\sum_{i= 1}^{n_R^0} \bar{b}^0_i \bar{a}^0_i \right) 
\end{align}

where the first terms in square brackets only contain non-zero modes and the last two terms sum over left and right handed zero modes. The path-integral formulation is then given by exponentiating the above Lagrangian and integrating on the measure:
\begin{align}
 \left(\prod_{n\neq 0} d\bar{b}_n d\bar{a}_n\right) \left(\prod_{n\neq 0} d{b}_n d{a}_n\right)d\bar{b}^0_1\dots d\bar{b}^0_{n^0_R}d\bar{a}^0_1\dots d\bar{a}^0_{n^0_R}d{b}^0_1\dots d{b}^0_{n^0_L}d{a}^0_1\dots d{a}^0_{n^0_L}
\end{align}

where $\prod_{n\neq 0} $ simply reminds us that this part of the measure does not contain any zero-mode. We can then derive the equations of motion for the grassmann coefficients:
\begin{align}
i\lambda_n a_n = -m \bar{b}_n, &\quad i\lambda_n b_n = m\bar{a}_n, ~~\lambda_n\neq 0 \label{massivemodes} \\
m\bar{b}^0_i = 0, &\quad m\bar{a}^0_i = 0, ~~ i=1,\ldots, n_{R}^0
\end{align}

We see that for massive fermions, the zero-modes are constrained to vanish. We must however recall that once we solve for the ``primed'' spinors, we will be interested in the massless limit, leaving therefore these zero-modes unconstrained. We can nevertheless integrate out the `massive modes' using (\ref{massivemodes}) so that to obtain our new second-order Lagrangian (after an appropriate rescaling and dropping an overall constant from the path-integral):
\begin{align}
\int d^4x \mathcal{L} = -\sum_{\lambda_n\neq 0 } \left(\lambda_n^2+m^2\right)b_n a_n  - m^2\left(\sum_{i= 1}^{n_L^0}b^0_i a^0_i +\sum_{i= 1}^{n_R^0} \bar{b}^0_i \bar{a}^0_i \right) 
\end{align}

We can then consider the massless limit we are interested in:
\begin{align}
\int d^4x \mathcal{L} = -\sum_{\lambda_n\neq 0 }\lambda_n^2 b_n a_n  
\end{align}

If we rewrite the Lagrangian in terms of the spinor fields, this is indeed equivalent to:
\begin{align}
\mathcal{L} = -(\bar{D} \chi )^T( D \xi) 
\end{align}

We see that the zero-modes have disappeared from the Lagrangian. This can be understood as they automatically satisfy the massless reality condition (\ref{realDano}) by means of them being zero-modes. However, we did not integrate over them\footnote{Another way to think about this is to remember that the path-integral of a theory whose quadratic operator contains zero-modes is singular. This is the case of the massless limit of Dirac theory, and to circumvent this singularity, the zero-modes can be left ``unsolved''.}, they still are present in the measure, which is now given by:
\begin{align}
\left(\prod_{n\neq 0} d{b}_n d{a}_n\right)d\bar{b}^0_1\dots d\bar{b}^0_{n^0_R}d\bar{a}^0_1\dots d\bar{a}^0_{n^0_R}d{b}^0_1\dots d{b}^0_{n^0_L}d{a}^0_1\dots d{a}^0_{n^0_L}
\end{align}

This, in turn, implies that for the anomaly the change in the measure is not given by (\ref{changemeas1}), but rather by:
\begin{align}
\begin{split}
\prod_{n=1}^{N}  db'_nda'_nd\bar{b'}^0_1\dots d\bar{b'}^0_{n^0_R}d\bar{a'}^0_1\dots d\bar{a'}^0_{n^0_R} &= \prod_{n=1}^{N} db_n da_nd\bar{b}^0_1\dots d\bar{b}^0_{n^0_R}d\bar{a}^0_1\dots d\bar{a}^0_{n^0_R} \\&\times \exp \Big[ 2i\alpha \left( \sum _{n=0}^N \{ left\}  -n^0_R \right) \Big] \label{changemeas2}
\end{split}
\end{align}

where for simplicity, we have again considered a constant phase $\alpha$ and denoted by $\sum _{n=1}^N \{ left\}$ the sum over unprimed spinor modes. All in all, if we gather the change in the measure, and the change in the Lagrangian, we obtain:
\begin{align}
\begin{split}
\delta Z &= \int \prod_{n=0}^{N}  db_nda_n \prod_{i=1}^{n^0_R}d\bar{b}^0_i d\bar{a}^0_{i} ~\Big[ 2i\alpha \left(n^0_L-n^0_R\right) \Big] ~ \exp\left[\sum_{\lambda_n\neq 0 }^N b_n a_n\lambda_n^2 \right] 
\end{split}
\end{align}

Which leads to the usual
\begin{align}
\int d^4x \mathcal{A}(x) = 2(n^0_L - n^0_R) =2( \dim\ker D - \dim\ker D^\dagger ) = 2 {\rm ind}~ D
\end{align}

We therefore see that at the level of the path-integral, one has to be careful in calculating the anomaly. This problem does not arise when the latter is computed pertubatively since triangle diagrams ``do not care'' about the modes expansion and automatically take into account the contribution from both chiralities.

\subsection{Fermion number anomaly in Weyl theory}

We now calculate the anomalous conservation of the fermionic number in the theory of one Weyl fermion. This result then generalises to $N$ Weyl fermions and the axial anomaly can be seen as a subcase of the general result. We will proceed as in the previous case, using a mode expansion of the first-order Lagrangian and then integrating out the modes of the primed fermions to achieve the second-order Lagrangian.

The Lagrangian for a Weyl fermions is given by:
\begin{align}\begin{split}
\mathcal{L}&=\im \lambda^\dagger (D\lambda)^T - \frac{m}{2}\left(\lambda
^T \lambda + \lambda^\dagger (\lambda^\dagger)^T\right) \\
 &=\im \lambda^\dagger_{A'} D^{A'}{}_A\lambda^A - \frac{m}{2}\left(\lambda^A \lambda_A + \lambda^\dagger_{A'} \lambda^{\dagger\,A'}\right)
\end{split}\end{align}

This leads to the equations of motion:
\begin{align}
\im (D\lambda) =-m\lambda^\dagger, \quad \im D_{A'}{}^A\lambda_A = -m\lambda^\dagger_{A'}
\end{align}

The kinetic term can be diagonalised when the fermions are expanded into modes:
\begin{align}
\lambda^\dagger(x) = \sum_n \bar{a}_n \hat\phi_{R\,n}(x) ,\quad \lambda(x) = \sum_n a_n \phi_{L\,n}(x)
\end{align}

with $D\phi_L = \lambda \phi_R$ as before. This then leads to (in the massless case):
\begin{align}
\int d^4x \mathcal{L}= \sum_n i\lambda_n \bar{a}_n a_n
\end{align}

whereas the measure becomes:
\begin{align}
 \left(\prod_{n\neq 0} d\bar{a}_n\right) \left(\prod_{n\neq 0}  d{a}_n\right)d\bar{a}^0_1\dots d\bar{a}^0_{n^0_R}d{a}^0_1\dots d{a}^0_{n^0_L}
\end{align}

and it is easy to read off the anomaly from the zero modes of the left and right handed fermions to be\footnote{Recall that $\dim\ker D^\dagger = \dim\ker \bar{D}^\dagger$.}
\begin{align}
\int d^4x \mathcal{A}(x) = n^0_L - n^0_R = \dim\ker D - \dim\ker D^\dagger =  {\rm ind}~ D
\end{align}

As we are interested in the second order formulation, let us see what happens with the mass terms. When expanded on the above modes, they become:
\begin{align}\begin{split}
\frac{m}{2}\int d^4 x\left(\lambda
^T \lambda + \lambda^\dagger (\lambda^\dagger)^T\right) &= \frac{m}{2}\sum_{m,n} a_m \Lambda^L_{mn}a_n + \bar a_m \Lambda^R_{mn}\bar a_n
\end{split}\end{align}

where $\Lambda^{L,R}$ are antisymmetric matrices given by:
\begin{align}
\Lambda^L_{mn} = \int d^4x \phi_{L\,m}\phi_{L\,n}, \quad \Lambda^R_{mn} = \int d^4x \hat\phi_{R\,m}\hat\phi_{R\,n}
\end{align}

For later purposes, we can separate the contribution to the mass terms containing zero modes:
\begin{align}\label{massterms}\begin{split}
\frac{m}{2}\sum_{m,n \neq 0} \left(a_m \Lambda^{'L}_{mn}a_n + \bar a_m \Lambda^{'R}_{mn}\bar a_n\right) &+  m\sum_{i,\{ n\neq 0\}} \left(a_i^0 \Lambda^L_{in}a_n + \bar a_i^0 \Lambda^R_{in}\bar a_n\right)\\& + \frac{m}{2}\sum_{i,j} \left(a_i^0 \Lambda^L_{ij}a_j^0 + \bar a_i^0 \Lambda^R_{ij}\bar a_j^0\right)
\end{split}\end{align}

where from now on, we denote by a prime the part of the matrix $\Lambda$ that does not contain the zero-modes subspaces. The massive Lagrangian can be rewritten as:
\begin{align}
\int d^4 x \mathcal{L} = \sum_{n\neq 0} i \bar a_n  \lambda_n a_n -\frac{m}{2}\sum_{m,n} a_m \Lambda^L_{mn}a_n + \bar a_m \Lambda^R_{mn}\bar a_n
\end{align}

with the equations of motion
\begin{align}
i\lambda_n a_n =m\sum_{m} \Lambda^R_{nm}\bar a_m  , \quad \lambda_n \neq 0 \\
  m\sum_{m} \Lambda^R_{im}\bar a_m =0, \quad i=1,\ldots, n^0_R \label{eqnzero}
\end{align}

It should be noticed here that the equations of motion mix the zero and non-zero modes of the operators. Furthermore, in the massive case, we see that the matrix $\Lambda^R_{mn}$ has $n^0_R$ ``null directions''. This corresponds to the previous case in which the massive equations of motion simply constrained the zero modes to vanish. However, as before, we only want to invert the equations of motion for the non-zero barred modes, we then rewrite:
\begin{align}
i\lambda_n a_n -m \sum_i  \Lambda^R_{ni}\bar a_i^0  =m\sum_{m\neq 0} \Lambda^{'R}_{nm}\bar a_m  , \quad \lambda_n \neq 0 
\end{align}

We then have:
\begin{align}
i\sum_{n\neq 0} \left(\Lambda^{'R}\right)^{-1}_{mn}\lambda_n a_n -m \sum_{i\,\{n\neq 0\}} \left(\Lambda^{'R}\right)^{-1}_{mn}  \Lambda^R_{ni}\bar a_i^0  =m \bar a_m  , \quad \lambda_m \neq 0 
\end{align}

Integrating out one barred fermion in the first sum in (\ref{massterms}), we have:
\begin{align}
\frac{m}{2}\sum_{m,n \neq 0} \bar a_m \Lambda^{'R}_{mn}\bar a_n \rightarrow \frac{1}{2}\sum_{n\neq 0} i \bar a_n  \lambda_n a_n - \frac{m}{2}\sum_{i,\{ n\neq 0\}}  \bar a_i^0 \Lambda^R_{in}\bar a_n
\end{align}

Repeating the same manipulation on the kinetic term and on the remaining mass term involving the barred non-zero modes, we obtain (after an appropriate rescaling of the fields):
\begin{align}\begin{split}
\int d^4x \mathcal{L}&= -\frac{1}{2}\sum_{m,n\neq 0} \left( i\lambda_n a_n + m \sum_i \bar{a}^0_i \Lambda^R_{in} \right) \left( \Lambda^{'R}\right)^{-1}_{nm}  \left( i\lambda_m a_m - m \sum_j \Lambda^R_{mj}\bar{a}^0_j   \right) \\& -\frac{m^2}{2}\sum_{m,n} a_m \Lambda^L_{mn}a_n -\frac{m^2}{2}\sum_{i,j}  \bar a_i^0 \Lambda^R_{ij}\bar a_j^0
\end{split}\end{align}

Notice that in the massive case, if we use the constraint (\ref{eqnzero}), then the contribution from the barred zero-modes disappears from the Lagrangian. Here, as we are interested in the case of a massless Weyl fermion, our Lagrangian finally becomes:
\begin{align}
\int d^4x \mathcal{L}&= \frac{1}{2}\sum_{m,n\neq 0}  \lambda_n a_n  \left(\Lambda^{'R}\right)^{-1}_{nm}  \lambda_m a_m 
\end{align}

which is the equivalent of 
\begin{align}\begin{split}
{\cal L} &=  -\frac{1}{2}(D \lambda )^T( D \lambda)\\
&= -({D}^{A'}{}_{ A}\lambda^A )(  D_{A'}{}^{B} \lambda_B ) \label{lagweyl}.
\end{split}\end{align}

with the extra minus sign coming from the ``transpose'' which effectively amounts to flipping an internal spinor contraction carried out with the antisymmetric epsilon metric, and we have:
\begin{align}
\left(\Lambda^{'R}\right)^{-1}_{nm} = \int d^4x \phi_{R\,n} \phi_{R\,m}, \quad m,n \neq 0
\end{align}

In the path integral for the theory, the measure becomes, after integrating out the non-zero modes:
\begin{align}
\left(\prod_{n\neq 0}  d{a}_n\right)d\bar{a}^0_1\dots d\bar{a}^0_{n^0_R}d{a}^0_1\dots d{a}^0_{n^0_L}
\end{align}

As in the case of Dirac fermions, we see that both the left and right handed zero modes have been left untouched and hence contribute to the anomaly. We can read off the contribution to the anomaly coming from the measure to be:
\begin{align}
\int d^4x \mathcal{A}^{meas} = n^0_L - n^0_R + \sum_{n\neq 0}\{left\}
\end{align}

where $\sum_{n\neq 0}\{left\}$ is the sum over non zero modes. As before, this should cancel with the contribution coming from the non-invariance of the Lagrangian. We have:
\begin{align}
\begin{split}
\int \prod_{n=0}^{N} da_n ~\Big[ (-i\alpha) (-1)&\sum_{m,n\neq 0}  \lambda_n a_n  \left(\Lambda^{'R}\right)^{-1}_{nm}  \lambda_m a_m  \Big] \\ &\hspace*{1.5cm}\times \exp\left[-\frac{1}{2}\sum_{m,n\neq 0}  \lambda_n a_n  \left(\Lambda^{'R}\right)^{-1}_{nm}  \lambda_m a_m \right]  \\
&=\int \prod_{n=0}^{N}  da_n ~\Big[ (-i\alpha)  \sum _{\lambda_n\neq 0 }^N 1 \Big] \\ &\hspace*{1.5cm}\times \exp\left[-\frac{1}{2}\sum_{m,n\neq 0}  \lambda_n a_n  \left(\Lambda^{'R}\right)^{-1}_{nm}  \lambda_m a_m\right] 
\end{split}
\end{align}
%\sum_{n\neq 0}\{ left\} 
Where we used:
\begin{align}\begin{split}
\int d \theta  \left(- \theta^T M \theta \right) \exp\left[- \frac{1}{2}\theta^T M \theta  \right] &= 2 \frac{\partial}{\partial t}\left.\int d \theta  \exp\left[- \frac{t}{2}\theta^T M \theta  \right]\right|_{t=1} \\&=2 \frac{\partial}{\partial t}t^{n/2}\left.\int d \xi \exp\left[- \frac{1}{2}\xi^T M \xi \right]\right|_{t=1} \\ &=n \int d \theta  \exp\left[- \frac{1}{2}\theta^T M \theta \right]
\end{split}\end{align}

where we changed variables $\xi =\sqrt{t}\theta$ and in the last line, we used that the latter is the identity operation at $t=1$. In the case we are interested in, $n$ is equal to the number of non-zero modes (modes appearing in Lagrangian), hence the result. Finally, using the contribution from the non-invariance of the Lagrangian to cancel the extra term in the transformation of the measure, we have:
\begin{align}
\int d^4x \mathcal{A} = n^0_L - n^0_R 
\end{align}

which is the usual Atiyah-Singer index theorem for the anomaly in the theory of one Weyl fermion.

\section{Discussion}
First of all, the main result obtained in this chapter is that the anomaly (chiral or fermion number) that arises in the first-order formalism can be equivalently derived using the second-order formalism. When it is calculated perturbatively, we have shown that, although the integral that has to be computed is different, there exists a shift in the loop momentum that maps the calculation in one formalism to the the calculation in the other. 
We then have developed the tools necessary to the computation of the anomaly using non-perturbative methods for two-component spinors and, again, have shown that the same result is obtained. However, it must be emphasised that the approach taken here is that the second-order Lagrangian is not fundamental but rather is obtained from its first-order counterpart. Indeed, as we have seen during the derivation, we have had to take into account the contribution of the right-handed zero modes corresponding to the primed fermions that were integrated out. This is due to the fact that the path-integral over an operator containing zero modes is singular and cannot be carried out without paying a particular attention to these. 

Furthermore, notice that nowhere in our derivation we have used the reality conditions that are imposed on the unprimed fermions. Indeed, we have simply worked with a Lagrangian in which the mass was set to zero after having integrated out the primed spinors, and with a general mode expansion of the remaining fields. The same calculation can be carried out after having imposed sharply the reality conditions. This leads for example to:
\begin{align}
D\xi = 0 \rightarrow \sum_{n} a_n D\phi_{L\,n} = \sum_{n\neq 0} a_n\lambda_n \phi_{R\,n} = 0 ~\leftrightarrow~ a_n = 0, ~\forall n \neq 0
\end{align}

From this, it follows that the Lagrangian vanishes and the measure only contains the integration over the zero modes that have been left unsolved by the above constraint. This immediately leads to the same result as we have derived, but in a much quicker way\footnote{Even though this method is quicker, the method we have presented makes the derivation more easily generalisable.}. 

There are some subtleties in the implementation of the reality conditions at the level of the action and in the counting of zero modes. Indeed, because the Weyl action in the second-order formalism is only considered as a limit after the reality conditions (massive) have been imposed, the modes that are left unsolved are only the left-handed zero modes and the ``massive'' left-handed modes.

% that are not in the kernel of the right-handed Dirac operator either 

%(for which we have solved the massive theory with no right-handed zero modes), $i.e.$:
%\begin{align}
%\phi_{L\,n} ~ \quad {\rm s.t.}\quad ~ D^\dagger (D \phi_{L\,n}) \sim  D^\dagger \phi_{R\, n} \neq 0
%\end{align}
%(the same happens when the reality conditions are not imposed sharply at the mode level, for all modes)
In the above derivation we have circumvented the problem by imposing the reality conditions on the (right-handed) massive modes only (and therefore keeping the contribution of the right-handed zero modes explicit for clarity). Now, because massive modes of either chirality are in one-to-one correspondence, the count of zero-modes in this subspace is not affected by the former's reality conditions. However, as we have seen before in the case of zero modes, this is not true any longer. Had we solved the massive reality conditions for the right-handed zero modes, these would have been constrained to vanish. Since at the field level, this is what has been done, we have effectively removed these modes from the theory. This, by extension, affects the number of left-handed zero modes\footnote{Because the right-handed zero modes are constrained to vanish, they are formally left-handed zero modes as well. Therefore, in the count of non trivial left-handed zero modes, there are $n^0_R$ less modes.} there are $n^0_L-n^0_R$ zero modes remaining. Hence giving the appropriate result but from a purely second-order perspective (the same conclusion would be obtained had we considered the second-order theory fundamental since the above argument relies only on the functional properties of the operators). We would like to emphasise that this is only an intuitive argument, a more rigorous explanation has not been developed for this thesis and is left for further exploration. 

%At the time when this thesis was redacted, the consequences on the non-perturbative calculation of the anomaly in the other approach in which the second-order formalism is taken as fundamental were not explored. At this point, it seems like a fundamental second-order theory of massless fermions fails to reproduce the same non-perturbative results as the first-order theory does, while their perturbative derivation is preserved. We do not know whether this is a ``proof'' that for fermions, although their second-order formalism is more practical as far as perturbative calculations are concerned, their first-order formalism is ``more'' fundamental.

%Could possibly take the approach that it is fundamental and calculate perturbatively and off-shell the contribution of the triangle diagrams.

%\input{TensorReduction}

\chapter{Unitarity}\label{chapuni}

\section{Introduction}
As we mentioned in the Introduction, we will only consider perturbative unitarity of the S-matrix. In order to prove perturbative unitarity, we shall use the so-called Largest-Time Equation (LTE), which we will explicitly construct for a scalar field theory. 

The organisation of this chapter is as follows: in the first section we discuss what it means for the S-matrix to be unitary. In the second section, we introduce the concept of anti-propagator, develop the LTE for scalars and show how this is related to unitarity in a simple example. In the third section, we summarise the main ingredients of the second-order theory for fermions that are needed for the proof, we discuss their propagators, derive the consequences of the reality conditions and finally develop the proof of unitarity. The last section is a discussion of the results we obtain.

\section{General remarks about unitarity}

In quantum field theory, the S-matrix (operator) can be written in the interaction picture:
\begin{align}
S= \mathcal{T}e^{i\int d^4x \mathcal{L}_{int}(x)}
\end{align}

where $\mathcal{T}$ denotes time-ordering\footnote{Notice that for interactions' Lagrangians containing derivatives, the latter is not equal to minus the interacting Hamiltonian, however the mismatch corresponds to non-covariant terms that are exactly cancelled in correlation functions by the chronological products of derivative terms. Therefore, the operator that is considered in the aforementioned situation is the covariant time-ordering operator.}. Taking matrix elements between physical states\footnote{In this chapter $\langle f | i \rangle \equiv \delta_{fi}$, were both states are seen as free asymptotic states, and therefore should not to be confused with the similar quantity that appears in the LSZ reduction formula.}, unitarity of the S-matrix reads:
\begin{align}
\langle f | i \rangle = \sum_{{\rm phys}~n} \langle f | S| n \rangle \langle n |S^\dagger | i \rangle = \sum_{{\rm phys}~n} \langle f | S^\dagger| n \rangle \langle n |S | i \rangle
\end{align}

where the sum runs only over physical intermediate states, and the latter are eigenstates of the free Hamiltonian. For practical calculations, one splits the S-matrix into the identity plus a transition operator $T$:
\begin{align}
S=1+iT
\end{align}

The unitarity equation in terms of the latter operator then reads:
\begin{align}
\langle f |(iT)| i \rangle +\langle f |(iT)^\dagger| i \rangle= -\sum_{{\rm phys}~n} \langle f | (iT)^\dagger| n \rangle \langle n |(iT)| i \rangle
\end{align}

Each of the terms appearing in the equation corresponds to a transition amplitude that can be computed by means of Feynman diagrams. In order to do so, the S-matrix exponential factor is expanded and we obtain, for instance,
\begin{align}
\langle f | (iT)| i \rangle = \langle f | \sum _{m=1}^\infty \frac{i^m}{m!}\int d^4x_1 \cdots d^4x_m \left[ \mathcal{T}\mathcal{L}_{int}(x_1) \cdots \mathcal{L}_{int}(x_m)\right] | i \rangle \label{amp}
\end{align}

This ought to be compared to:
\begin{align}
\langle f | (iT)^\dagger| i \rangle = \langle f | \sum _{m=1}^\infty \frac{(-i)^m}{m!}\int d^4x_1 \cdots d^4x_m \left[ \mathcal{T}^\dagger \mathcal{L}_{int}^\dagger(x_1) \cdots \mathcal{L}^\dagger_{int}(x_m)\right] | i \rangle \label{camp}
\end{align}
The first thing to notice is that for this amplitude we get an extra minus sign for each vertex. Let us now have a look at the terms inside the brackets. The operator $\mathcal{T}^\dagger$ corresponds to an anti-chronological time ordering, $i.e.$:
\begin{align}
\mathcal{T}^\dagger \mathcal{O}_1(x)\mathcal{O}_2(y)= \mp \theta(x^0-y^0)\mathcal{O}_2(y)\mathcal{O}_1(x) +\theta(y^0-x^0)\mathcal{O}_1(x)\mathcal{O}_2(y)
\end{align}
where the minus sign would be assigned to anti-commuting operators. Using Wick's theorem for correlation functions, this anti-chronological ordering will lead to anti-propagators, to be discussed below, connecting vertices. Finally, the reader should have noticed that we did not use the hermicity of the Lagrangian in the brackets. Indeed, if the Lagrangian were hermitian, we would obtain exactly the same Feynman rules for (\ref{amp}) and (\ref{camp}), replacing propagators by anti-propagators, up to an extra sign in the vertices. However, as we will shortly recall below, in the case of second-order fermions, the Lagrangian is not hermitian, and therefore we will have to work out the consequences of such a difference. It must be noted that it is usually assumed that a non-hermitian Lagrangian leads to a non-unitary theory, but we will show that this is not necessarily the case.

\section{Largest-time equation for scalars}
As we discussed in the introduction, we will develop a proof of unitarity using the LTE. It only relies on the decomposition of the Feynman propagator into a sum of forward and backward propagators with theta functions, and on a simple combinatorics argument. Because it does not directly depend on the dynamics of a system, and because it will turn out to be useful, we first develop it in the case of a scalar field.

\subsection{Propagators}

The propagator for a massive scalar field $\varphi(x)$ is given by:
\begin{align}
\Delta_F(x-y) \equiv \langle 0| \mathcal{T}\varphi(x) \varphi(y) |0\rangle = \int \frac{d^4p}{(2\pi)^4}\frac{-i}{p^2+m^2-i\varepsilon}e^{ip(x-y)}
\end{align}
with $p^2 = -p_0^2 + p_i^2$. It can alternatively be written as:
\begin{align}\begin{split}
\Delta_F(x-y)   &\equiv \theta(x^0-y^0)\Delta^+(x-y)+ \theta(y^0-x^0) \Delta^-(x-y) \\
&= \int \frac{d^3p}{(2\pi)^3}\left[\frac{e^{ip(x-y)}}{2\omega}\theta(x^0-y^0) + \frac{e^{-ip(x-y)}}{2\omega}\theta(y^0-x^0) \right]
\end{split}\end{align}

with $px = -\omega t + \vec{k}\cdot \vec{x}$ and 
\begin{align}\begin{split}
\Delta^\pm(x-y)= \int \frac{d^3p}{(2\pi)^3}\frac{e^{\pm ip(x-y)}}{2\omega}
\end{split}\end{align}

We see that $\Delta^+(x-y)$ describes an energy (particle) flow\footnote{We consider the positive energy particle to have a phase factor $\exp (-i\omega t)$.} from $y$ to $x$ whereas $\Delta^-(x-y)$ describes a flow from $x$ to $y$. On the other hand, the anti-propagator is given by:
\begin{align}
\Delta^*_F(x-y) \equiv \langle 0| \mathcal{T}^\dagger\varphi(x) \varphi(y) |0\rangle = \int \frac{d^4p}{(2\pi)^4}\frac{+i}{p^2+m^2+i\varepsilon}e^{-ip(x-y)}
\end{align}
We see that it is the complex conjugate of the ordinary propagator ($p\rightarrow-p$ change of variables). The important factor here is the $+i\varepsilon$ in the denominator. This implies that the poles will have opposite imaginary parts compared to the poles in the propagator, and therefore, when writing the anti-propagator as a three-dimensional integral, the contour will have to be closed in the opposite half-planes in the complex plane. This leads to:
\begin{align}\begin{split}
\Delta^*_F(x-y)   &\equiv \theta(y^0-x^0)\Delta^+(x-y)+ \theta(x^0-y^0) \Delta^-(x-y) \\
&= \int \frac{d^3p}{(2\pi)^3}\left[\frac{e^{ip(x-y)}}{2\omega}\theta(y^0-x^0) + \frac{e^{-ip(x-y)}}{2\omega}\theta(x^0-y^0) \right]
\end{split}\end{align}

In the following, we will need the following identities between propagators:
\begin{align}
&~~\Delta_F(x) = \theta(x^0)\Delta^+(x)  +\theta(-x^0)\Delta^-(x)\nonumber \\
&~~\Delta^*_F(x) = \theta(-x^0)\Delta^+(x)  +\theta(x^0)\Delta^-(x)\nonumber \\
&\begin{array}{ll}
(\Delta ^+(x))^* = \Delta^-(x) &;~\Delta^+(-x) = \Delta^-(x)\\
(\Delta_F(x))^* = \Delta^*_F(x) & ;~\Delta_F(-x) = \Delta_F(x) \\
\end{array}
\end{align}

\subsection{Largest-time equation}

Now that we have introduced the different types of propagators and their relations with respect to complex conjugation, we can derive a result known as the largest-time equation (LTE). Consider a Feynman integrand with $N$ interaction vertices $x_1, \ldots, x_N$ in Minkowski spacetime. We denote this function of the vertices points as $F(x_i)$, and the corresponding Feynman amplitude is obtained by integrating over all $x_i$ and adding emission and absorption factors for the external wave functions. For a scalar field (with appropriately normalised interactions), $F(x_i)$ is given by a factor of $i$ for each vertex and by propagators $\Delta_F(x_i-x_j) \equiv \Delta_F(ij)$ for each line joining the vertices $x_i$ and $x_j$.

The establishment of the LTE goes as follows: for each integrand $F(x_i)$ containing $N$ vertices, there are $2^N-1$ other functions of the $x_i$ that can be constructed. This is done by applying the following algorithm:
\begin{enumerate}
\item  Draw circles around vertices. There are $2^N$ possible combinations, the diagram with no circles being the original $F(x_i)$.\\
\item For each circle, swap the sign in the vertex $i \rightarrow -i$. \\
\item For uncircled connected vertices, write $\Delta_F(ij)$. If both vertices are circled, write $\Delta^*_F(ij)$. If $x_i$ is circled and $x_j$ is not, write $\Delta^+(ij)=\Delta^-(ji)$.% and if $x_i$ is not circled but $x_j$ is, then write $\Delta^-(ij)=\Delta^+(ji)$.
\end{enumerate}

After having constructed the $2^N$ graphs, we shall derive an equation for their sum. Consider the set of points $x_i$ contained in the Feynman integrand $F(x_i)$. In some reference frame, there will be a point, $x_\ell$, which has the largest time of all the points: $x_\ell^0 > x_i^0,~\forall i\neq \ell$. Consider now a graph belonging to the whole set of possible integrands where $x_\ell$ is uncircled and another point $x_s$ connected to it is also uncircled. Among the other $2^{N-1}$ graphs, there is another one which has the same circling pattern as the former, except for $x_\ell$ which is this time circled. Then, the sum over the two graphs is zero. Following our construction rules, the Feynman propagator $\Delta_F(\ell s)$ is replaced by $\Delta^+(\ell s)$ and the sign in the $x_\ell$ vertex is swapped. The statement is therefore true because we have:
\begin{align}
\Delta_F(\ell s) = \Delta^+(\ell s), \quad {\rm if} ~ x_\ell^0 > x_s^0
\end{align}

Similarly, if $x_\ell$ is uncircled and another point $x_s$ connected to it is circled, there is another graph which has the same circling pattern, except for $x_\ell$ which is circled. The sum vanishes since
\begin{align}
\Delta^-(\ell s) = \Delta^*_F(\ell s), \quad {\rm if} ~ x_\ell^0 > x_s^0
\end{align}

This generalises straightforwardly to the case in which $x_\ell$ is connected to several other vertices. The propagators will be equal, and only the sign swap in the vertex will make the sum cancel. \\
\medskip
It is then easy to see that the sum over all $2^N$ graphs will give zero. Indeed, we can group them in $2^{N-1}$ pairs whose graphs sum up to zero, see (Fig.\ref{LTEexample}).
\begin{figure}[H]\begin{center}
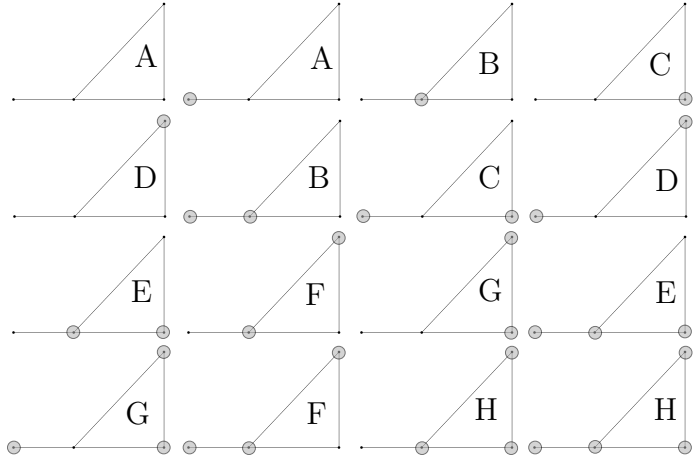\caption{Construction of the $2^4$ diagrams contributing to the LTE. Here, $x_\ell$ is taken to be the point at the far left of the diagram. Diagrams are grouped two by two in order to show how they cancel pairwise.}\label{LTEexample}
\end{center}\end{figure}

Finally, the Largest-Time Equation is written as:
\begin{align}
F(x_i) + F^*(x_i) = -\underline{F}(x_i)
\end{align}

where $F^*(x_i)$ denotes the complex conjugated integrand (all vertices and propagators are complex conjugated), and $\underline{F}(x_i)$ is the sum over the $2^{N}-2$ remaining graphs where there is at least one vertex of each type.

In the next subsection we explain how this is related to unitarity.

\subsection{Unitarity from the cutting rules}

We have just seen how we can derive an equation for the Feynman integrand appearing in a scalar field amplitude by simply using combinatorics. In order to obtain the corresponding S-matrix, one needs to multiply the corresponding diagram by plane waves for the emission and absorption of particles and integrate over the different vertices $x_i$:
\begin{align}
\langle f |iT|i \rangle = \int \left( \prod_{i=1}^N d^4 x_i \right)\left( \prod_{j=1}^{N_i} e^{ip_jx_{m_j}}\right)\left( \prod_{k=1}^{N_f} e^{-ip_kx_{m_k}}\right) ~F(x_i)
\end{align}
%\begin{align}
%\langle f |iT|i \rangle = \int \left( \prod_{i=1}^N d^4 x_i \right)\left( \prod_{j=1}^{N_i} \frac{e^{ip_jx_{m_j}}}{\sqrt{2\omega_j}}\right)\left( \prod_{k=1}^{N_f} \frac{e^{-ip_kx_{m_k}}}{\sqrt{2\omega_k}}\right) ~F(x_i)
%\end{align}

where $p_j$ denotes incoming momenta to a subset of vertices $x_{m_j}$ and $p_k$ denotes outgoing momenta from $x_{m_k}$. These factors are the same for all diagrams appearing in the LTE, the only difference is in the propagators involved. Notice also that there will in general be more terms in the RHS of the LTE than in the sum over physical states in the unitary relation. Indeed, for every {\it cut} propagator, $i.e.$ a propagator that is replaced by a sum over states in the unitary relation, there will be two different diagrams in the LTE: one where the vertex on the left is circled and the one on the right uncircled, and another diagram where the circles are swapped. In terms of the sum over states in $\sum_{{\rm phys}~n} \langle f | (iT)^\dagger| n \rangle \langle n |(iT)| i \rangle$, if the particle is emitted at $x$ by $T$ and absorbed at $y$ by $T^\dagger$, we obtain:
\begin{align}
\sum_{{\rm phys}~n}| n \rangle \langle n | \rightarrow \int \frac{d^3p}{(2\pi)^3}\frac{e^{-ip(x-y)}}{2\omega_p} = \Delta^+(y-x)
\end{align}

whereas in the LTE we will get two terms involving either $\Delta^+(y-x)$ or $\Delta^-(y-x)
$. We will see how the extra terms cancel due to energy conservation. As an example, consider the following scalar field interaction:
\begin{align}
\mathcal{L}_{int}= g_1 \varphi\varphi_1\varphi_2 + g_2 \varphi \varphi_3\varphi_4
\end{align}

and the $12\rightarrow 34$ tree level scattering:
\begin{align}
\mathcal{M}^{tree} = g_1g_2\int d^4x d^4y \varphi_1(x)\varphi_2(x)\varphi_3(y)\varphi_4(y) \left[ i \Delta_F(x-y)i   \right]
\end{align}

The LTE reads:
\begin{align}
i \Delta_F(x-y)i + (-i) \Delta_F^*(x-y)(-i)  =- \left[i \Delta^-(x-y)(-i) +(-i) \Delta^+(x-y)i   \right]
\end{align}
It can readily be checked using the decomposition of the Feynman (anti-)propagator in terms of the forward and backward propagators and that $\theta(t)+\theta(-t)=1$. Let us now derive the unitarity relation deriving from the LTE.

\begin{enumerate}
\item Assume energy flows from $x$ to $y$ and replace $\varphi_i$ by emission/absorption factors. We will have for the energy plane waves:
\begin{align}
e^{-ix^0(E_1+E_2)}e^{+iy^0(E_3+E_4)}
\end{align}

\item Integration over $x$ and $y$ will lead to two momentum conservation delta functions. For the term involving $\Delta^-(x-y)$ we have:
\begin{align}\begin{split}
\int dx^0 dy^0~& e^{-ix^0(E_1+E_2)}e^{+iy^0(E_3+E_4)} e^{+i\omega(x^0-y^0)} \\& \sim \delta(E_1+E_2 - \omega)\delta( \omega- E_3-E_4 ), \quad E_i,~\omega >0
\end{split}\end{align}
whereas for the term involving $\Delta^+(x-y)$, we have:
\begin{align}\begin{split}
\int dx^0 dy^0 ~&e^{-ix^0(E_1+E_2)}e^{+iy^0(E_3+E_4)} e^{-i\omega(x^0-y^0)} \\& \sim \delta(E_1+E_2 + \omega)\delta( \omega+ E_3+E_4 ), \quad E_i,~\omega >0
\end{split}\end{align}
which can never be satisfied for strictly positive energies.

\item We see that energy conservation kills the extra term in the RHS of the LTE, and we are left with a factor that corresponds to the sum over states in the unitary relation.
\begin{align}
\sum_{{\rm phys}~n}| n \rangle \langle n | = \Delta^-(x-y)
\end{align}
\end{enumerate}

To summarise the algorithm: write the Feynman integrand corresponding to a given process and write its corresponding LTE. The LHS is identical to the unitarity relation's LHS when integrated over vertices with plane waves factors inserted. The LTE tells us that this is equal to the sum over some cut diagrams, some of which have to be removed using energy conservation. Finally, show that the remaining diagrams correspond exactly to the RHS of the unitary relation using the fact that a sum over physical states is replaced by a forward or a backward propagator.
\medskip

The last step of the algorithm can be further simplified if one adds from the beginning an extra diagrammatic rule encoding the energy flow in a diagram. In a Feynman (anti-)propagator, energy flows both ways as it corresponds to a virtual particle. However, for cut propagators which correspond effectively to a physical sum over states, energy can only flow in one direction. As we have seen, in $\Delta^+(x-y)$ ($\Delta^-(y-x)$), energy flows from $y$ to $x$. Henceforth, for every diagram linking a circled to an uncircled vertex, we draw an arrow pointing towards the circle denoting energy flow. We now give the so-called {\it cutting rules} for unitarity:

\begin{enumerate}
\item Given a Feynman amplitude, draw all possible graphs obtained by circling a subset (all) of the vertices.
\item In addition, for every line linking two different types of vertices, draw an arrow pointing towards the circled vertex.
\item Choose a direction for the overall energy flow.
\item If a graph contains at least one vertex whose links are all incoming (outgoing), it vanishes.
\item Draw now a cut through all directed lines but not through undirected lines.
\item The only non-vanishing graphs are those for which energy flows from one side of the cut to the other side, and for which each side of the cut contains only one type of vertices.
\item The remaining graphs are in one-to-one correspondence with the terms appearing in the unitarity equation for that given diagram.
\end{enumerate}

This demonstrates that for a scalar field, perturbative unitarity of the S-matrix follows directly from the LTE. Note however, that the LTE (and thus unitarity) holds here diagram by diagram. This is more than is required, as all we need for perturbative unitarity to hold is that the S-matrix is unitary at a given order in the coupling constant(s). This will be important for the problem of second-order fermions as we will now see.

\section{Unitarity of second-order fermions}
We now focus on the theory of second-order fermions. We will give a short recap of the formalism for self-consistency of the chapter, then we will have a second look at two scattering processes that were calculated in previous chapters, we will explore the consequences of the reality conditions, and finally we will link the construction of the LTE for a scalar field to the theory of spinor fields.
\subsection{Brief reminder of second-order fermions}
Let us shortly summarise Section \ref{DiracRules}. The Lagrangian for second-order Dirac Quantum Electrodynamics is given by (we simply consider the fermionic part):
\begin{align}
{\cal L} =-{2} D_{A'}{}^{ A} \chi_A  D^{A'B} \xi_B - m^2 \chi^A\xi_A\label{lagchiral2U}
\end{align}
with 
\begin{align}
D_\mu \xi = (\partial_\mu - \im e A_\mu)\xi, \qquad
 D_\mu \chi = (\partial_\mu + \im eA_\mu)\chi,
\end{align}
where we included the electromagnetic coupling $|e|\ll 1$. Being not hermitian, the theory is supplemented with reality conditions:
\begin{align}
\xi^{\dagger\,A'} = -\frac{\im\sqrt{2}}{m}  D^{ A'A}\chi_A, \qquad \chi^{\dagger\,A'} = -\frac{\im\sqrt{2}}{m}  D^{ A'A}\xi_A. \label{real}
\end{align}
The status of the reality conditions can be clarified by the inclusion of source terms for the fields which are integrated out.
 The Lagrangian can be expanded so that:
 \begin{align}
{\cal L}= \mathcal{L}_0 + \mathcal{L}_{int}
\end{align}

with
\begin{align}
 \mathcal{L}_0 = -\partial^\mu \chi^A \partial_\mu \xi_A - m^2 \chi^A\xi_A, 
\end{align}

and
\begin{align}
 \mathcal{L}_{int} = 2ieA^{AA'}\left( \chi_A (\partial_{A'}{}^{B}\xi_B) +  (\partial_{A'}{}^{B}\chi_B ) \xi_A \right) - e^2 A^B{}_{B'}A^{B'}{}_{B}\chi^A\xi_A
\end{align}

To extract the propagator for the spinor fields, let us rewrite the free part of their Lagrangian as:
\begin{align}
i \mathcal{L}_{Dirac} = \chi_A\left[ i \epsilon^{AB} \left( -\square +m^2\right)\right]\xi_B
\end{align}

Then the inverse of the quadratic operator is:
\begin{align}
\langle 0| T\{\xi_A(p)\chi_B(-p)\}|0 \rangle \equiv S_F(p)_{AB} = \frac{-i}{p^2+m^2}\epsilon_{AB}
\end{align}
where, the field $\xi_A$ sits at the end of the directed line. Similarly, the Feynman rules for the vertices can be obtained by considering the momentum space version of the Lagrangian and setting, by convention, all our particles as incoming. Then the field can be expanded in positive-frequency modes and we obtain for the  vertices (all momenta incoming):
\begin{align}
\langle 0| A^{A'}{}_{A}(q) \chi_B(p)\xi_C(k)|0 \rangle &\rightarrow 2\im e \left[k_C{}^{A'}\epsilon_{BA} + p_B{}^{A'}\epsilon_{CA} \right] \\
\langle 0| A^{A'}{}_{A}(q_1) A^{B'}{}_{B}(q_2) \chi_C(p)\xi_D(k)|0 \rangle &\rightarrow -2\im e^2 \epsilon^{A'B'}\epsilon_{AB}\epsilon_{CD}
\end{align}

\subsection{Propagators}

In the case of second-order fermions, the propagator can be rewritten as:
\begin{align}
S_F(x-y)= \mathbb{I}\cdot \Delta_F(x-y)
\end{align}

where $\mathbb{I}$ is the identity over the unprimed spinors space. This implies that, as before, we have:
\begin{align}
S_F(x-y)= \mathbb{I}\cdot \Delta_F(x-y) =\mathbb{I}\cdot \left(\theta(x^0-y^0)\Delta^+(x-y) +\theta(y^0-x^0)\Delta^-(x-y)   \right) 
\end{align}

Because of this property of second-order fermionic propagators, we also have:
\begin{align}
S_F^\dagger(x-y)= -\mathbb{\bar I}\cdot \Delta_F^*(x-y) = -\left( S_F(x-y)\right)^*
\end{align}

where this time $\mathbb{\bar I}$ is the identity over primed spinors space, and we used $\mathbb{I}^\dagger = -\mathbb{\bar I}$ due to the antisymmetry of the spinor space metric. This minus sign will be important below. We see that as in the scalar field case, the anti-propagator is related to the complex conjugate of the propagator. This is not the case for the first-order propagator. We will, from here on, drop the identity matrices in the propagators and work with the scalar field notation $\Delta$. We therefore have as before:
\begin{align}
&~~\Delta_F(x) = \theta(x^0)\Delta^+(x)  +\theta(-x^0)\Delta^-(x)\nonumber \\
&~~\Delta^*_F(x) = \theta(-x^0)\Delta^+(x)  +\theta(x^0)\Delta^-(x)\nonumber \\
&\begin{array}{ll}
(\Delta ^+(x))^* = \Delta^-(x) &;~\Delta^+(-x) = \Delta^-(x)\\
(\Delta_F(x))^* = \Delta^*_F(x) & ;~\Delta_F(-x) = \Delta_F(x) \\
\end{array}
\end{align}

And this propagators satisfy the same LTE than we derived above. However, as we will now see, it does not apply straightforwardly to fermions as we are in fact working with a non-hermitian Lagrangian.

\subsection{Two warm-up examples}

The fact that the Lagrangian for second-order fermions is non-hermitian implies that in the unitary equation, the Feynman rules used in 
\begin{align}
\langle n | (iT)|m \rangle  ~~~ {\rm or}~~~\langle n | (iT)^\dagger |m \rangle
\end{align}

 not only differ by their dependence in the propagators, but the vertices are also different. Moreover, if in the first case we use polarisations for unprimed spinors, in the latter, we need to use primed polarisations. We will now see in two simple examples how the reality conditions (\ref{real}), when imposed linearly on the external (physical) states of each amplitude in the unitary equation, lead to the perturbative unitarity of the S-matrix.

\subsubsection*{Compton Scattering}
We first consider the simple tree-level example of Compton scattering, see Section \ref{seccompton} for more details. We recall that the amplitude can be split into its different channel contributions:
\begin{align}\begin{split}
\mathcal{M}_s^{A'}{}_{A}{}_{B}{}^{C'}{}_{C}{}_{D}&=-\frac{4e^2i}{s+m^2}\left( -\epsilon_{AC}k_{2B}{}^{A'}k_{4D}{}^{C'}+\epsilon_{AB}q_{C}{}^{A'}k_{4D}{}^{C'}\right. \\ & \hspace*{3cm} \qquad\left.-\epsilon_{CD}q_{A}{}^{C'}k_{2B}{}^{A'}-\frac{1}{2}m^2\epsilon^{A'C'}\epsilon_{AB}\epsilon_{CD}\right)   \end{split}
\end{align}\begin{align}
\begin{split}\mathcal{M}_u^{A'}{}_{A}{}_{B}{}^{C'}{}_{C}{}_{D}&=-\frac{4e^2i}{u+m^2}\left( +\epsilon_{AC}k_{2B}{}^{C'}k_{4D}{}^{A'}-\epsilon_{BC}p_{A}{}^{C'}k_{4D}{}^{A'}\right. \\ & \hspace*{3cm} \qquad\left. -\epsilon_{AD}p_{C}{}^{A'}k_{2B}{}^{C'}-\frac{1}{2}m^2\epsilon^{A'C'}\epsilon_{BC}\epsilon_{AD}\right)   \end{split} \\
\mathcal{M}&=\mathcal{M}_s + \mathcal{M}_u
\end{align}

Unitarity in the s-channel means:
\begin{figure}[H]\begin{center}
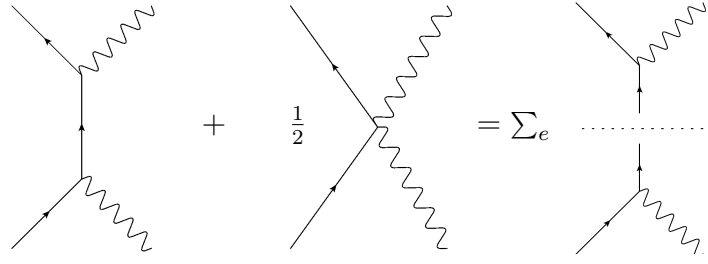\caption{s-channel contribution to Compton scattering.}\label{s-c}
\end{center}\end{figure}

where in the LHS it is understood that we sum over the diagrams and their complex conjugates. In order to compute the latter, we need the complex conjugated Feynman rules derived from the hermitian conjugate of the Lagrangian and the extra sign in the S$^\dagger$-matrix expansion. The quantity to consider is:
\begin{align}\begin{split}
-i\int d^4x \mathcal{L}^\dagger &= -i \int d^4x\left( -{2} D^{AA'} \xi_{A'}^\dagger  D_{A}{}^{B'} \chi^\dagger_{B'} - m^2 \xi_{A'}^\dagger \chi^{\dagger\, A'}\right)\\&= +i \int d^4x\left(-{2} D_{A}{}^{ A'} \xi_{A'}^\dagger  D^{AB'} \chi^\dagger_{B'} - m^2 \xi^{\dagger\, A'} \chi^{\dagger}_{ A'}\right)\label{lagchiraldagger}
\end{split}\end{align}

with now
\begin{align}
D_\mu \xi^\dagger = (\partial_\mu + \im e A_\mu)\xi^\dagger, \qquad
 D_\mu \chi^\dagger = (\partial_\mu - \im eA_\mu)\chi^\dagger
\end{align}

and where in the last line we rewrote the conjugated Lagrangian with the exact same index structure as the original Lagrangian. We therefore see that the momentum space Feynman rules will be exactly the same in the conjugated case:
\begin{align}
\langle 0| T^\dagger\{\chi^\dagger_{A'}(p)\xi^\dagger_{B'}(-p)\}|0 \rangle \equiv S^\dagger_F(p)_{A'B'} = \frac{-i}{p^2+m^2}\epsilon_{A'B'}
\end{align}

and
\begin{align}
\langle 0| A^{A}{}_{A'}(q) \xi^\dagger_{B'}(p)\chi^\dagger_{C'}(k)|0 \rangle &\rightarrow 2\im e \left[k_{C'}{}^{A}\epsilon_{B'A'} + p_{B'}{}^{A}\epsilon_{C'A'} \right]\\
\langle 0| A^{A}{}_{A'}(q_1) A^{B}{}_{B'}(q_2) \xi^\dagger_{C'}(p)\chi^\dagger_{D'}(k)|0 \rangle &\rightarrow -2\im e^2 \epsilon^{AB}\epsilon_{A'B'}\epsilon_{C'D'}  
\end{align}

Therefore the amplitudes are the same as in the usual case, except for the $SL(2,\mathbb{C})$ index structure of the representations that is swapped with its complex conjugated counterpart.

\medskip

Using these facts, the complex conjugate of the s-channel amplitude is given by:
\begin{align}\begin{split}
\mathcal{M}^*_s&=-\frac{4e^2i}{s+m^2}\left( -\epsilon_{A'C'}k_{2B'}{}^{A}k_{4D'}{}^{C}+\epsilon_{A'B'}q_{C'}{}^{A}k_{4D'}{}^{C}\right. \\ & \hspace*{3cm} \qquad\left. -\epsilon_{C'D'}q_{A'}{}^{C}k_{2B'}{}^{A}-\frac{1}{2}m^2\epsilon^{AC}\epsilon_{A'B'}\epsilon_{C'D'}\right)    \end{split}
\end{align}

where now $B',~D'$ denote primed fermions. This amplitude is obtained straightforwardly from the complex conjugate of the aforementioned Feynman rules. So far, we have written the amplitudes in momentum space, however to make sense of the unitarity equation, we rewrite:
\begin{align}\begin{split}
\mathcal{M}_s&=-4e^2i\int \frac{d^4q}{(2\pi)^4}\frac{e^{-iq(x-y)}}{q^2+m^2-i\varepsilon} F(k_i,q) \\
\mathcal{M}^*_s&=-4e^2i\int \frac{d^4q}{(2\pi)^4}\frac{e^{-iq(x-y)}}{q^2+m^2+i\varepsilon}F^*(k_i,q)   \end{split}
\end{align}

where $F$ corresponds the the integrand properly contracted with external polarisations:
\begin{align}
\epsilon^B(k_2),\quad \epsilon^D(k_4), \quad \epsilon^A{}_{A'}(k_1), \quad \epsilon^C{}_{C'}(k_3)
\end{align}

We see that the only difference comes from the fact that the second amplitude contains anti-propagators and that $F^*$ might in general differ from $F$ as they are projections on the polarisations coming from different spaces. In order to understand which polarisations are used in $F^*$, we need to use the mode decomposition of the primed and unprimed fermions and compare them. For unprimed spinors, the rules are given in Section \ref{DiracRules}. We simply recall their mode decomposition:
\begin{align}
 \xi_A(x)  &=  \sum_s \int d\Omega_k \Big(  u_A(s) a_k(s)  e^{\im kx} + v_A(s) c_k^\dagger(s) e^{-\im kx}\Big), \\
  \chi_A(x)  &=  \sum_s \int d\Omega_k \Big(  u_A(s) c_k(s)  e^{\im kx} + v_A(s) a_k^\dagger(s) e^{-\im kx}\Big)
\end{align}

For the primed fermions, using the linearised reality condition for external states:
\begin{align}
\chi^{\dagger\, A'} = i\frac{\sqrt{2}}{m}\partial^{A'}{} _{A}\xi^A, \quad \xi^{\dagger\,  A'} = i\frac{\sqrt{2}}{m}\partial^{A'}{} _{A}\chi^A
\end{align}

and the decomposition into modes of the primed fermions:
\begin{align}
 \xi_{A'}^\dagger (x)  &=  \sum_s \int d\Omega_k \Big(  u^\dagger_{A'}(s) a_k^\dagger(s)  e^{-\im kx} + v^\dagger_{A'}(s) c_k(s) e^{\im kx}\Big),  \\
  \chi^\dagger_{A'}(x)  &=   \sum_s \int d\Omega_k \Big(  u^\dagger_{A'}(s) c_k^\dagger(s)  e^{-\im kx} + v^\dagger_{A'}(s) a_k(s) e^{\im kx}\Big), 
\end{align}

we have:
\begin{align}
\frac{\sqrt{2}}{m}p^{A'}{}_A v^A(s) = u^{\dagger\, A'}(s), \quad \frac{\sqrt{2}}{m}p^{A'}{}_A u^A(s) = -v^{\dagger\, A'}(s)\label{realpol}
\end{align}

Now, if we consider the same external states in terms of creation and annihilation operators, but this time with the primed fermions decomposition, we have:
\begin{align}
|{\bf p}, s ,\pm e \rangle &\rightarrow e^{\im px} v^{\dagger\, A'}(s)|0\rangle \\
\langle{\bf p}, s,\pm e | & \rightarrow \langle 0 |  e^{-\im px} u^{\dagger\, A'}(s)
\end{align}

Using the reality condition (\ref{realpol}), these states can be rewritten as:
\begin{align}
 e^{\im px} v^{\dagger\, A'}(s)|0\rangle  &=  -\frac{\sqrt{2}}{m}p^{A'}{}_A \left(e^{\im px}u^A |0\rangle \right) \label{newpol1} \\
\langle 0 |  e^{-\im px} u^{\dagger\, A'}(s) &= \left(\langle 0 | v^A(s) e^{-\im px}\right) \frac{\sqrt{2}}{m}p^{A'}{}_A \label{newpol2}
\end{align}

We therefore see that the reality conditions imply that we can replace the primed polarisations using the reality equation independently of the state. This will allow us to easily compare numerators in the unitarity equation.
\medskip

We shall now show that we have $F=-F^*$, in other words, that the integrand in the amplitude is purely imaginary once it has been projected on the external polarisations. We can compute the numerators as follows:
\begin{align}\begin{split}
F(k_i, q)&= \epsilon^D_4\epsilon_3^C{}_{C'}\left( -\epsilon_{AC}k_{2B}{}^{A'}k_{4D}{}^{C'}+\epsilon_{AB}q_{C}{}^{A'}k_{4D}{}^{C'} \right. \\ & \hspace*{3cm} \qquad\left.-\epsilon_{CD}q_{A}{}^{C'}k_{2B}{}^{A'}-\frac{m^2}{2}\epsilon^{A'C'}\epsilon_{AB}\epsilon_{CD}\right) \epsilon^B_2\epsilon_1^A{}_{A'}  \\
&=(\epsilon_4 k_4)^{C'}(\epsilon_2k_2)^{A'} (\epsilon_1 \epsilon_3)_{A'C'} + (\epsilon_4 k_4)^{C'}(\epsilon_1\epsilon_2)^{A'} ( \epsilon_3 q)_{C'A'} \\ & \hspace*{3cm} \qquad +(\epsilon_4 \epsilon_3)^{C'}(\epsilon_2k_2)^{A'} ( \epsilon_1 q)_{A'C'}+ \frac{m^2}{2}(\epsilon _2 \epsilon_1)^{C'}(\epsilon_4 \epsilon_3)_{C'}
\end{split}\end{align}

where notation is as follows:
\begin{align}
(\lambda k )^{A'} \equiv \lambda^A k_A{}^{A'} , \quad (kp)_{A'B'} \equiv k_{A'}{}^A p_{AB'}, \quad (\lambda \chi) \equiv \lambda^A\chi_A
\end{align}
for any $\lambda, \chi \in (1/2,0)$ and $k,p \in (1/2,1/2)$, and for commuting spinors these inner products are antisymmetric. Also, $\epsilon_i \equiv \epsilon(k_i)$. The second numerator is given by:
\begin{align}\begin{split}
F^*(k_i, q) &=\epsilon^{*\, D'}_4\epsilon_3^{C'}{}_{C} \left( -\epsilon_{A'C'}k_{2B'}{}^{A}k_{4D'}{}^{C}+\epsilon_{A'B'}q_{C'}{}^{A}k_{4D'}{}^{C}\right. \\ & \hspace*{3cm} \qquad\left. -\epsilon_{C'D'}q_{A'}{}^{C}k_{2B'}{}^{A}-\frac{m^2}{2}\epsilon^{AC}\epsilon_{A'B'}\epsilon_{C'D'}\right)  \epsilon^{*\,B'}_2\epsilon_1^{A'}{}_{A} \\
&= \left( \frac{\sqrt{2}}{m}\epsilon^D_4 k_{4D}{}^{D'}\right)\epsilon_3^{C'}{}_{C} \left( \cdots \right) \left(-\frac{\sqrt{2}}{m}k_{2B}{}^{B'}\epsilon^{B}_2\right)\epsilon_1^{A'}{}_{A}  \\
&= \epsilon^{ D}_4\epsilon_3^{C'}{}_{C} \left( \epsilon^{AC}k_{2BA'}k_{4DC'}+\delta^{A}_{B}q^{C}{}_{A'}k_{4DC'} -\delta^{C}_{D}q^{A}{}_{C'}k_{2BA'}\right. \\ & \hspace*{3cm} \qquad\left.+\frac{m^2}{2}\epsilon_{A'C'}\delta^{A}_{B}\delta^{C}_{D}\right) \epsilon^{B}_2\epsilon_1^{A'}{}_{A} \\
&=(-1)\left((\epsilon_4 k_4)^{C'}(\epsilon_2k_2)^{A'} (\epsilon_1 \epsilon_3)_{A'C'} + (\epsilon_4 k_4)^{C'}(\epsilon_1\epsilon_2)^{A'} ( \epsilon_3 q)_{C'A'}\right. \\ & \hspace*{3cm} \qquad\left.+(\epsilon_4 \epsilon_3)^{C'}(\epsilon_2k_2)^{A'} ( \epsilon_1 q)_{A'C'}+ \frac{m^2}{2}(\epsilon _2 \epsilon_1)^{C'}(\epsilon_4 \epsilon_3)_{C'}\right)\label{fstarcompton}
\end{split}\end{align}

where in the second equality, we have used (\ref{newpol1}-\ref{newpol2}). In the third, the fact the the external fermions are on-shell ($i.e.$ $k^2=-m^2$). The fourth line, simply gives $F=-F^*$ as foretold. The LHS of the unitary equation is then given by:
\begin{align}\begin{split}
\mathcal{M}_s+ \mathcal{M}^*_s &= -4e^2i\int \frac{d^4q}{(2\pi)^4}\left(\frac{1}{q^2+m^2-i\varepsilon} -\frac{1}{q^2+m^2+i\varepsilon} \right)e^{-iq(x-y)}F(k_i,q) \\
&=4e^2 \int \frac{d^4q}{(2\pi)^4}\left(\Delta_F(q)+ \Delta^*_F(q)\right)e^{-iq(x-y)}F(k_i,q)
\end{split}\end{align}

Now, using
\begin{align}
\frac{1}{a \mp i\varepsilon} =   {\rm P}\left(\frac{1}{a} \right) \pm i\pi \delta(a)
\end{align}

The two principal values cancel and the imaginary parts add up to give:
\begin{align}\begin{split}
\mathcal{M}_s+ \mathcal{M}^*_s &= 4e^2\int \frac{d^4q}{(2\pi)^4}(2\pi) \delta(\omega^2-q_0^2) e^{-iq(x-y)}F(k_i,q) \\
&= 4e^2\int \frac{d^3q}{(2\pi)^3}\frac{1}{2\omega}\int dq_0 \left(\delta(\omega+q_0)+\delta(\omega-q_0)\right) e^{-iq(x-y)}F(k_i,q) \\
&= 4e^2\int \frac{d^3q}{(2\pi)^3}\frac{1}{2\omega}\left(e^{+i\omega(x^0-y^0)}F(k_i,q)+ e^{-i\omega(x^0-y^0)} F(k_i,\bar q)\right) e^{-i\vec{q}(\vec x-\vec y)}
\end{split}\end{align}

with in the last line we have $q_\mu = (-\omega, \vec q)$ and $\bar q_\mu = (\omega, \vec q)$. This is almost it, what remains to be done is adding the plane wave factors for the external particles and integrating over the position of the vertices:
\begin{align}\begin{split}
\int d^4x d^4y \left(\mathcal{M}_s+ \mathcal{M}^*_s \right)e^{i(k_1+k_2)x} e^{-i(k_3+k_4)y}
\end{split}\end{align}

Let us simply consider the time integral over the plane waves, we have:
\begin{align}\begin{split}
&\int dx^0 dy^0\left( e^{-ix^0 (E_1+E_2-\omega)} e^{-iy^0 (\omega - E_3-E_4 )} + e^{-ix^0 (E_1+E_2+\omega)} e^{-iy^0 (\omega + E_3+E_4) }\right)  \\
=&(2\pi)^2\Big(\delta(E_1+E_2-\omega)\delta(E_4+E_3-\omega)+\delta(E_1+E_2+\omega)\delta(E_3+E_4+\omega)\Big)
\end{split}\end{align}
where $k_{i\,\mu}= (-E_i, \vec{k}_i)$, and because the energies are positive, only the first term contributes, leading to the LHS of the unitarity equation:
\begin{align}
{\rm LHS} = 4e^2 \int \frac{d^3q}{(2\pi)^3}\frac{1}{2\omega}F(k_i,q) (2\pi)^4 \delta^{(4)}(k_1+k_2-q)(2\pi)^4 \delta^{(4)}(k_3+k_4-q)\label{LHS}
\end{align}

It now remains to compute the RHS. It will be given by, schematically:
\begin{align}
-\int d^4x d^4y \sum_{q} \left( \epsilon_1 \epsilon_2 V_3 \epsilon_q \right) \left( \epsilon^*_q V^*_3 \epsilon_3^* \epsilon_4^*\right)e^{i(k_1+k_2)x} e^{-i(k_3+k_4)y}
\end{align}

Before writing the integrand, we must make sense of the sum over states implied in the equation. We have:
\begin{align}
\sum_{spins} \epsilon^{A}\epsilon^{*A'}(k)= \frac{\sqrt{2}}{m}k^{AA'}, \quad \sum_{\vec{q}}\equiv \int \frac{d^3 q}{(2\pi)^3}\frac{1}{2\omega} \label{sumpol}
\end{align}

where the integration measure corresponds to the Lorentz invariant one. So that we write, for a one-fermion intermediate state:
\begin{align}
\sum_n |n\rangle\langle n|\equiv \sum_q \epsilon^A_q\epsilon^{*A'}_qe^{-iq(x-y)} = \int \frac{d^3 q}{(2\pi)^3}\frac{1}{2\omega}e^{-iq(x-y)}\frac{\sqrt{2}}{m}q^{AA'}
\end{align}
with $q_\mu = (-\omega, \vec q)$. The position space integrals can then be carried out very simply to lead to the same delta functions as in (\ref{LHS}). We are left with the contraction of the vertices and polarisation states. We have:
\begin{align}
-\int \frac{d^3 q}{(2\pi)^3}\frac{1}{2\omega}\epsilon^B_2\epsilon_1^{A}{}_{A'}  V_3^{A'}{}_{ABE} \frac{\sqrt{2}}{m}q^{EE'} V_3^{*C}{}_{C'D'E'} \epsilon_3^{C'}{}_{C} \epsilon^{*D'}_4
\end{align}

We only need to show that the integrand is equal to $4e^2 F(k_i,q)$ and then the unitarity equation would be proven for this case. We have, using again (\ref{newpol1}-\ref{newpol2}) and the fact that $q$ is now on-shell, for the contraction of the two amplitudes:
\begin{align}\begin{split}
&(-1)2ei\left( k_2^{A'}{}_B\epsilon_{EA}- q^{A'}{}_E\epsilon_{BA}\right)\frac{\sqrt{2}}{m}q^{EE'} \\ & \hspace*{3cm} \qquad 2ei \left(q^C{}_{E'}\epsilon_{D'C'}-k_4^C{}_{D'}\epsilon_{E'C'}\right)\left(-\frac{\sqrt{2}}{m}k_4^{D'}{}_D\right)\times{\rm pol} \\
=&4e^2\left( k_2^{A'}{}_B\epsilon_{EA}- q^{A'}{}_E\epsilon_{BA}\right)\left(-\epsilon^{EC}k_{4C'D}+q^E{}_{C'}\delta^C_D\right)\times{\rm pol}\\
=& 4e^2 \left(-k_2^{A'}{}_B\delta^C_Ak_{4C'D} + q^{A'C}\epsilon_{AB} k_{4C'D} + q_{AC'}\delta^C_D k_2^{A'}{}_B - \frac{m^2}{2}\delta^{A'}_{C'}\epsilon_{AB}\delta^C_D\right)\times{\rm pol} \\
=&4e^2 \left((\epsilon_2 k_2)^{A'}(\epsilon_1 \epsilon_3)_{A'C'}(\epsilon_4 k_4)^{C'} + (\epsilon_2 \epsilon_1)_{A'} (\epsilon_4 k_4)_{C'} (q \epsilon_3)^{A'C'} \right. \\ & \hspace*{3cm} \qquad\left.+ (\epsilon_2 k_2)^{A'}(\epsilon_4 \epsilon_3)^{C'} (\epsilon_1 q)_{A'C'} - \frac{m^2}{2}(\epsilon_2\epsilon_1)_{C'}(\epsilon_4 \epsilon_3)^{C'} \right)\\
=&4e^2 F(k_i, q)\label{numrhs}
\end{split}\end{align}

This implies that, upon integration over vertices, we have the unitarity equation (Fig.\ref{s-c}):
\begin{align}
\boxed{\begin{array}{rl}{\rm LHS}& = 4e^2 \int \frac{d^3q}{(2\pi)^3}\frac{1}{2\omega}F(k_i,q) (2\pi)^4 \delta^{(4)}(k_1+k_2-q)(2\pi)^4 \delta^{(4)}(k_3+k_4-q) \\ &={\rm RHS} \end{array}}\label{unitar}
\end{align}

Therefore proving unitarity for this process at this order in the coupling constant. It should be noted, that the main ingredient is the possibility to convert the complex conjugated amplitude into an usual one using the reality conditions on the external primed states and the fact the the sum over intermediate states also gives a reality condition-like factor. The rest follows from the decomposition of the propagators into forward and backward propagations.

\subsubsection*{Photon two-point function}
Let us now consider a case in which the fermions appear inside a loop. The easiest case is the photon two-point function, or charge renormalisation diagram that we discussed in Section \ref{seccharge}. Recall that the amplitude is given by:
\begin{align}\begin{split}
\mathcal{M}(k)= &(-1)4e^2 \left(\epsilon_k^A{}_{A'}\epsilon_{k}^{*B}{}_{B'}\right)\int d^Dx d^Dy~ e^{ikx} e^{-iky}\\& \int\frac{d^D p}{(2\pi)^D}\frac{d^D q}{(2\pi)^D} e^{-i(q+p)(x-y)}\frac{\left[- p^{A'}{}_{B}q^{B'}{}_{A} - q^{A'}{}_{B}p^{B'}{}_{A} +m^2 \epsilon^{A'B'}\epsilon_{AB}  \right]}{\left[p^2+m^2-i\varepsilon \right]\left[q^2+m^2 -i\varepsilon\right]}
\end{split}
\end{align}

In the LHS of the unitarity equation (Fig.\ref{loopcut}), there is a sum over the aforementioned amplitude and the amplitude calculated using conjugated Feynman rules.

\begin{figure}[H]\begin{center}
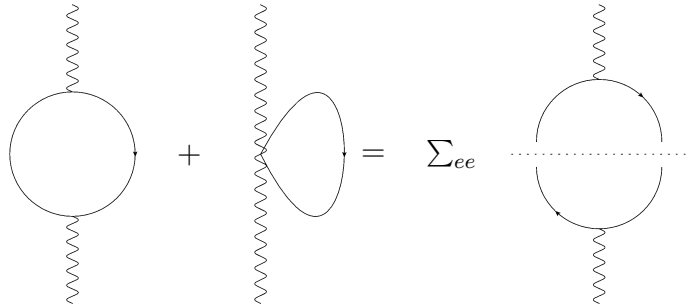\caption{Unitarity equation at one-loop.}\label{loopcut}
\end{center}\end{figure}

The latter is given by:
\begin{align}\begin{split}
\mathcal{M}^*(k)= &(-1)4e^2 \left(\epsilon_k^{A'}{}_{A}\epsilon_{k}^{*B'}{}_{B}\right)\int d^Dx d^Dy~  e^{ikx} e^{-iky}\\& \int\frac{d^D p}{(2\pi)^D}\frac{d^D q}{(2\pi)^D} e^{-i(q+p)(x-y)}\frac{\left[ -p^{A}{}_{B'}q^{B}{}_{A'} - q^{A}{}_{B'}p^{B}{}_{A'} +m^2 \epsilon^{AB}\epsilon_{A'B'}  \right]}{\left[p^2+m^2+i\varepsilon \right]\left[q^2+m^2 +i\varepsilon\right]} 
\end{split}
\end{align}

Notice that the sum of the amplitudes can again be written as a function of the scalar propagators:
\begin{align}\begin{split}
\mathcal{M}+\mathcal{M}^*= (-1)4e^2& \int d^Dx d^Dy  \int\frac{d^D p}{(2\pi)^D}\frac{d^D q}{(2\pi)^D} e^{-i(q+p-k)(x-y)} \\ & \hspace*{2cm} \qquad\times \left(-\Delta_F(p)\Delta_F(q)-\Delta^*_F(p)\Delta^*_F(q)\right) F(q,p,k)
\end{split}
\end{align}

where $F$ is again a function of the momenta with no poles. Because of this, the decomposition of the propagators appearing in the integral in terms of forward and backward propagations is still valid, since it only relies on the pole structure in the energy plane. Let us first explicitly show it in momentum space by considering the zeroth component integrals for the first term in the brackets in the above formula:
\begin{align}\begin{split}
\int\frac{d p_0}{(2\pi)}\frac{d q_0}{(2\pi)} e^{-i(q+p)_0(x-y)^0}
\Delta_F(p)\Delta_F(q) F(q,p,k)
\end{split}
\end{align}

We have for the first integral:
\begin{align}\begin{split}
\int\frac{d p_0}{(2\pi)} e^{-ip_0(x-y)^0}
&\Delta_F(p) F(q,p,k)\\& = \left[\frac{e^{i\omega_p (x-y)^0}}{2\omega_p}\theta_{yx}F(q, p,k) + \frac{e^{-i\omega_p(x-y)^0}}{2\omega_p}\theta_{xy}F(q, \bar p,k) \right]
\end{split}
\end{align}

with $\theta_{xy} \equiv \theta (x^0-y^0)$. Let us now consider the second integral over the first term in the square brackets:
\begin{align}\begin{split}
\int\frac{d q_0}{(2\pi)} e^{-iq_0(x-y)^0}
&\Delta_F(q) \theta_{yx}F(q, p,k) \\ &= \left[\frac{e^{i\omega_q (x-y)^0}}{2\omega_q}\theta_{yx}\theta_{yx} F( q, p,k) + \frac{e^{-i\omega_q(x-y)^0}}{2\omega_q}\theta_{xy}\theta_{yx}F(\bar q,  p,k) \right] \\
 &= \left[\frac{e^{i\omega_q (x-y)^0}}{2\omega_q}\theta_{yx} F( q, p,k) \right]
\end{split}
\end{align}

where in the second line we used $\theta_{xy}\theta_{yx} = 0$ and $\theta_{yx}\theta_{yx}=\theta_{yx}$. Similarly the second terms yields:
\begin{align}\begin{split}
\int\frac{d q_0}{(2\pi)} e^{-iq_0(x-y)^0}
\Delta_F(q)\theta_{xy}F(q, p,k) &= \left[\frac{e^{-i\omega_q(x-y)^0}}{2\omega_q}\theta_{xy}F(\bar q,   \bar p,k)\right]
\end{split}
\end{align}

So that:
\begin{align}\begin{split}
\int\frac{d p_0}{(2\pi)}\frac{d q_0}{(2\pi)} &e^{-i(q+p)_0(x-y)^0}
\Delta_F(p)\Delta_F(q) F(q,p,k)\\&= \left[\frac{e^{-i(\omega_q+\omega_p) (x-y)^0}}{(2\omega_q)(2\omega_p)}\theta_{xy}F(\bar q,  \bar p,k)+\frac{e^{i(\omega_q+\omega_p) (x-y)^0}}{(2\omega_q)(2\omega_p)}\theta_{yx} F( q, p,k)\right]
\end{split}
\end{align}

Finally, the antipropagators integral is:
\begin{align}\begin{split}
\int\frac{d p_0}{(2\pi)}\frac{d q_0}{(2\pi)}&e^{-i(q+p)_0(x-y)^0}
\Delta_F^*(p)\Delta^*_F(q) F(q,p,k)\\&= \left[\frac{e^{-i(\omega_q+\omega_p) (x-y)^0}}{(2\omega_q)(2\omega_p)}\theta_{yx}F(\bar q,  \bar p,k)+\frac{e^{i(\omega_q+\omega_p) (x-y)^0}}{(2\omega_q)(2\omega_p)}\theta_{xy} F( q, p,k)\right]
\end{split}
\end{align}

and adding the two  contributions gives:
\begin{align}\begin{split}
 \left[\frac{e^{-i(\omega_q+\omega_p) (x-y)^0}}{(2\omega_q)(2\omega_p)}F(\bar q,  \bar p,k)+\frac{e^{i(\omega_q+\omega_p) (x-y)^0}}{(2\omega_q)(2\omega_p)}F( q, p,k)\right]
\end{split}
\end{align}

where we used $\theta_{xy}+\theta_{yx}=1$. If we now consider the integration over vertices, including the external plane waves factors, only the second term contributes to the delta functions, as before, and we get for the LHS of the unitarity equation:
\begin{align}\begin{split}
\mathcal{M}+\mathcal{M}^*= 4e^2&  \int d\Omega_pd\Omega_q \left((2\pi)^D\delta^{(D)}(p+q-k)\right)^2 F(q,p,k)
\end{split}
\end{align}

where $d\Omega_p = \frac{d^{D-1} p}{(2\pi)^{D-1}(2\omega_p)}$ is the Lorentz invariant measure. It only remains to check that the RHS of the equation reproduces the same result. The sum over internal states with momenta $q$ and $p$ will yield the same integrals, furthermore, the integration over vertices will yield the same delta functions, we only need to check that the numerator is the same. Schematically:
\begin{align}
 -\left( \epsilon_k  V_3 \epsilon_p \epsilon_q \right) \left(   \epsilon^*_q \epsilon^*_p V^*_3 \epsilon_{k}^* \right) \stackrel{?}{=} 4e^2 F(q,p,k)
\end{align}

We have using (\ref{sumpol}.i) again:
\begin{align}\begin{split}
&(-1)(-2ei)^2( -p^{A'}{}_{E}\epsilon_{FA}- q^{A'}{}_{F}\epsilon_{EA}) \frac{2}{m^2}p^{EE'}q^{FF'} (p^{B}{}_{E'}\epsilon_{F'B'}+ q^{B}{}_{F'}\epsilon_{E'B'}) \\
=&4e^2 ( -p^{A'}{}_{E}\epsilon_{FA}- q^{A'}{}_{F}\epsilon_{EA})(-\epsilon^{EB}q^{F}{}_{B'} - \epsilon^{FB}p^{E}{}_{B'})\\
=&4e^2 ( -p^{A'B}q_{AB'}-q^{A'B}p_{AB'}- m^2\delta^B_A\delta^{A'}_{B'})
\end{split}\end{align}

So that, when contracting with the polarisations:
\begin{align}
4e^2 \epsilon_k^{A}{}_{A'} \left[- p^{A'}{}_{B}q^{B'}{}_{A} - q^{A'}{}_{B}p^{B'}{}_{A} +m^2 \epsilon^{A'B'}\epsilon_{AB}  \right] \epsilon_k^{*B}{}_{B'} =  4e^2 F(q,p,k)
\end{align}
and therefore:
\begin{align}
\boxed{ {\rm LHS} = 4e^2 \int d\Omega_pd\Omega_q \left((2\pi)^D\delta^{(D)}(p+q-k)\right)^2 F(q,p,k)= {\rm RHS}}
\end{align}

Hence proving unitarity for this process.

\subsection{Reality conditions and amplitude numerators}\label{realityampli}

In the two examples that we have explicitly developed, the proof of unitarity relied extensively on the fact that all contributions to the LHS and to the RHS share the same numerator. This is not at all an obvious fact, as in general the contributions to the unitarity equation are a mixture of amplitudes calculated using different Feynman rules (recall that our Lagrangian is not Hermitian). However, we discovered that by making use of the reality conditions (\ref{newpol1},\ref{newpol2}) and of (\ref{sumpol}), we can project the amplitudes derived using the hermitian-conjugated Lagrangian onto ``unprimed'' amplitudes. Without this trick it would, \`a priori, not be true that the theory is unitary. Therefore we should derive a general result involving amplitudes numerators, before proving in the next subsection that this and the scalar field theory LTE is enough to ensure unitarity of second-order fermions.

\medskip

In order to do so, we will need a compact notation for the numerators. We will henceforth use an index-free notation for the Feynman rules. We have for the vertices with $p_1$ incoming and $p_2$ outgoing:
\begin{align}
V^\mu_3= (-2ei)(\theta^\mu p_1 +p_2\theta^\mu )=V^{*\mu}_3 , \quad V_4^{\mu\nu} = (-2e^2i)\left(\theta^{\mu}\theta^{\nu}+\theta^{\nu}\theta^{\mu}\right)=V_4^{*\,\mu\nu}
\end{align}

where we wrote the quartic vertex so that the contribution to the two channels over the symmetrisation over photon lines is explicit. The ``propagators'' are simply given by:
\begin{align}
-i\mathbb{I}, \quad -i\mathbb{\bar I}
\end{align}

Notice that from now on, the order in which we contract vertices matters. We follow the usual rules for fermions: we climb up the charge arrow, and in the above vertex $p_2$ corresponds to that momentum. Also now, using unprimed spinors and following a fermionic line, the latter starts with a polarisation spinor $u_A$ and finishes (at the tip of the arrow) with another spinor $v^A$. Because of this, the reality conditions (\ref{newpol1},\ref{newpol2}) become:
\begin{align}
v^{\dagger}|0\rangle  &= \frac{\sqrt{2}}{m}p\cdot u |0\rangle  \label{newpol3} \\
\langle 0 |  u^{\dagger} &= \langle 0 | v\cdot p  \frac{\sqrt{2}}{m} \label{newpol4}
\end{align}

where $p \equiv p_\mu\theta^\mu$ and the extra sign in (\ref{newpol3}) comes from the modified spinor contractions. Finally:
\begin{align}
\sum_{spins} \epsilon\epsilon^\dagger(k) = \frac{\sqrt{2}}{m}k
\end{align}

Using this notation we shall derive results for a string of primed or unprimed fermions (or fermionic line) with one polarisation spinor at each end or with a sum over states inserted in the middle, as well as for (cut) single-loops. Indeed, the interaction vertices are such that these are the only two possible cases to consider. Also, we will restrict ourselves to the `s-channel' amplitudes only, as the proof is the same for other channels. Our convention is that all of the external photons are incoming.

\subsubsection*{Fermionic lines}
We first compare a string of unprimed fermions (uncut) to the same string of primed fermions. This case corresponds to the LHS of the unitarity equation. We will prove the results by induction, not taking into account what follows the photon line. First, for one external photon, we have:
\begin{align}\begin{split}
\mathcal{N}^{\mu} &= v(p_1) (-2ei)(\theta^\mu p + p_1\theta^\mu) u(p) 
\end{split}\end{align}
\begin{align}\begin{split}
\mathcal{N}^{*\mu} &= u^\dagger(p_1) (-2ei)(\theta^\mu p + p_1\theta^\mu) v^\dagger(p) = \frac{2}{m^2}v(p_1) (-2ei)p_1(\theta^\mu p + p_1\theta^\mu)pu(p) \\& =  - v(p_1) (-2ei)(\theta^\mu p + p_1\theta^\mu) u(p) =- \mathcal{N}^\mu
\end{split}\end{align}

which satisfies the unitarity equation trivially. For two photons, this amplitude can either be extended with a propagator and another cubic vertex, or replaced with the quartic vertex (in both cases, we choose the s-channel):
\begin{align}\begin{split}
&\mathcal{N}^{\mu_2\mu_1} \\&= v(p_2)\left[-i(-2ei)^2(\theta^{\mu_2} p_1 + p_2\theta^{\mu_2})(\theta^{\mu_1} p + p_1\theta^{\mu_1})+(-2e^2i)(s_1+m^2)\theta^{\mu_2}\theta^{\mu_1}\right] u(p)\\
&= v(p_2)(4e^2i)\left[\theta^{\mu_2} p_1\theta^{\mu_1} p  + p_2\theta^{\mu_2}\theta^{\mu_1} p +\theta^{\mu_2} p_1p_1\theta^{\mu_1}\right. \\ & \hspace*{3cm} \qquad\left. + p_2\theta^{\mu_2}p_1\theta^{\mu_1}-\theta^{\mu_2}(p_1p_1+\frac{m^2}{2})\theta^{\mu_1}\right] u(p)\\
&= v(p_2)(4e^2i)\left[\theta^{\mu_2} p_1\theta^{\mu_1} p  + p_2\theta^{\mu_2}\theta^{\mu_1} p  + p_2\theta^{\mu_2}p_1\theta^{\mu_1}-\theta^{\mu_2}\frac{m^2}{2}\theta^{\mu_1}\right] u(p)
\end{split}\end{align}

where we multiplied the quartic vertex by the inverse propagator with $s_1 = p_1^2 = 2 p_1 p_1$, so that it effectively comes with the same denominator in the amplitude. For the conjugated amplitude we have:
\begin{align}\begin{split}
\mathcal{N}^{*\mu_2\mu_1} &= u^\dagger(p_2)(4e^2i)\left[\theta^{\mu_2} p_1\theta^{\mu_1} p  + p_2\theta^{\mu_2}\theta^{\mu_1} p  + p_2\theta^{\mu_2}p_1\theta^{\mu_1}-\theta^{\mu_2}\frac{m^2}{2}\theta^{\mu_1}\right] v^\dagger(p)
\end{split}\end{align}

%In order to prove the result for any number of photons we will prove the following: 
In this case, we can explicitly carry through the amplitude the reality condition that we impose on $v^\dagger(p)$, which in turn will transform the last spinor $u^\dagger(p_2)$ in its unprimed equivalent. We have:
\begin{align}\begin{split}
&\mathcal{N}^{*\mu_2\mu_1} \\&= v^\dagger(p_2)(4e^2i)\left[\theta^{\mu_2} p_1\theta^{\mu_1} p  + p_2\theta^{\mu_2}\theta^{\mu_1} p  + p_2\theta^{\mu_2}p_1\theta^{\mu_1}-\theta^{\mu_2}\frac{m^2}{2}\theta^{\mu_1}\right] \frac{\sqrt{2}}{m}p u(p) \\
&= u^\dagger(p_2)\frac{\sqrt{2}}{m}(4e^2i)\left[\theta^{\mu_2} p_1\theta^{\mu_1} pp  + p_2\theta^{\mu_2}\theta^{\mu_1} pp  + p_2\theta^{\mu_2}p_1\theta^{\mu_1}p-\theta^{\mu_2}\frac{m^2}{2}\theta^{\mu_1}p\right]  u(p)\\
&= u^\dagger(p_2)\frac{\sqrt{2}}{m}(4e^2i)\left[-\frac{m^2}{2}\theta^{\mu_2} p_1\theta^{\mu_1}   - p_2\theta^{\mu_2}\frac{m^2}{2}\theta^{\mu_1} + p_2\theta^{\mu_2}p_1\theta^{\mu_1}p-\frac{m^2}{2}\theta^{\mu_2}\theta^{\mu_1}p\right]  u(p)\\
&= u^\dagger(p_2)\frac{\sqrt{2}}{m}(4e^2i)\left[p_2p_2\theta^{\mu_2} p_1\theta^{\mu_1}   - p_2\theta^{\mu_2}\frac{m^2}{2}\theta^{\mu_1} + p_2\theta^{\mu_2}p_1\theta^{\mu_1}p+p_2p_2\theta^{\mu_2}\theta^{\mu_1}p\right]  u(p)\\
&= u^\dagger(p_2)p_2\frac{\sqrt{2}}{m}(4e^2i)\left[p_2\theta^{\mu_2} p_1\theta^{\mu_1}   - \theta^{\mu_2}\frac{m^2}{2}\theta^{\mu_1} + \theta^{\mu_2}p_1\theta^{\mu_1}p+p_2\theta^{\mu_2}\theta^{\mu_1}p\right]  u(p)\\
&= -v(p_2)(4e^2i)\left[p_2\theta^{\mu_2} p_1\theta^{\mu_1}   - \theta^{\mu_2}\frac{m^2}{2}\theta^{\mu_1} + \theta^{\mu_2}p_1\theta^{\mu_1}p+p_2\theta^{\mu_2}\theta^{\mu_1}p\right]  u(p)\\
&=-\mathcal{N}^{\mu_2\mu_1}
\end{split}\end{align}

where we have used the fact that $p_2$ is on-shell, and that (\ref{newpol3}-\ref{newpol4}) imply:
\begin{align}
u^\dagger(p)p\frac{\sqrt{2}}{m}= -v(p), \quad \frac{\sqrt{2}}{m}p v^\dagger(p)= -u(p)
\end{align}

We have obtained the same result as in (\ref{fstarcompton}), but using this time an index-free notation.

%In order to be able to prove a general result involving $n$ external photons, we need an induction formula. At order $n \geq 1$ the numerator of the amplitude will be given by (excluding the external polarisations):
%
%
%\begin{align}
%\mathcal{N}^{\mu_n \cdots \mu_1} = V_3^{\mu_n}(-p_n, p_{n-1}) \cdot i\cdot  \mathcal{N}^{\mu_{n-1} \cdots\mu_1} + (-2e^2i)\theta^{\mu_n}(s_{n-1}^2+ m^2)\theta^{\mu_{n-1}}\cdot i \cdot \mathcal{N}^{\mu_{n-2} \cdots \mu_1}
%\end{align}
%
%with, for $n=1$, $i\cdot  \mathcal{N}^{\mu_0} \equiv 1$ and $\mathcal{N}^{\mu_{-1}}=0$. This can also be rewritten as (for $n\geq 2$): 
%
%\begin{align}
%\mathcal{N}^{\mu_n \cdots \mu_1} = \mathcal{N}^{\mu_n \mu_{n-1} } \cdot i\cdot  \mathcal{N}^{\mu_{n-2} \cdots\mu_1} + V_3^{\mu_n}(-p_n, p_{n-1}) \cdot i\cdot  (-2e^2i)\theta^{\mu_{n-1}}(s_{n-2}^2+ m^2)\theta^{\mu_{n-2}}\cdot i \cdot \mathcal{N}^{\mu_{n-3} \cdots \mu_1}
%\end{align}
%with the same rules as before for $n=2$. 

We will show:
\begin{align}
\frac{\sqrt{2}}{m}p_n \mathcal{N}^{*\,\mu_n \cdots \mu_1}p \frac{\sqrt{2}}{m} {=} - \mathcal{N}^{\mu_n \cdots \mu_1} \label{numTTdag}
\end{align}

In the Appendix, we showed (\ref{1to2})\footnote{In this chapter, we use the notation $\mathcal{A}$ to denote the amplitude that has not yet been projected over the fermions polarisations.} 
\begin{align}
 S_n \mathcal{A}_{(1)}^{\mu_n \dots \mu_1}  S  =m D_n \mathcal{A}_{(2)}^{\mu_n \dots \mu_1}  D  
\end{align}

where the amplitude on the left is either from unprimed to unprimed or from primed to primed and on the right, it is an amplitude built using  second-order Feynman rules (or equivalently their anti-chronological conjugated rules). $D_i$ stands for either $\Delta_F$ or $-\Delta_F^*$. And $S_i$ stands for the first-order fermionic propagator. If we consider unprimed fermions external states, we have:
\begin{align}\begin{split}
 v_{(1)}(p_n)(\sqrt{2}p_n +m) D_n &\mathcal{A}_{(1)}^{\mu_n \dots \mu_1}D  (-\sqrt{2}p +m)u_{(1)}(p)\\&  = mv_{(1)}(p_n) D_n \mathcal{A}_{(2)}^{\mu_n \dots \mu_1}  D u_{(1)}(p) 
\end{split}\end{align}
 
If on the LHS we consider the primed to primed amputated amplitude, we have:
\begin{align}
- m^2 v_{(1)}^{\dagger}(p_n)\mathcal{A}_{(1)}^{\mu_n \dots \mu_1} u_{(1)}^{\dagger}(p)  =  m^2 v_{(2)}(p_n) \mathcal{A}_{(2)}^{\mu_n \dots \mu_1}  u_{(2)}(p) 
\end{align}

where we have rescaled the second-order polarisation spinors. Consider now primed fermions external states:
\begin{align}\begin{split}
 v_{(1)}^{\dagger}(p_n)(-\sqrt{2}p_n +m) D_n&\mathcal{A}_{(1)}^{\mu_n \dots \mu_1}D  (\sqrt{2}p +m)u_{(1)}^{\dagger}(p) \\& = mv_{(1)}^{\dagger}(p_n) D_n \mathcal{A}_{(2)}^{\mu_n \dots \mu_1}  D u_{(1)}^{\dagger}(p) 
\end{split}\end{align}

 with the the primed to primed amputated amplitude on the LHS:
\begin{align}
 +m^2 v_{(1)}^{\dagger}(p_n)\mathcal{A}_{(1)}^{\mu_n \dots \mu_1} u_{(1)}^{\dagger}(p)  =  m^2v_{(2)}^{\dagger}(p_n) \mathcal{A}_{(2)}^{\mu_n \dots \mu_1}   u_{(2)}^{\dagger}(p) 
\end{align}

So that we have:
\begin{align}
v_{(2)}(p_n) \mathcal{A}_{(2)}^{\mu_n \dots \mu_1}  u_{(2)}(p)  = - v_{(2)}^{\dagger}(p_n) \mathcal{A}_{(2)}^{\mu_n \dots \mu_1}   u_{(2)}^{\dagger}(p) 
\end{align}

Which is equivalent to:
\begin{align}
\mathcal{M}^{\mu_n \dots \mu_1}   = - \mathcal{M}^{*\mu_n \dots \mu_1}   
\end{align}

since the Feynman rules used on the RHS are the same for $\langle f|(iT)|i \rangle$ and $\langle f|(iT)^\dagger|i \rangle$, and by extension this leads to (\ref{numTTdag}).

\medskip

Now, consider the same string of fermions, but with a sum over states inserted somewhere in the middle, and with the rest of the amplitude calculated using the conjugated Feynman rules. This would correspond to the RHS of the unitarity equation. Again, we would like to show that the numerator of this quantity is proportional to the numerator of the full (conjugated) amplitude. In order to do so, let us first state a result concerning the vertices that are involved in the diagrams. As we saw before, for any number of connected photons $n\geq2$, we will have several diagrams (restricting ourselves without loss of generality to the s-channel) contributing to the amplitude. More precisely, for any two cubic vertices connected by a propagator, there will be another diagram where such contribution is replaced by a (s-channel contribution of a) quartic vertex, and the rest of the amplitude is identical. Specifically, we will have:
\begin{align}\begin{split}
\mathcal{A}^{\mu_2\mu_1} & =\left[(-2ei)^2(\theta^{\mu_2} p_1 + p_2\theta^{\mu_2})(-i)(\theta^{\mu_1} p + p_1\theta^{\mu_1})+(-2e^2i)(s_1+m^2)\theta^{\mu_2}\theta^{\mu_1}\right]\\
&= (4e^2i)\left[\theta^{\mu_2} p_1\theta^{\mu_1} p  + p_2\theta^{\mu_2}\theta^{\mu_1} p  + p_2\theta^{\mu_2}p_1\theta^{\mu_1}-\theta^{\mu_2}\frac{m^2}{2}\theta^{\mu_1}\right]\\
&=\left[(-2ei)^2(\theta^{\mu_2} p_1 + p_2\theta^{\mu_2})(-i)(\theta^{\mu_1} p + p_1\theta^{\mu_1})\right]_{s_1=-m^2}
\end{split}\end{align}

and it will always be the case that the quartic vertex contribution effectively sets on-shell, in the numerator, the momentum propagating between the cubic vertices. So that, any numerator for a string of fermions can be written as:
\begin{align}\begin{split}
\mathcal{A}^{\mu_n\cdots \mu_1} & = (-i)^{n-1} \left[\prod_{i=n}^{1}  V_3^{\mu_i}(-p_{i},p_{i-1})\right]_{s_i=-m^2}\label{onshellnum}
\end{split}\end{align}
where by $\prod_{i=n }^{1}$, we mean the ordered contraction of cubic vertices, starting from the outgoing fermion:
\begin{align}\begin{split}
\prod_{i=n}^{1}  V_3^{\mu_i}(-p_{i},p_{i-1})= (-2ei)(\theta^{\mu_n} p_{n-1} +p_n\theta^{\mu_n} )\cdots(-2ei)(\theta^{\mu_1} p +p_1\theta^{\mu_1} ) 
\end{split}\end{align}

Now that we have this result, we can look at the full numerator:
\begin{align}\begin{split}
\mathcal{\tilde N}^{\mu_n \cdots \mu_1} &= v(p_n)\mathcal{A}^{\mu_n \cdots \mu_{q+1}}\frac{\sqrt{2}}{m}q \mathcal{A}^{*\mu_q \cdots \mu_1} v^\dagger(p)\\
 &= v(p_n)\mathcal{A}^{\mu_n \cdots \mu_{q+1}}\frac{\sqrt{2}}{m}q \mathcal{A}^{*\mu_q \cdots \mu_1}p\frac{\sqrt{2}}{m} u(p)\\
 &= -v(p_n)\mathcal{A}^{\mu_n \cdots \mu_{q+1}} \mathcal{A}^{\mu_q \cdots \mu_1} u(p)\\
 &= (-i) v(p_n)\mathcal{A}^{\mu_n \cdots \mu_{q+1}}(-i) \mathcal{A}^{\mu_q \cdots \mu_1} u(p)\\
 &= (-i)\mathcal{ N}^{\mu_n \cdots \mu_1}
 \end{split}\end{align}

where in the third line we used (\ref{numTTdag}) and in the last line the fact that any amplitude can be built from lower order amplitudes with on-shell numerators (\ref{onshellnum}). Moreover, the results does not depend on the index $q$, and therefore is valid for any cut along the string. Notice, that we recover the result from the first example where we had:
\begin{align}
\mathcal{N} = (-4e^2i) F(q,k_i), \quad \mathcal{N}^* = (+4e^2i) F(q,k_i), \quad \mathcal{\tilde N} = (-4e^2) F(q,k_i)
\end{align}

We therefore see that the numerators are proportional to each other. We will use this result in the next subsection, but before we will prove similar results in the case of loops.

\subsubsection*{Loops}

First of all, if we consider a closed fermion loop (LHS of the unitarity equation), we have trivially:
\begin{align}
\mathcal{N}^{\mu_n\cdots \mu_1}=(-1){\rm Tr} \left( \mathcal{A}^{\mu_n\cdots \mu_1} \right)= (-1){\rm Tr} \left( \mathcal{A}^{*\mu_n\cdots \mu_1} \right)= \mathcal{N}^{*\mu_n\cdots \mu_1}
\end{align}
since the traces are independent of the type of fermions involved inside the loop. On the RHS, however, we will have something more complicated. Suppose the loop is cut through the propagators labelled by $k,~l$, then 
\begin{align}
\begin{split}
\mathcal{\tilde N}^{\mu_n\cdots \mu_1} &= \mathcal{A}^{\mu_{k+1}\cdots \mu_{l}} v(p_k)v(p_l)u^\dagger(p_l)u^\dagger(p_k) \mathcal{A}^{*\mu_{l+1}\cdots \mu_{k}} \\
&= \mathcal{A}^{\mu_{k+1}\cdots \mu_{l}} \frac{2}{m^2}p_l p_k \mathcal{A}^{*\mu_{l+1}\cdots \mu_{k}} \\
&= (-)i^2{\rm Tr}\left(\mathcal{A}^{\mu_{k+1}\cdots \mu_{l}}(-i)^2 \mathcal{A}^{\mu_{l+1}\cdots \mu_{k}} \right)\\
&=(-1)\mathcal{N}^{\mu_n\cdots \mu_1}
\end{split}
\end{align}

which again reproduces the result we obtained in the loop example.

\subsection{Unitarity of spinors from scalar field theory}

We will now use the results obtained about the numerators in order to factor out the fermionic dependence (numerators) from the unitarity equation, and deal instead with a purely scalar case. As before, we will treat separately the case of a string of fermions with $V\geq 2$ connected photons, and the fermionic loop. 

\subsubsection*{String of fermions}

In order to conclude our proof of unitarity, let us rewrite the most general amplitude for a string of fermions as:
\begin{align}
\langle f | (iT) | i\rangle = \int \prod_{i=1}^V d^4 x_i e^{ipx_1}e^{-iqx_V} \int \prod_{j=1}^{I}\frac{d^4p_j}{(2\pi)^4}e^{ip_j(x_{j+1}-x_{j})}\Delta_F(p_j) \mathcal{N}\left(\{p_j\}, q, p\right)
\end{align}

with $V$ the number of vertices, $I=V-1$ the number of internal lines (propagators), $x_1$ the vertex attached to the incoming fermion with momentum $p$, $x_V$ the vertex attached to the outgoing fermion with momentum $q$, and where we have stripped off a factor of $(-i)^I$ from the numerator (as it was defined above) in order to have the exact expression for the Feynman propagators. Also, we have omitted the spacetime indices on the numerator. Let us now define a $4I$ dimensional momentum variable as well as a $4I$ dimensional spacetime variable:
\begin{align}
P^I = \bigoplus_{j=1}^{I}p_j, \quad X^I = \bigoplus_{j=1}^{I}(x_{j+1}-x_j)
\end{align}

as well as a propagator function:
\begin{align}
\mathcal{D}(P) = \prod_{j=1}^{I}\Delta_F(p_j)
\end{align}

The amplitude can then be rewritten as (dropping for a moment the $q$ and $p$ dependence of the numerator):
\begin{align}
\langle f | (iT) | i\rangle = \int \prod_{i=1}^V d^4 x_i e^{ipx_1}e^{-iqx_V} \int\frac{d^{4I}P}{(2\pi)^{4I}}e^{iPX}\mathcal{D}(P) \mathcal{N}\left( P\right)
\end{align}

The second integral is now simply rewritten as a $4I$ dimensional Fourier transform of a product of functions. Using the convolution theorem, we have:
\begin{align}
\int\frac{d^{4I}P}{(2\pi)^{4I}}e^{iPX}\mathcal{D}(P) \mathcal{N}\left( P\right) = \int d^{4I}Z ~\mathcal{D}(Z) \mathcal{N}\left( X-Z\right)
\end{align}

where $\mathcal{D}(Z)$ and  $\mathcal{N}\left( Z\right)$ are the Fourier transforms into position space of their respective momentum space quantities. The full amplitude is now given by:
\begin{align}
\langle f | (iT) | i\rangle = \int \prod_{i=1}^V d^4 x_i e^{ipx_1}e^{-iqx_V}\int d^{4I}Z ~\mathcal{D}(Z) \mathcal{N}\left( X-Z ; p,q \right)
\end{align}

Let us look more carefully at the propagator function in position space:
\begin{align}
\mathcal{D}(Z) =\int \prod_{j=1}^{I}\frac{d^4p_j}{(2\pi)^4}e^{ip_j(z_{j+1}-z_{j})}\Delta_F(p_j) =\prod_{j=1}^{I} \Delta_F(z_{j+1}-z_{j})
\end{align}

This is nothing but the first term in the LTE for scalar fields stripped off the vertices factors! The same can be applied to the conjugated amplitude to obtain:
\begin{align}\begin{split}
\langle f | &(iT) ^\dagger | i\rangle\\&= \int \prod_{i=1}^V d^4 x_i e^{ipx_1}e^{-iqx_V} \int \prod_{j=1}^{I}\frac{d^4p_j}{(2\pi)^4}e^{ip_j(x_{j+1}-x_{j})}\left(-\Delta_F^*(p_j)\right) \mathcal{N}^*\left(\{p_j\}, q, p\right) \\
& = \int \prod_{i=1}^V d^4 x_i e^{ipx_1}e^{-iqx_V}\int d^{4I}Z ~\mathcal{\bar D}(Z) \mathcal{N}\left( X-Z ; p,q \right)
\end{split}\end{align}

where we used $\mathcal{N}^*= -\mathcal{N}$, and 
\begin{align}
\mathcal{\bar D}(Z) =(-1)\int \prod_{j=1}^{I}\frac{d^4p_j}{(2\pi)^4}e^{ip_j(z_{j+1}-z_{j})}\left(-\Delta_F^*(p_j)\right) =(-1)^{I+1}\prod_{j=1}^{I} \Delta^*_F(z_{j+1}-z_{j})
\end{align}

Now, the LHS of the unitarity equation becomes:
\begin{align}\begin{split}
\langle f | (iT) | i\rangle+\langle f | (iT) ^\dagger | i\rangle
& = \int \prod_{i=1}^V d^4 x_i e^{ipx_1}e^{-iqx_V} \\ & \hspace*{1cm} \qquad \times\int d^{4I}Z ~\left(\mathcal{D}(Z)+\mathcal{\bar D}(Z) \right)\mathcal{N}\left( X-Z ; p,q \right)
\end{split}\end{align}

Let us look at the sum over the propagator functions:
\begin{align}
\mathcal{D}(Z)+\mathcal{\bar D}(Z)  =\prod_{j=1}^{I} \Delta_F(z_{j+1}-z_{j})+(-1)^{I+1}\prod_{j=1}^{I} \Delta^*_F(z_{j+1}-z_{j})
\end{align}

This expression is very similar to the LHS of the scalar LTE for a string of scalars, indeed we have in the scalar case:
\begin{align}
i\prod_{j=1}^{I}i \Delta_F(z_{j+1}-z_{j})+(-i)\prod_{j=1}^{I}(-i) \Delta^*_F(z_{j+1}-z_{j})= i^{I+1}\left(\mathcal{D}(Z)+\mathcal{\bar D}(Z)\right)
\end{align}

On the other hand, the RHS is given by:
\begin{align}\begin{split}
-\sum_{p=1}^{I}\left[  \left(\prod_{j=1}^{p-1}i \Delta_F(z_{j+1}-z_{j})\right) i\Delta^+(z_{p+1}-z_p) \left(\prod_{i=p+1}^{I}(-i)\Delta^*_F(z_{i+1}-z_{i})\right)(-i)\right. \\ 
+\left. \left(\prod_{j=1}^{p-1}(-i) \Delta_F^*(z_{j+1}-z_{j})\right) (-i)\Delta^-(z_{p+1}-z_p) \left(\prod_{i=p+1}^{I}i\Delta_F(z_{i+1}-z_{i})\right)(i) \right]
\end{split}\end{align}

where the sum over $p$ denotes all possible cuts. We can factor out the $i$ factors so that:
\begin{align}\begin{split}
-(-i)^{I+1}\sum_{p=1}^{I}(-1)^p \left[ \left(\prod_{j=1}^{p-1} \Delta_F(z_{j+1}-z_{j})\right) \Delta^+(z_{p+1}-z_p) \left(\prod_{i=p+1}^{I}\Delta^*_F(z_{i+1}-z_{i})\right)\right. \\ 
+\left. (-1)^{I+1}\left(\prod_{j=1}^{p-1} \Delta_F^*(z_{j+1}-z_{j})\right) \Delta^-(z_{p+1}-z_p) \left(\prod_{i=p+1}^{I}\Delta_F(z_{i+1}-z_{i})\right) \right]
\end{split}\end{align}

and we define the above quantity as:
\begin{align}
-(-i)^{I+1} \sum_{p=1}^{I}(-1)^p  \left( \mathcal{D}^+_p(Z) + \mathcal{D}^-_p(Z)\right)
\end{align}

The LTE for the scalar fields gives us therefore:
\begin{align}
i^{I+1}\left(\mathcal{D}(Z)+\mathcal{\bar D}(Z)\right) = -(-i)^{I+1} \sum_{p=1}^{I}(-1)^p  \left( \mathcal{D}_p^+(Z) + \mathcal{D}_p^-(Z)\right)
\end{align}

 or 
\begin{align}
\left(\mathcal{D}(Z)+\mathcal{\bar D}(Z)\right) = (-1)^{I} \sum_{p=1}^{I}(-1)^p  \left( \mathcal{D}_p^+(Z) + \mathcal{D}_p^-(Z)\right)
\end{align}

We can plug this back in our unitarity equation:
\begin{align}\begin{split}
\langle f | (iT) | i\rangle&+\langle f | (iT) ^\dagger | i\rangle\\
& = \int \prod_{i=1}^V d^4 x_i e^{ipx_1}e^{-iqx_V} \\ & \hspace*{1cm} \qquad \times\int d^{4I}Z ~(-1)^{I} \sum_{p=1}^{I}(-1)^p \left( \mathcal{D}_p^+(Z) + \mathcal{D}_p^-(Z)\right)\mathcal{N}\left( X-Z ; p,q \right)
\end{split}\end{align}

but because the energy flows from $x_1$ to $x_V$, the only non-vanishing contribution is\footnote{To see it explicitly, we can rewrite this amplitude as it was originally given.}:
\begin{align}\begin{split}
\langle f | (iT) | i\rangle&+\langle f | (iT) ^\dagger | i\rangle\\& = \int \prod_{i=1}^V d^4 x_i e^{ipx_1}e^{-iqx_V} \\ & \hspace*{1.3cm} \qquad \times\int d^{4I}Z ~  \sum_{p=1}^{I}(-1)^{I+p}  \mathcal{D}_p^+(Z) \mathcal{N}\left( X-Z ; p,q \right)
\end{split}\end{align}

Let us now check the RHS. We have:
\begin{align}\begin{split}
 -\sum_{{\rm phys}~n} \langle f | &(iT)^\dagger| n \rangle \langle n |(iT)| i \rangle \\ & = -\sum_{p=1}^I(-i)(-1)^{I+p}\int \prod_{i=1}^V d^4 x_i e^{ipx_1}e^{-iqx_V} \\
  &\left(\int \prod_{j < p}\frac{d^4p_j}{(2\pi)^4}e^{ip_j(x_{j+1}-x_{j})}\Delta_F(p_j)
  \times \int \frac{d^3p_p}{(2\pi)^3(2\omega_p)}e^{ip_p(x_{p+1}-x_{p})} \right.\\ & \hspace*{1.3cm} \qquad \times\left.\int \prod_{i > p}\frac{d^4p_i}{(2\pi)^4}e^{ip_i(x_{i+1}-x_{i})}\Delta^*_F(p_i)  \mathcal{\tilde N}\left(\{p_j, p_i\}, p_p, q, p\right)\right)\\
  & = \sum_{p=1}^I(-1)^{I+p}\int \prod_{i=1}^V d^4 x_i e^{ipx_1}e^{-iqx_V} \\
  &\left(\int \prod_{j < p}\frac{d^4p_j}{(2\pi)^4}e^{ip_j(x_{j+1}-x_{j})}\Delta_F(p_j)
  \times \int \frac{d^3p_p}{(2\pi)^3(2\omega_p)}e^{ip_p(x_{p+1}-x_{p})} \right.\\ & \hspace*{1.3cm} \qquad \times\left.\int \prod_{i > p}\frac{d^4p_i}{(2\pi)^4}e^{ip_i(x_{i+1}-x_{i})}\Delta^*_F(p_i)  \mathcal{N}\left(\{p_j, p_i\}, p_p, q, p\right)\right)  
\end{split}\end{align}

where we used $\mathcal{\tilde N}=(-i)\mathcal{ N}$ and when we factor out $(-i)^I$ from the numerator there is one extra factor to cancel the former (with the overall minus sign), and there is a factor of $(-1)^{I+p}$ for the extra minus sign in each anti propagator. In order to proceed as before, we need to rewrite the sum over intermediate states as a four-dimensional integral:
\begin{align}
\int \frac{d^3p}{(2\pi)^3(2\omega_p)}e^{\pm ip(x-y)} f(p) = \int \frac{d^4p}{(2\pi)^4}e^{\pm ip(x-y)} (2\pi)\theta(\mp p_0)\delta(p^2+m^2)f(p)
\end{align}

so that we rewrite the RHS as:
\begin{align}\begin{split}
 -\sum_{{\rm phys}~n} \langle f | &(iT)^\dagger| n \rangle \langle n |(iT)| i \rangle\\ & = \sum_{p=1}^I(-1)^{I+p}\int \prod_{i=1}^V d^4 x_i e^{ipx_1}e^{-iqx_V} \\ & \hspace*{0cm} \qquad \times\left(\int \frac{d^{4I}P}{(2\pi)^{4I}}e^{iPX}\underbrace{\prod_{j < p}\Delta_F(p_j)\Delta^+(p_p) \prod_{i > p}\Delta^*_F(p_i)}_{\mathcal{D}^+_p(P)}~ \mathcal{ N}\left(P, q, p\right)\right)\\
  & = \int \prod_{i=1}^V d^4 x_i e^{ipx_1}e^{-iqx_V}\sum_{p=1}^I \int d^{4I}Z~ {\mathcal{D}^+_p(Z)}~ \mathcal{ N}\left(Z-X, q, p\right) 
\end{split}\end{align}

and therefore we have proved the unitarity equation for a string of fermions:
\begin{align}
\boxed{\begin{array}{rl}{\rm LHS}& =\int \prod_{i=1}^V d^4 x_i e^{ipx_1}e^{-iqx_V}\sum_{p=1}^I (-1)^{I+p} \int d^{4I}Z~ {\mathcal{D}^+_p(Z)}~ \mathcal{ N}\left(Z-X, q, p\right)\\ &={\rm RHS} \end{array}}
\end{align}

\subsubsection*{Fermionic loop}

%We consider this time a $2\rightarrow N-2$ one-loop amplitude (in order to have at least one non-vanishing delta function):

% e^{ik_1x_1+ik_2x_2}e^{-i\sum_{j=3}^{V} k_j x_j}
We are left with the case of a fermionic loop, let us therefore rewrite the most general amplitude for a loop:
\begin{align}
\langle f | (iT) | i\rangle = \int \prod_{i=1}^V d^4 x_i \int \prod_{j=1}^V\frac{d^4p_j}{(2\pi)^4}e^{ip_j(x_{j+1}-x_{j})}\Delta_F(p_j) \mathcal{N}\left(\{p_j\}\right)
\end{align}

with $x_{V+1}=x_1$ and this time the number of internal lines and of external particles $I=N=V$. We do not write explicitly the photon polarisations or plane waves as this loop could be connected to a more general amplitude. As before, we define a generalised momentum and position which are this time $4V$ dimensional:
\begin{align}
P^I = \bigoplus_{j=1}^{V}p_j, \quad X^I = \bigoplus_{j=1}^{V}(x_{j+1}-x_j)
\end{align}

as well as a propagator function:
\begin{align}
\mathcal{D}(P) = \prod_{j=1}^{V}\Delta_F(p_j)
\end{align}

The amplitude can then be rewritten as:
\begin{align}
\langle f | (iT) | i\rangle = \int d^{4V}X\int\frac{d^{4V}P}{(2\pi)^{4V}}e^{iPX}\mathcal{D}(P) \mathcal{N}\left( P\right)
\end{align}

As before, the second integral is now simply rewritten as a $4V$ dimensional Fourier transform of a product of functions. Using the convolution theorem, we have:
\begin{align}
\int\frac{d^{4V}P}{(2\pi)^{4V}}e^{iPX}\mathcal{D}(P) \mathcal{N}\left( P\right) = \int d^{4V}Z ~\mathcal{D}(Z) \mathcal{N}\left( X-Z\right)
\end{align}

where $\mathcal{D}(Z)$ and  $\mathcal{N}\left( Z\right)$ are the Fourier transforms into position space of their respective momentum space quantities. The full amplitude is now given by:
\begin{align}
\langle f | (iT) | i\rangle = \int d^{4V}X d^{4V}Z ~\mathcal{D}(Z) \mathcal{N}\left( X-Z \right)
\end{align}

Let us look more carefully at the propagator function in position space:
\begin{align}
\mathcal{D}(Z) =\int \prod_{j=1}^{V}\frac{d^4p_j}{(2\pi)^4}e^{ip_j(z_{j+1}-z_{j})}\Delta_F(p_j) =\prod_{j=1}^{V} \Delta_F(z_{j+1}-z_{j})
\end{align}

Again, this is nothing but the first term in the LTE for scalar fields stripped off the vertices factors. The same can be applied to the conjugated amplitude to obtain:
\begin{align}\begin{split}
\langle f | (iT) ^\dagger | i\rangle&= \int \prod_{i=1}^V d^4 x_i \int \prod_{j=1}^{V}\frac{d^4p_j}{(2\pi)^4}e^{ip_j(x_{j+1}-x_{j})}\left(-\Delta_F^*(p_j)\right) \mathcal{N}^*\left(\{p_j\}, q, p\right) \\
& = \int d^{4V}X d^{4V}Z ~\mathcal{\bar D}(Z) \mathcal{N}\left( X-Z \right)
\end{split}\end{align}

where we used $\mathcal{N}^*= \mathcal{N}$ for loops, and 
\begin{align}
\mathcal{\bar D}(Z) =\int \prod_{j=1}^{V}\frac{d^4p_j}{(2\pi)^4}e^{ip_j(z_{j+1}-z_{j})}\left(-\Delta_F^*(p_j)\right) =(-1)^{V}\prod_{j=1}^{I} \Delta^*_F(z_{j+1}-z_{j})
\end{align}

Now, the LHS of the unitarity equation becomes:
\begin{align}\begin{split}
\langle f | (iT) | i\rangle&+\langle f | (iT) ^\dagger | i\rangle
\\& = \int d^{4V}X d^{4V}Z ~\left(\mathcal{D}(Z)+\mathcal{\bar D}(Z) \right)\mathcal{N}\left( X-Z \right)
\end{split}\end{align}

Let us look at the sum over the propagator functions:
\begin{align}
\mathcal{D}(Z)+\mathcal{\bar D}(Z)  =\prod_{j=1}^{V} \Delta_F(z_{j+1}-z_{j})+(-1)^{V}\prod_{j=1}^{V} \Delta^*_F(z_{j+1}-z_{j})
\end{align}

This expression is very similar to the LHS of the scalar LTE for a string of scalars, indeed we have in the scalar case:
\begin{align}
(-1)\prod_{j=1}^{V}i \Delta_F(z_{j+1}-z_{j})+(-1)\prod_{j=1}^{V}(-i) \Delta^*_F(z_{j+1}-z_{j})= (-1)i^{V}\left(\mathcal{D}(Z)+\mathcal{\bar D}(Z)\right)
\end{align}

In order to write the RHS, recall that in this case the LTE implies that one can only have one sequence of propagators followed by one sequence of antipropagators (linked by $\Delta^\pm$). Indeed, we only consider single cuts that split the loop into two pieces. In order to write an expression for this sum over cuts, notice that the expression in terms of (anti-)propagators only depends on the distance between the two cut propagators, and on the position of one of the two cuts. Therefore we define the distance between the first cut propagator and the second:
\begin{align}
\delta^i \equiv \wp(p_{i+\delta^i})-\wp(p_{i})
\end{align}

where $\wp(p_i)=i$ is the position of the $i^{th}$ propagator, and the subscript $i$ in $\delta^i$ labels the first cut propagator we consider. Then, we can write the RHS as:
\begin{align}\begin{split}
-\sum_{i=1}^{V}\sum_{\delta^i=1}^{V-1} i&\Delta^+(z_{i+1}-z_i) \left(\prod_{j=i+1}^{i-1+\delta^i} (-i)\Delta^*_F(z_{j+1}-z_j)\right)\\ & \hspace*{1.3cm} \qquad \times(-i)\Delta^-(z_{i+1+\delta^i}-z_{i+\delta^i})\left(\prod_{j=i+1+\delta^i}^{i-1+V} i\Delta_F(z_{j+1}-z_j)\right)
\end{split}\end{align}

Notice that the number of (anti-)propagators in the sum does not depend on the position of the first cut propagator (by symmetry of the loop) but only on $\delta^i$. Moreover, for $\delta^i=1$, there is no antipropagator, and for $\delta^i=V-1$, no propagator. We can factor out the explicit $i$ dependence:
\begin{align}\begin{split}
(-1)i^V\sum_{i=1}^{V}\sum_{\delta^i=1}^{V-1}(-1)^{\delta^i} &\Delta^+(z_{i+1}-z_i) \left(\prod_{j=i+1}^{i-1+\delta^i} \Delta^*_F(z_{j+1}-z_j)\right)\\ & \hspace*{1.3cm} \qquad \times\Delta^-(z_{i+1+\delta^i}-z_{i+\delta^i})\left(\prod_{j=i+1+\delta^i}^{i-1+V} \Delta_F(z_{j+1}-z_j)\right)
\end{split}\end{align}

%\begin{align}\begin{split}
%-\sum_{1\leq p < q \leq V}\left[  \left(\prod_{j=1}^{p-1}i \Delta_F(z_{j+1}-z_{j})\right) i\Delta^+(z_{p+1}-z_p) \left(\prod_{i=p+1}^{q-1}(-i)\Delta^*_F(z_{i+1}-z_{i})\right) \right. \\ 
%\times (-i)\Delta^-(z_{q+1}-z_q) \left(\prod_{j=q+1}^{V}i \Delta_F(z_{j+1}-z_{j})\right)\\
%+\left(\prod_{j=1}^{p-1}(-i) \Delta_F^*(z_{j+1}-z_{j})\right) (-i)\Delta^-(z_{p+1}-z_p) \left(\prod_{i=p+1}^{q-1}i\Delta_F(z_{i+1}-z_{i})\right)\\
%\left. \times i\Delta^+(z_{q+1}-z_q) \left(\prod_{j=q+1}^{V}(-i) \Delta^*_F(z_{j+1}-z_{j})\right) \right]
%\end{split}\end{align}

%
%\begin{align}\begin{split}
%(-1)i^V\sum_{\substack{ 1\leq p , q \leq V \\ p\neq q}}\left[  \left(\prod_{j=1}^{p-1} \Delta_F(z_{j+1}-z_{j})\right) \Delta^+(z_{p+1}-z_p) \left(\prod_{i=p+1}^{q-1}\Delta^*_F(z_{i+1}-z_{i})\right) \right. \\ 
%\times \left. \Delta^-(z_{q+1}-z_q) \left(\prod_{j=q+1}^{V} \Delta_F(z_{j+1}-z_{j})\right)\right]
%\end{split}\end{align}

and we define the above quantity as:
\begin{align}
(-1)i^V\sum_{i=1}^{V}\sum_{\delta^i=1}^{V-1}(-1)^{\delta^i}\mathcal{D}_{\delta^i}(Z) 
\end{align}

The LTE for the scalar fields gives us therefore:
\begin{align}
(-1)i^{V}\left(\mathcal{D}(Z)+\mathcal{\bar D}(Z)\right)= (-1)i^V\sum_{i=1}^{V}\sum_{\delta^i=1}^{V-1}(-1)^{\delta^i}\mathcal{D}_{\delta^i}(Z) 
\end{align}

or 
\begin{align}
\left(\mathcal{D}(Z)+\mathcal{\bar D}(Z)\right) = \sum_{i=1}^{V}\sum_{\delta^i=1}^{V-1}(-1)^{\delta^i}\mathcal{D}_{\delta^i}(Z) 
\end{align}

We can plug this back in our unitarity equation:
\begin{align}\begin{split}
\langle f | (iT) | i\rangle&+\langle f | (iT) ^\dagger | i\rangle
\\& =\int d^{4V}X d^{4V}Z ~\sum_{i=1}^{V}\sum_{\delta^i=1}^{V-1}(-1)^{\delta^i}\mathcal{D}_{\delta^i}(Z) \mathcal{N}\left( X-Z \right)
\end{split}\end{align}

for particles that are all incoming, there cannot be only one circled vertex in the RHS of the LTE (otherwise the delta function is not satisfied at that vertex), therefore, there must at least be one anti-propagator. Hence, in $\mathcal{D}_{\delta^i}(Z)$, we must have $\delta^i>1$. The only non-vanishing contribution is:
\begin{align}\begin{split}
\langle f | (iT) | i\rangle&+\langle f | (iT) ^\dagger | i\rangle
\\& = \int d^{4V}X d^{4V}Z ~ \sum_{i=1}^{V}\sum_{\delta^i=2}^{V-1}(-1)^{\delta^i}\mathcal{D}_{\delta^i}(Z) \mathcal{N}\left( X-Z \right)
\end{split}\end{align}

Let us now check the RHS. We have:
\begin{align}\begin{split}
 -&\sum_{i=1}^{V}\sum_{\delta^i=2}^{V-1}(-1)^{\delta_i-1}\int \prod_{j=1}^V d^4 x_j  \left(\int  \frac{d^3p_i}{(2\pi)^3(2\omega_i)}e^{ip_i(x_{i+1}-x_{i})} \right. \\
  &\hspace*{1cm}\left.\times \int\prod_{j=i+1}^{i-1+\delta^i}\frac{d^4p_j}{(2\pi)^4}e^{ip_j(x_{j+1}-x_{j})}\Delta^*_F(p_j)\int  \frac{d^3p_{i+\delta^i}}{(2\pi)^3(2\omega_{i+\delta^i})}e^{-ip_{i+\delta^i}(x_{i+1+\delta^i}-x_{i+\delta^i})}  \right. \\
  &\hspace*{1cm}\left.\times\int \prod_{j=i+1+\delta^i}^{i-1+V}\frac{d^4p_i}{(2\pi)^4}e^{ip_j(x_{j+1}-x_{j})}\Delta_F(p_j)  \mathcal{\tilde N}\left(\{p_j\}\right)\right)\\
 = &   \sum_{i=1}^{V}\sum_{\delta^i=2}^{V-1}(-1)^{\delta^i}\int \prod_{j=1}^V d^4 x_j  \left(\int  \frac{d^3p_i}{(2\pi)^3(2\omega_i)}e^{ip_i(x_{i+1}-x_{i})} \right. \\
  &\hspace*{1cm}\left. \times\int\prod_{j=i+1}^{i-1+\delta^i}\frac{d^4p_j}{(2\pi)^4}e^{ip_j(x_{j+1}-x_{j})}\Delta^*_F(p_j)  \int  \frac{d^3p_{i+\delta^i}}{(2\pi)^3(2\omega_{i+\delta^i})}e^{-ip_{i+\delta^i}(x_{i+1+\delta^i}-x_{i+\delta^i})}  \right. \\
  &\hspace*{1cm}\left.\times\int \prod_{j=i+1+\delta^i}^{i-1+V}\frac{d^4p_i}{(2\pi)^4}e^{ip_j(x_{j+1}-x_{j})}\Delta_F(p_j)  \mathcal{ N}\left(\{p_j\}\right)\right)\\
\end{split}\end{align}

where we used $\mathcal{\tilde N}=(-1)\mathcal{ N}$ and when we factor out $(-i)^V$ from the numerator there are two extra factors to cancel the former. Moreover for each antipropagator, there is a factor of $(-1)$ `unused', so that in the end there is an extra factor of $(-1)^{\delta^i}$. In order to proceed as before, we need to rewrite the sum over intermediate states as a four-dimensional integral:
\begin{align}
\int \frac{d^3p}{(2\pi)^3(2\omega_p)}e^{\pm ip(x-y)} f(p) = \int \frac{d^4p}{(2\pi)^4}e^{\pm ip(x-y)} (2\pi)\theta(\mp p_0)\delta(p^2+m^2)f(p)
\end{align}

so that we rewrite the RHS as:
\begin{align}\begin{split}
 -\sum_{{\rm phys}~n} \langle f | (iT)^\dagger| n \rangle \langle n |(iT)| i \rangle & = \sum_{i=1}^{V}\sum_{\delta^i=2}^{V-1}(-1)^{\delta^i}\int d^{4V}X\\  
 &\hspace*{-4cm}\times\left(\int \frac{d^{4I}P}{(2\pi)^{4I}}e^{iPX}\underbrace{\Delta^+(p_i)\prod_{j=i+1}^{i-1+\delta^i}\Delta^*_F(p_j)\Delta^-(p_{i+\delta^i})
  \prod_{k=i+1+\delta^i}^{i-1+V}\Delta_F(p_k)}_{\mathcal{D}_{\delta^i}(P)}~ \mathcal{ N}\left(P\right)\right)\\
  & = \sum_{i=1}^{V}\sum_{\delta^i=2}^{V-1}(-1)^{\delta^i}\int d^{4V} dX d^{4I}Z~ \mathcal{D}_{\delta^i}(Z)~ \mathcal{ N}\left(Z-X\right) 
\end{split}\end{align}

and therefore we have proved the unitarity equation for a fermionic loop:
\begin{align}
\boxed{{\rm LHS} =\int d^{4V}X d^{4V}Z ~ \sum_{i=1}^{V}\sum_{\delta^i=2}^{V-1}(-1)^{\delta^i}\mathcal{D}_{\delta^i}(Z) \mathcal{N}\left( X-Z \right)= {\rm RHS} }
\end{align}

\section{Discussion}

In this chapter we proved that the second-order Dirac theory is indeed unitary even though its Lagrangian is not hermitian. After having developed the tools necessary for the proof in the case of a scalar field, we have shown that the unitarity of the theory relies deeply on the fact that external (physical) states are subject to a set of reality conditions. Without these, it would have been impossible to link the complex conjugated amplitudes to their original counterparts. Furthermore, the structure of the interactions are also constrained by, this time, a fully non-linear set of equations. Indeed, if we consider the theory as a construction arising from the first-order formalism, the form of the vertices is dictated by the gauge-covariant Dirac equation. This, in turn, forces the quartic vertex to be how it is, and therefore allows for the necessary cancellations and factorisations in the unitary equation. Notice that if, instead, we had considered the second-order theory as fundamental, the quartic vertex would have been included in a Noether construction of the (locally) gauge-invariant Lagrangian. Together with the constraint on external states, the quartic vertex ensure the unitarity of the theory.

We argue that unitarity holds because of the right balance of interactions and constraints, the reader might wonder what would happen if one were to add an additional quartic interaction to the Lagrangian, such as:
\begin{align}
\mathcal{L}_\lambda \sim {\lambda}(\xi \chi)^2
\end{align}

Because the fermions have mass dimension one in the second-order formalism (and their propagator is scalar), this interaction is power-counting renormalisable\footnote{If there is only the above interaction present in the theory, the superficial degree of divergence is $D=4-N_f$, with $N_f$ the external number of fermions. This is to be compared to the usual $D= 4-N_f +P_f$ where $P_f$ is the number of propagators.}, besides having a tree-level amplitude $\sim (s/m^2)^2$ which is a sign of its effective character. However, once can see that in order to ensure the unitarity of the theory (assuming the same reality conditions on the external states $\partial \xi \sim m \chi^\dagger$), one must also include:
\begin{align}
\mathcal{L}_{\lambda^*} \sim {\lambda^*}\frac{1}{m^4}(\partial \xi \partial \chi)^2
\end{align}
which is non-renormalisable\footnote{$D= 4L$ since for each internal propagator there will be two derivatives from the vertices in the numerator}. We therefore see that in the second-order formalism, there is a clear tension between renormalisability and unitarity. The fact that the above interactions (when required to be unitary) are non-renormalisable is intuitively understood as they arise from a $(\bar\Psi \Psi)^2$ term in the first-order formalism, which is known to be non-renormalisable.

Aside from the case of second-order fermions that was considered here, this proof of unitarity can be generalised to other non-hermitian theories that have a scalar-type propagator. Indeed, for such a theory to be unitary, some reality conditions are needed. Using a generalisation of the results of Section \ref{realityampli}, we can see that this will impose stringent constraints on the numerators already at tree-level. For example, in a spinor-helicity formalism, one must require the theory to produce the same helicity configurations using both sets of (usual or conjugated) Feynman rules. At this time, it seems that it is sufficient for a theory that is non-hermitian to be $PT$ symmetric, as a future line of research, it would be worth exploring this in the context of the constraints numerators must satisfy.
\vfill

\chapter{Unification}\label{sec:unific}
\section{Introduction}
We have seen in Chapter \ref{chapSM} that our second-order Lagrangian (\ref{SML}) is much simpler and more compact than the first-order Lagrangian. Moreover, in joining together the barred fermions, we have made explicit an approximate symmetry of the theory. In this spirit, we will discuss in this chapter different ``unifications'' in the second-order formalism. We do not claim that the groups that we will introduce below represent new GUTs, but rather that the Lagrangian of the SM can be written using higher approximate symmetries in a much simpler way. Section \ref{sec:pati} is based on work done in \cite{Espin:2013wia}, whereas the other sections contain mainly ongoing research. The main result of this chapter can be found in Section \ref{gguni}. For an introduction to group theory, we refer the reader to \cite{Georgi:1999wka,Ramond:2010zz} for a physicist's perspective, while \cite{dynkin2000selected} is a comprehensive mathematical reference. We mainly use the same conventions as in \cite{Georgi:1999wka}.

\section{${\rm SO}(8)$ unification}
As we have seen, each generation of fermions has 16 components. Furthermore, in our second-order Lagrangian where the weak $SU(2)$ is frozen, the fermionic representations appearing are real. Indeed, for each $SU(3)$ triplet, there is an anti-triplet with opposite electromagnetic charge and the same occurs for the singlets. This allows us to look for a 16 dimensional real representation of a gauge group that is broken down to $SU(3)\times U(1)$, thereby escaping the embedding constraints imposed by the weak charges. The simplest group containing a 16-dimensional real irreducible representation is $SO(9)$ with its spinor representation.  However, $SO(8)$ contains two real 8-dimensional representations, we will therefore start with the latter.
We want to find an embedding $SU(3)\times U(1) \subset SO(8)$. The Dynkin diagram for this group is

\begin{tabular}{>{$}r<{$}m{4cm}m{1cm}}

D_4 \equiv \mathfrak{so}(8) 
&
  \begin{dynkin}
    \foreach \x in {1,...,2}
    {
        \dynkindot{\x}{0}
    }
    \dynkindot{2.8}{.8}
    \dynkindot{2.8}{-.8}
    \dynkinline{1}{0}{2}{0}
    \dynkinline{2}{0}{2.8}{.8}
    \dynkinline{2}{0}{2.8}{-.8}
  \end{dynkin}
& {}
\\

\end{tabular}

If we denote its simple roots by $\alpha_i$, $i=1,\ldots,4$, a useful representation for the latter is:
\begin{align}
\alpha_i = e_i-e_{i+1}, \quad \alpha_4 = e_3+e_4, \quad e_i\cdot e_j = \delta_{ij}
\end{align}

Its Cartan matrix is given by:
\begin{align}
A_{ij}=\left( \begin{array}{cccc} 2 & -1 & 0 & 0 \\ -1 & 2 & -1 & 0 \\ 0 & -1 & 2 & 0 \\ 0 & -1 & 0 &2\\
\end{array} \right)
\end{align}

so that the simple roots in the Dynkin basis are given by:
\begin{align}
\left(\alpha_i\right)_j = A_{ij}
\end{align}

A suitable embedding is obtained by considering the $SU(3)$ subgroup generated by $\alpha_1$ and $\alpha_2$ (the root on the left and the middle root). As for which representation is needed, $SO(8)$ has no 16 dimensional irreducible representation, however it has three 8 dimensional representations (triality).

\renewcommand*{\arraystretch}{1}
The spinor irreps of $SO(8)$ are given by the decomposition of their highest-weight in terms of Dynkin indices: 
\begin{align}
{\bf 8}_{(0010)} = \left( \begin{array}{cccc} 0 & 0 & 1 & 0 \\
\end{array} \right)\quad {\bf 8}_{(0001)} = \left( \begin{array}{cccc} 0 & 0 & 0 & 1 \\
\end{array} \right)
\end{align}

Our choice of embedding leads us to consider the two spinor representations ${\bf 8}_{(0010)}+{\bf 8}_{(0001)}$. The first one decomposes under $SU(3)$ as:
\begin{align}
{\bf 8}_{(0010)} \rightarrow \begin{array}{cccc}
{\bf 3}\subset &\left( \begin{array}{cccc} 1 & 0 & 0 & -1 \\ \end{array} \right) &\left( \begin{array}{cccc} -1 & 1 & 0 & -1 \\ \end{array} \right)  &\left( \begin{array}{cccc} 0 & -1 & 1 & 0 \\ \end{array} \right)  \\
{\bf \bar{3}}\subset  &\left( \begin{array}{cccc} 0 & 1 & -1 & 0 \\ \end{array} \right) &\left( \begin{array}{cccc} 1 & -1 & 0 & 1 \\ \end{array} \right)  &\left( \begin{array}{cccc} -1 & 0 & 0 & 1 \\ \end{array} \right)  \\
{\bf 1}\subset & \left( \begin{array}{cccc} 0 & 0 & 1 & 0 \\ \end{array} \right) & & \\
{\bf 1} \subset &\left( \begin{array}{cccc} 0 & 0 & -1 & 0 \\ \end{array} \right)  & & \\
\end{array}
\end{align}

and the second as:
\begin{align}
{\bf 8}_{(0001)} \rightarrow \begin{array}{cccc}
{\bf 3}\subset &\left( \begin{array}{cccc} 1 & 0 & -1 & 0 \\ \end{array} \right) &\left( \begin{array}{cccc} -1 & 1 & -1 & 0 \\ \end{array} \right)  &\left( \begin{array}{cccc} 0 & -1 & 0 & 1 \\ \end{array} \right)  \\
{\bf \bar{3}}\subset  &\left( \begin{array}{cccc} 0 & 1 & 0 & -1 \\ \end{array} \right) &\left( \begin{array}{cccc} 1 & -1 & 1 & 0 \\ \end{array} \right)  &\left( \begin{array}{cccc} -1 & 0 & 1 & 0 \\ \end{array} \right)  \\
{\bf 1} \subset & \left( \begin{array}{cccc} 0 & 0 & 0 & 1 \\ \end{array} \right) & & \\
{\bf 1} \subset &\left( \begin{array}{cccc} 0 & 0 & 0 & -1 \\ \end{array} \right)  & & \\
\end{array}
\end{align}

This has been derived in the following way: we seek an embedding of the SM unbroken gauge group inside $SO(8)$ and the latter can be found by considering the reduced Dynkin diagram obtained after omitting the last two roots in $D_4$. The new diagram describes an embedding $SU(3)\times U(1) \subset SO(8)$ where the $SU(3)$ sector is generated by the first two simple roots. Now, in order to see how our spinor representation transforms under the subgroup, one needs simply remember that for $SU(3)$ the (anti-)triplet representation is obtained by a successive lowering of the highest-weight state by the two simple roots.

We still have to check the electric charges of those particles. In order to do so, we need to find the charge generator $Q$. The latter is a linear combination of Cartan's generators of $SO(8)$ and as such can be written as\footnote{In the Chevalley basis where $\left[E_\alpha, E_{-\alpha}\right]= H_\alpha$.}:
\begin{align}
Q = \left( \begin{array}{cccc} a & b & c & d \\
\end{array} \right)
\end{align}

For the $U(1)$ subgroup to commute with $SU(3)$, we have to require that the charges within an (anti-) triplet are equal. Using that the charge of a state $\lambda$ is given by:
\begin{align}
 Q\cdot \lambda = q_\lambda
\end{align}

This allows us to define a charge operator 
\begin{align}
Q = -\frac{1}{3}\left( \begin{array}{cccc} 1 & 2 & 3 & 0 \\
\end{array} \right)
\end{align}

and we obtain the branching rule 
\begin{align}
{\bf 8}_{(0010)}+{\bf 8}_{(0001)} = {\bf 3}_{\frac{2}{3}} + {\bf 3}_{-\frac{1}{3}}  +{\bf \bar{3}}_{-\frac{2}{3}}  +{\bf \bar{3}}_{\frac{1}{3}}  + {\bf 1}_{-1} +{\bf 1}_0+{\bf 1}_0+{\bf 1}_{1}
\end{align}

Later on, we will construct the embedding of the SM into $SO(9)$ in the same way.

\section{${\rm SU}(2)\times {\rm SU}(4)$ unification} \label{sec:pati}

We now construct a Pati-Salam-like unification \cite{Pati:1974yy}. Consider the second-order Lagrangian in the form (\ref{SML}) that we recall here:
\begin{align}\label{SML2}
\mathcal{L}_{SM,f} =  -2\mathcal{D}\bar Q^i\mathcal{D} {Q_i}-{2} \mathcal{D} \bar L^i \mathcal{D} {L_i} - \rho^2 \left(\Lambda_q \bar Q\right)^i Q_i  - \rho^2 \left(\Lambda_l \bar L\right)^i L_i
\end{align}

Here the quarks' kinetic term contains a sum over the three colour indices. Spelling this out we have the following kinetic term:
\begin{align}
-2 (\mathcal{D}\bar Q^i)^r(\mathcal{D} {Q_i})^r -2(\mathcal{D}\bar Q^i)^g(\mathcal{D} {Q_i})^g -2(\mathcal{D}\bar Q^i)^b(\mathcal{D} {Q_i})^b-2 D\bar{L}^i DL_i
\end{align}

where $r,g,b$ are the three colours and $(\mathcal{D}Q_i)^*$ denotes the projection of a triplet on a particular colour index. As in the Pati-Salam unification, it seems legitimate at the level of this Lagrangian to define the leptons as the fourth ``colour''. This suggests that we put all of the SM fermions into two multiplets:
\begin{align}\label{uni-fermions}
S_i :=\left( \begin{array}{cccc} u_i^r & u_i^g & u_i^b & \nu_i \\ d_i^r & d_i^g & d_i^b & e_i \end{array}\right), \qquad 
\bar{S}_i :=\left( \begin{array}{cc} \bar{u}_i^r & \bar{d}_i^r\\ \bar{u}_i^g & \bar{d}_i^g \\ \bar{u}_i^b & \bar{d}_i^b \\ \bar{\nu}_i & \bar{e}_i \end{array}\right)
\end{align}
We can then rewrite the Lagrangian in terms of $S_i,~\bar{S}_i$, using covariant derivatives appropriate for each field. However, what seems to spoil this picture is the different electric charges of the quarks and leptons. As we have seen above, the embedding into a larger gauge-group is not always straightforward. The same problem arises in the usual Pati-Salam treatment, where it is solved by using a non-trivial embedding of the SM gauge group into a larger group, see e.g. \cite{Baez:2009dj} for a nice exposition. 

Thus, to understand what is happening with the electric charges, we need to understand how the SM symmetry group sits inside some larger gauge group. If for $SO(8)$ we have carried out the construction at the level of the representation in a purely algebraic way, we will here do it at the level of the action of the group on the states. Note that the symmetry group that is unbroken in the Lagrangian (\ref{SML2}) is ${U}(1)\times{SU}(3)$. The weak ${ SU}(2)$ no longer acts on our fermions, as they are all ${SU}(2)$-invariant objects. However, there is a leftover from this gauge group in the form of the massive gauge field (also ${ SU}(2)$-invariant) that acts on doublets $Q_i, L_i$ and does not act on $\bar{Q}_i, ~\bar{L}_i$. So, the group ${SU}(2)$ is broken, but the fact that the fermions come as doublets tells us that it was there. Similarly, now that we put in (\ref{uni-fermions}) leptons on the same footing as the quarks, it appears that there is an ${ SU}(4)$ behind this construction. So, we take ${ SU}(2)\times {SU}(4)$ as the GUT gauge group that acts on multiplets $S_i,~\bar{S}_i$, and look for an embedding of ${ U}(1)\times{ SU}(3)$ into it. The sought embedding is given by
\begin{align}
{U}(1)\times { SU}(3) \ni \{\alpha, h\} \to \left\{ \left( \begin{array}{cc} \alpha^3 & 0\\ 0 & \alpha^{-3}\end{array} \right) , \left( \begin{array}{cc} \alpha h & 0 \\ 0 & \alpha^{-3} \end{array} \right)\right\}\in {SU}(2)\times {SU}(4)
\end{align}
where in the top-right corner of the second matrix, we have a $3\times 3$ matrix, and therefore (for example for $h=1$) $\alpha \equiv \alpha \mathbb{I}_3$. Let us check how this works out for the charges. According to this prescription the ${ U}(1)$ acts on the up quarks as $u_i\to \alpha^{1+3} u_i$, which corresponds to the correct electric charge of $4/6=2/3$ for $\alpha= e^{1/6}$. Similarly, for the down quarks we have $d_i\to \alpha^{1-3} d_i$, which gives the correct electric charge of $-2/6=-1/3$. For the neutrino we have $\nu_i\to \alpha^{3-3} \nu_i$, which gives zero electric charge, and for the electrons $e_i\to \alpha^{-3-3} e_i$, which gives the electric charge $-1$. This gives all the correct quantum numbers of the unbarred fermions. For the barred ones it is clear that we simply have to use the hermitian conjugate representation of ${SU}(2)\times { SU}(4)$ (but not of the Lorentz group, because the barred fermions are still unprimed 2-component spinors). 

We can now write the kinetic terms for all the fermions in a very compact form
\begin{align}\label{unif-kin}
-{2} \mathcal{D}\bar{S}^i \mathcal{D}S_i
\end{align}
Here $\mathcal{D}$ is the covariant derivative relevant for each multiplet. The ${SU}(3)$ and ${U}(1)$ connections are present in both $\mathcal{D}S_i$ and $\mathcal{D}\bar{S}^i$ in a symmetric way, with the hermitian conjugate connections appearing in $\mathcal{D}\bar{S}^i$. This is because the representation of the unbroken gauge group of the SM is indeed real as mentioned earlier. However, the massive ${ SU}(2)$ gauge field appears asymmetrically in that $\mathcal{D}\bar{S}^i$ is diagonal in the isospin indices, while $\mathcal{D}S_i$ is not, see Section \ref{sec:inter}.
Let us now discuss the mass terms. These can again be written in terms of $\bar{S}_i, S_i$ as
\begin{align}\label{unif-mass}
-\rho^2 \, \bar{S}_i \Lambda^{ij} S_j
\end{align}
The mass matrices $\Lambda^{ij}$ appearing here are complicated objects. Each of them is an $8\times 8$ block matrix that consists of 4 different entries $\Lambda^{ij}_u, \Lambda^{ij}_d, \Lambda^{ij}_\nu,\Lambda^{ij}_e$. It thus breaks ${ SU}(2)$ symmetry completely, while the ${SU}(4)$ is broken down to ${ U}(1)\times {SU}(3)$. 

Overall, the sum of two terms (\ref{unif-kin}), (\ref{unif-mass}) gives the Lagrangian:
\begin{align}\label{unif-tot}
-{2} \mathcal{D}\bar{S}^i \mathcal{D}S_i-\rho^2 \, \bar{S}_i \Lambda^{ij} S_j
\end{align}

We note that the unification described here is different from the Pati-Salam model, as no second ${ SU}(2)$ has been used. This seems natural in the second-order formalism in which the weak ${ SU}(2)$ has been frozen from the beginning by using physical variables. In our framework the unbarred doublets simply transform under the hermitian conjugate representation of ${SU}(2)\times {SU}(4)$ and therefore the same $SU(2)$ acts on both barred and unbarred particles. Furthermore, in the Pati-Salam unification the representations that are needed are complex because both the usual (left-)weak and right-weak $SU(2)$s are unbroken and act differently on the doublets. Here we have the following (real) reducible representation:
\begin{align}
S\oplus \bar S = ({\bf 2} ,{\bf 4}) \oplus ({\bf 2}, {\bf \bar{4}}) \in SU(2) \times SU(4)
\end{align}

whereas in Pati-Salam, the representation used is:
\begin{align}
S\oplus \bar S = ({\bf 2},{\bf 1},{\bf 4}) \oplus ({\bf 1},{\bf 2}, {\bf \bar{4}}) \in SU(2)\times SU(2) \times SU(4)
\end{align}

which is complex as clearly we have $\overline{ ({\bf 2},{\bf 1},{\bf 4})}\neq ({\bf 1},{\bf 2}, {\bf \bar{4}})$ (with ${\bf \bar{2}} \sim{\bf 2}$). Notice now that ${SU}(2)\times {SU}(4)\sim { SO}(3)\times { SO}(6)\subset {SO}(9)$. Below we discuss the direct embedding of ${ U}(1)\times {SU}(3)\subset SO(9)$, and in the last section we will see how the former can be constructed.

\section{${\rm SO}(9)$ unification}

We  give an algebraic derivation of the particle content of the aforementioned 16-dimensional spinor irrep of $SO(9)$ by making use of Dynkin diagrams and of the highest weight construction. The Dynkin diagram of $SO(9)$ belongs to the $C_n$ series for (rank) $n=4$:

\medskip
\renewcommand*{\arraystretch}{1.5}
\begin{tabular}{>{$}r<{$}m{4cm}}

C_4 \equiv \mathfrak{so}(9) 
&
  \begin{dynkin}
    \dynkinline{1}{0}{2}{0};
    \dynkinline{2}{0}{3}{0};
    \dynkindoubleline{4}{3};
    \foreach \x in {1,...,3}
    {
        \dynkindot{\x}{0}
    }
    \dynkinddot{4}{0}
  \end{dynkin}
\\

\end{tabular}

\medskip
\noindent If we denote its simple roots by $\alpha_i$, $i=1,\ldots,4$, a useful representation for the latter is:
\begin{align}
\alpha_i = e_i-e_{i+1}, \quad \alpha_4 = e_4, \quad e_i\cdot e_j = \delta_{ij}
\end{align}

Its Cartan matrix is given by:
\begin{align}
A_{ij}=\left( \begin{array}{cccc} 2 & -1 & 0 & 0 \\ -1 & 2 & -1 & 0 \\0 & -1 & 2 & -2\\0 & 0 & -1 & 2\\
\end{array} \right)
\end{align}

so that the simple roots in the Dynkin basis are given by:
\begin{align}
\left(\alpha_i\right)_j = A_{ij}
\end{align}

\renewcommand*{\arraystretch}{1}
The spinor irrep of $SO(9)$ is given by the decomposition of its highest-weight in terms of Dynkin indices: 
\begin{align}
{\bf 16} = \left( \begin{array}{cccc} 0 & 0 & 0 & 1 \\
\end{array} \right)
\end{align}

We can therefore construct the whole irrep in the Dynkin basis. We do it here explicitly:
\begin{figure}[H]\begin{center}
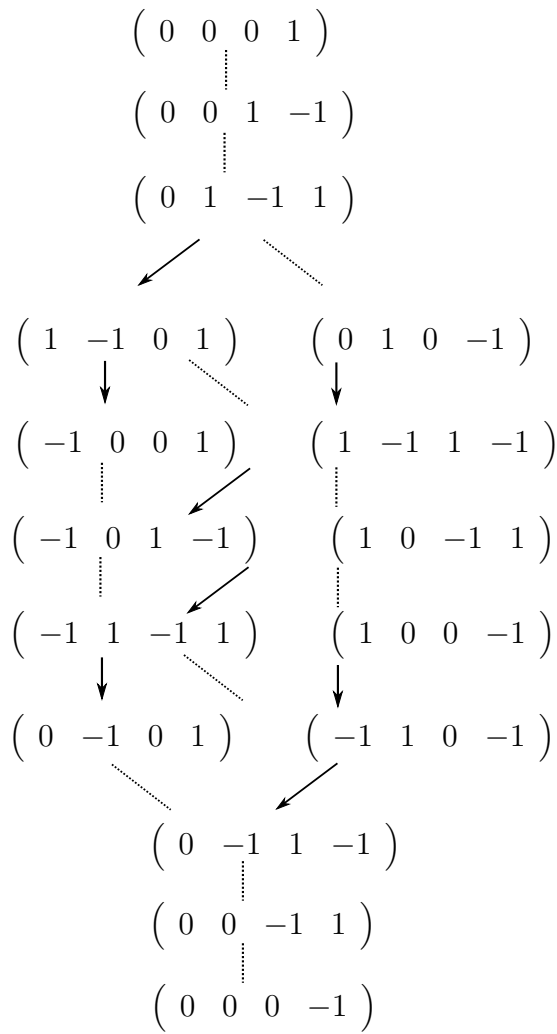 
\caption{Highest-weight construction of the spinor representation of $SO(9)$ in the Dynkin basis.}
\end{center}\end{figure}

where we denoted by an arrow the action of the simple roots $\alpha_1$ and $\alpha_2$ and by dotted lines the others. As before, we seek an embedding of the SM unbroken gauge group inside $SO(9)$. Again, a good embedding can be found by considering the reduced Dynkin diagram obtained after omitting the last two roots in $C_4$. The new diagram describes an embedding $SU(3)\times U(1) \subset SO(9)$ where the $SU(3)$ sector is generated by the first two simple roots. Now, in order to see how our spinor representation transforms under the subgroup, one needs simply remember that for $SU(3)$ the (anti-)triplet representation is obtained by a successive lowering of the highest-weight state by the two simple roots. It is easy to see that states linked by arrows decompose appropriately under $SU(3)$ therefore, we see that the 16 decomposes (under $SU(3)$) as follows:
\begin{align}
{\bf 16} = {\bf 3} + {\bf 3} + {\bf \bar{3}} +{\bf \bar{3}} + {\bf 1 }+{\bf 1 }+{\bf 1 }+{\bf 1 }
\end{align}
which maps into the SM particle content as far as color is concerned. We still have to check the electric charges of those particles. In order to do so, we need to find the charge generator $Q$. Again:
\begin{align}
Q = \left( \begin{array}{cccc} a & b & c & d \\
\end{array} \right)
\end{align}

For the $U(1)$ subgroup to commute with $SU(3)$, we have to require that the charges within an (anti-) triplet are equal. In order to fix our conventions, we have the following triplets:
\begin{align}
\begin{array}{l}\left( \begin{array}{cccc} 1 & 0 & -1 & 1 \\
\end{array} \right)\\ \left( \begin{array}{cccc} -1 & 1 & -1 & 1 \\
\end{array} \right)\\ \left( \begin{array}{cccc} 0 & -1 & 0 & 1 \\
\end{array} \right) \end{array}, \quad \begin{array}{c}\left( \begin{array}{cccc} 1 & 0 & 0 & -1 \\
\end{array} \right)\\ \left( \begin{array}{cccc} -1 & 1 & 0 & -1 \\
\end{array} \right)\\ \left( \begin{array}{cccc} 0 & -1 & 1 & -1 \\
\end{array} \right) \end{array}
\end{align}
while the other two 3 dimensional weight subspaces are anti-triplets and the remaining four states are singlets. Using that the charge of a state $\lambda$ is given by:
\begin{align}
 Q\cdot \lambda = q_\lambda
\end{align}

We obtain, on the one hand:
\begin{align}
Q = -\frac{1}{3}\left( \begin{array}{cccc} 1 & 2 & 3 & 0 \\
\end{array} \right)
\end{align}
which can be understood as $Q= -\frac{1}{3}(H^1+2H^2+3H^3)$, $H^i$ being the Cartan generators, and on the other hand:
\begin{align}
{\bf 16}= {\bf 3}_{\frac{2}{3}} + {\bf 3}_{-\frac{1}{3}}  +{\bf \bar{3}}_{-\frac{2}{3}}  +{\bf \bar{3}}_{\frac{1}{3}}  + {\bf 1 }_{-1} +{\bf 1 }_0+{\bf 1 }_0+{\bf 1 }_{1}
\end{align}
which corresponds to the particle content of the SM. Therefore, we see that the fermionic states of the (second-order) SM can be gathered into a single $SO(9)$ irrep.

\section{${\rm SO}(10)$ unification}

It is known that all left-handed fermions of the SM can be put into a single 16-dimensional irreducible representation of ${SO}(10)$, see e.g. \cite{Srednicki:2007qs}, section 97 or \cite{Baez:2009dj}. So, our unprimed fermions $\bar{S}^i, S^i$ can be combined into a single multiplet of ${SO}(10)$. This ${ SO}(10)$ is obtained from the already encountered ${SU}(2)\times { SU}(4)$ by adding another ${ SU}(2)$ that mixes $\bar{S}^i$ and $S^i$. Putting these groups together we have ${ SU}(2)\times{SU}(2)\times{ SU}(4)\sim {SO}(4)\times {SO}(6)\subset { SO}(10)$. The Lagrangian one obtains is of the same schematic form (\ref{SML2}), now with a single 16-dimensional fermionic multiplet ${\mathcal F}^i$. We will not explore this theory any further, since for our purposes it is enough to consider the above $SO(9)$ irrep.
%as there seems to be no benefit in putting all fermions into a single multiplet from the point of view of reducing the complexity of the Lagrangian. At the same time, one has to introduce ${\rm dim}({\rm SO}(10))-{\rm dim}({SU}(2)\times{ SU}(4))=45-18=27$ more gauge fields in the ${\rm SO}(10)$ unification as compared to the ${\rm SU}(2)\times {\rm SU}(4)$ framework. 

We also note that it is the ${SO}(10)$ unification scheme that incorporates the ${SU}(5)\subset{ SO}(10)$ viewpoint \cite{Georgi:1974sy} on the SM fermions. The second-order version of the ${ SU}(5)$ model is also possible, but we refrain from spelling out the details as it has no immediate added interest.
% given that this model requires ${\rm dim}({SU}(5))-{\rm dim}({\rm SU}(2)\times{\rm SU}(4))=6$ more gauge fields than ${\rm SU}(2)\times {\rm SU}(4)$ based GUT. It is thus less economical than the model described in the previous subsection. 

\section{Gauge-gravity unification}\label{gguni}
Earlier, we saw that the unprimed fermions of the SM (16 per generation) sit nicely into the 16-dimensional spinor irrep of $SO(9)$. However, a higher ``unification'' can occur if we consider gravity as a gauge theory. The idea of reformulating gravity in terms of new variables dates back to \cite{Ashtekar:1986yd}, where an $SU(2)$ connection was introduced to account for the gravitational degrees of freedom of the theory. More recently, a novel description of gravity as diffeomorphism invariant gauge theory was developed, see \cite{Krasnov:2011up,Krasnov:2011pp,Krasnov:2012pd}. It seems therefore natural, in our efforts to seek the nicest irrep possible for the fermions, to take their spinor index into account. Indeed, as it is well known, the Lorentz group $SO(1,3)$ is isomorphic (up to complexification) to $SU(2)_L\times SU(2)_R$ and all the field representations of the former can be built using the latter. Fermions are no exception to this rule and unprimed spinors live in the $({\bf 2} ,{\bf 1})$ irrep, while primed fermions live in the $({\bf 1},{\bf 2} )$\footnote{In this chapter we use the dimensional notation for irreps. The representation $({\bf 2} ,{\bf 1})$ would be denoted $(\frac{1}{2}, 0)$ in a spin notation.}. Nevertheless the two irreps are linked by complex conjugation, and based on the gauge-theoretical approach to gravity, it is not a leap of faith to consider the extended symmetry group for the SM representations to be
\begin{align}
G_{SM+GR} = SO(9) \times SO(3)_{GR}
\end{align}
 
where we used to well known isomorphism $SO(3)\sim SU(2)$. The representation we need is (for the unprimed fermions only at this point):
\begin{align}
\mathcal{F}= ({\bf 16},{\bf 2})
\end{align}

where $\mathcal{F}$ denotes the fermions multiplet, and we have now a 32 dimensional object that takes into account the doubling of the original 16 due to the fact that spinors are $SO(3)_{GR}$ doublets. Once this has been set, the simplest embedding that can be considered is the obvious:
\begin{align}
SO(9) \times SO(3)_{GR} \subset SO(12)
\end{align}

This embedding is the most natural, because $SO(12)$ possesses a 32-dimensional (pseudo-)real spinor representation. Not only that, but it actually has two 32-dimensional spinor irreps, so that the second can account for the primed fermions if needed! It turns out that this embedding exists, and after a ``breaking''
\begin{align}
  SO(12)~ \rightarrow ~ SO(9) \times SO(3)_{GR}
\end{align}

its spinor irreps transform in the right manner under its subgroup:
\begin{align}
{\bf 32} = ({\bf 16},{\bf 2}), \quad {\bf 32'} = ({\bf 16},{\bf 2})
\end{align}

In what follows we will be particularly interested in a more direct embedding of the SM symmetry group, even though the embedding we have just described can be considered as an intermediate step. More precisely we want:
\begin{align}\begin{split}
SO(12) &\supset  SO(9) \times SO(3)_{GR}\\&  \supset SO(6)_c \times SO(3)_{W}\times SO(3)_{GR} \sim SU(4)_c \times SU(2)_W \times SU(2)_{GR}
\end{split}\end{align}

As far as representations are concerned we want:
\begin{align}
{\bf 32 } = ({\bf 4},{\bf 2},{\bf 2}) \oplus ({\bf \bar{4}},{\bf 2},{\bf 2})
\end{align}

This would correspond to the colour-weak representations of Section \ref{sec:pati} with the spinor index taken into account. In this section, we will explicitly construct the representation using a Clifford Algebra that generates $SO(12)$ and its subsequent ladder operators that generate the $SU(6)\subset SO(12)$ subgroup. This will allow us to construct and name all the states that appear in the 32 in terms of their SM quantum numbers. Moreover, using the the $SU(6)$ subgroup that we have just mentioned, we will be able to represent this states as antisymmetric tensors (forms) living in a 6-dimensional complex space. This brings along many investigation opportunities that we will mention in the Discussion, but before considering the full $SO(12)$ algebra we will shortly focus on a toy-model that describes a neutrino-electron doublet.

\subsection{Quick reminder: ${\rm SO}(2n)$ spinor irreps}
We would like to give a short reminder on the way spinor representations of $SO(2n)$ can be built using Clifford Algebras (CA). We will focus on the ``even'' orthogonal groups as these will be the main focus in this section. In order to build their spinor irrep(s), we construct the following CA:
\begin{align}
\left\{\Gamma_I , \Gamma_J \right\}= 2 \delta_{IJ}, \quad I,J = 1,\ldots, 2n+1
\end{align}

The algebra generators are then constructed as:
\begin{align}
M_{ij} = \frac{1}{4i}\left[\Gamma_i, \Gamma_j \right], \quad i,j = 1,\ldots, 2n\label{genCA}
\end{align}
and the last gamma matrix that was not used commutes with all of the generators:
\begin{align}
\Gamma_{2n+1} = (-i)^n\prod_{i=1}^{2n} \Gamma_i, \quad \left[ \Gamma_{2n+1}, M_{ij}\right] =0 , ~\forall i,j
\end{align}

This implies that the $SO(2n)$ (Dirac) spinor representations are in fact reducible, and the above gamma matrix acts as a projector onto two (Weyl) irreps:
\begin{align}
P^{\pm} = \frac{1}{2}(1\pm\Gamma_{2n+1}) \rightarrow D^n, ~ D^{n-1}
\end{align}

where $D^n, ~ D^{n-1}$ denote the two irreps and both have dimension $2^{n-1}$. Furthermore, depending on the value of $n$, they are either (pseudo-)real or complex conjugate to each other. All we need to know here is that for $SO(6)$ they are complex and for $SO(12)$ they are pseudo-real. 

Using the aforementioned CA, we can build the two irreps in the following way. First, define a set of (pairs of) ladder operators:
\begin{align}
A_s \defeq \frac{1}{2}(\Gamma_{2s-1}-i\Gamma_{2s}),\quad A_s^\dagger \defeq \frac{1}{2}(\Gamma_{2s-1}+i\Gamma_{2s})\label{laddersun}
\end{align}
with $s= 1, \ldots, n$. They form a set of canonically normalised ladder operators, in the sense that:
\begin{align}
\left\{A_s, A_r \right\}=0= \left\{A_s^\dagger, A_r^\dagger \right\}, \quad \left\{A_s, A_r^\dagger \right\}= \delta_{rs}
\end{align}

Furthermore, the set of creation operators $\left\{ A_s^\dagger\right\}$ generates a $SU(n)\subset SO(2n)$ subalgebra, and with respect to the former, they are tensor operators that transform according to its defining representation\footnote{Precisely: $T_a = \sum_{rs}A^\dagger_r \left[ T_a\right]^{(n)}_{rs}A_s$ where $T_a^{(n)}$ denotes the defining representation's generators of $SU(n)$ and $\left[ T_a, A^\dagger _s\right] = \sum_{r}A^\dagger_r \left[ T_a\right]^{(n)}_{rs}$.}. Now, it only remains to construct the spinor irreps of $SO(2n)$. We mentioned that there are two such irreps, and these are given by their fundamental weights. We could use the simple roots to construct all the states appearing in the irreps, however, we decide to take a different approach that is closer in spirit to what is usually done for physical systems. Therefore, consider the lowest weight (or vacuum). Let us denote it by:
\begin{align}
|0 \rangle,  \quad A_s |0 \rangle  = 0,~\forall s
\end{align}
Because each pair of ladder operators acts on a separable spin $1/2$ subspace, we can also use a spin up-down notation to label the states. Therefore, we have equivalently:
\begin{align}
|0 \rangle\equiv  \bigotimes_{n}|- \rangle 
\end{align}

We will use both notations in the two examples that we work out below, however let us, for now, focus on the properties of the states. Because the creation operators can also be seen as tensor operators, the first excited states are just the defining representation of $SU(n)$:
\begin{align}
A^\dagger_s|0 \rangle \sim {\bf n}
\end{align}
Next, two ``particle'' states will transform as a rank-two antisymmetric tensor because the creation operators anti-commute:
\begin{align}
A^\dagger_sA^\dagger_r|0 \rangle \sim ({\bf n}\otimes {\bf n})_{a}
\end{align}

Similarly, we generate all the higher-rank antisymmetric tensors up to $n$. We therefore have:
\begin{align}
D^{n}\oplus D^{n-1} = \sum_{m=0}^{n} \left[ m \right ]
\end{align}

where $\left[ m \right ]$ denotes the rank-$m$ antisymmetric irrep. Equivalently, these spinors irreps can be thought of as $m-$forms in $n$ complex dimensions, so that:
\begin{align}
D^{n}\oplus D^{n-1} = \Lambda\mathbb{C}^n
\end{align}

Finally, because the last gamma matrix $\Gamma_{2n+1}$ anticommutes with the creation operators, it commutes with an even number of them, and therefore (as we have already mentioned) the Dirac spinor splits into two Weyl irreps. In terms of forms, we have:
\begin{align}
D^{n}\oplus D^{n-1} = \Lambda^{e}\mathbb{C}^n \oplus \Lambda^{o}\mathbb{C}^n
\end{align}
 where $e,~o$ stand for even and odd, and which irrep is matched to which forms' subspace depends on the value of $n$. We will make the details explicit in the two examples we discuss below.

%{\rm SO}(9) \times{\rm SO}(3)_{GR}
\subsection{Toy model: ${\rm SO}(3)_{W}\times{\rm SO}(3)_{GR}\subset {\rm SO}(6)$ }\label{toyso6}

Let us now focus our attention on a simple toy model. Consider a single generation weak doublet consisting of a neutrino ($q=0$) and an electron ($q=-1$). Their weak isospin is $+1/2$ and $-1/2$ respectively. Furthermore, we consider a second-order description of their dynamics, so that their field representation is a doublet of unprimed spinors. In matrix form, the particle content can be described as:
\begin{align}
\chi^a_A \equiv \left(\begin{array}{cc} \nu_1 & \nu_2 \\ e_1 & e_2\end{array} \right)
\end{align}
where $A=1,2$ is their spinor index. In terms of groups and representations, the symmetry group is ${SU}(2)_{W}\times{SU}(2)_{GR}\sim {SO}(3)_{W}\times{SO}(3)_{GR}$ and their irrep is:
\begin{align}
\chi \sim ({\bf 2}, {\bf 2})
\end{align}
We would like to consider an embedding of that group into $SO(6)\sim SU(4)$ and because of the dimension of the representation, it is natural to look at (one of) the Weyl spinors of $SO(6)$:
\begin{align}
{\bf 4} = ({\bf 2}, {\bf 2}) \quad {\rm or} \quad {\bf \bar{4}} = ({\bf 2}, {\bf 2})\label{embconst}
\end{align}

At this point, we could choose to work with the first irrep without loss of generality. Below we will see that both arise naturally in our construction.

We would like to construct explicitly the generators of the subgroup starting from the generators of $SO(6)$. In order to do so, notice:
\begin{align}
{\bf 6} = ({\bf 4}\otimes {\bf 4})_a
\end{align}

where ${\bf 6}$ is the vector irrep of $SO(6)$. Using the constraint we want to impose (\ref{embconst}), we have:
\begin{align}
{\bf 6} = ({\bf 3}, {\bf 1}) \oplus ({\bf 1}, {\bf 3})
\end{align}
where on the LHS we have an $SO(6)$ irrep and on the RHS we have ${SO}(3)_{W}\times{SO}(3)_{GR}$ representations. The above equation means that a 6-dimensional vector decomposes into two 3-dimensional vectors. Let us see what this tells us about the generators. As we have done above, we have now in hands a construction of the $SO(6)$ generators using 6 gamma matrices:
\begin{align}
M_{ij}= \frac{1}{4i}\left[\Gamma_i, \Gamma_j \right], \quad i,j=1,\ldots,6
\end{align}
and we have an extra matrix $\Gamma_7$ from which we can build a projector. The above embedding leads us to consider the following generators for the subalgebra:
\begin{align}
J_{ij} &= M_{ij},~~i,j= 1,2,3 \rightarrow SO(3)_W \\
K_{ij} &= M_{ij},~~i,j= 4,5,6 \rightarrow SO(3)_{GR}
\end{align}
This (explicit) embedding has all the right properties we have stated above. In order to see it, we can explicitly construct all the states. As it is usually done for $SU(2)$ algebras, we construct two ``spin'' ladder operators and a measurable operator. It is usual to choose the three Cartan generators of $SO(6)$ to be $M_{12},~M_{34},~M_{56}$ and therefore, we fix $M_{12} \equiv J^3,~~M_{56} \equiv K^3$, the rest of the generators easily follow. To summarise:
\begin{align}
\begin{split}
\left\{ \begin{array}{l} J^3 \equiv M_{12} \\ J^2 \equiv M_{31} \\ J^1 \equiv M_{23}\end{array}\right., \quad J^{\pm}= \frac{1}{\sqrt{2}}(J^1\pm i J^2)
\end{split}\\
\begin{split}\left\{ \begin{array}{l} K^3 \equiv M_{56} \\ K^2 \equiv M_{45} \\ K^1 \equiv M_{64}\end{array}\right., \quad K^{\pm}= \frac{1}{\sqrt{2}}(K^1\pm i K^2)
\end{split}
\end{align}
with, for example:
\begin{align}
\left[ J^3, J^\pm\right] = \pm J^\pm, \quad \left[J^+, J^- \right] = J^3
\end{align}

We would like to have an expression for these generators in terms of the $SU(3)$ ladder operators (\ref{laddersun}). Using (\ref{genCA}) and the inverse transformation:
\begin{align}
\Gamma_{2s-1} = A_s+A^\dagger_s, \quad \Gamma_{2s} = i\left( A_s- A_s^\dagger\right)
\end{align}
we can obtain a general formula for the generators:
\begin{align}
M_{2s,2r} &= \frac{i}{2}\left( A_sA_r +A^\dagger_sA^\dagger_r - A^\dagger_sA_r -A_s A^\dagger_r\right)\\
M_{2s-1,2r-1}&= \frac{1}{2i}\left(A_sA_r+A^\dagger_sA^\dagger_r + A^\dagger_sA_r +A_s A^\dagger_r\right)\\
M_{2s,2r-1}&= \frac{1}{2}\left(A_sA_r -A^\dagger_sA^\dagger_r - A^\dagger_sA_r +A_s A^\dagger_r\right)\label{genladder}
\end{align}

We therefore have:
\begin{align}
\begin{split}
\left\{ \begin{array}{l} J^3 = N_1 -\frac{1}{2} \\ J^+ =\frac{1}{\sqrt{2}}\Sigma_2 A_1^\dagger \\ J^- =\frac{1}{\sqrt{2}}A_1\Sigma_2 \end{array}\right. \end{split}\\
\begin{split}\left\{ \begin{array}{l} K^3 = N_3 -\frac{1}{2} \\ K^+ =\frac{i}{\sqrt{2}}\Delta_2 A_3^\dagger \\ K^- =\frac{i}{\sqrt{2}}A_3\Delta_2\end{array}\right. \end{split}
\end{align}

where we defined:
\begin{align}
N_i\defeq A_i^\dagger A_i, \quad \Sigma_2 \defeq A_2+ A^\dagger_2, \quad \Delta_2 \defeq A_2- A_2^\dagger
\end{align}

First of all, notice that all the generators are bilinears in the ladder operators. This implies that once we are in a spinor irrep (${\bf 4}$ or ${\bf \bar{4}}$), we stay in it (as expected). Then notice that, although both $SU(2)$s have a mix of ladder operators (1,2 and 2,3), the weak group acts ``mainly'' on the first subspace, while the Lorentz group acts on the third. The second direction is ``broken'' at the level of the subgroup.

We now want to construct the states. In order to do so, we need a vacuum state. In the previous subsection, we saw that the state:
\begin{align}
|0\rangle = | - - -\rangle
\end{align}

is annihilated by all ladder operators and can therefore be taken as vacuum state. However, we are here interested in a particular subalgebra, $i.e.$, we want a vacuum that is annihilated by a particular set of generators. In this case, there are two:
\begin{align}
|0 \rangle _{e}= | - - -\rangle, \quad |0\rangle _{o}=| - + -\rangle, \quad A_s |0 \rangle _{e,o} = 0,~s=1,3
\end{align}

where $e,~o$ stands for even or odd vacuum, and the parity of the latter is dictated by its $\Gamma_{7}$ value and is even for a state belonging to $\Lambda^e\mathbb{C}^3$ and odd otherwise. Moreover, we have in our case:
\begin{align}
|0 \rangle _{e}\in {\bf \bar{4}}= D^2 , \quad |0\rangle _{o}\in {\bf {4}}= D^3
\end{align}

As it is expected, using our subalgebra we can construct two different irreps, corresponding to the freedom of choice (\ref{embconst}). Using the quantum numbers that we want the particles to have, it is then easy to see that we have either:
\begin{align}
{\bf \bar{4}}= \left(\begin{array}{cc} \nu_1 & \nu_2 \\ e_1 & e_2\end{array} \right) = \left(\begin{array}{cc} | + - +\rangle & | + + -\rangle\\ | - + +\rangle& | - - -\rangle\end{array} \right)
\end{align}

or
\begin{align}
{\bf {4}}= \left(\begin{array}{cc} \nu_1 & \nu_2 \\ e_1 & e_2\end{array} \right) = \left(\begin{array}{cc} | + + +\rangle & | + - -\rangle\\ | - - +\rangle& | - + -\rangle\end{array} \right)
\end{align}

We could conclude with these two equations as now we know explicitly how both the embedding group and its subgroup act on each state, and we have therefore explicitly constructed the embedding. It is however interesting to (equivalently) express everything in terms of $m-$forms in three complex dimensions. In order to do so, let us denote:

\begin{align}
A^\dagger_1 |0\rangle \sim dt, \quad A^\dagger_2 |0\rangle \sim du, \quad A^\dagger_3 |0\rangle \sim dv
\end{align}

Using these definitions, our original spinor can be rewritten in two different ways:
\begin{align}
\tilde\chi' &= \tilde\nu_1(p)dt \wedge dv +\tilde \nu_2(p)dt\wedge du +\tilde e_1(p) du\wedge dv + \tilde e_2(p)\\
\chi &= \nu_1(p)dt \wedge du\wedge dv +  \nu_2(p)dt +  e_1(p)  dv +  e_2(p)du 
\end{align}
where $p=(t,u,v)$ is a point in $\mathbb{C}^3$ and where the tilde denotes another possible function for the corresponding state. In this notation, we used the fact that a scalar product is well defined for forms and that they admit a complete basis upon which they can be decomposed. These two representations can equivalently describe the particle content that we wished to have in our theory. Nonetheless, it seems natural at this point to choose one, because the notation is more elegant, we can simply choose the spinor irrep ${\bf 4}$:
\begin{align}
\chi&= \nu_1(p)dt \wedge du\wedge dv +  \nu_2(p)dt +  e_1(p)  dv +  e_2(p)du 
\end{align}

However, it seems a bit brutal to throw away an entire irrep that we have just constructed. Indeed, recall that in Chapter \ref{chapSM} we saw that in addition to the weak doublet, there is an additional pair or barred fermions with opposite quantum numbers:
\begin{align}
\bar \nu_A, \quad \bar e_A
\end{align}

If, as we did in the second-order formulation of the SM, we build a doublet out of them and give them the correct weak isospin quantum numbers, we can rewrite for the ${\bf \bar 4}$:
\begin{align}
\bar \chi &= \bar e_1(p)dt \wedge dv + \bar e_2(p)dt\wedge du +\bar \nu_1(p) du\wedge dv + \bar \nu_2(p)
\end{align}

This is beautifully unified, as we know that in $SO(6)$, we have:
\begin{align}
{\bf \bar 4}\otimes {\bf 4}={\bf 15}^{adj} \oplus {\bf 1}
\end{align}

so that we can build a scalar out of these two irreps. Let us look in our case at the scalar we obtain:
\begin{align}
\bar\chi \chi = \left( \bar \nu_2\nu_1 - \bar \nu_1\nu_2  +\bar e_2e_1 -\bar e_1e_2 \right) dt\wedge du\wedge dv
\end{align}
where we allowed ourselves a change of phase in the definition of the wavefunctions to obtain the correct relative sign. This is nothing but the mass term we would like to write for a fermion, $e.g.$:
\begin{align}
\bar{e}^Ae_A = \bar e_2e_1 -\bar e_1e_2
\end{align}

 At this point we could try to construct a Lagrangian in terms of these quantities, it seems however that they live in a three dimensional space so that the interpretation that is given to the forms must be carefully thought about. This is still a current research interest.
After this warm-up, we consider the full SM symmetry group unified with gravity.

\subsection{The real deal: ${\rm SU}(4)_c \times {\rm SU}(2)_W \times {\rm SU}(2)_{GR}\subset {\rm SO}(12) $}

In this final subsection, we would like to consider all of the SM fermionic section (one generation). There are 16 unprimed fermions, and each of them can be seen as a $SU(2)_{GR}$ doublet. This amounts to a total of 32 ``particles''. To be as general as possible, we also consider the primed fermions, which also add up to 32 independent objects. We will treat the weak $SU(2)_W$ as frozen, however, both barred and unbarred fermions are seen as weak doublets as was done in Chapter \ref{chapSM}. We will therefore talk about the approximate symmetry of the (free Lagrangian of the) SM. The smallest group with two 32-dimensional spinor irreps is $SO(12)$ and we wish to describe the embedding:
\begin{align}
{SU}(4)_c \times { SU}(2)_W \times {SU}(2)_{GR}\subset SO(6)_c \times SO(6)_{WG}\subset  SO(12)
\end{align}

So that:
\begin{align}
{\bf 32} = ({\bf 4}, {\bf 4}) \oplus ({\bf \bar{4}}, {\bf \bar{4}}), \quad {\bf 32}' = ({\bf 4}, {\bf \bar{4}}) \oplus ({\bf \bar{4}},{\bf 4})
\end{align}

where the embedding is such that $\overline{\bf 32}  ={\bf 32} $, $\overline{{\bf 32}'}  ={\bf 32}' $ and ${\bf 12} \in {\bf 32} \otimes {\bf 32}'\ni ({\bf 6}, {\bf 1})\oplus ({\bf 1}, {\bf 6})$. This embedding can be understood as follows: $SO(12)$ admits an $SU(6)$ subalgebra generated by 6 pairs of ladder operators. Its Dirac spinor representation will split into two Weyl spinors as before, and we have:
\begin{align}
{\bf 32} = \Lambda^e \mathbb{C}^6, \quad {\bf 32}' = \Lambda^o \mathbb{C}^6
\end{align}

As far as $SO(6)$ irreps are concerned, recall:
\begin{align}
{\bf 4} = \Lambda^{o}\mathbb{C}^3, \quad {\bf \bar 4} = \Lambda^{e}\mathbb{C}^3
\end{align}

Thus, we have:
\begin{align}
({\bf 4}, {\bf 4}) \in \Lambda^{e}\mathbb{C}^6, \quad ({\bf 4}, {\bf \bar{4}}) \in \Lambda^o \mathbb{C}^6
\end{align}
and similarly for the conjugate representations. 

Furthermore, to see that we can embed all of the SM particle content with the right quantum numbers, it is sufficient to check that we can embed two spinor irreps of $SO(6)$ that transform independently, in other words, that 3 pairs of ladder operators generate the first $SO(6)_c$ spinor and the remaining three generate the weak-gravity $SO(6)_{WG}$. Indeed, if this is true, then Sections \ref{sec:pati} and \ref{toyso6} allow us to conclude.

Let us now then proceed with the details of the embedding. We have for the generators:
\begin{align}
A_{ij} &= M_{ij}, ~~i,j = 1,\ldots,6 ~\rightarrow ~ SO(6)_{WG} \\
B_{ij} &= M_{ij}, ~~i,j = 7,\ldots,12 ~\rightarrow ~ SO(6)_{c}
\end{align}

At the next ``breaking'' step we have:
\begin{align}
J_{ij} &= A_{ij}, ~~i,j = 1,\ldots,3 ~\rightarrow ~ SO(3)_{W} \\
K_{ij} &= A_{ij}, ~~i,j = 4,\ldots,6 ~\rightarrow ~ SO(3)_{GR}
\end{align}

But let us focus on the first two subalgebras. It is easy to see that:
\begin{align}
\left[A_{ij}, B_{kl} \right]=0
\end{align}

Also, the Cartan generators of $SO(12)$ being $M_{1,2},~M_{3,4},~M_{5,6},~M_{7,8},~M_{9,10},~M_{11,12}$, we see that they are equally split into each $SO(6)$. Finally, using (\ref{genladder}), it is easy to see that $A_{ij}$ will be given by bilinears of $A_s,~A_s^\dagger$, $s=1,2,3$, while $B_{ij}$ will be given by bilinears of $A_s,~A_s^\dagger$, $s=4,5,6$. This nice factorisation happens because $SO(12)$ belongs to the $SO(4m)$ type of algebras. At this stage, the construction of the embedding becomes trivial because we already know which representations can appear on each subspace: namely the two Weyl spinors of each $SO(6)$ as mentioned earlier. We therefore have as expected:
\begin{align}\begin{split}
{\bf 32}&= \left( (\Lambda^e \mathbb{C}^3+\Lambda^o \mathbb{C}^3), (\Lambda^e \mathbb{C}^3+\Lambda^o \mathbb{C}^3) \right)_{e} \\
&= \left( \Lambda^e \mathbb{C}^3, \Lambda^e \mathbb{C}^3 \right) \oplus\left( \Lambda^o \mathbb{C}^3, \Lambda^o \mathbb{C}^3 \right)
\end{split}\end{align}

and similarly for ${\bf 32}'$. \\
\medskip

We could conclude this chapter here, but let us carry on for a bit and try to construct a Lagrangian for such representations. As we did before, let us denote the first set of generators for the (colour) forms to be $dx,~dy,~dz~$, whereas we have as before the weak-gravity space generated by $dt,~du,~dv$ with the $dt$ subspace corresponding to the weak isospin quantum number and the $dv$ subspace to the spin. The exact dictionary stating which particle corresponds to which form in this 6-dimensional complex space can be obtained as was done in Section \ref{toyso6}, but we will refrain to do so here as no further insight can be gained from it. It is simply a generalisation of Sections \ref{sec:pati} and \ref{toyso6}. All we need for the argument below is that we choose to embed all the unprimed spinors in ${\bf 32}$ and all the primed spinors in ${\bf 32}'$. We will take the point of view that we take seriously the interpretation given to us by the forms representations and hence, we will try to build a Lagrangian on a 6-dimensional complex space. Using:
\begin{align}
 {\bf 32}\otimes {\bf 32}& = {\bf 1} \oplus \Lambda^2\mathbb{R}^{12}\oplus \Lambda^4\mathbb{R}^{12}\oplus \Lambda^{sd}\mathbb{R}^{12}\\
 {\bf 32}'\otimes {\bf 32}'& = {\bf 1} \oplus \Lambda^2\mathbb{R}^{12}\oplus \Lambda^4\mathbb{R}^{12}\oplus \Lambda^{asd}\mathbb{R}^{12}\\
 {\bf 32}\otimes {\bf 32}'& = {\bf 12} \oplus \Lambda^3\mathbb{R}^{12}\oplus \Lambda^5\mathbb{R}^{12}\label{kinetickronecker}
\end{align}

where $\Lambda^{(a)sd}\mathbb{R}^{12}$ denotes the rank-6 antisymmetric tensor that satisfies a real (anti-)self-duality property. We see that mass terms can be built as expected. Indeed, let us denote the spinor irreps:
\begin{align}
|\mathcal{F}\rangle \equiv \mathcal{F}= {\bf 32}, \quad |\overline{\mathcal{F}}\rangle\equiv \overline{\mathcal{F}}= {\bf 32}'
\end{align}

Then there is a natural pairing that gives us a scalar quantity. In terms of forms, it is given by their usual scalar product so that:
\begin{align}
\langle \mathcal{F}|\mathcal{F}\rangle  +\langle \overline{\mathcal{F}}|\overline{\mathcal{F}}\rangle \sim \mathcal{M}(p)dt\wedge dx\wedge dy\wedge dz\wedge du\wedge dv
\end{align}
where $\mathcal{M}(p)$ is a (real) scalar mass function that contains mass terms for all fermions and $p=(t,x,y,z,u,v)$ is a point in a 6-dimensional space and with
\begin{align}
\langle \mathcal{F}| \equiv \star \mathcal{F}, \quad \langle \overline{\mathcal{F}}| \equiv \star \overline{\mathcal{F}}
\end{align}

is the Hodge dual of the form\footnote{In 6 dimensions, the Hodge dual of an even form is an even form. This is why the singlet representation appears in the Kronecker product ${\bf 32}\otimes {\bf 32}$ (and similarly with the other spinor irrep).}. Let us now assume that we build a kinetic term for the fermions using (\ref{kinetickronecker}). On the RHS we have the 12-dimensional (real) vector; the rank-3 (real) antisymmetric tensor with dimension 220; and the rank-5 (real) antisymmetric tensor with dimension 792. For a first-order Dirac-type Lagrangian, we need a map:
\begin{align}
\langle \overline{\mathcal{F}}| \mathcal{D} |\mathcal{F}\rangle  \rightarrow {\bf 1}
\end{align}

 where $\mathcal{D}$ is a covariant derivative on the space of forms. In order to get a singlet, we can for example consider the covariant derivative to be given simply by a 12-(real)-dimensional $SO(12)$ vector because:
 \begin{align}
 \left({\bf 12}\otimes {\bf 12}\right)_s \ni {\bf 1} \label{vecscalar}
 \end{align}
 
where $s$ denotes a symmetric Kronecker product. This would be it for a simple non-interacting Lagrangian in 6-complex dimensions as, indeed we have that the simple partial derivative is a (complex) 6-dimensional vector\footnote{In a 6-dimensional complex space, we would surely expect a kinetic operator of the type $\partial + \bar{\partial}\sim {\bf 6} \oplus \overline{{\bf 6}} = {\bf 12}$.}. However, we expect the covariant derivative to include gauge-fields as well. Let us assume that they transform according to the adjoint representation and that they carry a vector index as well, therefore, we write:
\begin{align}
\mathcal{D}\in {\bf 12} \otimes {\bf 66}^{adj} = {\bf 12} \oplus {\bf 220} \oplus {\bf 560} 
\end{align} 

where the last 560-dimensional representation corresponds to the fundamental weight $(110000)$. We then have:
\begin{align}
\langle \overline{\mathcal{F}}| \mathcal{D} |\mathcal{F}\rangle  = \left({\bf 12}\oplus {\bf 220}\oplus {\bf 792 }\right)\otimes \left({\bf 12} \oplus {\bf 220} \oplus {\bf 560} \right)
\end{align}

It remains to construct a scalar out of this Kronecker product. In addition to (\ref{vecscalar}), we also have:
\begin{align}
\left({\bf 220}\otimes {\bf 220}\right)_s \ni {\bf 1}
\end{align}

All the other Kronecker products do not generate singlets, therefore we can define our kinetic term to be:
\begin{align}
\left.{\rm Tr}\right|_{SO(12)}\langle \overline{\mathcal{F}}| \mathcal{D} |\mathcal{F}\rangle  ={\bf 1}
\end{align}

Where $\left.{\rm Tr}\right|_{SO(12)}$ denotes the usual trace over the algebra, and the trace of any non-trivial ($i.e.$ other than the singlet) representation of the algebra vanishes. We see that the only contribution coming from the covariant derivative corresponds to:
\begin{align}
\mathcal{D}= {\bf 12} \oplus {\bf 220}
\end{align}
where the ${\bf 12}$ can be seen as the contribution from the partial derivative, and the ${\bf 220}\equiv \Lambda^3\mathbb{R}^{12}$ as the representation in which the gauge-fields live. In terms of forms in 6-dimensional complex space, we have:
\begin{align}\begin{split}
{\bf 220}=\Lambda^3\mathbb{R}^{12} &= \left(\Lambda^1\mathbb{C}^6 \otimes \Lambda^2\mathbb{C}^6\right) \oplus \star \left(\Lambda^1\mathbb{C}^6 \otimes \Lambda^2\mathbb{C}^6\right) \oplus \Lambda^3\mathbb{C}^6 \oplus \Lambda^3\mathbb{C}^6\\
&= (\overline{\bf 70} \oplus {\bf 20}) \oplus ({\bf 70} \oplus {\bf 20}) \oplus {\bf 20}\oplus {\bf 20}
\end{split}\end{align}
with $ \star \left(\Lambda^1\mathbb{C}^6 \otimes \Lambda^2\mathbb{C}^6\right)=\left(\Lambda^5\mathbb{C}^6 \otimes \Lambda^4\mathbb{C}^6\right)$. This can be obtained by looking at the decomposition ${\bf 220} = \left( {\bf 32}\otimes {\bf 32}'\right)_a$, with ${\bf 32} = \Lambda^{e}\mathbb{C}^6$, and with the additional constraint that we want a rank-3 antisymmetric tensor.

Finally, if in terms of 6-dimensional representations, we assume:
\begin{align}
\mathcal{D} \equiv (\partial + \overline{\partial}) + (\mathcal{A} + \overline{\mathcal{A} })
\end{align}

We have:
\begin{align}
\mathcal{A} = {\bf 70} \oplus {\bf 20} \oplus {\bf 20}
\end{align}

Let us summarise what we have achieved here. We have seen that we can embed the whole SM symmetry group and gravity seen as a gauge theory inside $SO(12)$. Within this group, its two spinor representations encapsulate both primed and unprimed fermions in a unified way, so that they have the right ``low-energy'' quantum numbers. Due to the equivalence between the spinor representations of $SO(2n)$ and forms in $n$-complex-dimensions, we have seen that there is a possibility to reinterpret the states of the SM taking seriously this equivalence. We therefore schematically constructed, using basic representation theory, a possible Lagrangian (first-order in this case) for fermions in 6-dimensions. A more detailed construction of the latter is an ongoing research project.

\section{Discussion}

In this chapter we have explored the different unification patterns that can be explored using (mainly) a second-order approach to fermionc Lagrangians. Indeed, the content of these theories differs from the usual first-order content as we do not need to take into account half of the particles (primed fermions). Furthermore, since in the second-order SM the weak symmetry is frozen, smaller unifying groups can be found as we have seen with the $SO(8)$ and $SO(9)$ examples. Nevertheless, it is always possible to consider different approximate symmetries, where left- and right-handed particles are taken as representation of the same weak group. In that sense, these are not GUTs, but rather (once the Lagrangian has been written down) approximate symmetries of the theory. Finally, we have seen that gravity can also be included in the discussion, once its description as a gauge theory has been established. In this case, it is interesting to see that we can construct higher-dimensional Lagrangians that might be able to reproduce the ``low-energy'' content of the SM. It seems that the two extra dimensions that correspond to the gravity degrees-of-freedom ought to be integrated-out (by means of compactification or any other suitable mechanism), thus recovering a four-dimensional spacetime. As mentioned earlier, this is ongoing research.

\chapter*{Conclusion}
\addcontentsline{toc}{chapter}{Conclusion}
\pagestyle{fancy}
%\addtolength{\headwidth}{\marginparsep} %these change header-rule width
%\addtolength{\headwidth}{\marginparwidth}
\lhead{ \bfseries \nouppercase{Conclusion}}
\chead{} 
\rhead{} 
\lfoot{\small\scshape PhD Thesis} 
\cfoot{\thepage} 
\rfoot{\footnotesize Johnny Espin} 
\renewcommand{\headrulewidth}{.3pt} 
\renewcommand{\footrulewidth}{.3pt}
\setlength\voffset{-0.25in}
\setlength\textheight{648pt}

The aim of this thesis was to demonstrate that a second-order formulation of fermionic field theories is indeed possible. In order to do so, we took a bottom-up approach to the construction of the formalism. We first developed the tools necessary for the free-field theory of second-order fermions in Chapters \ref{chapMaj} and \ref{chapDirac}. There we saw that the first-order field equations play a special role in this new formalism. Indeed, they are now seen as reality conditions: a constraint that should be imposed at the level of the mode-decomposition so as to kill half of the solutions of the second-order Klein-Gordon equation. As far as the free-theory is concerned, switching from a first- to a second-order formalism is merely a change of interpretation. The main differences appear when one considers interacting theories, as we saw in Chapters \ref{chapED} and \ref{chapSM}. 

In the second-order formalism, the complexity that arose in the first-order propagators and number of fields present in the theory is shifted to the interaction vertices. Nevertheless, due to the fact that the new Lagrangians that were obtained only contain unprimed two-component spinors, the theories appear to be more compact. Moreover, and more particularly in the case of the Standard Model, this led to an entire reformulation of both its bosonic and fermionic sector. There, we were led to combine the ${SU}(2)$ singlets into ``doublets'', as well as to define ${SU}(2)$-invariant combinations from the fermion doublets and the Higgs. Similarly, the angular part of the Higgs field was absorbed into the gauge fields to produce ${SU}(2)$-frozen massive gauge fields. This reformulation of the SM allows for a more direct description of its physical content.

%The obtained Lagrangian is quite compact, see (\ref{L-intr}) and can be used as the starting point for concrete computations of SM scattering amplitudes. As we already emphasized in the Introduction, it is likely to be worth converting the arising computation rules into a computer code, as the second order formalism leads to faster performance than the one based on the first order Lagrangian. 

Another interesting aspect of the new SM Lagrangian is that the Higgs field appears non-polynomially. The analogy between the Higgs field and the conformal factor of the metric was already emphasised in \cite{Faddeev:2008qc}. Indeed, consider a Weyl transformation $g_{\mu\nu} \to \rho^2 g_{\mu\nu}$ with, recall, $\rho^2$ the modulus squared of the Higgs field. Under such a conformal rescaling the spinor metric $\epsilon_{AB}$ transforms as $\epsilon_{AB}\to\rho \epsilon_{AB}, \epsilon^{AB}\to\rho^{-1}\epsilon^{AB}$. The Dirac operator changes as $D_{AA'} \chi_B \to D_{AA'}\chi_B - (\partial_{BA'} \log\rho) \chi_A$, see $e.g.$ \cite{Penrose:1987uia}, formula (5.6.15). Then, if we define the transformation rule for second-order spinors to be $S_A \to \rho^{-1} S_A$, similarly for $\bar{S}_A$, the quantity $\epsilon^{AB} D_{AA'} S_B$ transforms homogeneously $\epsilon^{AB} D_{AA'} S_B\to \rho^{-2} \epsilon^{AB} D_{AA'} S_B$, with the covariant derivative remaining unchanged. This implies that under such a transformation
\begin{align}
\sqrt{g} \,D\bar{S} DS \to \frac{1}{\rho} \sqrt{g} \, D\bar{S} DS,
\end{align}

where we have taken into account that $\sqrt{g}\to \rho^4 \sqrt{g}$, and there is an extra factor of $1/\rho$ coming from the contraction of the primed spinor indices. Similarly, 
\begin{align}
\sqrt{g}\, \bar{S} S \to \rho \sqrt{g}\, \bar{S} S.
\end{align}
Thus, we see that, as already observed in \cite{Faddeev:2008qc} for the bosonic sector, the Higgs field $\rho$ enters the fermionic Lagrangian (\ref{SML}) as the conformal factor of a transformation $g_{\mu\nu}\to\rho^2 g_{\mu\nu}$ (there the Higgs field was reabsorbed into the covariant derivative so as to give a canonical kinetic term). It would be interesting to understand the implications of a second-order formulation for fermions on the modifications of the Higgs effective potential in the context of frame-independence of General Relativity. This new formalism could possibly affect different Higgs-inflation scenarios that depend on the parametrisation of the potential \cite{Bezrukov:2009db,Bezrukov:2010jz,Bezrukov:2014ina}.

\bigskip

In the third part of this thesis, we developed the tools necessary to carry out perturbative calculations in the two simplest theories: Dirac and Majorana-Weyl Electrodynamics. As we have mentioned above, in the case of second-order fermions, the complexity of the first-order propagator is shifted to the interaction vertices. As a matter of fact, the former becomes a simple scalar-type propagator, while the latter now contain derivatives (cubic vertex). Furthermore, there is now a four-valent interaction, which although very simple (it takes the form of an identity operator over vector and spinor representations), is of the uttermost importance. Indeed, the effect of integrating-out half of the fermions (in our convention, primed spinors) is that these excitations are effectively set on-shell. In perturbation theory, the propagating internal primed fermions that are constrained to satisfy their first-order field equation ``resonate'' so as to contract their propagator into a quartic vertex. We saw explicitly how this happened in Chapter \ref{chaptertree} for the case of Compton scattering. Therefore, we see that including this quartic vertex ensures that the scattering amplitudes that are computed using the second-order formalism are equivalent to the results obtained using the usual first-order formulation. Nonetheless, we also saw that (at tree level), this new quartic vertex could be forgotten as long as a new set of rules to construct the scattering amplitudes was followed. More details on this procedure are described in Appendix \ref{ApxRules}. In Chapter \ref{chapterreno}, we have also developed a few textbook examples of loop calculations in order to emphasise that, even at loop level, the equivalence between the formalisms is maintained. There the quartic vertex plays the same role as at tree-level.

It must be noticed that perturbative calculations in the second-order formalism are much more economical than in the first-order framework. This is not only due to the fact that we are dealing with half the number of fields, but also to the fact that we are working with two-dimensional spinor irreps (as compared to the four component Dirac spinors). The algebra of gamma matrices has effectively been taken care of, so that all that is left to calculate is spinor contractions. In the case of two-component first-order spinor Lagrangians, the algebra computations are also simpler, however, there one has to deal with a huge amount of Feynman diagrams. In our formalism, the presence of extra diagrams containing quartic vertices is much less cumbersome.

\bigskip

Finally, in the last part of this work we covered some advanced field-theoretical aspects that are specific to the second-order formalism. A deeper and more specific discussion about each of these topics can be found in their respective chapters, we summarise here the main points. In Chapter \ref{chapterano}, we explored the calculation of the anomalies that could arise in this framework. The final result is equivalent to its first-order counterpart, however, both the perturbative and non-perturbative calculations have to be carefully conducted. 

In Chapter \ref{chapuni}, we showed that the second-order theory is unitary even though its Lagrangian is not Hermitian. Any other result would have been a hard blow to the formalism. Moreover, the results and methodology derived in this thesis may lead to new insights on the development of different theories described by complex fields upon which reality conditions need to be imposed. Such an example would be an $SU(2)$ connection description of gravity, for which no definitive answer has been found yet.

The last results we presented in Chapter \ref{sec:unific} can be tied to the ideas that were first introduced in Chapter \ref{chapSM}. Indeed, after a very compact Lagrangian for the fermionic section of the SM was obtained, it seemed natural to explore this direction further. This led to the results in the former chapter. There we showed how a different interpretation of the particle content of the SM could lead to new unification patterns. In addition to this, once gravity is included in the game, interesting new aspects coming from the symmetry groups appear. Indeed, spinor irreps of $SO(2n)$ groups admit a representation as forms in $n$ complex dimensions. With gravity being added to the picture, it seems natural to look at four (when available) of those dimensions as spacetime, while the rest can be thought of as compactified or ``frozen'' or ``UV'' dimensions. This idea is currently being investigated.

\bigskip

To conclude this thesis, we hope this work will motivate new research directions in the framework of second-order fermionic field theories. A few aspects were discussed here, while possible future lines of research were also mentioned. Nevertheless, we believe the topic to be very vast, and possibly tied to many other research areas such as Scattering Amplitudes, Twistors or $PT$ symmetric theories to just mention a few.

\part{Appendices}
\appendix
\pagestyle{fancy}
%\addtolength{\headwidth}{\marginparsep} %these change header-rule width
%\addtolength{\headwidth}{\marginparwidth}
\lhead{ \bfseries \nouppercase{\leftmark}}
\chead{} 
\rhead{} 
\lfoot{\small\scshape PhD Thesis} 
\cfoot{\thepage} 
\rfoot{\footnotesize Johnny Espin} 
\renewcommand{\headrulewidth}{.3pt} 
\renewcommand{\footrulewidth}{.3pt}
\setlength\voffset{-0.25in}
\setlength\textheight{648pt}
\chapter{Two-component spinors}\label{appendixA}

% WORK WITH INPUTS!!!!!!

\section{${\rm SO}(1,3)\sim {\rm SL}(2,\C)$}\label{MinkoSpinors}

The Lorentz group ${SO}(1,3)$ acts on Minkowski spacetime $M^{1,3}$, with signature convention $(-,+,+,+)$\footnote{and orientation $\epsilon^{0123}=+1$}, via:
$$ x^\mu \to \Lambda^\mu{}_\nu x^\nu, \qquad x^\mu=(t,x,y,z)\in M^{1,3}.$$
Let us form from the components of $x^\mu$ an Hermitian $2\times 2$ matrix\footnote{The normalisation is chosen for later convenience.}
\begin{align}\label{X}
X=\frac{1}{\sqrt{2}} \left(\begin{array}{cc} t-z & x+i y \\ x-i y & t+z \end{array}\right), \qquad X^\dagger=X
\end{align}

It is not hard to see that:
\begin{align}
2 ~{\rm det}(X) = t^2-x^2-y^2-z^2=-\eta_{\mu\nu}x^\mu x^\nu, \qquad \eta_{\mu\nu}={\rm diag}(-1,1,1,1)
\end{align}
Moreover, one can show that any Hermitian $2\times 2$ matrix is of this form for some $t,x,y,z$. Thus, Minkowski spacetime $M^{1,3}$ can be identified with the space of Hermitian $2\times 2$ matrices:
\begin{align}
 M^{1,3} \sim X\in {\rm Mat}(2\times 2): X^\dagger=X
 \end{align}

Consider now the group ${SL}(2,\C)$ of complex $2\times 2$ matrices with unit determinant:
\begin{align}{SL}(2,\C) \ni g=\left( \begin{array}{cc} a & b \\ c& d \end{array} \right), \qquad ad-bc=1
\end{align}

This group acts on the space of Hermitian matrices via:
\begin{align}\label{11-g}
X\to g X g^\dagger. 
\end{align}
It is clear that this action preserves ${\rm det}(X)$ (because ${\rm det}(g)=1$), and preserves the space of Hermitian $2\times 2$ matrices. Thus, this gives a norm-preserving action of ${SL}(2,\C)$ on Minkowski spacetime, and thus an embedding of ${SL}(2,\C)$ into ${SO}(1,3)$. Since the element $g=-1\in {SL}(2,\C)$ is sent to the identity in ${ SO}(1,3)$, this embedding can be seen to be a $2\to 1$ covering map.

\section{Spinors}

There are two inequivalent fundamental 2-dimensional representations of the group ${SL}(2,\C)$. Both are isomorphic as vector spaces, but with a different action of ${SL}(2,\C)$. Consider a column with entries being two arbitrary complex numbers:
\begin{align} \xi = \left( \begin{array}{c} \xi_1 \\ \xi_2 \end{array} \right)
\end{align}

Such a column is called a {\it spinor}. There is a natural action of ${SL}(2,\C)$ on $\xi$ given by:
\begin{align}\label{unprimed}
\xi \to g \xi \qquad g\in { SL}(2,\C)
\end{align}

Spinors on which the Lorentz group ${SL}(2,\C)$ acts as above are called {\it unprimed}. However, there is yet another natural action of ${SL}(2,\C)$ on the space of spinors. This is given by:
\begin{align}\label{primed}
\bar{\xi} \to g^* \bar{\xi} \qquad g\in {  SL}(2,\C)
\end{align}

where $g^*$ is the matrix consisting of complex conjugates of $a,b,c,d$. Spinors on which ${  SL}(2,\C)$ acts this way are called {\it primed}. In order to avoid confusion we shall label primed spinors by a symbol with a bar over it, which is why the spinor in the above formula is denoted by $\bar{\xi}$. Unprimed and primed spinors constitute two fundamental (inequivalent) representations of ${  SL}(2,\C)$. The corresponding representation spaces (spaces of unprimed and primed spinors) are denoted by:
\begin{align} {\rm unprimed} \quad \xi\in V^{(1/2,0)}, \qquad {\rm primed}\quad \bar{\xi}\in V^{(0,1/2)}
\end{align}

This notation has to do with the fact that a general (finite dimensional) representation of the Lorentz group ${  SL}(2,\C)$ is specified by two half-integers $j,j'$. The corresponding representation space is denoted by $V^{(j,j')}$. Thus, the representations realising the unprimed and primed spinors are the simplest possible ones.

We note that the above definitions of the action of ${  SL}(2,\C)$ on unprimed and primed spinors imply that the complex conjugate of an unprimed spinor is a primed one.

\subsection*{Index notation}

It turns out to be very convenient to introduce a certain spinor index notation. In this notation we represent the matrix $g\in {  SL}(2,\C)$ by $g_A{}^B$, where $A,B=1,2$ are the spinor indices. An unprimed spinor is denoted by $\xi_A$, and the action of  ${  SL}(2,\C)$ on $\xi$ is given by:
\begin{align} \xi_A \to g_A{}^B \xi_B
\end{align}

The index notation is developed in order to avoid thinking which of these two indices corresponds to rows and which to columns, and so this is left unspecified (even though is not hard to deduce from the fact that $\xi_A$ was a column). Similarly for the primed spinors, we introduce a new type of spinor indices $A',B'=1,2$, so that a primed spinor is denoted by $\bar{\xi}^{A'}$. The action of  ${  SL}(2,\C)$ on primed spinors is then:
\begin{align}\bar{\xi}^{A'} \to \bar{\xi}^{B'} (g^\dagger)_{B'}{}^{A'}
\end{align}

Here $g^\dagger=(g^T)^*$ is the Hermitian conjugate of $g$.

\subsection*{The spinor metric}

The determinant condition ${\rm det}(g)=1$ can be rewritten as:
\begin{align}
1= \frac{1}{2} \epsilon^{AB} g_A{}^C g_B{}^D \epsilon_{CD}
\end{align}

where $\epsilon^{AB}=-\epsilon^{BA}$ and similarly $\epsilon_{AB}=-\epsilon_{BA}$ are anti-symmetric tensors. This is easily shown to be equivalent to:
\begin{align}\epsilon^{AC} g_A{}^B g_C{}^D = \epsilon^{CD}
\end{align}

Thus, $\epsilon^{AB}$ is an ${  SL}(2,\C)$-invariant metric in $V^{(1/2,0)}$, and the following bilinear form in the space of unprimed spinors is  ${  SL}(2,\C)$-invariant:
\begin{align}(\xi,\lambda) \equiv (\xi\lambda) :=- \epsilon^{AB} \xi_A \lambda_B
\end{align}

The minus sign is introduced here for future convenience (we will later develop index free notations where it will disappear). Note that this bilinear form is anti-symmetric, and so the norm squared of any spinor is zero:
\begin{align}(\xi,\xi)=0
\end{align}

Similarly, one introduces an  ${  SL}(2,\C)$-invariant metric $\epsilon_{A'B'}$ in the space of primed spinors, as well as an  ${  SL}(2,\C)$-invariant bilinear pairing:
\begin{align}
[\bar{\xi},\bar{\lambda}] \equiv [\bar{\xi}\bar{\lambda}]:= \bar{\xi}^{A'} \bar{\lambda}^{B'}\epsilon_{A'B'}
\end{align}

Note that we denoted the primed inner product by square brackets, while the unprimed spinor product was denoted by the usual round brackets. This will be convenient below.

\subsection*{Raising and lowering of spinor indices}

Using the invariant metrics in $V^{(1/2,0)},V^{(0,1/2)}$ one can define an operation of raising and lowering of indices. Thus, to raise an index of an unprimed spinor we define:
\begin{align}
\xi^A:= \epsilon^{AB}\xi_B
\end{align}

To lower a spinor index we need the inverse of the metric $\epsilon^{AB}$, which we define via:
\begin{align}\epsilon_{AB}: \epsilon^{AB} \epsilon_{AC} = \delta_C{}^B
\end{align}

where $\delta_A{}^B$ is the Kronecker delta. We would like the spinor $\xi^A$ obtained above with its index lowered to be the original spinor $\xi_A$, which is achieved via the following definition:
\begin{align}\xi_A:= \xi^B\epsilon_{BA}
\end{align}

Note that in these formulas the upper index to the left is contracted with a lower index to the right, which is the rule how these formulas can be memorised. 

One defines similar raising and lowering of primed indices via $\epsilon_{A'B'}$ and its inverse $\epsilon^{A'B'}$ defined via:
\begin{align}\epsilon^{A'B'}: \epsilon^{A'B'}\epsilon_{A'C'} = \delta_{C'}{}^{B'}
\end{align}

Thus, we have:
\begin{align}\bar{\xi}_{A'}:=\bar{\xi}^{B'}\epsilon_{B'A'}, \qquad \bar{\xi}^{A'} := \epsilon^{A'B'} \bar{\xi}_{B'}
\end{align}

Note that the operation of raising-lowering a pair of spinor indices is now not innocuous:
\begin{align} \xi^A \lambda_A = - \xi_A \lambda^A
\end{align}

Now that we understand how the spinor indices can be raised and lowered, we can write down a formula that summarises the effect of the complex conjugation on an unprimed spinor. Indeed, from (\ref{primed}) we see that the complex conjugate of an unprimed spinor transforms under ${  SL}(2,\C)$ as a primed one. We write:
\begin{align}(\xi_A)^* = (\xi^*)_{A'}
\end{align}

Thus, the rule is that under the complex conjugation the unprimed spinor index gets replaced by a primed one, whose symbol is the original symbol with a sign of complex conjugation added.

\subsection*{The soldering form}

If we write the Hermitian $2\times 2$ matrix $X$ in the spinor index notation as $X_A{}^{A'}$, it can be written as a linear combination of $2\times 2$ matrices times the components of the 4-vector $x^\mu$. This defines the following object:
\begin{align}X_A{}^{A'}= \theta_{\mu A}{}^{A'} x^\mu
\end{align}

The object $\theta_{\mu A}{}^{A'}$ is called {\it the soldering form}. It provides an isomorphism between the space of Hermitian $2\times 2$ matrices $X_A{}^{A'}$ and Minkowski spacetime whose elements are $x^\mu$. Note that the matrix $X_A{}^{A'}$ transforms (\ref{11-g}) as a vector in the irreducible representation space $V^{(1/2,1/2)}$. Thus, the usual 4-vectors $x^\mu$ are seen to form an irreducible representation of the Lorentz group more complicated than the spinor representations. You need two spinors (unprimed and a primed one) to get a vector!

Note that the soldering form is Hermitian. This property is best written for the soldering form with both its indices raised:
\begin{align}(\theta_{\mu}^{AA'})^* = \theta_\mu^{AA'}
\end{align}

Now, computing the determinant of $X_A{}^{A'}$ we get:
\begin{align}{\rm det}(X)=  \frac{1}{2} \theta_{\mu A}{}^{A'} \theta_{\nu B}{}^{B'} \epsilon^{AB} \epsilon_{A'B'} x^\mu x^\nu
\end{align}

On the other hand, the same determinant is equal to half of $-\eta_{\mu\nu} x^\mu x^\nu$. Thus, we see that the following relation between the Minkowski metric and the soldering form holds:
\begin{align}\label{metric}
\eta_{\mu\nu} = - \theta_{\mu A}{}^{A'} \theta_{\nu B}{}^{B'} \epsilon^{AB} \epsilon_{A'B'}
\end{align}

\subsection*{The spinor basis}

It is very convenient to introduce in each space $V^{(1/2,0)},~V^{(0,1/2)}$ a certain spinor basis. Since each space is (complex) 2-dimensional we need two basis vectors for each space. Let us denote these by:
\begin{align}o_A, \iota_A \in V^{(1/2,0)}, \qquad o^{A'}, \iota^{A'}\in V^{(0,1/2)}
\end{align}

Note that we shall assume that the basis in the space of primed spinors is the complex conjugate of the basis in the space $V^{(1/2,0)}$:
\begin{align}\iota^{A'}=(\iota^A)^*, \qquad o^{A'}=(o^A)^*
\end{align}

The basis vectors are pronounced as ``omicron'' and ``iota''. Since the norm of every spinor is zero, we cannot demand that each of the basis vectors is normalised. However, we can demand that the product between the two basis vectors in each space is unity. Thus, the basis vectors satisfy the following normalisation:
\begin{align}\iota^A o_A = 1, \qquad \iota^{A'} o_{A'}=1
\end{align}

Of course, a spinor basis in each space $V^{(1/2,0)},~V^{(0,1/2)}$ is only defined up to an ${  SL}(2,\C)$ rotation. Any ${  SL}(2,\C)$ rotated basis gives an equally good basis, and it can be seen that any two bases can be related by a (unique) ${  SL}(2,\C)$ rotation. 

Once a spinor basis is introduced, we have the following expansion of the $\epsilon_{AB}$ symbol:\begin{align}
\epsilon_{AB}=o_A \iota_B -\iota_A o_B
\end{align}

A similar formula is also valid for $\epsilon_{A'B'}$.

\subsection*{The soldering form in the spinor basis}

The following explicit expression for the soldering form $\theta_{\mu A}{}^{A'}$ in terms of the basis one-forms $t_\mu,x_\mu,y_\mu,z_\mu$ and the spinor basis vectors $o^A, o^{A'}, \iota^A, \iota^{A'}$ can be obtained:
\begin{align}\begin{split}
\theta_\mu^{AA'} = \frac{t_\mu}{\sqrt{2}} ( o^A o^{A'} + \iota^A \iota^{A'}) &+ \frac{z_\mu}{\sqrt{2}}  ( o^A o^{A'} - \iota^A \iota^{A'}) \\& + \frac{x_\mu}{\sqrt{2}}  ( o^A \iota^{A'} + \iota^A o^{A'})+\frac{i y_\mu}{\sqrt{2}}  ( o^A \iota^{A'} - \iota^A o^{A'}) \label{theta-xyz}
\end{split}\end{align}

Here we have given a formula for the soldering form with its both spinor indices raised. This expression encodes the same information as in the formula (\ref{X}) for $X_A{}^{A'}$ in terms of a matrix. However, one now never needs to think about what corresponds to column and what to a row, and can manipulate with the spinor objects in a completely algorithmic (algebraic) fashion, which is convenient. Note that the above expression is explicitly Hermitian.

\subsection*{A doubly null tetrad}

Collecting the components in front of equal spinor combinations in the above formula for the soldering form we can rewrite it as:
\begin{align}\label{theta}
\theta_\mu^{AA'} =l_\mu o^A o^{A'}+ n_\mu \iota^A \iota^{A'}+ m_\mu o^A \iota^{A'} + \bar{m}_\mu \iota^A o^{A'}
\end{align}

where
\begin{align} l_\mu = \frac{t_\mu+z_\mu}{\sqrt{2}}, \qquad n_\mu = \frac{t_\mu-z_\mu}{\sqrt{2}}, \qquad m_\mu = \frac{x_\mu+i y_\mu}{\sqrt{2}}, \qquad \bar{m}_\mu = \frac{x_\mu-i y_\mu}{\sqrt{2}}
\end{align}

Note that $l,~n$ are real one-forms, while $\bar{m}_\mu = m_\mu^*$. The above collection of one-forms is known as a {\it doubly null tetrad}. Indeed, it is easy to see that all four one-forms introduced above are null, $e.g.$ $l^\mu l_\mu=0$. The only non-zero products are:
\begin{align}
l^\mu n_\mu = -1, \qquad m^\mu \bar{m}_\mu=1
\end{align}

Thus, the Minkowski metric can be written in terms of a doubly null tetrad as:
\begin{align} \eta_{\mu\nu}= -l_\mu n_\nu - n_\mu l_\nu + m_\mu \bar{m}_\nu + \bar{m}_\mu m_\nu
\end{align}

which can also be verified directly by substituting (\ref{theta}) into the formula (\ref{metric}) for the metric.

\subsection*{Grassmann spinors}

We have seen that the norm of any spinor is zero. This is due to the fact that the spinor metric $\epsilon^{AB}$ is anti-symmetric and so:
\begin{align} (\xi,\xi) = - \epsilon^{AB} \xi_A \xi_B = \xi_2 \xi_1 - \xi_1 \xi_2 = 0
\end{align}

However, this is so if the spinor components commute: $\xi_1\xi_2 = \xi_2\xi_1$, which is the case for ordinary spinors. Let us now introduce a new type of spinors, whose components {\it anti-commute}:
\begin{align}\label{anti-comm} 
\xi_1 \xi_2 = -\xi_2 \xi_1
\end{align}

Numbers that anti-commute are known from algebra, and form the so-called Grassmann algebra. In the case of two-component spinors we introduce a Grassmann algebra generated by two anti-commuting generators $\xi_1, \xi_2$. These are required to anti-commute with each other (\ref{anti-comm}), but also with themselves:
\begin{align}\xi_1 \xi_1 = - \xi_1 \xi_1
\end{align}

and similarly for $\xi_2$. This, in particular implies that each generator $\xi_1,\xi_2$ is {\it nilpotent}:
\begin{align}\xi_1 \xi_1 = 0, \qquad \xi_2\xi_2 = 0
\end{align}

Going back to spinor index notations, we define a Grassmann-valued (unprimed) spinor $\xi_A$ so that it satisfies:
\begin{align}\xi_A \xi_B = -\xi_B \xi_A
\end{align}

i.e., anti-commutes with itself. For such a spinor its norm (squared):
\begin{align}
(\xi,\xi)=\xi^A \xi_A =2\xi_2 \xi_1 \not= 0
\end{align}

If we have a collection of Grassmann-valued spinors $\xi_A, \lambda_A, \ldots$ these anti-commute with themselves and between each other:
\begin{align}
\xi_A \lambda_B = -\lambda_B \xi_A
\end{align}

Primed Grassmann spinors are defined analogously. They anti-commute with themselves, with other primed spinors, as well as with unprimed spinors. 

Finally, let us define the action of the Hermitian conjugation on the Grassmann-valued spinors (we cannot talk about complex conjugation anymore, as Grassmann-valued spinors are not numbers; we need Hermitian conjugation instead). We have:
\begin{align} (\xi_A \lambda_B)^\dagger = (\lambda_B)^\dagger (\xi_A)^\dagger = (\lambda^\dagger)_{B'} (\xi^\dagger)_{A'}
\end{align}

Thus, the Hermitian conjugation acts on spinors as on operators, in that the Hermitian conjugates of all operators are taken in the opposite order, as one is used to in quantum mechanics.

\subsection*{Index-free notations}

As is usual in the 2-component spinor literature, we shall sometimes use an index-free notation:
\begin{align}
\lambda^A \xi_A := \lambda \xi\equiv (\lambda\xi), \qquad (\lambda^\dagger)_{A'} (\xi^\dagger)^{A'} = \lambda^\dagger \xi^\dagger\equiv[\lambda^\dagger\xi^\dagger]
\end{align}

Thus, if no indices are shown in a fermionic contraction, this means that the {\it natural} contraction is used, i.e. unprimed spinors are contracted as in $\lambda^A \xi_A$ and primed spinors are contracted in an opposite way as $(\lambda^\dagger)_{A'} (\xi^\dagger)^{A'}$. This is a natural convention, for we have:
\begin{align}
(\lambda\xi)^\dagger = \xi^\dagger \lambda^\dagger
\end{align}

Sometimes, when more than two spinors are present in an index-free formula, it is necessary to put brackets around spinors to make it clear which pairs are contracted. Then we use round brackets to denote contractions of unprimed spinors and square brackets for contractions of primed spinors.

\subsection*{Self-dual two-forms}

The following self-dual two-forms play a very important role in the second-order formulation of fermions. They are defined as:
\begin{align}
\Sigma^{AB} = \frac{1}{2} \theta^{A}{}_{A'}\wedge \theta^{BA'}\label{sd2f}
\end{align}

Explicitly, in terms of the null tetrad and the spinor basis we get:
\begin{align}\label{Sigma}
\Sigma^{AB} = l\wedge m \, o^A o^B + \bar{m}\wedge n \, i^A i^B + (l\wedge n - m\wedge \bar{m}) i^{(A} o^{B)}
\end{align}
 
This formula can be used to derive all the necessary identities involving the self-dual two forms.

\subsection*{${\rm SU}(2)$ spinors}\label{su2spinors}

We will need ${  SU}(2)$ spinors when we consider the Hamiltonian formulation of any of our fermionic theories. Our conventions here are reminiscent of those in Appendix A of  \cite{Ashtekar:1991hf}, but there are some differences. In particular, we use a Hermitian tetrad, while the convention in  \cite{Ashtekar:1991hf} is that the tetrad is anti-Hermitian. 

Let us first consider ordinary, non-Grassmann-valued spinors. To define ${  SU}(2)$ spinors we need a Hermitian positive-definite form on spinors. This is a rank 2 mixed spinor $G_{A'A}$: $\bar{G}_{A'A}=G_{A'A}$, such that for any spinor $\lambda^A$ we have $\bar{\lambda}^{A'} \lambda^A G_{A'A}>0$. Here $\bar{\lambda}^{A'}$ is the complex conjugate of $\lambda^A$. We can define the ${  SU}(2)$ transformations to be those ${  SL}(2,\C)$ ones that preserve the form $G_{A'A}$. Then $G_{A'A}$ defines an anti-linear operation $\star$ on spinors via:
\begin{align}
(\lambda^\star)^A:=G^{AA'}\bar{\lambda}_{A'}
\end{align}

We require that the anti-symmetric rank 2 spinor $\epsilon_{AB}$ is preserved by the $\star$-operation:
\begin{align}
(\epsilon^\star)^{AB}=\epsilon^{AB}
\end{align}

which implies the following normalisation condition:
\begin{align}\label{G-norm}
G_{AA'} G^{A'}{}_{B}=\epsilon_{AB}
\end{align}

Using the normalisation condition we find that $(\lambda^{\star\star})^A=-\lambda^A$ or:
\begin{align}
\star^2=-1
\end{align}

Thus, the $\star$-operation so defined is similar to a ``complex structure'', except for the fact that it is anti-linear:
\begin{align}
(\alpha \lambda^A +\beta \eta^A)^\star= \bar{\alpha} (\lambda^\star)^A+\bar{\beta} (\eta^\star)^A
\end{align}

We note that using the $\star$-operation we can rewrite the positive-definite quantity $\bar{\lambda}^{A'} \lambda^A G_{A'A}$ as follows:
\begin{align}
\bar{\lambda}^{A'} \lambda^A G_{A'A} = \lambda_A (\lambda^\star)^A  >0
\end{align}

Now for the purpose of 3+1 decompositions to be carried out below, we need to introduce a special Hermitian form that arises once a time vector field is chosen. We can then consider the zeroth component of the soldering form:
\begin{align}
\theta_{0}^{AA'}\equiv \theta_\mu^{AA'} \left( \frac{\partial}{\partial t}\right)^\mu =\frac{1}{\sqrt{2}} \left( o^A o^{A'} + \iota^A \iota^{A'}\right)
\end{align}

It is Hermitian, and so we can use a multiple of $\theta_0^{AA'}$ as $G^{AA'}$. It remains to satisfy the normalisation condition (\ref{G-norm}). This is achieved by:
\begin{align}\label{herm-form}
G^{AA'} := \sqrt{2} \theta_0^{AA'}
\end{align}

We then define the spatial soldering form via:
\begin{align}\label{sigma-sp}
\sigma^{i\, AB}:= G^{AA'} \theta^{i\, B}{}_{A'}
\end{align}

which is automatically symmetric $\sigma^{i\,AB}=\sigma^{i\,(AB)}$ because its anti-symmetric part is proportional to the product of the time vector with a spatial vector, which is zero. Explicitly, in terms of the spinor basis introduced above we have:
\begin{align}\label{sigma-small}
\sigma^{i\,AB}= -m^i o_A o_B + \bar{m}^i \iota_A \iota_B+\frac{z^i}{\sqrt{2}}(\iota_A o_B+o_A\iota_B )
\end{align}

The action of the $\star$-operation on the basis spinors is as follows:
\begin{align}
(o^\star)^A =  \iota^A, \qquad (\iota^\star)^A =  -o^A
\end{align}

It is then easy to see from (\ref{sigma-small}) that the spatial soldering form so defined is anti-Hermitian with respect to the $\star$ operation:
\begin{align}
(\sigma^{i\,\star})^{AB}=-\sigma^{i\,AB}
\end{align}

The following property of the product of two spatial soldering forms holds:
\begin{align}\label{sigma-ident}
\sigma^i_A{}^B\sigma^j_B{}^C=\frac{1}{2}\delta^{ij} \epsilon_A{}^C-\frac{i}{\sqrt{2}}\epsilon^{ijk} \sigma^k_A{}^C.
\end{align}

Below we will also often use the following related quantities :
\begin{align}\label{T}
T^i_A{}^B:=i \sqrt{2} \sigma^i_A{}^B
\end{align}

which have the following nicer algebra:
\begin{align}\label{T-algebra}
T^i_A{}^B T^j_B{}^C=-\delta^{ij} \epsilon_A{}^C+ \epsilon^{ijk} \, T^k_A{}^C
\end{align}

Now, using the Hermitian form (\ref{herm-form}), we extend the $\star$-operation defined above to Grassmann-valued spinors. Thus, we define a new operation on Grassmann-valued spinors which is a combination of the usual Hermitian conjugation $\dagger$ acting on a Grassmann-valued fermion with the operation of converting the primed index into an unprimed one:
\begin{align}
(\lambda^\star)^A := G^{AA'} (\lambda^\dagger)_{A'}\label{starconjug}
\end{align}

This operation is of importance when we discuss the 3+1 decomposition of the standard Weyl and Dirac actions.

%\chapter{Soldering-Form and Fierz Identities}
%
%\section{Fierz identities}
%
%\section{Identities in D dimensions}

\chapter{Pauli and gamma matrices}

\section{Soldering-form v. Pauli and gamma matrices}

It would be useful for the unacquainted reader to have a dictionary between commonly used quantities such as the Pauli and gamma matrices, and the soldering for which is extensively used throughout this thesis. In order to do so, we recall briefly some properties of the former. Dirac gamma matrices satisfy the Clifford algebra:
\begin{align}
\{ \gamma^\mu, \gamma^\nu \}= -2\eta^{\mu\nu}
\end{align}

so that $(\gamma^0)^2=1,~(\gamma^i)^2=-1$. We can then choose extra hermicity constraints on the matrices such that these still satisfy the algebra. They are:
\begin{align} \left( \gamma^\mu\right)^\dagger=\gamma^0 \gamma^\mu \gamma^0
\end{align}

The matrix $\gamma_5$ is defined:
\begin{align}
\gamma_5 = i\gamma^0\gamma^1\gamma^2\gamma^3,\quad \left( \gamma_5\right)^\dagger=\gamma_5
\end{align}

and finally the generators of the Lorentz group:
\begin{align}
S^{\mu\nu} = \frac{i}{4}[ \gamma^\mu, \gamma^\nu ]
\end{align}

The Pauli matrices are generators of the spin $1/2$ representation of $SU(2)$ and are defined by their algebra:
\begin{align}
\sigma^i\sigma^j = \delta^{ij} + i\epsilon^{ijk}\sigma^k
\end{align}

These are related to the gamma matrices through their extension to four dimensions:
\begin{align}
\sigma^\mu =(1, +\sigma^i), \quad \bar{\sigma}^\mu = \epsilon^T (\sigma^\mu)^* \epsilon = (1,-\sigma^i)
\end{align}

They satisfy:
\begin{align}
\sigma^\mu\bar{\sigma}^\nu + \sigma^\nu\bar{\sigma}^\mu = -2\eta^{\mu\nu},\quad \bar\sigma^\mu {\sigma}^\nu + \bar\sigma^\nu{\sigma}^\mu = -2\eta^{\mu\nu}
\end{align}

which leads to:
\begin{align}
\gamma^\mu = \left(\begin{array}{cc}0 & \sigma^\mu \\ \bar\sigma^\mu &0  \end{array}\right)
\end{align}
This is the Weyl or chiral representation of the gamma matrices which allows us to decompose Dirac spinors into two irreducible spinors. The Lorentz generators in each representation are then given by:
\begin{align}
\sigma^{\mu\nu} = \frac{i}{2} \sigma^{[\mu} \bar\sigma^{\nu ]},\quad \bar\sigma^{\mu\nu} = \frac{i}{2} \bar\sigma^{[\mu} \sigma^{\nu ]}
\end{align}

Finally, these quantities can be connected to the soldering form and self-dual two-forms. We introduce an index notation for the sigma matrices:
\begin{align}
\sigma^\mu \defeq (\sigma^\mu )_{AA'}, \quad \bar\sigma^\mu \defeq (\bar\sigma^\mu )^{A'A}
\end{align}

We have then:
\begin{align}
 (\bar\sigma^\mu )^{A'A} = \epsilon^{B'A'}(\sigma^\mu)^*_{B'B}\epsilon^{BA}
\end{align}

And their algebra can written as:
\begin{align}
(\sigma^\mu\bar{\sigma}^\nu)_{A}{}^{B} = -\eta^{\mu\nu}\delta_{A}{}^{B} - 2i (\sigma^{\mu\nu} )_{A}{}^{B},\quad (\bar\sigma^\mu{\sigma}^\nu)^{A'}{}_{B'} = -\eta^{\mu\nu}\delta^{A'}{}_{B'} - 2i (\bar\sigma^{\mu\nu} )^{A'}{}_{B'}
\end{align}

It is now possible to identify:
\begin{spacing}{1.5}
\begin{align}
\begin{array}{rclrcl}
(\bar\sigma^\mu )^{A'A} &\equiv & \sqrt{2}\theta^{\mu\,A'A}, & (\sigma^\mu )_{AA'} &\equiv & \sqrt{2}\theta^\mu _{AA'} \\
- i (\sigma^{\mu\nu} )_{A}{}^{B} &\equiv &\Sigma^{\mu\nu} {}_{A}{}^{B}   ,&- i (\bar\sigma^{\mu\nu} )^{A'}{}_{B'} &\equiv & \bar\Sigma^{\mu\nu A'}{}_{B'} \\
\end{array}
\end{align}
\end{spacing}

So that:
\begin{align}
\gamma^\mu =\sqrt{2}\left(\begin{array}{cc}0 & \theta^\mu _{AA'}  \\   \theta^{\mu\,A'A}  &0  \end{array}\right)
\end{align}

\section{Gamma matrices algebra}

When dealing with gamma matrices, we work within the mainly plus signs signature, and within a dimensional regularisation framework (whenever allowed). We therefore have spacetime indices denoted by greek letters than run from 0 to $D-1$, $e.g.$: $\mu,\nu=0,\ldots,D-1$. On the other hand, the internal spinor space can be taken as four dimensional. We then have:
\begin{align}
\left\{ \gamma^\mu, \gamma^\nu \right\} = -2 \eta^{\mu\nu} \mathbb{I}_4, \quad \gamma^\mu \gamma_\mu = -D\mathbb{I}_4
\end{align}

We then have the following identities that we need in our calculations:
\begin{align}
\gamma^\mu \gamma^\alpha \gamma^\beta \gamma_\mu &= 4 \eta^{\alpha\beta} \mathbb{I}_4 - (D-4) \gamma^\alpha \gamma^\beta \\
{\rm Tr}\left(\gamma^\alpha\gamma^\mu  \gamma^\beta \gamma_\mu \right)&= -4(D-2) \eta^{\alpha\beta} \\
{\rm Tr} \left(\gamma^\alpha\gamma^\mu  \gamma^\beta \gamma^\nu \right)&=  \left( -4\eta^{\mu\nu}\eta^{\alpha\beta}+4\eta^{\mu\alpha}\eta^{\nu\beta} +4\eta^{\nu\alpha}\eta^{\mu\beta}\right)\\
{\rm Tr} \left(\gamma^5\gamma^\mu\gamma^\nu  \gamma^\alpha\gamma^\beta \right)&=  4i\epsilon^{\mu\nu\alpha\beta}
\end{align}

where $S_{(\mu\nu)}$ is a symmetric tensor. For our purposes, we will not need any other identity involving the gamma matrices.

\chapter{Euclidean space}

\section{Four-dimensional Euclidean space}
In order to work with a path-integral formulation, we need to regularise the integration by switching to an Euclidean signature. Therefore, we continue analytically all our quantities into Euclidean space such that $x_4= ix^0,~\partial_4= -i\partial_0,~\gamma_4= i\gamma_0$ and $A_4= iA^0$. In terms of gamma matrices, we now have:
\begin{align}
\{ \gamma^\mu, \gamma^\nu \}= -2\delta^{\mu\nu}, \quad \left( \gamma^\mu\right)^\dagger=-\gamma^\mu , \quad \gamma_5 = -\gamma^1\gamma^2\gamma^3\gamma^4,\quad \left( \gamma_5\right)^\dagger=\gamma_5
\end{align}

The only calculation that will be carried out using Euclidean space and gamma matrices will be the anomaly in Section \ref{anodiraceuclid}. More details about the formalism can be found there. In this appendix we will focus on developing the Euclidean formalism for two-components spinors.

\section{Euclidean space and two-component spinors}\label{appendix_euclidean}

%Before going further, we need to define the Euclidean version of the theory and hence a set of self-adjoint operators which eigenmodes will be used in the regularisation of the path integral.

We repeat the main steps of Section \ref{MinkoSpinors}, this time using an Euclidean signature for our space. Let us shortly recall that we can form from the (complex) components of $x^\mu$ a $2\times 2$ matrix:
\begin{align}\label{Xc2}
X=\frac{1}{\sqrt{2}} \left(\begin{array}{cc} t-z & x+i y \\ x-i y & t+z \end{array}\right)
\end{align}

We earlier saw that for a real $x^\mu$, the matrix is Hermitian $X^\dagger= X$ and that any Hermitian $2\times 2$ matrix is of this form for some $t,x,y,z$. We also saw that we have:
\begin{align}
2 ~{\rm det}(X) = t^2-x^2-y^2-z^2=-\eta_{\mu\nu}x^\mu x^\nu, \qquad \eta_{\mu\nu}={\rm diag}(-1,1,1,1)
\end{align}
Thus, Minkowski spacetime $M^{1,3}$ can be identified with the space of Hermitian $2\times 2$ matrices:
\begin{align}
 M^{1,3} \sim X\in {\rm Mat}(2\times 2): X^\dagger=X
 \end{align}

We know define the Euclidean version of the above formulas. In order to do so, we must first define an Euclidean spinor conjugation. Above we had the usual Hermitian conjugation that sent a spinor to its complex conjugate:
\begin{align}
w^A = \alpha\iota^A + \beta o^A \mapsto
 \bar{w}^{A'} = \bar\alpha\iota^{A'} + \bar\beta o^{A'} 
\end{align}

and that defines an isomorphism $SO(1,3) \sim SL(2,\mathbb{C})/\mathbb{Z}_2$. In this case, we want the isomorphism to be $SO(4) \sim SU(2)\times SU(2)/\mathbb{Z}_2$, so that our conjugation should not mix the two $SU(2)$ spinors subspaces and be invariant over the latter. Therefore, we seek two maps
\begin{align}
w^A = \alpha\iota^A + \beta o^A \mapsto \hat{w}^A, \qquad \hat{w}^A w_A = \alpha\bar\alpha+\beta\bar\beta \\
z^{A'} = \gamma\iota^{A'} + \delta o^{A'} \mapsto \hat{z}^{A'}, \qquad \hat z_{A'}{z}^{A'} = \gamma\bar\gamma+\delta\bar\delta
\end{align}

So that the two $SU(2)$ inner products are left invariant. Using $\iota^A o_A = 1 = \iota^{A'} o _{A'}$, we obtain:
\begin{align}
w^A = \alpha\iota^A + \beta o^A \mapsto \hat{w}^A= \bar\beta \iota^A + \bar\alpha(- o^A) \\
z^{A'} = \gamma\iota^{A'} + \delta o^{A'} \mapsto \hat{z}^{A'}= \bar\delta(-\iota^{A'})+ \bar\gamma o^{A'}
\end{align}

We see that our new conjugation is also antilinear but does not interchange primed and unprimed indices. Also, it is easy to see that:
\begin{align}
\hat{\hat{w}}^A= - w^A , \quad \hat{\hat{z}}^{A'}= - z^{A'} 
\end{align}

Finally, one can show that (see below) imposing $\hat{X} = X$ is equivalent to requiring that $x^\mu = (it, x, y, z)$ should be real. Calculating
\begin{align}
X^A{}_{A'}\hat{X}^{A'}{}_A = (it)^2+x^2+y^2+z^2=\delta_{\mu\nu}x^\mu x^\nu, \qquad \delta_{\mu\nu}={\rm diag}(1,1,1,1)
\end{align}

shows that Euclidean space $\mathbb{R}^4$ can be identified with the space of $2\times 2$ matrices that are self-adjoint under the above conjugation:
\begin{align}
\mathbb{R}^4 \sim X\in {\rm Mat}(2\times 2): \hat X=X
 \end{align}

We can make explicit the isomorphism by considering the group ${  SU}(2)$ of unitary transformations:
\begin{align}{  SU}(2) \ni g=\left( \begin{array}{cc} \alpha & -\bar\beta \\ \beta& \bar\alpha \end{array} \right), \qquad |\alpha|^2+|\beta|^2=1, \quad \alpha,\beta \in \mathbb{C}
\end{align}

This group acts on the space of self-adjoint matrices via:
\begin{align}
X\to L X R
\end{align}

where $L$ and $R$ act on two different $SU(2)$ representations (left and right handed). It is clear that this action preserves ${\rm det}(X)$ (because ${\rm det}(g)=1,~ g= L,R$), and preserves the space of self-adjoint $2\times 2$ matrices by construction. Thus, this gives a norm-preserving action of $SU(2)\times SU(2)$ on Euclidean spacetime, and thus an embedding of $SU(2)\times SU(2)$ into ${  SO}(4)$. Since flipping the sign in both rotations does not change the way the group acts, this embedding can be seen to be a $2\to 1$ covering map. As before, we write
\begin{align}X_A{}^{A'}= \theta_{\mu A}{}^{A'} x^\mu
\end{align}

The Euclidean version can be obtained from a Wick rotation of the Minkowski version by noting $\tau = x_4 = ix^0 = it, ~\tau \in \mathbb{R}$ and $\theta_4 = -i\theta_0$. This allows us to rewrite, as in the Minkowski case, the soldering form in terms of a doubly null tetrad:
\begin{align}\label{thetaE2}
\theta_\mu^{AA'} =\bar{n}_\mu o^A o^{A'}- n_\mu \iota^A \iota^{A'}+ m_\mu o^A \iota^{A'} + \bar{m}_\mu \iota^A o^{A'}
\end{align}

where
\begin{align} \bar{n}_\mu = \frac{-i\tau_\mu+z_\mu}{\sqrt{2}}, \qquad n_\mu = \frac{i\tau_\mu+z_\mu}{\sqrt{2}}, \qquad m_\mu = \frac{x_\mu+i y_\mu}{\sqrt{2}}, \qquad \bar{m}_\mu = \frac{x_\mu-i y_\mu}{\sqrt{2}}
\end{align}

Note that now $\bar{n}= n^*$ as $\bar{m}= m^*$. The only non-zero products are:
\begin{align}
\bar{n}^\mu n_\mu = 1, \qquad m^\mu \bar{m}_\mu=1
\end{align}

Thus, the Euclidean metric can be written in terms of a doubly null tetrad as:
\begin{align} \delta_{\mu\nu}= \bar n_\mu n_\nu + n_\mu \bar n_\nu + m_\mu \bar{m}_\nu + \bar{m}_\mu m_\nu,
\end{align}

Using the formula for the soldering form (\ref{thetaE2}) as well as:
\begin{align}
\hat{\iota}^A= -o^A ,\quad \hat{\iota}^{A'}= o^{A'}
\end{align}

and the antilinearity of the conjugation, it is now straightforward to verify $\hat X = X$. Furthermore, this can trivially be extended to the statement that for any real vector $v^\mu$, we have $\hat{V} = V$. The reader might have noticed that this conjugation is related to the $SU(2)$ conjugation that we encountered (\ref{starconjug}). Indeed, recall that
\begin{align}
G^{AA'} = o^Ao^{A'} + \iota^A\iota^{A'}
\end{align}

The conjugation (\ref{starconjug}) that makes use of this matrix precisely acts on the spinor basis as this ``newly'' introduced Euclidean conjugation. This is no coincidence as we are dealing, in Euclidean space, with $SU(2)$ spinors.

%\chapter{Grassmann integrals}

\chapter{Feynman rules: from first- to second-order formalism}\label{ApxRules}

In this appendix, we make explicit the equivalence of the two formalisms at the perturbative level using examples that have already been carried out in the main body of this work. We therefore remind the reader of the different Lagrangians and Feynman rules with which we are dealing and repeat a few simple calculations, this time in both formalisms, while explaining their link.

\section{Dirac fermions}
\subsection*{Dirac Lagrangian}
\subsubsection{Second-order}
Recall that the Lagrangian for the fermionic sector of second-order Quantum Electrodynamics is given by:
\begin{align}
{\cal L} =-{2} D_{A'}{}^{ A} \chi_A  D^{A'B} \xi_B - m^2 \chi^A\xi_A
\end{align}
with 
\begin{align}
D_\mu \xi = (\partial_\mu - i e A_\mu)\xi, \qquad
 D_\mu \chi = (\partial_\mu + i eA_\mu)\chi,
\end{align}
where we included the electromagnetic coupling $|e|\ll 1$. The Lagrangian can be expanded so that:
\begin{align}
{\cal L}= \mathcal{L}_0 + \mathcal{L}_{int}
\end{align}

with
\begin{align}
 \mathcal{L}_0 = -\partial^\mu \chi^A \partial_\mu \xi_A - m^2 \chi^A\xi_A, 
\end{align}
and
\begin{align}
 \mathcal{L}_{int} = 2ieA^{AA'}\left( \chi_A (\partial_{A'}{}^{B}\xi_B) +  (\partial_{A'}{}^{B}\chi_B ) \xi_A \right) - e^2 A^B{}_{B'}A^{B'}{}_{B}\chi^A\xi_A
\end{align}

\subsubsection{First-order}
The Lagrangian can be written in terms of two-component spinor fields, and reads:
\begin{align}\label{Dirac-L3}
{\cal L}_{\rm Dirac} = -i \sqrt{2} \chi^\dagger_{A'} D^{A'A} \chi_A -i \sqrt{2} \xi^\dagger_{A'} D^{A'A} \xi_A - m(\chi^A \xi_A + \chi^\dagger_{A'} \xi^{\dagger\,A'}).
\end{align}

with 
\begin{align}
 \mathcal{L}_{int} = \sqrt{2}eA^{AA'}\left(\chi^\dagger_{A'} \chi_A  -  \xi^\dagger_{A'} \xi_A \right) 
\end{align}

\subsection*{Propagator and Feynman Rules}

\subsubsection{Second-order}
To extract the propagator for the spinor fields, let us rewrite the free part of their Lagrangian as:
\begin{align}
i \mathcal{L}_{Dirac} = \chi_A\left[ i \epsilon^{AB} \left( -\square +m^2\right)\right]\xi_B
\end{align}
Then the inverse of the quadratic operator is:
\begin{align}
\langle 0| T\{\xi_A(p)\chi_B(-p)\}|0 \rangle \equiv D(p)_{AB} = \frac{-i}{p^2+m^2}\epsilon_{AB}
\end{align}
where, the field $\xi_A$ sits at the end of the directed line. The Feynman rules for the vertices are:
\begin{align}
\langle 0| A^{A'}{}_{A}(q) \chi_B(p)\xi_C(k)|0 \rangle &\rightarrow  2i e \left[k_C{}^{A'}\epsilon_{BA} + p_B{}^{A'}\epsilon_{CA} \right] \\
\langle 0| A^{A'}{}_{A}(q_1) A^{B'}{}_{B}(q_2) \chi_C(p)\xi_D(k)|0 \rangle &\rightarrow  -2i e^2 \epsilon^{A'B'}\epsilon_{AB}\epsilon_{CD}
\end{align}
%\begin{figure}[H]\begin{center}
%\input{cubic.pdf_tex} \quad \input{quartic.pdf_tex}\caption{Cubic and quartic vertices for the QED$_2$ interactions}
%\end{center}\end{figure}

\subsubsection{First-order}
In this case we have four propagators:
\begin{align}
\langle 0| T\{\xi_A(p)\xi^\dagger_{A'}(-p)\}|0 \rangle &\equiv -i\sqrt{2} \frac{p_{A'A}}{p^2+m^2}\\
\langle 0| T\{\chi_A(p)\chi^\dagger_{A'}(-p)\}|0 \rangle &\equiv -i\sqrt{2} \frac{p_{A'A}}{p^2+m^2} \\
\langle 0| T\{\xi^\dagger_{A'}(p)\chi^\dagger_{B'}(-p)\}|0 \rangle &\equiv -im \frac{\epsilon_{A'B'}}{p^2+m^2} \\ 
\langle 0| T\{\xi_A(p)\chi_B(-p)\}|0 \rangle &\equiv -im \frac{\epsilon_{AB}}{p^2+m^2} 
\end{align}

We follow \cite{Dreiner:2008tw} for Feynman diagrams' conventions: an outgoing arrow denotes a primed fermion whereas an incoming arrow denotes an unprimed one. Moreover, the rule for the spinor contractions is that we climb up the charge arrows. The sign in the momentum propagators denotes a momentum flow antiparallel to the contraction arrow (Fig.\ref{mmprop}).

\begin{figure}[H]  \centering\begin{center}
$\vcenter{\hbox{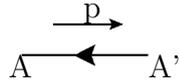}}$
    	\caption{Momentum propagator diagram}\label{mmprop}
\end{center}\end{figure}

In the case in which there is one or several mass propagators (two outgoing or two incoming arrows on the same propagator), we follow the direction dictated by the rest of the graph ($e.g.$ previous external fermion or momentum propagator). Notice that in the case where we have a mass propagator in between two momentum propagators, once we climb up the charge arrow of the first momentum propagator, the second one will be climbed down. This does not affect the calculation of the diagram since we are working with an explicit index notation and the ``charge arrow rule'' simply allows us to set signs consistently among diagrams. The interaction vertices with the current are given by:
\begin{align}\begin{split}
\langle 0| A^{A'}{}_{A}(q) \chi^{\dagger\,B'}(p)\chi^C(k)|0 \rangle &\rightarrow-i\sqrt{2}e~\epsilon^{A'B'}\delta_{A}{}^{C}\\ \langle 0| A^{A'}{}_{A}(q) \xi^{\dagger\,B'}(p)\xi^C(k)|0 \rangle &\rightarrow+i\sqrt{2}e~\epsilon^{A'B'}\delta_{A}{}^{C}
\end{split}\end{align}

\subsection*{Tree level and the quartic vertex}

We reconsider Compton-scattering, as we said, the first tree-level process in which the new quartic vertex comes into play. In Chapter \ref{chaptertree}, we saw that both formalisms led to the same result. This was most easily seen after we performed a simple trick that we recall here for this example. Define the reduced channel amplitudes:
\begin{align}
\mathfrak{M}_{s_i} \defeq (s_i+m^2)\mathcal{M}_{s_i}
\end{align}
where $m$ is the mass of the fermion in the channel. Then, the amputated amplitude for a two-fermions-two-photons process with momenta $k_i$ is:
\begin{align}\begin{split}
\mathcal{M}(s_1,s_2, \{k_i\})&= \frac{\mathfrak{M}_{s_1}(s_1,\{k_i\})}{(s_1+m^2)}+\frac{\mathfrak{M}_{s_2}(s_2,\{k_i\})}{(s_2+m^2)}+\mathcal{V}_4\\ &=  \frac{\mathfrak{M}_{s_1}(s_1=-m^2,\{k_i\})}{(s_1+m^2)}+\frac{\mathfrak{M}_{s_2}(s_2=-m^2,\{k_i\})}{(s_2+m^2)}\end{split} \label{rule1A}
\end{align}
where $\mathcal{V}_4$ is the quartic vertex. We will see that this trick is enough to show the equivalence of amplitudes at tree-level. In the following we will see how this rule generalises for loops involving the four-vertex.

\subsection*{Loops equivalence}
In Chapter \ref{chapterreno}, we computed the photon two-point function in the second-order formalism. We now compute the same quantity in the first-order two-component Dirac formalism. In that case, there are four diagrams: two in which both propagators incorporate the momentum contribution (the vertices are denoted by:$~\xi^\dagger\xi\rightarrow\xi\xi^\dagger,~\chi\chi^\dagger\rightarrow\chi^\dagger\chi$) and two in which both are mass insertions ($\xi^\dagger\xi\rightarrow\chi^\dagger\chi,~\chi^\dagger\chi\rightarrow\xi^\dagger\xi$). This immediately leads to:
\begin{align}\begin{split}
i\Pi^{(1)}&(k)^{A'}{}_{A}{}^{B'}{}_{B} \\&= (-1)4e^2 \int\frac{d^D p}{(2\pi)^D}\frac{\left[ p^{A'}{}_{B}(p+k)^{B'}{}_{A} + (p+k)^{A'}{}_{B}p^{B'}{}_{A} +m^2 \epsilon^{A'B'}\epsilon_{AB}  \right]}{\left[p^2+m^2 \right]\left[(p+k)^2+m^2 \right]}
\end{split}
\end{align}
%\begin{align}\begin{split}
%i\Pi^{\rm 1-loop}(k)^{A'}{}_{A}{}^{B'}{}_{B} &= (-1)4e^2 \int\frac{d^D p}{(2\pi)^D}\frac{\left[2 p^{A'}{}_{B}(p+k)^{B'}{}_{A} +m^2 \epsilon^{A'B'}\epsilon_{AB}  \right]}{\left[p^2+m^2 \right]\left[(p+k)^2+m^2 \right]}
%\end{split}
%\end{align}
%
%The first term in the brackets can be split into two and one part shifted with a reverted loop momentum to obtain:
This is the same amplitude as in the second-order formalism. Notice though, that we needed shift freedom obtained through dimensional regularisation in order to match them.

\bigskip

All in all, we come to the conclusion that the two formalisms can be matched into each other. Furthermore, it is worth stating that, in the second-order formalism, one could consider only diagrams containing cubic vertices with an additional rule (at tree level): wherever there is a contraction in the numerator leading to a propagating momentum squared, the latter is set on-shell. This is understood in the sense that the quartic vertex encodes the information that used to be carried by the primed fermions. Those propagators containing mass insertions of the primed spinors have been set on-shell and contracted to form quartic vertices. Hence, when we consider only unprimed propagating degrees of freedom in the cubic vertices, we obtain the correct amplitude, up to some resonances in the virtual particles\footnote{Recall that the reality conditions are $\xi^\dagger \sim \partial \xi $, so that an on-shell primed spinor in a mass insertion propagator will lead to a resonance $\partial^2\rightarrow p^2$.} that have to be accounted for. Below we will see how this happens in the case of non-trivial loop diagrams.

\section{Majorana-Weyl theory}
The photon two-point function calculation develops in the same way as for the Dirac fermion. The first non-trivial result arises when one considers the triangle anomaly diagrams. This calculation will mimic what was done in Chapter \ref{chapterano}.

\subsection*{First-order Lagrangian}
We start with a massive Majorana spinor coupled to an external vector field (we forget about gauge symmetry for a bit). Indeed, our aim is to link the first-order calculation to the second-order one and to expose the behaviour of the diagrams containing a quartic vertex. The Lagrangian is given by:
\begin{align}
{\cal L}_{\rm Maj} = -i \sqrt{2} \lambda^\dagger_{A'} D^{A'A} \lambda_A  - \frac{m}{2}(\lambda^A \lambda_A + \lambda^\dagger_{A'} \lambda^{\dagger\,A'}).
\end{align}

with
\begin{align}
 D^{A'A} \lambda_A  = \left(\partial - ieA  \right)^{A'A}\lambda_A
\end{align}

\subsection*{First-order perturbative calculation}
The anomaly can be computed in perturbation theory by means of Feynman diagrams. Indeed, one shows that the divergence of the current has a non-zero matrix element to create two photons, where the amplitude which is considered is given by:
\begin{align}
 \langle k_1,k_2 | j^{ A'}_{A}(x)|0\rangle &=  \epsilon^{*\, B}_{B'}(k_1)\epsilon^{*\, C}_{C'}(k_2)\mathcal{M}^{ABC}_{A'B'C'}(k_1,k_2)\nonumber\\
\langle k_1,k_2 | p\cdot j(x)|0\rangle &\neq 0
\end{align}

We briefly translate the two-component anomaly calculation of \cite{Dreiner:2008tw} into our notation. The Feynman rules are as follows:
\begin{align}
\langle 0| \lambda^\dagger_{A'}(p)\lambda_A(-p)|0 \rangle & \equiv\frac{-i\sqrt{2}p_{A'A}}{p^2+m^2}\\
\langle 0| T\{\lambda^\dagger_{A'}(p)\lambda^\dagger_{B'}(-p)\}|0 \rangle & \equiv -i \frac{\epsilon_{A'B'}}{p^2+m^2}\\
\langle 0| T\{\lambda_A(p)\lambda_B(-p)\}|0 \rangle & \equiv -i \frac{\epsilon_{AB}}{p^2+m^2} 
\end{align}
\begin{align}
\langle 0| A^{A'}{}_{A}(q)\lambda^{\dagger\, B'}(p)\lambda^C(k)|0 \rangle &\rightarrow i\sqrt{2}e~\epsilon^{A'B'}\delta_{A}{}^{C}
\end{align}
We have one momentum propagator and two mass-insertion propagators as well as one cubic vertex and the rule for contracting the indices are as above (climbing up the arrows and consistently contracting any mass insertion).

Taking into account the two orientations for the triangle diagrams, and denoting the incoming photons by spinor indices $(AA'),~(BB'),~(CC')$ and their massless momenta $k_1,~k_2,~k_3$:
\begin{align}\begin{split}
i\mathcal{M}^{(1)}(k_2,k_3)&= 8e^3\int\frac{d^4q}{(2\pi)^4}\frac{1}{(q-k_2)^2+m^2}\frac{1}{q^2+m^2}\frac{1}{(q+k_3)^2+m^2}
\\& \times\Bigg[\left(q^{B'}{}_{C}(q-k_2)^{A'}{}_{B}(q+k_3)^{C'}{}_{A}- q^{C'}{}_{B}(q-k_2)^{B'}{}_{A}(q+k_3)^{A'}{}_{C}\right) \\
&-\frac{m^2}{2}\epsilon^{B'C'}\left( \epsilon_{AB}(q+k_3)^{A'}{}_{C} +\epsilon_{AC}(q-k_2)^{A'}{}_{B}\right) \\
&-\frac{m^2}{2}\epsilon^{A'C'}\left(\epsilon_{AB}q^{B'}{}_{C}+\epsilon_{CB}(q-k_2)^{B'}{}_{A}\right)\\
&-\frac{m^2}{2}\epsilon^{A'B'}\left(-\epsilon_{AC}q^{C'}{}_{B}-\epsilon_{BC}(q+k_3)^{C'}{}_{A}\right)\Bigg] \label{amplitudeweyl1stm}
\end{split}\end{align}

This amplitude leads to the usual anomalous conservation of the current in the massless limit. Note that dimensional regularisation is not used here as it leads to some subtleties in the definition of the integral\footnote{In other words, $\gamma_5$ has to be carefully defined if dimensional regularisation were to be used.}. Before continuing the analysis of the amplitude, we will first derive the same amplitude in the second-order formalism.

\subsection*{Second-order Lagrangian}

We will now carry out the calculation for a Majorana fermion in a second-order formalism coupled to an external vector field. The Lagrangian in this case is given by:
\begin{align}
{\cal L} = - D_{A'}{}^{ A} \lambda_A  D^{A'B} \lambda_B - \frac{m^2}{2} \lambda^A\lambda_A\label{lagweyl2A}.
\end{align}
This should be supplemented with the reality conditions:
\begin{align}
m\lambda^{\dagger\, A'} = -i\sqrt{2}D^{ A'A}\lambda_A \label{realityweyl2}
\end{align}
The field equations that result from the above Lagrangian are
\begin{align}
{2} D_{A'}{}^{ A} D^{A'B} \lambda_B + m^2 \lambda^A=0
\end{align}
We see that the Lagrangian is not invariant under the usual $U(1)$ transformations
\begin{align}
 \delta\lambda= +ie\alpha \lambda
\end{align}
However, the field equations and the reality condition are in the massless limit (where the vector field can be considered as a gauge field). Furthermore, the current given by:
\begin{align}
j_{A}{}^{A'}=ie\left( \lambda_A  D^{A'B} \lambda_B -  D^{A'B} \lambda_B\lambda_A \right) = 2ie \lambda_A  D^{A'B} \lambda_B
\end{align}

is conserved on-shell in that limit. We will show that the amplitude for the triangle diagrams is equivalent in both formalisms up to some boundary terms that are fixed in the calculation of the anomaly (Chapter \ref{chapterano}, Appendix \ref{apxexplicit}).

\subsection*{Perturbative calculation in the second-order formalism}

The calculation is identical to the one carried out in Chapter \ref{chapterano}, we therefore refer the reader to the latter for further details. Let us simply recall:
\begin{align}\begin{split}
i\mathcal{M}(k_2,k_3)= 8e^3\int&\frac{d^4q}{(2\pi)^4}\Bigg[ \frac{\mathcal{I} + \mathcal{J} }{D(-k_2)D(k_3)D(0)  } + \frac{\mathcal{A}}{D(-k_2)D(k_3)} \\& + \frac{\mathcal{B}}{D(-k_2)D(0)}+\frac{\mathcal{C}}{D(0)D(k_3)}\Bigg] \label{amplitudeweylA}
\end{split}\end{align}

 with now  $D(k) = (q+k)^2+m^2$ and where the contribution from $\mathcal{I},~\mathcal{J},~\mathcal{A},~\mathcal{B},$ and $\mathcal{C}$ do not depend on the mass and are the same as before:
\begin{align}\begin{split}
\mathcal{I}= q^{B'}{}_{C}(q-k_2)^{A'}{}_{B}(q+k_3)^{C'}{}_{A}- q^{C'}{}_{B}(q-k_2)^{B'}{}_{A}(q+k_3)^{A'}{}_{C}
\end{split}\end{align}

These are the terms that, in the massless limit lead to the anomalous conservation of the current in a theory of one Weyl fermion. They correspond to the $m^2\rightarrow 0 $ limit of $i\mathcal{M}^{(1)}$ in (\ref{amplitudeweyl1stm}). In the second-order case, the triangle diagram yields an extra contribution:
\begin{align}\begin{split}
\mathcal{J}=&\frac{1}{2}q^2\epsilon^{B'C'}\left(\epsilon_{AB}(q+k_3)^{A'}{}_{C} +\epsilon_{AC}(q-k_2)^{A'}{}_{B}\right) \\
&+\frac{1}{2}(q+k_3)^2\epsilon^{A'C'}\left(\epsilon_{AB}q^{B'}{}_{C}+\epsilon_{CB}(q-k_2)^{B'}{}_{A}\right)\\
&+\frac{1}{2}(q-k_2)^2\epsilon^{A'B'}\left(-\epsilon_{AC}q^{C'}{}_{B}-\epsilon_{BC}(q+k_3)^{C'}{}_{A}\right)\end{split}
\end{align}

These terms arise from the contractions of momenta by propagators as it was the case when we computed the photon two-point function. They are expected to cancel out with terms arising from the quartic vertex:
\begin{align}
\mathcal{A}&= \frac{1}{4}\epsilon^{B'C'}\epsilon_{BC}(k_2+k_3)^{A'}{}_{A} \\
\mathcal{B}&=-\frac{1}{4} \epsilon^{A'C'}\epsilon_{AC}{k_2}^{B'}{}_{B} \\
\mathcal{C}&=-\frac{1}{4}\epsilon^{A'B'}\epsilon_{AB}{k_3}^{C'}{}_{C}
\end{align}

The difference between the massless and the massive case appears now. If before, the propagators could be cancelled simply by terms of the type $\sim q^2$, we now need an extra contribution from the mass squared. Therefore, we can combine these four terms in the following way: in the $\mathcal{J}$ term, one can add and substract an $m^2$ term. Then, we will have numerators such as $q^2+m^2$ that cancel one propagator and are added to $\mathcal{A},~\mathcal{B}$ or $\mathcal{C}$, and there will remain three terms proportional to $m^2$. The latter are written as:
\begin{align}\begin{split}
\mathcal{J}(p_i^2=-m^2) = &-\frac{1}{2}m^2\epsilon^{B'C'}\left(\epsilon_{AB}(q+k_3)^{A'}{}_{C} +\epsilon_{AC}(q-k_2)^{A'}{}_{B}\right) \\
&-\frac{1}{2}m^2\epsilon^{A'C'}\left(\epsilon_{AB}q^{B'}{}_{C}+\epsilon_{CB}(q-k_2)^{B'}{}_{A}\right)\\
&-\frac{1}{2}m^2\epsilon^{A'B'}\left(-\epsilon_{AC}q^{C'}{}_{B}-\epsilon_{BC}(q+k_3)^{C'}{}_{A}\right)\end{split}
\end{align}
where $p_i$ denotes the momentum flowing the propagators. These terms are in one-to-one correspondence with the terms in (\ref{amplitudeweyl1stm}). We are finally left with the quartic-vertex contributions with the additional terms coming from $\mathcal{J}$. We have:
\begin{align}
\mathcal{\tilde A}&= \frac{1}{2}\epsilon^{B'C'}\left( \frac{1}{2}\epsilon_{BC}(k_2+k_3)^{A'}{}_{A} +\epsilon_{AB}(q+k_3)^{A'}{}_{C} +\epsilon_{AC}(q-k_2)^{A'}{}_{B}\right)\\
\mathcal{\tilde B}&=\frac{1}{2} \epsilon^{A'C'}\left(-\frac{1}{2}\epsilon_{AC}{k_2}^{B'}{}_{B}+\epsilon_{AB}q^{B'}{}_{C}+\epsilon_{CB}(q-k_2)^{B'}{}_{A}\right) \\
\mathcal{\tilde C}&=\frac{1}{2}\epsilon^{A'B'}\left(-\frac{1}{2}\epsilon_{AB}{k_3}^{C'}{}_{C}-\epsilon_{AC}q^{C'}{}_{B}-\epsilon_{BC}(q+k_3)^{C'}{}_{A}\right)
\end{align}

For the second-order amplitude to be equal to the first-order one, these three quantities should vanish as is the case in the usual massless calculation (Appendix \ref{apxexplicit}), and therefore the constraint that has to be imposed is the same. Recall that we are not allowed to use dimensional regularisation: this implies that the terms proportional to the loop momentum can not be freely shifted as they diverge linearly. However, if we were able to use shift invariance to rewrite them, $e.g.$ for $\epsilon_{AB}(q+k_3)^{A'}{}_{C} +\epsilon_{AC}(q-k_2)^{A'}{}_{B}$, we would obtain:
\begin{align}\begin{split}
 & \frac{1}{2}\left(\epsilon_{AB}(q+k_3)^{A'}{}_{C} +\epsilon_{AC}(q-k_2)^{A'}{}_{B}\right) + \frac{1}{2}\left(\epsilon_{AB}(q+k_3)^{A'}{}_{C} +\epsilon_{AC}(q-k_2)^{A'}{}_{B}\right) \\ =& \frac{1}{2}\left(\epsilon_{AB}(q+k_3)^{A'}{}_{C} +\epsilon_{AC}(q-k_2)^{A'}{}_{B}\right) + \frac{1}{2}\left(\epsilon_{AB}(-q+k_2)^{A'}{}_{C} -\epsilon_{AC}(q+k_3)^{A'}{}_{B}\right)\\
=& -\frac{1}{2}\left( (k_3+k_2)^{A'}{}_C\epsilon_{AB}+ (k_2+k_3)^{A'}{}_{B}\epsilon_{AC}\right) \\
=& -\frac{1}{2}(k_2+k_3)^{A'}{}_A\epsilon_{BC}
\end{split}\end{align}

where in the second line we have shifted $q\rightarrow q+k_2-k_3$ and then $q\rightarrow -q$ so that their denominators coincide. All in all, each expression into brackets would individually cancel and the second-order amplitude would be equal to its first-order counterpart. The problem here arises due to the lack of dimensional regularisation that we used before to match the calculations. However, it was shown in Chapter \ref{chapterano} that there always exists a shift in the loop momentum in the second-order amplitude that matches another shift in the first-order case so that the physical content of both is equivalent\footnote{This is true when the fermion is coupled to a background field that satisfies transversality conditions.}. We could therefore make the following assumption: if, from the beginning, we only considered the triangle diagrams with no quartic vertices and set on-shell the contracted momenta in the numerators thereby extending (\ref{rule1A}) to loop diagrams, we would have obtained (without worrying about shift freedom and lack of dimensional regularisation schemes) the sought amplitude. This assumption can be considered as an extra Feynman rule of the second-order formalism. Its proof is beyond the scope of this thesis, we shall however see that when dimensional regularisation is allowed, it holds.

\section{Transition from first- to second-order diagrams}\label{apxtransition}
\subsection*{Index-free Feynman rules}
We recall the Feynman rules for two Dirac two-component spinors, however, we rewrite them in a more convenient way (as far as the index structure is concerned):
\begin{align}
\langle 0| T\{\chi_A(-p)\chi^{\dagger \, A'}(p)\}|0 \rangle &\equiv -i\sqrt{2} \frac{p_A{}^{A'}}{p^2+m^2}\\
 \langle 0| T\{\xi^A(-p)\xi^\dagger_{A'}(p)\}|0 \rangle &\equiv -i\sqrt{2} \frac{p^A{}_{A'}}{p^2+m^2} \\
\langle 0| T\{\xi^\dagger_{A'}(-p)\chi^{\dagger\,B'}(p)\}|0 \rangle &\equiv -im\frac{\delta_{A'}{}^{B'}}{p^2+m^2}\\
\langle 0| T\{\xi^A(-p)\chi_B(p)\}|0 \rangle &\equiv -im \frac{\delta^{A}{}_{B}}{p^2+m^2} 
\end{align}

where the momentum flows from the primed to the unprimed spinor (we will call this the positive direction flow). And the interaction vertices with the current as:
\begin{align}
\langle 0| A^\mu(q) \chi^{\dagger}_{B'}(p)\chi^C(k)|0 \rangle &\rightarrow+i\sqrt{2}e~\left(\theta^\mu\right)_{B'}^C\\
 \langle 0| A^\mu(q) \xi^{\dagger\,B'}(p)\xi_C(k)|0 \rangle &\rightarrow -i\sqrt{2}e~\left(\theta^\mu\right)^{B'}_C
\end{align}
We can conveniently rewrite all of these propagators in an index free notation as:
\begin{align}
\frac{-i(\sqrt{2}p+m)}{p^2+m^2}
\end{align}

with the momentum flowing in the same direction as the contraction arrows and $p \equiv p_\mu \theta^\mu$. Notice that these arrows correspond to the charge flow for one of the Dirac fermions and is opposite to it for the other ($\chi$ has the same charge as $\xi^\dagger$). Moreover, when the propagator is written in this way, the reader must remember that it is simply a convenient rewriting. Indeed, in the first-order two-component formalism, we have either a mass or a momentum insertion in the propagator, not both. The above notation is merely a compact and useful way of gathering many terms in one amplitude (see below). The vertices are given by:
\begin{align}
\pm ie \sqrt{2}\theta^\mu
\end{align}
where the sign depends on whether it is a $\chi$ or $\xi$ vertex. This two-component index free notation is so far ambiguous because it encapsulates too much information. Indeed, we à priori do not know with which spinor we are dealing. We therefore need ``extra'' Feynman rules. These will fall into two categories: open strings of fermions, $i.e.$, one incoming fermion going into $n$ photons and one outgoing fermion at tree level, and closed fermion loops.

\subsection*{String of Dirac fermions}

We will start the conversion of the first-order Feynman rules for the case of a string of fermions. Let us start by listing our rules:
\begin{enumerate}
\item We are interested in un-amputated amplitudes with one incoming and one outgoing unprimed spinor, and $n$ photons (that can be taken to be amputated). Hence, there will be two types of fermionic propagators connected to the amplitude:
\begin{align}
\frac{+i\sqrt{2}p}{p^2+m^2}
\end{align} 

or
\begin{align}
\frac{-im}{p^2+m^2}
\end{align} 

Depending on whether the amputated first-order amplitude starts with a primed or unprimed spinor.

\item Each internal fermionic line has also either type of propagator. We therefore consider, as previously stated, a compact notation\footnote{The change of sign in this abstract propagator is due to the momentum flowing from unprimed to primed spinor in the first external propagator.}
\begin{align}
S(p) = \frac{i(\sqrt{2}p-m)}{p^2+m^2}= D(p)(-\sqrt{2}p+m)
\end{align}

\item In order to fix the sign in the vertices, we consider the incoming unprimed spinor to be $\chi$ and the outgoing unprimed spinor to be $\xi$ (this fixes the charge flow). Therefore, we consider the default vertex (for an amplitude with no mass insertions, see below) to be 
\begin{align}
+ie\sqrt{2}\theta^\mu
\end{align}

and after a mass insertion, it becomes (until the next swap):
\begin{align}
-ie\sqrt{2}\theta^\mu
\end{align}
%
%at each $m$ insertion (operation that consists in swapping $\chi\leftrightarrow\xi$). For example, if the incoming un-amputated external propagator is an $m$ insertion, the first vertex will be a $\xi A^\mu\xi^\dagger$ vertex and given by $+ie\sqrt{2}\theta^\mu$.

\item Because an $m$ insertion swaps the spinor fields and therefore inverts the ``positive'' contraction direction\footnote{In the following $\xi \xi^\dagger$ propagators the momentum will flow from $\xi$ to $\xi^\dagger$, which is defined as the negative direction flow.}, the sign in the momentum propagators and in the vertices changes after an odd number of $m$ insertions until there is another $m$ insertion (if any) so that the number becomes even.
% } Not correct: integration by parts and anticommutation in the lagrangian yields the same sign

\item Because we are only interested in unprimed un-amputated amplitudes (these contain all the necessary information), a string of propagators and vertices can only contain an even number of $\theta$'s. The terms with an odd number of soldering forms can be neglected once the conversion to second-order amplitudes is finished as they correspond to primed to unprimed amplitudes. However, we will keep them in order to prove general conversation rules since, $e.g.$, the extra terms in a $n$ photons process, will contribute to the $n+1$ photons process.

\end{enumerate}

Now that we fixed our rules and conventions, we will go through a few examples and then give a general formula for the conversion. \\
\medskip \\
First, consider the simplest case consisting of a single vertex, with incoming momentum $p$ and outgoing $p_1 = p+k_1$. We have:
\begin{align}\begin{split}
\mathcal{M}(p, p_1) &= D_1(-\sqrt{2}p_1+m) ( ie\sqrt{2}\theta^\mu)(-\sqrt{2}p+m)D \\
&= D_1(-\sqrt{2}p_1+m) \left[(-2ei)\theta^\mu p - ie\sqrt{2} \theta^\mu m \right]D \\
&= D_1\left[(-\sqrt{2}p_1+m) (-2ei)\theta^\mu p - (+\sqrt{2}p_1+m) ie\sqrt{2} \theta^\mu m \right]D\\
&= D_1\left[(-2ei)(m\theta^\mu p +p_1\theta^\mu m)- \sqrt{2}((-2ie)p_1 \theta^\mu p + ie m\theta^\mu m) \right]D\\
&= D_1\left[m V^\mu_3- \sqrt{2}((-2ie)p_1 \theta^\mu p + ie m\theta^\mu m) \right]D
\end{split}\end{align}

%Not quite right
%\begin{align}\begin{split}
%\mathcal{M}(p, p_1) &= D_1(\sqrt{2}p_1+m) (\mp ie\sqrt{2}\theta^\mu)(\sqrt{2}p+m)D \\
%&= D_1(\sqrt{2}p_1+m) \left[(-2ei)\theta^\mu p + ie\sqrt{2} \theta^\mu m \right]D \\
%&= D_1\left[(\sqrt{2}p_1+m) (-2ei)\theta^\mu p + (-\sqrt{2}p_1+m) ie\sqrt{2} \theta^\mu m \right]D\\
%&= D_1\left[(-2ei)(m\theta^\mu p +p_1\theta^\mu m)+ \sqrt{2}((-2ie)p_1 \theta^\mu p + ie m\theta^\mu m) \right]D\\
%&= D_1\left[m V^\mu_3+ \sqrt{2}((-2ie)p_1 \theta^\mu p + ie m\theta^\mu m) \right]D
%\end{split}\end{align}

Notice, in the second line, the change in sign in the vertex due to the mass insertion. In the last line, the first term corresponds to the second-order cubic vertex with incoming $p$ and outgoing $p_1$, and the remaining terms map an unprimed spinor into a primed one and would be discarded if we were only interest in that amplitude.
\\
\medskip \\
Next we consider the string with two photons labelled by $k_1,~\mu$ and $k_2,~\nu$. There are here two diagrams to consider: in order to compute the amplitude, we need to consider the symmetrisation of the photons' external legs. We will try to extract a conversion formula from this amplitude instead of simply discarding the extra terms as we have done before.
\begin{align}\begin{split}
\mathcal{M}(p, p_1,p_2) &= S_2( ie\sqrt{2}\theta^\nu)  D_1\left[m V^\mu_3- \sqrt{2}((-2ie)p_1 \theta^\mu p + ie m\theta^\mu m)  \right]D  \\ 
&= S_2  D_1\left[( ie\sqrt{2}\theta^\nu)(m V^\mu_3)+ (-2ie)^2\theta^\nu p_1 \theta^\mu p \right. \\
&\hspace*{3cm} \quad \left.- (ie\sqrt{2}\theta^\nu)\sqrt{2} ie m\theta^\mu m \right]D \\ 
&= S_2  D_1\left[( ie\sqrt{2}\theta^\nu)(m V^\mu_3)+ (-2ie)^2(\theta^\nu p_1)( \theta^\mu p )\right. \\
&\hspace*{3cm} \quad \left.+ 2e^2\theta^\nu (m^2+p_1^2)\theta^\mu - 4e^2 (\theta^\nu p_1 )(p_1\theta^\mu) \right]D \\ 
&= S_2  \left[( ie\sqrt{2}\theta^\nu)D_1(-\sqrt{2}p_1+m)V^\mu_3+ (-2e^2i)\theta^\nu\theta^\mu  \right]D \\ 
\end{split}\end{align}

with $p_2 = p_1+k_2$ and we have made use of $p \equiv p_\mu\theta^\mu$ and $pp =p_\mu\theta^\mu \theta^\nu p_\nu = p^2/2$. Once we symmetrise the amplitude we obtain:
\begin{align}\begin{split}
\mathcal{M}^s(p, p_1,p_2) &= S_2  \left[( ie\sqrt{2}\theta^\nu)S_1 V^\mu_3+ (-2e^2i)\theta^\nu\theta^\mu  \right]D  ~~+(k_1,\mu\leftrightarrow k_2,\nu) \\ 
&= S_2  \left[( ie\sqrt{2}\theta^\nu)D_1(-\sqrt{2}p_1+m)V^\mu_3 ~~+(k_1,\mu\leftrightarrow k_2,\nu)\right]D 
\\
&\hspace*{6cm} \quad + S_2(-2e^2i)\eta^{\nu\mu}  D \\ 
&= S_2  \left[( ie\sqrt{2}\theta^\nu)S_1 V^\mu_3 ~~+(k_1,\mu\leftrightarrow k_2,\nu)\right]D 
+ S_2V_4^{\nu\mu} D \label{2g}
\end{split}\end{align}

This allows us to extract a conversion formula:
\begin{align}
\boxed{ (ie\sqrt{2}\theta^{\alpha_2})S_1  (ie\sqrt{2}\theta^{\alpha_1})S = ( ie\sqrt{2}\theta^{\alpha_2})S_1 V^{\alpha_1}_3 D + \bar{V}_4^{{\alpha_2}{\alpha_1}}D}\label{unsym}
\end{align}

where $\bar{V}_4^{{\alpha_2}{\alpha_1}}= (-2e^2i)\theta^{\alpha_2}\theta^{\alpha_1}$ is the unsymmetrised quartic vertex. Notice that in (\ref{unsym}) there is only one photon momentum ($k_1$) and therefore, to obtain the exact formula for two external photons containing the quartic vertex we need to consider three vertices (the third one being added $ad~hoc$):
\begin{align}
( ie\sqrt{2}\theta^{\alpha_3})S_2( ie\sqrt{2}\theta^{\alpha_2})S_1  ( ie\sqrt{2}\theta^{\alpha_1})S , 
\end{align}

apply (\ref{unsym}) twice, and then symmetrise over the ``physical'' $(k_1,\alpha_1\leftrightarrow k_2,\alpha_2)$. We then obtain:
\begin{align}\begin{split}
( ie\sqrt{2}\theta^{\alpha_3})\mathcal{M}^s(p, p_1,p_2) =&  ( ie\sqrt{2}\theta^{\alpha_3})S_2\left[V_3^{\alpha_2}D_1 V^{\alpha_1}_3 ~+(1\leftrightarrow 2) +V_4^{{\alpha_2}{\alpha_1}}\right]D \\
&\qquad+ \bar{V}_4^{\alpha_3 \alpha_2} D_1V_3^{\alpha_1} D
\end{split}\end{align}

where the terms in square brackets is the second-order amplitude and $ (1\leftrightarrow 2)\equiv (k_1,\alpha_1\leftrightarrow k_2,\alpha_2)$. We could have multiplied the above equation by the inverse of the third $ad~hoc$ vertex and, as before, simply discard terms with an odd number of soldering forms however, as presented the formula will be more useful when dealing with loops. 
 
The above formula is easily  generalised to $n$ photons:
\begin{align}
\boxed{\begin{array}{l}( ie\sqrt{2}\theta^{\alpha_{n+1}})S_n \mathcal{M}_{(1)}^{\alpha_n \dots \alpha_1} S  \\ \hspace*{2cm}=( ie\sqrt{2}\theta^{\alpha_{n+1}})S_n \mathcal{M}_{(2)}^{\alpha_n \dots \alpha_1} D+ V_4^{\alpha_{n+1}\alpha_n}D_{n-1} \mathcal{R}_{(2)}^{\alpha_{n-1} \dots \alpha_1}D  \end{array}}\label{sym}
\end{align}

where $V_4\mathcal{R}_{(2)}$ is the ``rest'' that contributes to the next amplitude only. All in all, when we consider the amplitudes we are interested in (recall we only want an even number of soldering forms in the RHS and that the last vertex is $ad~hoc$), we have:
\begin{align}
\left. S_n \mathcal{M}_{(1)}^{\alpha_n \dots \alpha_1}  S \right|_{unprimed} =m D_n \mathcal{M}_{(2)}^{\alpha_n \dots \alpha_1}  D  \label{1to2}
\end{align}

The dimensions of the amplitudes seem to mismatch, recall however that the first-order wavefunctions are related to the second-order ones by:
\begin{align}
\{ u_{(1)},~\bar{u}_{(1)}\} = \sqrt{m}~\cdot~ \{u_{(2)},~\bar{u}_{(2)}\}
\end{align}

Moreover, in the LHS, there is only either a momentum or mass propagator and the external spinors are onshell, $i.e.$:
\begin{align}
S u_{(1)} = m u_{(1)}D, \quad \bar{u}_{(1)} S_n = m D_n \bar{u}_{(1)}
\end{align}

All in all:
\begin{align}
\boxed{\left. \bar{u}_{(1)} \mathcal{M}_{(1)}^{\alpha_n \dots \alpha_1} u_{(1)} \right|_{full} = \bar{u}_{(2)}\mathcal{M}_{(2)}^{\alpha_n \dots \alpha_1}  u_{(2)} }
\end{align}

\subsection*{Dirac loops}

In order to find similar formulas for loops, we will consider (\ref{sym}) without the explicit symmetrisation. As before, we will use a bar notation to denote unsymmetrised second-order quantities. In the first-order formalism, a one-loop diagram for the $n+1$ photons amplitude is given by:
\begin{align}
{\rm Tr}\left[( ie\sqrt{2}\theta^{\alpha_{n+1}})S_n \mathcal{M}_{(1)}^{\alpha_n \dots \alpha_1} S \right]
\end{align}

which is given in the second-order formalism, after conversion by:
\begin{align}
{\rm Tr}\left[( ie\sqrt{2}\theta^{\alpha_{n+1}})S_n \bar{\mathcal{M}}_{(2)}^{\alpha_n \dots \alpha_1} D+ \bar V_4^{\alpha_{n+1}\alpha_n}D_{n-1} \bar{\mathcal{R}}_{(2)}^{\alpha_{n-1} \dots \alpha_1}D  \right]
\end{align}
with $D=D_{n+1}$. In this trace, only terms with an even number of soldering forms contribute as the trace of an odd number of these vanishes. We see that the second term in the brackets is directly related to the loop amplitude, whereas in the first one, only the momentum propagator in $S_n$ contributes. All in all:
\begin{align}\begin{split}
{\rm Tr}&\left[( ie\sqrt{2}\theta^{\alpha_{n+1}})S_n \mathcal{M}_{(1)}^{\alpha_n \dots \alpha_1} S \right] \\&\hspace*{1.5cm}= {\rm Tr}\left[(-2 ie)(\theta^{\alpha_{n+1}}p_n)D_n\bar{\mathcal{M}}_{(2)}^{\alpha_n \dots \alpha_1} D+ \bar{\mathcal{B}}_{(2)}^{\alpha_{n+1}\alpha_n  \dots \alpha_1}\right]
\end{split}\end{align}

where $\bar{\mathcal{B}}$ denotes a second-order formalism's loop diagram where the ``last'' vertex is quartic. \\ \medskip \\
Even though we have not symmetrised the diagram over its external photon states, as we are working with Dirac two-component fermions, there is still a piece missing the the diagram. Indeed, we must consider the same amplitude with the interchange $\chi \leftrightarrow \xi^\dagger$. Since there is always an even number of soldering forms and that the swap only affect these terms, there will not be any sign change in the individual terms. However, as we keep the charge flow equal and swap the fields, the contraction direction changes. Schematically a loop diagram is given by, if we keep the ``last'' vertex fixed and apply the swap on it:
\begin{align}
{\rm Tr}\left[(+ ie\sqrt{2}\theta^{\alpha_{n+1}})S_n \mathcal{M}_{(1)}^{\alpha_n \dots \alpha_1} S \right] +
{\rm Tr}\left[(- ie\sqrt{2}\theta^{\alpha_{n+1}})S \mathcal{M}_{(1)}^{\alpha_1 \dots \alpha_n} S_n \right]
\end{align}

where the amplitudes need not be equal. If we apply the same formulas as above to the second term we obtain for the latter:
\begin{align}\begin{split}
{\rm Tr}&\left[(- ie\sqrt{2}\theta^{\alpha_{n+1}})S \mathcal{M}_{(1)}^{\alpha_1 \dots \alpha_n} S_n \right] \\ & \hspace*{1.5cm}= {\rm Tr}\left[(-2 ie)(-\theta^{\alpha_{n+1}}p)D\bar{\mathcal{M}}_{(2)}^{\alpha_1 \dots \alpha_n}  D_n+ \bar{\mathcal{B}}_{(2)}^{\alpha_{n+1} \alpha_1 \dots \alpha_n }\right]
\end{split}\end{align}

where the additional minus sign comes from the fact that the propagators get an extra minus sign due to the direction change. The difference in the two first-order reversed amplitudes comes from the the ``last'' vertex, $i.e.$, how it connects to the rest of the amplitude. Once this vertex has been factored out, the rest is equal and simply contracted in an opposite direction. Therefore, using:
\begin{align}
{\rm Tr} \left[ \theta^\mu \theta^{\alpha_1}\theta^{\alpha_2}  \dots \theta^{\alpha_{n-1}} \theta^{\alpha_n}\right]= {\rm Tr} \left[ \theta^\mu \theta^{\alpha_n} \theta^{\alpha_{n-1}}\dots \theta^{\alpha_2} \theta^{\alpha_1}\right]
\end{align}

We have:
\begin{align}
 {\rm Tr}\left[(-2 ie)(\theta^{\alpha_{n+1}}p)D\bar{\mathcal{M}}_{(2)}^{\alpha_1 \dots \alpha_n}  D_n\right] =  {\rm Tr}\left[(-2 ie)(p\theta^{\alpha_{n+1}})D_n\bar{\mathcal{M}}_{(2)}^{\alpha_n \dots \alpha_1}  D\right]
\end{align}

and finally:
\begin{align}\begin{split}
{\rm Tr}\left[\mathcal{L}_{(1)}^{\alpha_{n+1} \dots \alpha_1}  \right] & = {\rm Tr}\left[V_3^{\alpha_{n+1}}D_n\bar{\mathcal{M}}_{(2)}^{\alpha_n \dots \alpha_1} D+ \bar{\mathcal{B}}_{(2)}^{\alpha_{n+1}\alpha_n  \dots \alpha_1}+ \bar{\mathcal{B}}_{(2)}^{\alpha_{n+1} \alpha_1 \dots \alpha_n }\right]\\&= {\rm Tr}\left[\mathcal{L}_{(2)}^{\alpha_{n+1} \dots \alpha_1}  \right] \label{loop}
\end{split}\end{align}

where $\mathcal{L}_{(i)}$ denotes the full one-loop diagram (counting both two-component spinors) written in either formalism. Notice however, that the second-order amplitude is written in terms of unsymmetrised quartic vertices $\bar{V}_4$. We will explain the apparent mismatch in a few examples below.

\subsection*{One-loop photon two-point function}
We will now construct explicitly the second-order formalism amplitudes from the first-order diagrams for the simple case of a two-point photon amplitude. We choose to fix the vertex $\mu$, there are then two diagrams (corresponding to one charge flow but both directions of contraction): one when the latter is a $\chi\chi^\dagger$ vertex and one when it is a $\xi^\dagger\xi$ vertex. The first diagram is given by:
\begin{align}\begin{split}
&(-1) \int \mathcal{D}\ell~ {\rm Tr} \left[(ie\sqrt{2}\theta^\mu)(-\sqrt{2}\ell +m) ( ie\sqrt{2}\theta^\nu)(-\sqrt{2}\ell_1+m) \right]D D_1 \\
=&(-1) \int \mathcal{D}\ell~ (-2ei)^2{\rm Tr} \left[(\ell_1\theta^\mu) (\ell\theta^\nu)-\frac{m^2}{2}(\theta^\nu\theta^\mu) \right]D D_1 \\
=&(-1) \int \mathcal{D}\ell~ (-2ei)^2{\rm Tr} \left[(\ell_1\theta^\mu) ((\ell\theta^\nu)+(\theta^\nu \ell_1)) \right]D D_1 +(-2e^2i){\rm Tr} \left[(\theta^\nu\theta^\mu)\right]D  \\
=&(-1) \int \mathcal{D}\ell~ (-2ei){\rm Tr} \left[(\ell_1\theta^\mu) V_3^\nu \right]D D_1 +{\rm Tr}\left[ \bar{V}_4^{\nu\mu}\right]D 
\end{split}\end{align}

where terms as $(\ell \theta^\nu)$ indicate that the momentum propagator $\ell$ attaches itself to the primed spinor in the the vertex $\theta^\nu$ (climbing up contraction arrows) and we used $p \equiv p_\mu\theta^\mu$ and $pp =p_\mu\theta^\mu \theta^\nu p_\nu = 1/2p^2$ as before to cancel the $m^2$ terms. Similarly, the second diagram is given by:
\begin{align}\begin{split}
&(-1) \int \mathcal{D}\ell~ {\rm Tr} \left[(-ie\sqrt{2}\theta^\mu)(\sqrt{2}\ell +m) (- ie\sqrt{2}\theta^\nu)(\sqrt{2}\ell_1+m) \right]D D_1 \\
=&(-1) \int \mathcal{D}\ell~ (-2ei)^2{\rm Tr} \left[(\ell_1\theta^\nu) (\ell\theta^\mu)-\frac{m^2}{2}(\theta^\nu\theta^\mu) \right]D D_1 \\
=&(-1) \int \mathcal{D}\ell~ (-2ei)^2{\rm Tr} \left[(\ell\theta^\mu) ((\ell_1\theta^\nu)+(\theta^\nu \ell)) \right]D D_1 +(-2e^2i){\rm Tr} \left[(\theta^\nu\theta^\mu)\right]D_1  \\
=&(-1) \int \mathcal{D}\ell~ (2ei)^2{\rm Tr} \left[(\theta^\mu\ell) ((\ell\theta^\nu)+(\theta^\nu \ell_1)) \right]D D_1 +(-2e^2i){\rm Tr} \left[(\theta^\nu\theta^\mu)\right]D_1 \\
=&(-1) \int \mathcal{D}\ell~ (-2ei){\rm Tr} \left[(\theta^\mu\ell) V_3^\nu \right]D D_1 +{\rm Tr}\left[ \bar{V}_4^{\nu\mu}\right]D_1
\end{split}\end{align}

All in all, when we sum up both contributions, we obtain:
\begin{align}\begin{split}
(-1) \int \mathcal{D}\ell~ {\rm Tr} \left[V_3^\mu V_3^\nu \right]D D_1 +{\rm Tr}\left[ \bar{V}_4^{\nu\mu}\right]D_1+{\rm Tr}\left[ \bar{V}_4^{\nu\mu}\right]D
\end{split}\end{align}

The reader can recognise (\ref{loop}) for $n=1$. We said earlier that the second-order Feynman rules give rise to the quartic vertex $V_4$ and not $\bar{V}_4$ and in our case the amplitude is given in terms of the latter. Notice however that:
\begin{align}
{\rm Tr}\left[ \bar{V}_4^{\nu\mu}\right] = \frac{1}{2}{\rm Tr}\left[ {V}_4^{\nu\mu}\right]
\end{align}

So that our amplitude can be written as:
\begin{align}\begin{split}
&(-1) \int \mathcal{D}\ell~ {\rm Tr} \left[V_3^\mu V_3^\nu \right]D D_1 +\frac{1}{2}{\rm Tr}\left[ {V}_4^{\nu\mu}\right]D_1+\frac{1}{2}{\rm Tr}\left[ {V}_4^{\nu\mu}\right]D \\&= (-1) \int \mathcal{D}\ell~ {\rm Tr} \left[V_3^\mu V_3^\nu \right]D D_1 +{\rm Tr}\left[ {V}_4^{\nu\mu}\right]D
\end{split}\end{align}

Which is what would be written using second-order Feynman rules.

\subsection*{Massive Majorana in a loop}
We give here an explicit construction of the second-order triangle diagrams for a loop consisting of massive Majorana fermions coupled to external vector fields (gauge fields in the massless limit).  For the former, there is only one kind of vertex given by $-ie\sqrt{2}\theta^\mu $ and the propagator is given by $\frac{-i(\sqrt{2}p+m)}{p^2+m^2} $ (when the momentum flows in the same direction as the contraction arrows). %In this case, there is one additional `sign' rule to take into account: the next $p$ insertion after an odd number of $m$ insertions is reversed as this corresponds to a swap in the flow. 
\\
We consider three incoming spin 1 particles with momenta $k_i$. In this index free notation, in order to obtain all the possible diagrams we simply need to fix one vertex (say $\mu$) and consider both orientations for the latter. Since we are solely interested in unprimed Feynman rules for the second-order formalism, each momentum insertion will contract with the vertex containing a primed spinor, and mass insertions will also contract when consisting of two primed spinors. All in all, the amplitude for the first orientation is given by:
\begin{align}\begin{split}
&(-1) \int \mathcal{D} p ~{\rm Tr}\left[ (-ie\sqrt{2}\theta^\mu ) (\sqrt{2}p_2+m) (ie\sqrt{2}\theta^\beta) \right. \\ &\hspace*{4cm}\quad \left.\times(\sqrt{2}p_1+m) (ie\sqrt{2}\theta^\alpha)(\sqrt{2}p+m) \right]DD_1D_2 \\
=&(-1) \int \mathcal{D} p ~(-2ie)^3{\rm Tr}\left[ (p\theta^\mu)(p_2\theta^\beta)(p_1\theta^\alpha) \right. \\ &\hspace*{2.7cm}\quad \left.+\frac{m^2}{2}\left( (p_2\theta^\beta)(\theta^\alpha\theta^\mu)+(p\theta^\mu)(\theta^\beta\theta^\alpha)-(\theta^\beta p_1)(\theta^\alpha\theta^\mu) \right) \right]DD_1D_2
\end{split}\end{align}

In order to convert the amplitude into second-order product of vertices, for each $m^2$ term, we add a $p_i^2$ term such as to cancel one denominator and obtain a quartic vertex (the momentum is chosen according to the $(\theta \theta)$ contraction) and subtract the same term but this time writen as $2p_i p_i$ in order to obtain a cubic vertex with the remaining terms. All in all, we obtain for this first amplitude:
\begin{align}\begin{split}
(-1) \int \mathcal{D} p ~&{\rm Tr}\left[ (-2ei)(p\theta^\mu)V_3^\beta V_3^\alpha \right]DD_1D_2 
 \\ &\hspace*{1.8cm}\quad +{\rm Tr}\left[(-2ei) (p\theta^\mu){\bar V}_4^{\beta\alpha} \right]DD_2 
+{\rm Tr}\left[ V_3^\beta {\bar V}_4^{\alpha\mu} \right]D_1D_2 
\end{split}\end{align}

with, for a momentum flow following the charge flow in the vertices:
\begin{align}\begin{split}
&V_3^\mu(k_{out},k_{in}) = (-2ei)(k_{out}\theta^\mu - \theta^\mu k_{in}) \\
&{\bar V}_4^{\alpha\beta} = 2e^2i (\theta^\alpha\theta^\beta) = \frac{1}{2}V_4^{\alpha\beta} + 2e^2i \Sigma^{\alpha\beta}
\end{split}\end{align}

The second amplitude is then written (remember that the momentum now mainly flows in the opposite direction to the charge):
\begin{align}\begin{split}
(-1) \int \mathcal{D} p ~&(-2ie)^3{\rm Tr}\left[ -(p_2\theta^\mu)(p\theta^\alpha)(p_1\theta^\beta) \right. \\ &\hspace*{0cm}\quad \left.-\frac{m^2}{2}\left( (p_2\theta^\mu)(\theta^\alpha\theta^\beta)+(p\theta^\alpha)(\theta^\beta\theta^\mu)-(\theta^\alpha p_1)(\theta^\beta\theta^\mu) \right) \right]DD_1D_2
\end{split}\end{align}

We now use the identity:
\begin{align}
{\rm Tr} \left[ \theta^\mu \theta^{\alpha_1}\theta^{\alpha_2}  \dots \theta^{\alpha_{n-1}} \theta^{\alpha_n}\right]= {\rm Tr} \left[ \theta^\mu \theta^{\alpha_n} \theta^{\alpha_{n-1}}\dots \theta^{\alpha_2} \theta^{\alpha_1}\right]
\end{align}

So that the amplitude can be rewritten as:
\begin{align}\begin{split}
(-1) \int \mathcal{D} p ~&{\rm Tr}\left[ (-2ei)(-\theta^\mu p_2)V_3^\beta V_3^\alpha \right]DD_1D_2 
 \\ &\hspace*{2cm}\quad +{\rm Tr}\left[(-2ei) (-\theta^\mu p_2){\bar V}_4^{\beta\alpha} \right]DD_2 
+{\rm Tr}\left[ V_3^\alpha {\bar V}_4^{\mu\beta} \right]D_1D 
\end{split}\end{align}

All together, the whole triangle diagram amplitude is given by:
\begin{align}\begin{split}
(-1) \int \mathcal{D} p ~&{\rm Tr}\left[ V_3^\mu V_3^\beta V_3^\alpha \right]DD_1D_2 
+{\rm Tr}\left[V_3^\mu{\bar V}_4^{\beta\alpha} \right]DD_2 
 \\ &\hspace*{3cm}\quad +{\rm Tr}\left[ V_3^\alpha {\bar V}_4^{\mu\beta} \right]D_1D +{\rm Tr}\left[ V_3^\beta {\bar V}_4^{\alpha\mu} \right]D_1D_2 
\end{split}\end{align}

\chapter{Explicit calculation of the anomaly}\label{apxexplicit}

We will now compute the amplitude of the divergence of the first current. This amounts to dot the amplitude with its momentum $ik_1=-i(k_2+k_3)$. The procedure is the same as in Chapter \ref{chapterano} and Appendix \ref{apxtransition}: using $-i(k_2+k_3)=-i(-(q-k_2)+(q+k_3))$, each term with a different denominator in the integrand comes with its shifted counterpart, so that the value of the integral is a boundary term (see below) which we give as:
\begin{align}\begin{split}
ik_1& \cdot i\mathcal{M}\\ &= \frac{e^3}{8\pi^2}(k_2+k_3)^{A}{}_{A'}\Bigg[ \epsilon^{B'C}(k_2-k_3)^{A'}{}_{(B}\epsilon_{C)A}- \epsilon^{A'C'} {k_2}^{B'}{}_{(A}\epsilon_{C)B} - \epsilon^{A'B'} {k_3}^{C'}{}_{(A}\epsilon_{B)C} \Bigg] \\
&= \frac{e^3}{8\pi^2}(k_2+k_3)^{A}{}_{A'}\Bigg[\left( -\Sigma_{DA} \right)^{B'C'}_{~B~C} (k_2-k_3)^{A'D} \\ &\hspace*{4cm} + \left(\Sigma_{DB} \right)^{A'C'}_{~A~C} {k_2}^{B'D} +  \left(\Sigma_{DC} \right)^{A'B'}_{~A~B} {k_3}^{C'D} \Bigg]
\end{split}\end{align}

We now make us of the fact that all the pairs of spinor indices we be contracted by either the momentum $(k_2+k_3)$ or the external polarisation vectors of the photons. Therefore, using
\begin{align}
\theta_{\mu\,AA'} \theta_{\nu\,B}{}^{A'} = -\frac{1}{2}\eta_{\mu\nu}\epsilon_{AB} +\Sigma_{\mu\nu\, AB}
\end{align}

and replacing the spinor indices by their Minkowski counterpart:
\begin{align}\begin{split}
ik_1 & \cdot i\mathcal{M}\\ &= \frac{e^3}{8\pi^2}(k_2+k_3)_\mu\Bigg[ - \Sigma^{\mu\beta}\cdot \Sigma^{\nu\alpha}(k_2-k_3)_\beta +\Sigma^{\mu\alpha}\cdot \Sigma^{\nu\beta}{k_2}_\beta+\Sigma^{\mu\nu}\cdot \Sigma^{\alpha\beta}{k_3}_\beta  \Bigg] \\
&= \frac{e^3}{4\pi^2}\Bigg[  \Sigma^{\mu\beta}\cdot \Sigma^{\nu\alpha} + \Sigma^{\mu\nu}\cdot \Sigma^{\alpha\beta} \Bigg]{k_2}_\mu{k_3}_\beta \\&= \frac{e^3}{4\pi^2}\Bigg[  \frac{1}{2}\left( \eta^{\mu\nu}\eta^{\alpha\beta}-\eta^{\mu\beta}\eta^{\nu\alpha}\right) - i\epsilon^{\mu\nu\alpha\beta} \Bigg]{k_2}_\mu{k_3}_\beta
\end{split}\end{align}

We consider the on-shell amplitude (so that $k_i^2 =0=\epsilon_i\cdot k_1$) and we obtain:
\begin{align}
ik_1 \cdot i\mathcal{M}_{\text{on-shell}} &=- \frac{ie^3}{4\pi^2} \epsilon^{\mu\nu\alpha\beta} {k_2}_\mu{k_3}_\beta\epsilon_\nu(k_2)\epsilon_\alpha(k_3)
\end{align}
Hence the divergence of the first current is anomalous. We should check whether it is the case for the divergence of, by symmetry, either of the remaining currents. Let us recall the amplitude:
\begin{align}\begin{split}
i\mathcal{M}(k_2,k_3)= 4e^3\int&\frac{d^4q}{(2\pi)^4}\Bigg[q^{B'}{}_{C}(q-k_2)^{A'}{}_{B}(q+k_3)^{C'}{}_{A} \\& \hspace*{2cm}- q^{C'}{}_{B}(q-k_2)^{B'}{}_{A}(q+k_3)^{A'}{}_{C}\Bigg]\\ &\times \frac{1}{(q-k_2)^2q^2(q+k_3)^2}+ (k_2;BB') \leftrightarrow (k_3;CC') \\&+
\frac{ie^3}{16\pi^2}\Bigg[ \left( \eta^{\mu\nu}\eta^{\alpha\beta} - \eta^{\mu\beta}\eta^{\nu\alpha} - 2i \epsilon^{\mu\nu\alpha\beta}\right){k_3}_\beta \\& \hspace*{2cm}+\left( \eta^{\mu\alpha}\eta^{\nu\beta} - \eta^{\mu\beta}\eta^{\nu\alpha} + 2i \epsilon^{\mu\nu\alpha\beta}\right){k_2}_\beta\Bigg]
\end{split}\end{align}

where it is understood that matching spinor indices are rewritten in terms of their spacetime counterpart. The terms in the second brackets, that correspond to the bubbles mentioned in Chapter \ref{chapterano}, are given by a boundary term as we saw in Appendix \ref{apxtransition}. The value of the latter is calculated straightforwardly using the methods presented in the two cited chapters.

In the computation of the divergence of the first current, it turned out that the terms that appear in the first brackets lead to a vanishing contribution due to Lorentz invariance, and only the remaining boundary terms contributed. Now, in calculating the divergence of the remaining two currents we then need to make some effort with the first two terms only. Furthermore the result will be symmetric under the exchange of momenta and indices $(k_2;\nu) \leftrightarrow (k_3;\alpha)$, so only one calculation is needed. We have then:
\begin{align}\begin{split}
ik_2& \cdot i\mathcal{M} = 2ie^3\int \frac{d^4q}{(2\pi)^4}\Bigg[ \left( \frac{(q+k_3)^{A'}{}_{C}q^{C'}{}_{A}}{q^2(q+k_3)^2} - \frac{(q-k_3)^{C'}{}_{A}q^{A'}{}_{C}}{q^2(q-k_3)^2}\right) \\  &\qquad \qquad + \left(\frac{(q-k_3)^{C'}{}_{A}(q+k_2)^{A'}{}_{C}}{(q+k_2)^2(q-k_3)^2}  - \frac{(q+k_3)^{A'}{}_{C}(q-k_2)^{C'}{}_{A}}{(q-k_2)^2(q+k_3)^2} \right) \Bigg]\\&\hspace*{4cm} - (AA' \leftrightarrow CC') \\  & \hspace*{4cm}+\frac{ie^3}{8\pi^2}\epsilon^{\mu\nu\alpha\beta}{k_2}_\nu{k_3}_\beta + \ldots
\end{split}\end{align}

Where the dots are terms that vanish onshell. The first quantity in brackets vanishes by Lorentz invariance, whereas in order to compute the second integral, it suffices to notice that it can be rewritten as:
\begin{align}
2ie^3\int\frac{d^4q}{(2\pi)^4}  \Big[I(q+S) - I(q) \Big]
\end{align}

and using the fact that for a quadratically divergent integral we have:
\begin{align}\begin{split}
\int\frac{d^4q}{(2\pi)^4} f(q+a)-f(a) &= \frac{i}{(2\pi)^4}\left(2\pi^2 a_{D'}{}^D \lim_{q\rightarrow \infty} q^ {D'}{}_D q^2 f_o(q)\right. \\ &\hspace*{2cm}\left. + \pi^2 a_{E'}{}^E a_{D'}{}^D \lim_{q\rightarrow \infty} q^{E'}{}_{E} q^2 \frac{\partial}{\partial q_{D'}{}^D}f_e(q)\right)
\end{split}\end{align}
where the LHS momenta have been continued to Euclidian space by Wick rotation and $f_o,~f_e$ are the odd and even components of the function $f$. In our case:
\begin{align}
I(q) = \frac{(q+k_3)^{A'}{}_{C}(q-k_2)^{C'}{}_{A}}{(q-k_2)^2(q+k_3)^2}, \quad S= k_2-k_3
\end{align}

Using:
\begin{align}
I_o(q)&= \frac{-q^{A'}{}_{C} {k_2}^{C'}{}_{A} +q^{C'}{}_{A}{k_3}^{A'}{}_{C}}{q^4} - \frac{2q^{C'}{}_{A} q^{A'}{}_{C} q\cdot (k_3-k_2)}{q^6}+\mathcal{O}(q^{-4})\\
\frac{\partial I_e(q)}{\partial q_{D'}{}^D} &= \frac{\epsilon^{A'D'}\epsilon_{DC} q^{C'}{}_{A} + \epsilon^{C'D'}\epsilon_{DA} q^{A'}{}_{C}}{q^4}- \frac{4 q^{A'}{}_{C} q^{C'}{}_{A} q^{D'}{}_{D}}{q^6} +\mathcal{O}(q^{-4})
\end{align}

so that the evaluation of the integral yields:
\begin{align}
\frac{e^3}{8\pi^2}\left( {k_3}^{A'}{}_{C}{k_2}^{C'}{}_{A}- {k_2}^{A'}{}_{C}{k_3}^{C'}{}_{A}\right) 
\end{align}

We now make use of the fact that all the pairs of spinor indices will be contracted by the external polarisation vectors of the photons, $i.e.$
\begin{align}
k^{A'}{}_{C} ~\varepsilon(k)^{C}{}_{C'} =\bar{\Sigma}^{\mu\nu\, A'}{}_{C'}k_\mu \epsilon_\nu,\quad k^{C'}{}_{A} ~\varepsilon(k)^{C}{}_{C'} ={\Sigma}^{\mu\nu\, C}{}_{A} k_\mu \epsilon_\nu
\end{align}

thus obtaining:
\begin{align}
{\text{on-shell}}  &\rightarrow \frac{e^3}{8\pi^2}\Bigg[ \bar{ \Sigma}^{\mu\nu}\cdot \bar{\Sigma}^{\alpha\beta}-\Sigma^{\mu\nu}\cdot \Sigma^{\alpha\beta}  \Bigg]k_{3\,\mu}\varepsilon_{3\,\nu}k_{2\,\alpha}\varepsilon_{1\,\beta} =- \frac{ie^3}{8\pi^2}\epsilon^{\mu\nu\alpha\beta}\varepsilon_{1\,\mu}\varepsilon_{3\,\alpha}k_{2\,\nu}k_{3\,\beta}
\end{align}

So that finally
\begin{align}
ik_2 \cdot i\mathcal{M} &= - \frac{ie^3}{8\pi^2}\epsilon^{\mu\nu\alpha\beta}k_{2\,\nu}k_{3\,\beta} +\frac{ie^3}{8\pi^2}\epsilon^{\mu\nu\alpha\beta}k_{3\,\beta} = 0 = ik_3 \cdot i\mathcal{M}
\end{align}

All in all, if we consider a shifted amplitude (as we saw the result is shift dependent), we obtain:
\begin{align}
ik_1 \cdot i\mathcal{M}(k_2,k_3;a)&= -\frac{ie^3}{4\pi^2}(1-4c)\epsilon^{\mu\nu\alpha\beta}{k_2}_\mu{k_3}_\beta \\
ik_2 \cdot i\mathcal{M}(k_2,k_3;a)&= -\frac{ie^3}{4\pi^2}(2c)\epsilon^{\mu\nu\alpha\beta}{k_2}_\nu{k_3}_\beta \\
ik_3 \cdot i\mathcal{M}(k_2,k_3;a)&= -\frac{ie^3}{4\pi^2}(2c)\epsilon^{\mu\nu\alpha\beta}{k_2}_\alpha{k_3}_\beta
\end{align}

So that we can remove the anomaly in two of the currents by choosing $c=0$ but there will always be one anomalous current. The symmetric choice would be to set $c= 1/6$ (which must be the case when we are dealing with three identical currents).

%
%\chapter{Group Theory}
%Follow Georgi
%
%%\section{Roots and Weights}
%%
%%\section{Simple roots}
%%
%%
%%\section{The Classical Groups}
%%
%%\section{Classification theorem and subalgebras}
%
%\section{Spinor representations}
%
%\section{Clifford algebras}
%
%\section{Embeddings}

\bibliographystyle{hieeetr}
\bibliography{ThesisBib}{}

\end{document}